\documentclass[authoryear]{elsarticle}
\usepackage{amsmath}
\usepackage{float}
\usepackage{subfigure}
\usepackage{amsfonts}
\usepackage{lineno}
\usepackage{caption}
\usepackage{subfigure}
\usepackage{xcolor}
\usepackage[linesnumbered,ruled]{algorithm2e}
\modulolinenumbers[5]
\usepackage{natbib}
\usepackage{lipsum}
\usepackage{cleveref}
\usepackage{chngpage}

\journal{Computational Statistics $\&$ Data Analysis}

\DeclareMathOperator*{\argmin}{argmin}

\usepackage[T1]{fontenc}

\makeatletter
\def\@author#1{\g@addto@macro\elsauthors{\normalsize%
    \def\baselinestretch{1}%
    \upshape\authorsep#1\unskip\textsuperscript{%
      \ifx\@fnmark\@empty\else\unskip\sep\@fnmark\let\sep=,\fi
      \ifx\@corref\@empty\else\unskip\sep\@corref\let\sep=,\fi
      }%
    \def\authorsep{\unskip,\space}%
    \global\let\@fnmark\@empty
    \global\let\@corref\@empty  
    \global\let\sep\@empty}%
    \@eadauthor={#1}
}
\makeatother

\begin{document}
\begin{frontmatter}

\title{Robust Boosting for Regression Problems}
\author{Xiaomeng Ju \corref{cor1}}

\author{Mat\'{i}as Salibi\'{a}n-Barrera }
\cortext[cor1]{Corresponding author}

\address{Department of Statistics \\ University of British Columbia, Canada}

\begin{abstract}
Gradient boosting algorithms construct a regression predictor using a linear
	combination of  ``base learners''.  Boosting also offers an approach to
	obtaining robust non-parametric regression estimators that are scalable to
	applications with many explanatory variables. The robust  boosting
	algorithm is based on a two-stage approach, similar to what is done for
	robust  linear regression: it first minimizes a robust residual scale
	estimator, and then improves it by optimizing a bounded
	loss function. Unlike previous robust boosting proposals  this approach
	does not require computing an ad-hoc residual scale estimator in each
	boosting iteration. Since the loss functions involved in this robust boosting
	algorithm are typically non-convex, a reliable initialization
	step is required, such as an $L_1$ regression tree, which is also fast to compute.
	A robust variable importance measure can also be calculated via a
	permutation procedure.  Thorough simulation studies and several data
	analyses show that, when no atypical observations are present, the
	robust boosting approach works as well as the standard gradient
	boosting with a squared loss. Furthermore, when the data contain
	outliers, the robust boosting estimator outperforms the alternatives in
	terms of prediction error and variable selection accuracy. 
\end{abstract}

\begin{keyword}
 Regression \sep Boosting \sep Robustness  \sep Ensemble methods \end{keyword} 
\end{frontmatter}

\let\thefootnote\relax\footnotetext{Contact:   xiaomeng.ju@stat.ubc.ca (Xiaomeng Ju);  Department of Statistics,  2207 Main Mall, University of British Columbia, Vancouver, BC,  V6T1Z4, Canada. 
}
\let\thefootnote\relax\footnotetext{\hspace{-1ex} An Appendix is provided as on-line supplementary material. 
}

\section{Introduction}

The goal of robust regression methods is to provide reliable estimates for the
unknown regression function and predictions for future observations,
even when the training data set may contain atypical
observations. These ``outliers'' need not be ``extreme'' or aberrant values,
but might simply be points following a different model from the one that
applies to the majority of the data.  It is well known that classical methods
can provide seriously misleading results when trained on such heteregenous or
contaminated data sets, and thus obtaining robust alternatives is of immediate
practical importance. 

In the robust regression literature much of the attention has been devoted to
linear models, and a rich variety of proposals exists for them.  For a review
see, for example, \citet{maronna2018robust}.  Fewer options are available for
robust non-parametric regression methods, and generally they have not been
developed as thoroughly as the linear ones. 
Among them we can mention: robust locally weighted regression
\citep{cleveland1979robust, wang19941},  local M-estimators
\citep{boentefraiman89}, local regression quantiles \citep{welsh96}, and the
proposals in \citet{hardletsy88}, \citet{hardle90}, and \citet{ohnychkalee07}.
Unfortunately, when applied to problems with several explanatory variables,
these methods suffer from the well known curse of dimensionality
\citep{esl2009,curseofdim}, requiring
training sample sizes that grow exponentially with the number of covariates.
An exception is the class of 
robust estimators for additive models recently proposed in 
\citep{boente2017robust}. However, selecting appropriate bandwidths for each of
the components of an additive model can be computationally prohibitive when a
moderate or large number of explanatory variables are involved. 

In this paper we study robust non-parametric regression estimators based on
gradient boosting \citep{Friedman2001}.  These ensemble methods are scalable to
high-dimensional data, and typically require selecting fewer tuning parameters
than additive models. 
Our proposal applies to practical situations where the
proportion of potential outliers in the training and validation sets is
unknown, and no parametric structure for the regression model is available.


Most robust boosting algorithms in the literature were designed for
classification problems \citep{kegl2003robust, rosset2005robust,
freund2009more, miao2015rboost, li2018boosting}.  Previous proposals to
robustify regression boosting methods replaced the squared loss with the absolute value
function, or with Huber's loss. 
\citet{lutz2008robustified} introduced several ways to robustify boosting,
primarily focusing on linear regression.  One of their
methods (\texttt{Robloss}) can be extended to 
the scope of our study (nonlinear regression) by replacing the simple linear regression base learners with other types of base learners. 

%
A different class of robust nonparametric regression methods is based on random
forests.  \citet{roy2012robustness} proposed variations of random forests that
use the median to aggregate trees and make the tree splits.
\citet{meinshausen2006quantile} introduced a quantile random forest which
models the full conditional distribution of the response variable, not just the
conditional mean.  \citet{li2017forest} established a connection between random
forests and locally weighted regression.  They suggested a forest-type
regression framework which includes  \citet{meinshausen2006quantile}'s quantile
random forest as a special case. It  is also applicable to other loss
functions, such as Huber's or Tukey's.


Our main concern with these proposals is that they use robust loss functions
that either may yield estimators with low efficiency (e.g. the $L_1$ loss), or
require an auxiliary residual scale estimator (e.g. Huber's and Tukey's loss
functions).  For the latter, previous proposals suggested using in each step a
robust scale estimator obtained with the residuals from the fit at the previous
iteration \citep{Friedman2001, lutz2008robustified}.  It is easy to see that
this changing scale estimator may not work well, since the scale can be
overestimated in early iterations and this might result in observations with
large residuals not being considered outlying.  In fact, our numerical
experiments confirm that this is the case even in relatively simple settings.
To address this problem, we propose a robust boosting method that directly
minimizes a robust scale of the residuals, and hence does not need to compute
an auxiliary residual scale estimator.  Although in principle our approach can
be used with any scale estimator, gradient boosting uses the partial 
derivatives of
the objective function. Hence, in this paper we focus on minimizing an
M-estimator of scale, for which we can compute the corresponding gradient.
This approach can be seen as an extension of S-estimators
\citep{rousseeuw1984robust} to this class of non-parametric regression
estimators. Moreover, this robust boosting estimator can be used (along with
its associated robust residual scale estimator) as the initial fit for a
boosting M-type estimator using Huber's or Tukey's loss. As in the case of linear
models, this second stage is expected to result in a robust estimator with
higher efficiency (less variability) than the S-estimator.

The remainder of this paper is organized as follows. Section
\ref{sec:methodology} introduces our proposed robust boosting regression
estimator, and a variable importance score based on permutations.  Section
\ref{sec:simu} reports the results of our simulation studies comparing our
method with previously proposed robust boosting algorithms, while results
obtained on benchmark data are discussed in Section \ref{sec:emp}.  Finally,
Section \ref{sec:conclusion} summarizes our findings and conclusions.

\section{Methodology}  \label{sec:methodology}

In this section we introduce our robust gradient boosting algorithm based on
minimizing an M-estimator of the scale of the residuals.  We also show how this
fit can be further improved with a second gradient boosting step based on a
bounded loss function. This second step reduces the variability of the fit
without affecting its robustness.  Finally, we discuss a variable importance
method based on the permutation importance scores used with random forest
\citep{breiman2001random}. 

\subsection{SBoost: gradient boosting based on M-scale estimators}
\label{sec:sboost}

Consider a training data set  $(\mathbf{x}_i,  y_i)$, where $i \in
\mathcal{I}_{\text{train}}$, $\mathbf{x}_i \in \mathbb{R}^p$ and $y_i \in
\mathbb{R}$.  The goal is to estimate the regression function $F$ that relates
the explanatory variables $\mathbf{x}$ to the response $y$.  
When second moments exist, the regression function $F$ 
is generally defined as 
$F = \arg \min_G E_{Y, \mathbf{X}} \left[ Y - G( \mathbf{X} ) \right]^2$, 
where the minimum is taken over a sufficiently rich set of functions $G$ (e.g. 
measurable and squared integrable functions). 
Following  \cite{Friedman2001}, one can take a slightly more general approach
and define the target regression function as the minimizer 
$F = \arg \min_G E_{Y, \mathbf{X}} L \left(Y,  G( \mathbf{X} ) \right)$
for a pre-specified loss function $L(a, b)$. Different choices for $L$ require
specific regularity conditions for such a minimum to exist. 
Gradient boosting assumes that the target
function $F$ can be approximated with a function $\hat{F}$
of the form: 
\begin{equation}  \label{eq:23} 
\hat{F}(\mathbf{x}) \, = \,  \sum_{t= 1}^T \alpha_{t} h_t(\mathbf{x}) \, , 
\end{equation}
where the functions $h_t : \mathbb{R}^p \to \mathbb{R}$ are called ``base learners''. 
If the base learners $h_t$ are restricted to
be functions of a single variable, 
equation \eqref{eq:23} above 
is akin to assuming an additive model. 
%
%
Gradient boosting can be seen as a step-wise iterative algorithm to minimize
the empirical risk 
\begin{equation*} 
\argmin_{F} \sum_{i \in \mathcal{I}_{\text{train}}} L(y_i, F(\mathbf{x}_i)) 
 \end{equation*}
over functions $F$ of the form \eqref{eq:23}.  More specifically, let $n$ be
the number of observations in the training set, $n =
|\mathcal{I}_{\text{train}}|$, and consider the objective function above as a
function of the $n$-dimensional vector $(F(\mathbf{x}_1), \ldots,
F(\mathbf{x}_n))^\top$.  At each step $t$ we compute the 
$i$-th coordinate of its gradient 
at the point obtained from the previous 
iteration 
$(\hat{F}_{t-1}(\mathbf{x}_1), 
\ldots, \hat{F}_{t-1}(\mathbf{x}_n))^\top$:
$$
g_{t,i} \,  =  \, 
\frac{ \partial L(y_{i}, b) }{\partial b} 
\Big \vert_{b = \hat{F}_{t-1}(\mathbf{x}_{i})} \, , 
\qquad i \in \mathcal{I}_{\text{train}} \, ,
$$
and form the gradient vector $\mathbf{g}_t = (g_{t,1}, g_{t,2}, \ldots,
g_{t,n})^\top$.  Similarly to the usual gradient descent algorithm that
sequentially adds to the current point the negative gradient vector of the
objective function, gradient boosting adds to the current function estimate  an
approximation to the negative gradient vector ($-\mathbf{g}_t$) using a base
learner $h_t$.  More specifically, at the $t$-th iteration $h_t$ is chosen to
minimize 
$$
\sum_{i \in \mathcal{I}_{\text{train}}} (h_t(\mathbf{x}_i) + g_{t,i})^2 \, ,
$$
over members $h_t$ of the family of base learners ($\mathcal{H}$).  We then set 
$$
\hat{F}_{t}(\mathbf{x}) = \hat{F}_{t-1}(\mathbf{x}) +  \alpha_t
h_t(\mathbf{x}),
$$
 where $\alpha_t$ is the optimal step size given by 
 \begin{equation}
 \alpha_t \ = \ \argmin_{\alpha} \, 
 \sum_{i \in \mathcal{I}_{\text{train}}} L(y_i,
 \hat{F}_{t-1}(\mathbf{x}_i) + \alpha h_t(\mathbf{x}_i))  \, .
 \label{eq:alpha-optimal}
 \end{equation}
Note that this algorithm requires an initial estimate $\hat{F}_0(\mathbf{x})$.
When $L$ is the squared loss this initial estimator is usually taken to be the
sample mean of the response variable in the training set. It is easy to see
that this simple choice may not work well when there are atypical observations
in the data set. 
In Section~\ref{sec:init} below
we present a better initialization approach for our
setting. 


%

Prior attempts to robustify boosting replaced the squared loss with 
Huber's loss, which requires the use of a preliminary scale
estimator of the residuals. However, to estimate the scale of the residuals
we need a robust regression estimator with which to compute the residuals. 
To avoid this problem we consider the following minimization problem
instead:
\begin{equation}
\label{s:objective}
    \hat{F} \, = \, \argmin_{F} \hat{\sigma}_n(F) \, , 
\end{equation}
where $F$ is as in
\eqref{eq:23}, and 
$\hat{\sigma}_n(F)$ is a robust estimator of the scale of the 
residuals $r_i(F) = y_i - F(\mathbf{x}_i)$, $i \in 
\mathcal{I}_{\text{train}}$. 
More specifically, in this paper we will focus on the case 
where $\hat{\sigma}_n$ is an M-estimator of scale given as the solution of
\begin{equation} \label{eq:s:1}
   \frac{1}{|\mathcal{I}_{\text{train}}|} 
	\sum_{i \in \mathcal{I}_{\text{train}}}
	\rho_0 \left( \frac{y_i - F(\mathbf{x}_i)}{\hat{\sigma}_n(F)}\right) = \kappa \, ,  
\end{equation}
where $\kappa > 0$, and 
$\rho_0 : \mathbb{R} \to [0, +\infty)$ is a symmetric 
function around zero. These 
two parameters 
control the robustness and efficiency of the resulting estimator 
\citep{maronna2018robust}. 
Note that if we take $\rho_0(u) = u^2$ this 
approach reduces to the usual  
boosting with the squared loss function. 

When $\rho_0$ is bounded
(and in that case we can assume without loss of generality that 
$\sup_t \rho_0(t) = 1$), using 
$\kappa = 1/2$ in \eqref{eq:s:1} yields a residual scale estimator with maximal breakdown point \citep{maronna2018robust}.
In what follows we use functions $\rho_0$ in Tukey's bisquare family 
(but our approach applies to any bounded
differentiable function $\rho_0$): 
\begin{align}
	\label{tukey-loss} \rho_{\text{Tukey},c}(u) &=  
\begin{cases}  1 -
		\left(1 - (\frac{u}{c})^2 \right)^3 &  \text{if} \ |u| \leq c \, ;
		\\  1  &  \text{if} \ |u| > c \, , 
\end{cases}
\end{align}
where $c > 0$ is a user chosen tuning constant. 

In the robustness literature it is customary to assume that 
non-outlying observations follow a specific regression model, 
e.g. $Y = F({\mathbf X}) + \sigma \varepsilon$, 
where $\sigma > 0$ is an unknown scale parameter and 
the residuals $\varepsilon$ have a fixed distribution $H$. 
In that case, the above residual M-scale estimator will be 
consistent if $\rho_0$ satisfies $E_{H} \rho_0 ( \varepsilon ) = \kappa$. A very 
common assumption is that, for clean data, $\varepsilon$ 
is Gaussian. In that case, the choices $c = 1.547$
in \eqref{tukey-loss} and $\kappa = 1/2$ in 
\eqref{eq:s:1} result in a consistent and highly robust
estimator. Similarly, highly robust 
and consistent estimators for other 
models can be obtained by choosing
the tuning constant $c$ in \eqref{tukey-loss}
as the solution to the equation $E_{H} \rho_0 ( \varepsilon ) = 1/2$. 
We refer the interested reader to \cite{maronna2018robust} for details. 


To compute the \texttt{SBoost} estimator, we follow the same approach used for
gradient boosting, namely: at each step $t$ we calculate the gradient of the
objective function \eqref{s:objective} (considered as a function of the $n$ fitted values), and
approximate the negative gradient with a base learner.  The $i$-th
coordinate of the gradient at step $t$ is given by:
$$
g_{t,i} \,  = \, \Bigl.
\frac{ \partial \hat{\sigma}_n(F) }{ \partial F(\mathbf{x}_{i})}
\Bigr|_{F(\mathbf{x}_{i}) = \hat{F}_{t-1}(\mathbf{x}_i)}
\, , \qquad 
i \in \mathcal{I}_{\text{train}} \,,
$$
where $\hat{\sigma}_n(F)$ is defined in  \eqref{eq:s:1} .

The above can be computed explicitly by 
differentiating both sides of equation (\ref{eq:s:1})
with respect to $F(\mathbf{x}_\ell)$,
and noting that 
\begin{align*} 
0 \, = \, 
\sum_{i \in \mathcal{I}_{\text{train}}} 
\frac{\partial}{\partial F(\mathbf{x}_\ell)} \left[
\rho_0 \left( \frac{y_i - F(\mathbf{x}_i)}{ \hat{\sigma}_n(F)} \right)
	\right]
	\, , \qquad \ell \in 
\mathcal{I}_{\text{train}} \, .
\end{align*}
We have
\begin{align*}
 0 \, = \, 
\sum_{i \in \mathcal{I}_{\text{train}}} 
\psi_0 \left( \frac{y_i - F(\mathbf{x}_i)}{\hat{\sigma}_n(F)}\right) 
	\frac{ (y_i - F(\mathbf{x}_i))g_{t,\ell}}{\hat{\sigma}_n^2(F)}  
	+ \psi_0 \left( \frac{y_j - F(\mathbf{x}_j)}{\hat{\sigma}_n(F)} \right) 
	\frac{1}{\hat{\sigma}_n(F)},  \label{eq:sboost:abc}
\end{align*}
where $\psi_0 = \rho_0^{\prime}$.
Solving for $g_{t,\ell}$ in the equation above we get
$$
g_{t,\ell} \ = -\ C_{t} \, \psi_0 \left( 
\frac{y_\ell - \hat{F}_{t-1}(\mathbf{x}_\ell)}{
\hat{\sigma}_n (\hat{F}_{t-1})}
\right)  \, , 
$$
where the constant 
$$C_{t} =  
 \left[\sum_{i \in \mathcal{I}_{\text{train}}} 
 \psi_0 \left(\frac{y_i - \hat{F}_{t-1}(\mathbf{x}_i)}{
	 \hat{\sigma}_n (\hat{F}_{t-1}) } \right)  
\left( \frac{ y_i - 
\hat{F}_{t-1}(\mathbf{x}_i) }{ 
\hat{\sigma}_n(\hat{F}_{t-1}) } \right)
\right]^{-1}\, . 
$$ 
The resulting \texttt{SBoost} algorithm 
is shown in Algorithm \ref{SBoost}. 
 \begin{algorithm}[tp]
    \SetKwInOut{Input}{Input}
    \SetKwInOut{Output}{Output}
     \SetKwInOut{Initialize}{Initialize}
    \Input{A data set $(\mathbf{x}_i, y_i),  i \in \mathcal{I}_{\text{train}}$  \\
    The number of iterations $T_1$   \\ 
    The class of base learners $\mathcal{H}$ \\
    Initialization $\hat{F}_0(\mathbf{x})$ }
     \For{ $t = 1:T_1$}
     {
       $\hat{\sigma}_n (\hat{F}_{t-1}) =  \{\sigma: \frac{1}{| \mathcal{I}_{\text{train}}|}\sum_{i \in \mathcal{I}_{\text{train}}} \rho_0 \left( \frac{y_i - \hat{F}_{t-1}(\mathbf{x}_i)}{\sigma}\right) = \kappa\}$ \\
	 $C_t =  \left[\sum_{i \in \mathcal{I}_{\text{train}}} \psi_0 \left(\frac{y_i - \hat{F}_{t-1}(\mathbf{x}_i)}{\hat{\sigma}_n (\hat{F}_{t-1} )}\right) \left(\frac{y_i - \hat{F}_{t-1}(\mathbf{x}_i)}{\hat{\sigma}_n (\hat{F}_{t-1} )}\right)\right]^{-1}$ \\
        $g_{t,\ell} = -C_t\psi_0 \left(\frac{y_{\ell}- \hat{F}_{t-1}(\mathbf{x}_{\ell})} {\hat{\sigma}_n (\hat{F}_{t-1})} \right)$  \\
        $h_t = \argmin_{h \in \mathcal{H}}  \sum_{i \in |\mathcal{I}_{\text{train}}|}  (g_{t,i} + h(\mathbf{x}_i))^2$ \\
        $\alpha_t = \argmin_{\alpha} \hat{\sigma}\left(\hat{F}_{t-1} + \alpha h_t\right)$ \\
        $\hat{F}_{t}(\mathbf{x}) = \hat{F}_{t-1}(\mathbf{x}) +  \alpha_t h_t(\mathbf{x})$ 
        }
      \Output{$\hat{F}_{T_1}(\mathbf{x})$,       $\hat{\sigma}(\hat{F}_{T_1})$}  
          \caption{SBoost}
          \label{SBoost}
\end{algorithm} 

\subsubsection{The initial fit} \label{sec:init}

Gradient boosting algorithms require an initial fit 
$\hat{F}_0(\mathbf{x})$ 
to calculate the gradient vector for the first iteration. 
The classical gradient boosting algorithm is 
generally initialized with a constant fit
$\hat{F}_0(\mathbf{x}) \equiv \alpha_0$
where $\alpha_0$ solves
$$
\alpha_0 \, = \, \argmin_{a \in \mathbb{R}} \,
\sum_{i \in \mathcal{I}_{\text{train}} } L(y_i, a) \, .
$$

Since the loss function in Algorithm \ref{SBoost} is non-convex, the choice of an
initial fit for \texttt{SBoost} will typically  be more important
than for the usual gradient boosting with a squared loss function. 
A natural initial fit for a robust version of gradient boosting would be
to take $\alpha_0 = \mbox{median}(y_1, \ldots, y_n)$.  However, we have found
that this choice may not work well when the response variable includes a wide
range of values.   The median initialization does not  consider any explanatory
variables  and can be far from the global optimum.  Due to the non-convexity of
the objective function and the gradient descent like approach that boosting
takes,  we may end up reaching a local optimum with a large objective value.
Intuitively the problem is that in those cases many ``good''  observations may
have large residuals $y_i - \alpha_0$ that result in  zero initial gradients
$g_{1,i} = 0$, and  the algorithm may fail to update $\hat{F}(\mathbf{x}_i)$ in
the following iterations.  

Instead, we recommend using a regression tree computed with the $L_1$ loss
\citep{breiman2017classification}. This tree, which we call \texttt{LADTree},
at each split  minimizes the average absolute value of the residuals.  This
initial fit is expected to be robust to outliers in the training set, it is
able to handle data of mixed types (quantitative, binary and categorical), and
it is scalable to data with a large number of explanatory variables. 

\subsection{RRBoost}  \label{sec:rrboost}

Although our numerical experiments confirm the robustness 
of the \texttt{SBoost} fit 
(see Section \ref{sec:results}), 
they also show a very low efficiency when the data 
do not contain outliers. Mimicking what is done to 
fix this issue in the case of S-estimators for linear 
regression, 
we propose to refine the \texttt{SBoost} estimator by using it 
as the initial fit for further gradient boosting iterations
based on 
a bounded loss function that is closer to the squared 
loss function than $\rho_0$ 
in \eqref{eq:s:1} (e.g. 
a Tukey's bisquare  function with a larger tuning constant). 
Moreover, in this second step we can use the 
scale estimator associated with the final 
\texttt{SBoost} fit ($\hat{\sigma}_n(\hat{F}_{T_1})$). 

This 
two-step procedure, which we call \texttt{RRBoost}, can be
summarized as follows:
\begin{itemize}
 \item \textbf{Stage 1:}  compute an $S$-type boosting estimator (\texttt{SBoost})
	 with  high robustness but possibly low   efficiency; 
 \item \textbf{Stage 2:}  compute an $M$-type boosting estimator initialized
	 at the function and scale estimators obtained in Stage 1. 
\end{itemize}

With this approach we avoid the problem of many robust boosting proposals in the
literature that require a preliminary residual scale estimator.  Since it is
known that highly robust  S-estimators for linear regression have low efficiency
\citep{hossjer92}, the second step attempts to improve the efficiency of
\texttt{SBoost}, similarly to what MM estimators do for linear regression
models \citep{yohai1987high}.

The M-type boosting fit is given by 
\begin{equation}\label{eq:rrboost}
\hat{F} = \argmin_{F}  \sum_{ i \in \mathcal{I}_{\text{train}}} 
	\rho_1 \left( \frac{y_i - F(\mathbf{x}_i)}{\hat{\sigma}_n} \right) \, ,
\end{equation}
where  $F$ is of the form \eqref{eq:23}, and $\hat{\sigma}_n =
\hat{\sigma}(\hat{F}_{T_1})$ is the residual scale estimator obtained from
\texttt{SBoost}.  In finite dimensional regression models, the shape 
of the function $\rho_1$ determines the robustness and
efficiency of the resulting estimator with respect to the assumed model for the clean data. 
In what follows we will use a Tukey's bisquare 
function~\eqref{tukey-loss} for $\rho_1$.
When the non-outlying observations are assumed to follow a 
Gaussian regression model, the choice 
$c = 4.685$ yields a highly-efficient robust estimator with 
maximal breakdown point 
\citep{maronna2018robust}.
%

The algorithm for the second stage starts from the \texttt{SBoost} function
 estimator $\hat{F}_{T_1}(\mathbf{x})$, and sequentially adds base learners
 $h_1(\mathbf{x})$, \ldots, $h_{T_2}(\mathbf{x})$ that approximate the
 vector of gradients at each step. At the $t$-th iteration, the $i$-th coordinate of the 
gradient of the objective function above is given by 
\begin{align*} 
g_{t,i} &= 
	\frac{\partial}{\partial  F(\mathbf{x}_i)}  \rho_1 \left( \frac{y_i  -
	F(\mathbf{x}_i)}{
\hat{\sigma}_n} \right)  \notag \\
& = - \frac{1}{\hat{\sigma}_n} \psi_1 \left( \frac{y_i - F(\mathbf{x}_i)}{
	\hat{\sigma}_n} \right) \, , 
\qquad i \in \mathcal{I}_{\text{train}} \, ,
\label{eq:rrboost.grad}
 \end{align*} 
 where $\psi_1 = \rho_1'$. 
The resulting \texttt{RRBoost} procedure is described in Algorithm \ref{code-rrboost}. 
 
Many boosting algorithms also include a ``shrinkage'' option \citep{Friedman2001}. 
The idea is to reduce the size of the function updates by multiplying 
the optimal step size $\alpha_t$ in \eqref{eq:alpha-optimal}
by a small fixed constant $\gamma \in (0, 1)$. 
Shrinkage is expected to reduce the impact of each base learner and
approximate more finely the search path in gradient descent \citep{telgarsky2013margins}.  A small $\gamma$ typically improves the predictive accuracy, but requires a larger number of iterations and thus 
increases the computational cost of the algorithm. Our software 
implementation includes a shrinkage option, which corresponds 
to replacing line 7 in Algorithm~\ref{SBoost} with 
    $$\hat{F}_{t}(\mathbf{x}) \ = \ \hat{F}_{t-1}(\mathbf{x}) + 
    \gamma \, \alpha_t \,  h_t(\mathbf{x}) \, , $$
   and line 5 in Algorithm \ref{code-rrboost} with 
      $$\hat{F}_{T_1 + t}(\mathbf{x}) \ = \ \hat{F}_{T_1 + t-1}(\mathbf{x}) +  
      \gamma \, \alpha_t \, h_t(\mathbf{x}) \, .$$ 
We ran \texttt{RRBoost} with shrinkage parameter $\gamma = 0.1,0.3,0.5,$ and $1$ in one of our simulated settings. As expected, we observed improvement in prediction error as $\gamma$ decreases at the cost of requiring a significantly increased number of iterations to reach early stopping time.  For more details we refer the interested reader to the Appendix in the on-line supplementary material.       
 \vspace{0.3cm}
\begin{algorithm}[tp]
    \SetKwInOut{Input}{Input}
    \SetKwInOut{Output}{Output}
     \SetKwInOut{Initialize}{Initialize}
      \SetKwInOut{Stagea}{Stage 1}
     \SetKwInOut{Stageb}{Stage 2}
    \Input{A data set $(\mathbf{x}_i, y_i),  i \in \mathcal{I}_{\text{train}}$  \\
       The number of stage 1 iterations $T_1$   \\ 
       The number of stage 2 iterations $T_2$   \\ 
    The class of base learners $\mathcal{H}$ }
     \Stagea{Run SBoost for $T_1$ iterations and obtain: \\
$ \hat{F}_{T_1}(\mathbf{x}_i)$, and  $ \hat{\sigma}_n$ }
     \Stageb{}
     \hspace{1cm}
     \For{ $t = 1:T_2$}
     {
            \hspace{1cm} $g_{t,i} = - \frac{1}{\hat{\sigma}_n }
	    \psi_1 \left(\frac{y_{i}  - \hat{F}_{T_1 + t-1}(\mathbf{x}_{i})}{\hat{\sigma}_n} \right)$ \\
             \hspace{1cm} $h_t = \argmin_{h \in \mathcal{H}}  \sum_{i \in \mathcal{I}_{\text{train}}} (g_{t,i} + h(\mathbf{x}_i))^2$ \\
                   \hspace{1cm}  $\alpha_t = \argmin_{\alpha} \sum_{i \in \mathcal{I}_{\text{train}}} \rho_1 \left( \frac{y_i  - \hat{F}_{T_1 + t-1}(\mathbf{x}_i)  - \alpha h_t(\mathbf{x}_i)}{\hat{\sigma}_n} \right)$ \\
            \hspace{1cm} $\hat{F}_{T_1 + t}(\mathbf{x}) = \hat{F}_{T_1 + t-1}(\mathbf{x}) +   \alpha_t h_t(\mathbf{x})$ 
        }
      \Output{$\hat{F}_{T_1 + T_2}(\mathbf{x})$} 
          \caption{RRBoost algorithm}
            \label{code-rrboost}
\end{algorithm}

 \subsection{Early stopping}  \label{sec:early}

 The boosting algorithm is intrinsically greedy: 
 at each step it attempts to maximize the reduction in the value of the loss
 function evaluated on the training set. Consequently, overfitting the training
 set at the expense of an increased generalization error (i.e. a poor
 performance on future data) is a practical concern. To mitigate this problem
 we propose to use an early stopping rule based on monitoring the performance
 of the predictor on a validation set. This approach can be seen as a form of
 regularization. 
 
 We define the early stopping time of each stage of \texttt{RRBoost} as the 
 iteration
 that achieves the lowest loss function value on a validation set
 ($\mathcal{I}_{\text{val}}$). Let 
 $T_{1, \text{max}}$ and $T_{2, \text{max}}$ be the maximum number of
 iterations allowed in \texttt{RRBoost} (Algorithm~\ref{code-rrboost}).  In the first stage (\texttt{SBoost}), we refer to (\ref{eq:s:1})
 and define the validation loss: 
  $$\text{L}_{1,
 \text{val}}(t) = \hat{\sigma}_{\text{val}}(\hat{F}_t),  $$  
 where $\hat{\sigma}_{\text{val}}(\hat{F}_t)$ satisfies
 $$ \frac{1}{|\mathcal{I}_{\text{val}}|}
 \sum_{i \in \mathcal{I}_{\text{val}}} \rho_0 \left(\frac{y_i - \hat{F}_t(\mathbf{x}_i)}{\hat{\sigma}_{\text{val}}(\hat{F}_t)}\right) =  \kappa. $$
  The early stopping time for SBoost is 
   $$T_{1, \text{stop}}= \argmin_{t \in \{1,...,T_{1,\text{max}}\}} \text{L}_{1,\text{val}}(t).$$
 In the second stage, referring to (\ref{eq:rrboost}) we define the validation loss: 
   $$\text{L}_{
 \text{2,val}}(t) = \frac{1}{|\mathcal{I}_{\text{val}}|}\sum_{i \in \mathcal{I}_{\text{val}}} \rho_1 \left( \frac{ y_i - \hat{F}_{t}(\mathbf{x}_i)}{\hat{\sigma}_{\text{val}}(\hat{F}_{T_{1,\text{stop}}})} \right), $$ 
 and the early stopping time 
    $$T_{2, \text{stop}}= \argmin_{t \in \{T_{1,\text{max}},...,T_{2,\text{max}}\}} \text{L}_{2,\text{val}}(t).$$

In practice, when running boosting for $T_{1,\text{max}} + T_{2,\text{max}}$ iterations is too costly, one can terminate the boosting iterations when the validation error ceases to improve after several iterations.

\subsection{Robust variable importance} \label{robust-imp}

Variable importance methods 
typically rank the available explanatory variables in terms of their
contribution to the 
prediction strength of the fit
(see also \cite{Fisher2019etal} for a more general
discussion). 
The variable importance 
measure suggested by \cite{Friedman2001} for gradient boosting 
is based on averaging the improvement of a squared error at each tree split over
all iterations, which cannot be generalized to arbitrary loss functions and 
is limited to tree base learners.   Motivated by the permutation variable importance 
criterion used in random forests \citep{breiman2001random}, 
here we suggest a suitable robust modification 
that applies to \texttt{RRBoost}. 

Permutation importance measures 
are typically based on randomly reordering the values of a explanatory variable in the validation data, and 
recording the resulting change in prediction accuracy.   
The idea is that a variable is ``important''  if breaking its relationship with 
the response (by randomly shuffling its values) increases the prediction error.  
Since we assume that our validation data may be contaminated, a robust prediction 
accuracy metric is needed to reduce the potential impact of outliers on  
our prediction error quantification. 
To this end, we propose to trim the prediction errors 
that deviate from the median by more than 3 MAD (median absolute deviations). 
More specifically, let $\tilde{\mathcal{I}}_{\text{val}}$ be the indices of the observations 
in the validation set that are not flagged as outliers: 
\begin{equation} 
     \tilde{\mathcal{I}}_{\text{val}} = \{i: i \in \mathcal{I}_{\text{val}} \ \text{and} \  \left| \hat{F}(\mathbf{x}_i) - y_i - 
     \hat{\mu}_r 
     \right| < 3 \hat{\sigma}_r \} \label{trim:eq1} \, ,  
 \end{equation}
 where $\hat{\mu}_r = \text{median}(\hat{F}(\mathbf{x}_j) - y_j)$ and 
\begin{equation}
        \hat{\sigma}_r  = \text{MAD}(  \hat{F}(\mathbf{x}_i) - y_i, i \in \mathcal{I}_{\text{val}}) \, , \label{trim:eq2} 
    \end{equation}
with $\text{MAD}(r_i) = 1.438 \, \text{median}_{1 \le i \le n} | r_i - 
     \hat{\mu} |$, and $\hat{\mu} = \text{median}_{1 \le j \le n} r_j$. 
Finally, to evaluate the prediction error of our predictor on the validation set we 
use the trimmed root-mean-square error 
and define the importance of variable $j$ as: 
$$\text{VI}(\mathbf{x}_{.j}) = \sqrt{\frac{1}{|\tilde{\mathcal{I}}_{\text{val}}|}\sum_{i \in \tilde{\mathcal{I}}_{\text{val}}} (\hat{F}_{\pi_j}\left(\mathbf{x}_i) - y_i \right)^2} - \sqrt{\frac{1}{|\tilde{\mathcal{I}}_{\text{val}}|}\sum_{i \in \tilde{\mathcal{I}}_{\text{val}}} (\hat{F}\left(\mathbf{x}_i) - y_i \right)^2},$$
where $\hat{F}(\mathbf{x})$ is fitted using the validation data, and  $\hat{F}_{\pi_j}(\mathbf{x})$ is fitted using validation data with the $j$-th variable permuted.

\section{Simulation studies}  \label{sec:simu}

We performed extensive numerical experiments to assess the performance 
of \texttt{RRBoost} and compare it with that of other 
proposals in the literature. 
In particular, we study the prediction properties of different 
regression boosting 
methods when the training and validation sets contain outliers, 
and also when no atypical observations are present. 

\subsection{Set up} \label{set-up}

Consider a data set $D_n = \{(\mathbf{x}_i, y_i), i=1, \ldots, n \}$, 
consisting of explanatory variables  $\mathbf{x}_i \in \mathbb{R}^p$
and 
responses $y_i \in \mathbb{R}$ that follow the model
\begin{equation*}\label{eq:111}
y_i = g(\mathbf{x}_i) + C\epsilon_i \, , \qquad i = 1, \ldots, n \, ,
\end{equation*}
where the errors $\epsilon_i$ are independent from  the explanatory variables
$\mathbf{x}_i$, and $C > 0$ is a constant that controls the signal-to-noise
ratio (SNR): \begin{equation*} \label{SNR}
\text{SNR} = \frac{\text{Var}(g(\mathbf{x}))}{\text{Var}{(C\epsilon)}} \, .
\end{equation*} 
In our experiments, we considered three $g$ functions: 
\begin{align*}
g_1(\mathbf{x}) &= \ 2 \, \mathbf{x}_{.1} -2\mathbf{x}_{.2} +  8 \, \left (\mathbf{x}_{.3} - \frac{1}{2}
\right)^2 + \exp(\mathbf{x}_{.4}) + 0.5 \, \cos(8\pi \mathbf{x}_{.5}) \, \exp(2\mathbf{x}_{.5}) \, , \\
g_2(\mathbf{x}) &= \ 5 \, \left(\mathbf{x}_{.1}^2 + \left(\mathbf{x}_{.2}\mathbf{x}_{.3} - \frac{1}{\mathbf{x}_{.2}\mathbf{x}_{.4}}\right)^2\right)^{\frac{1}{2}} \, , \quad \mbox{ and} \\
g_3(\mathbf{x}) &=\, \left[ \sum_{j=1}^{5} \Bigl( 1 + (-1)^j 0.8 \mathbf{x}_{.j} + \sin(6\mathbf{x}_{.j})
\Bigr) \right] \, \sum_{j=1}^{3} \left(1+ \mathbf{x}_{.j}/3 \right) \, .
\end{align*}
The first 
function ($g_1$) is additive, and we expect simple base learners to work well in this case. 
For $g_2$ we modified a function used in \cite{friedman1991multivariate} (see also \cite{ridgeway1999boosting} and \cite{sigrist2018gradient}), 
while $g_3$ was adapted from \cite{buhlmann2003boosting}.

We considered the following error distributions: 
\begin{itemize}
    \item $D_0$: \textbf{Clean} - $\epsilon \sim N(0, 1)$;
    \item $D_1(\alpha)$: \textbf{Symmetric gross errors} 
		
		-  $\epsilon \sim 
		(1 - \alpha) N(0, 1) + 0.5 \alpha N(20,0.1^2) +  0.5 \alpha N(-20,0.1^2)$; 
		
%
    \item $D_2 (\alpha)$: \textbf{Asymmetric gross errors} 
		-  $\epsilon \sim 
		(1 - \alpha) N(0, 1) + \alpha N(20,0.1^2)$; 
	\item $D_3$: \textbf{Skewed errors} - $\epsilon =  W -  \exp(1/2)$, where $W$ has a 
	standard log-normal distribution; and
	\item $D_4$: \textbf{Heavy-tailed symmetric errors} - $\epsilon \sim T_1$, a Student's T with 1 degree of freedom. 
\end{itemize}
For the contaminated settings $D_1$ and $D_2$, we considered rates of contamination equal to 10\% and 20\% (which correspond to $\alpha =10\%$ and $\alpha = 20\%$ respectively).  The skewed error distribution in 
$D_3$ was centered (by  
subtracting $\sqrt{e}$) so that its mean is equal to 0.  

In each case, we constructed training, validation, and test sets.
Test sets were always generated with uncontaminated data ($D_0$) and had $1000$ observations.
Each estimator was fitted using the training and validation sets, 
while the test sets were reserved to evaluate the performance of the different estimators. 
The training sets were of size 300 and 3000, corresponding to data of small and moderate sizes.  
The corresponding validation sets had 200 and 2000 observations, respectively. 
We used $p = 10$ and $p = 400$ explanatory variables. 
The combination of training size 300 and $p = 400$ corresponds to a  
``high-dimensional'' case where the size of the training set is smaller than 
the number of features.  
The signal-to-noise ratio was set to  $\text{SNR} = 6$ when $p = 10$ and $\text{SNR} = 10$  when $p = 400$. 

The variables ($\mathbf{x}_{.1},..., \mathbf{x}_{.p})$ were simulated to have uniform marginal distributions
between 0 and 1, except for the case $g = g_2$, where 
$\mathbf{x}_{.2}$ and $\mathbf{x}_{.4}$  in $g_2$ appear in the
denominator term, and followed a ${\cal U}(1,2)$ distribution. 
We considered three types of correlation structures for
the variables in $\mathbf{x}$: 
\begin{itemize}
\item $S_0$: Independent components; 
\item  $S_1$: Decreasing correlation structure: 
$\mbox{cor}(\mathbf{x}_{.i}, \mathbf{x}_{.j}) = 0.8^{|i-j|}$; and 
\item $S_2$: Block correlation structure for active explanatory variables
\citep{zou2005regularization}:
$$Cor(\mathbf{x}_{.i}, \mathbf{x}_{.j}) = 
\begin{cases}
0.8 &  \mathbf{x}_{.i} \ \text{and} \ \mathbf{x}_{.j} \  \text{are in the same block};\\
0 &  \text{otherwise},
\end{cases}
$$
where the blocks were defined for each regression function $g$ as follows:
\begin{itemize}
        \item for $g_1$ and $g_3$: Block 1: $\left[ \mathbf{x}_{.1}, \mathbf{x}_{.2}, \mathbf{x}_{.3}\right]$,  
        Block 2: $\left[ \mathbf{x}_{.4}, \mathbf{x}_{.5} \right]$, and 
        \item for $g_2$: Block 1: $\left[ \mathbf{x}_{.1}, \mathbf{x}_{.2} \right]$,
        Block: 2: $\left[ \mathbf{x}_{.3}, \mathbf{x}_{.4} \right]$. 
\end{itemize}
\end{itemize}

For each possible 
combination of regression function, error distribution, 
variable correlation structure, sample size, and 
number of features, we conducted 100 independent experiments 
and compared the following non-parametric predictors: 
\begin{itemize}
    \item \texttt{L2Boost:} gradient boosting with squared error loss \citep{Friedman2001};
    \item \texttt{LADBoost:} gradient boosting with absolute error loss \citep{Friedman2001};
    \item \texttt{MBoost:}  gradient boosting with Huber's loss \citep{Friedman2001};
    \item \texttt{Robloss:} Robloss boosting \citep{lutz2008robustified};
    \item \texttt{SBoost:} SBoost (see Section \ref{sec:sboost}) intialized with \texttt{LADTree};
    \item \texttt{RRBoost:}  RRBoost (see Section \ref{sec:rrboost}) intialized with \texttt{LADTree}.
\end{itemize}
All methods under comparison were computed using decision trees of 
the maximum depth equal to 1, 2, and 3 as  base learners (note that a 
tree of depth 1 is often called a ``decision stump'').   
In order to facilitate comparison and manage the computational effort of our
experiment, we did not use a shrinkage parameter. 

\texttt{MBoost} and  \texttt{Robloss} both used Huber's loss and 
the residual scale estimator is re-computed at each iteration. 
For \texttt{MBoost} we followed \citep{Friedman2001} and used
the 90\% quantile of the residuals of the previous iteration, 
while \texttt{Robloss} used the MAD of residuals  \citep{lutz2008robustified}.  
For \texttt{RRBoost}, we used the early stopping rules discussed in Section \ref{sec:early} with $T_{1, \text{max}}= 500$ and $T_{2, \text{max}}= 1000$ for $d = 1$ trees and $T_{1, \text{max}}= 300$ and $T_{2, \text{max}}= 500$  for $d = 2$ and $d = 3$ trees. For \texttt{L2Boost} and  \texttt{LADBoost}, we tracked their loss functions evaluated on the validation set ($L_{\text{val}}(t)$)  and calculated the early stopping time ($T_{stop}$): 
$$T_{stop}= \argmin_{t \in
\{1,...,T_{\text{max}}\}} \text{L}_{\text{val}}(t) \, , 
$$ where the maximum number of iterations $T_{\text{max}} = 1500$ for $d = 1$ trees and $T_{\text{max}} = 800$ for $d = 2$ and $d = 3$ trees.   For \texttt{MBoost} and \texttt{Robloss},  since the tuning constant of their loss functions varies in each iteration, we used average absolute deviation (AAD) of the validation residuals for early stopping and calculated 
$$T_{stop}= \argmin_{t \in
\{1,...,T_{\text{max}}\}} \text{AAD}_{\text{val}}(t) \, , 
$$ 
where
$$\text{AAD}_{\text{val}}(t)  =  \frac{1}{|\mathcal{I}_{\text{val}}|} \sum_{i \in{ \mathcal{I}}_{\text{val}}} \left|
 \hat{F}_{t}(\mathbf{x}_i) - y_i \right| \,, $$
 with $T_{\text{max}} = 1500$ for $d = 1$ trees and $T_{\text{max}} = 800$ for $d = 2$ and $d = 3$ trees. 
 
To initialize \texttt{RRBoost} we considered \texttt{LADTree}s with depth 0, 1, 2, 3 and 4, and the minimum number of observations per node set at 10, 20 or 30 (note that an \texttt{LADTree} of depth 0 corresponds to using the median of the responses as the initial fit).  Of the 13 different combinations  we chose the one that performs the best on the validation set.  As the validation set may also contain outliers, we first fitted a depth 0 tree and trimmed as potential outliers those observations with residuals 
deviating from their median by more than 3 times
their MAD.

All our simulations were run using the \texttt{R} programming environment \citep{RLang}. Regression trees were fit using the \texttt{rpart} package,
including the  \texttt{LADTree}s and the base learners. The code  used in our experiments is publicly available on-line at \texttt{https://github.com/xmengju/RRBoost}.

\subsection{Performance metrics}
We compared different methods in terms of  their predictive performance and variable selection accuracy.  For predictive performance,  we recorded the root-mean-square error (RMSE) evaluated on the clean test set ($\mathcal{I}_{\text{test}}$) at the early stopping time ($T_{\text{stop}}$):  
 $$\text{RMSE}(T_{\text{stop}}) =
 \sqrt{\frac{1}{|\mathcal{I}_{\text{test}}|}\sum_{i \in
 \mathcal{I}_{\text{test}}} (\hat{F}_{T_{\text{stop}}}(\mathbf{x}_i) - y_i)^2}.
 $$
 
For variable selection accuracy, we  measured the proportion of variables recovered \citep{kazemitabar2017variable}.   Specifically, let $M$ be the set of indices of explanatory variables that were used to generate the data ($M = \{1, 2, \ldots, 5\}$). After ranking all $p$ features  using the robust permutation variable importance index,  we recorded the proportion of true variables in the  top $|M|$ of the ordered list:
\begin{equation*}
	\label{var:imp}
\frac{1}{|M|} \left| \{j, \text{VI}(\mathbf{x}_{.j}) >= \text{VI}(\mathbf{x}_{.(|M|)}) \} \cap  M\right|, 
\end{equation*}
where $\text{VI}(\mathbf{x}_{.(|M|)})$ is an order statistics, denoting the $|M|$th largest value of  $\text{VI}(\mathbf{x}_{.1}), ..., \text{VI}(\mathbf{x}_{.p})$. 

\subsection{Results} \label{sec:results}

There are 756 possible combinations of 
regression function, error distribution, 
feature correlation structure, sample size,  
number of features and depth of base learners. All our results can be found in the on-line supplementary material. 
Here we discuss in detail three representative settings, and below give 
general conclusions from our study.  
Consider the following three cases:
\begin{itemize}
\item Setting 1: $g = g_1$, $S = S_0$, $n = 300$, $p = 10$; 
\item Setting 2: $g = g_2$, $S = S_1$, $n = 3000$, $p = 400$;  and 
\item Setting 3:  $g = g_3$, $S = S_2$, $n = 300$, $p = 400$. 
\end{itemize}
These settings were chosen to represent different regression functions
$g$, correlation structures for the features, and data sizes. 

The depths of the tree learners that resulted in 
better performances for these three cases were: 
$d = 1$ for Settings 1 and 3, and $d = 2$  for Setting 2. 

\begin{figure}[tp]
  \centering
    \makebox[\linewidth][c]{
	\subfigure[No outliers]{\includegraphics[width=0.6\textwidth, page = 1]{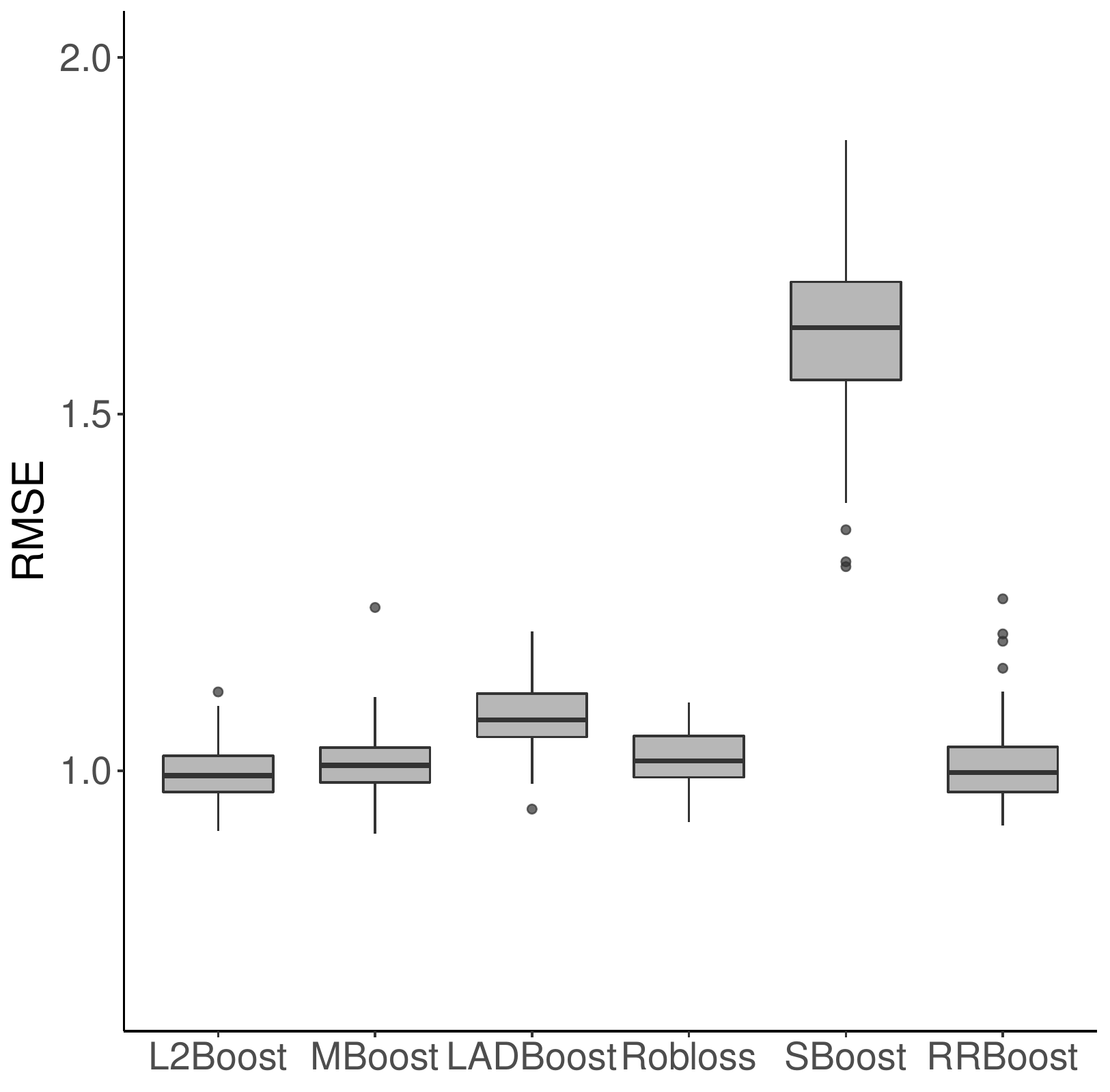}}
	\subfigure[Symmetric gross error]{\includegraphics[width=0.6\textwidth, page = 2]{R2_figs_simulation/setting_1_cov_x_indep_max_depth_1_n_300_p_10}}
	}
	  \makebox[\linewidth][c]{
	\subfigure[Asymmetric gross error]{\includegraphics[width=0.6\textwidth, page = 3]{R2_figs_simulation/setting_1_cov_x_indep_max_depth_1_n_300_p_10}}
		\subfigure[Skewed and heavy-tailed distribution]{\includegraphics[width=0.6\textwidth, page = 4]{R2_figs_simulation/setting_1_cov_x_indep_max_depth_1_n_300_p_10}}
		}
 \caption{Boxplots of RMSEs on the test sets for 100 independent runs using data generated from Setting 1($g = g_1$, $S = S_0$, $n = 300$, $p = 10$).  Panel (a) corresponds to clean training sets with Gaussian errors ($D_0$);  panel (b) corresponds to symmetric gross error contaminated ($D_1$) data; panel (c) corresponds to asymmetric gross error contaminated ($D_2$) data; and panel (d) corresponds to skewed distributed ($D_3$) and heavy-tailed distributed ($D_4$) data. In panels (b) and (c), the yellow and blue boxplots correspond to 10\% and 20\% outliers, respectively.  In panel (d), the yellow bloxplots correspond to data with log-normaly distributed errors ($D_3$) and the blue boxplots correspond to data with heavy-tailed distributed errors ($D_4$).}
\label{figure:setting1}
\end{figure}

\begin{figure}[tp]
  \centering
    \makebox[\linewidth][c]{
	\subfigure[No outliers]{\includegraphics[width=0.6\textwidth, page = 1]{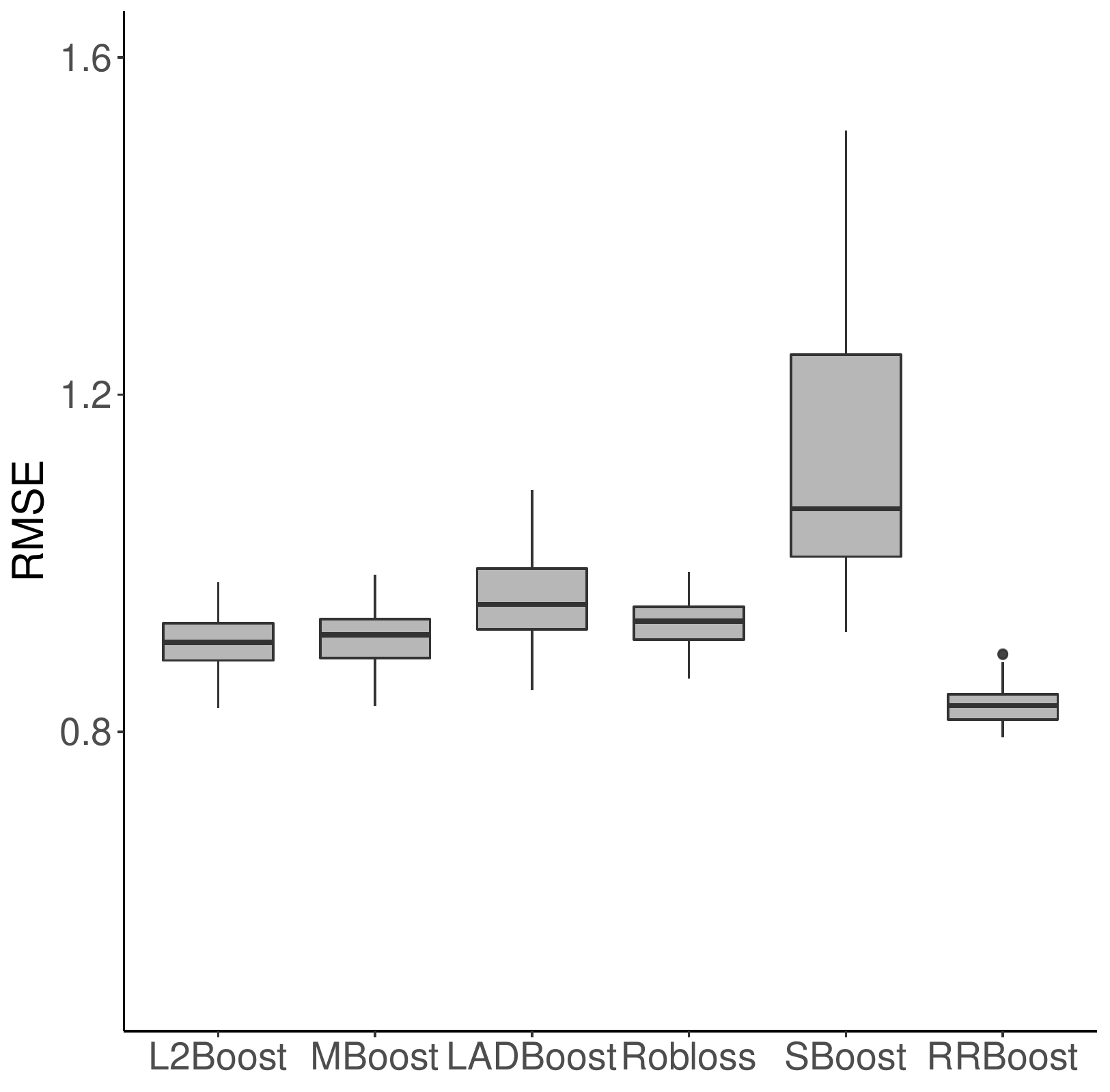}}
	\subfigure[Symmetric gross error]{\includegraphics[width=0.6\textwidth, page = 2]{R2_figs_simulation/setting_2_cov_x_dep1_max_depth_2_n_3000_p_400}}
	}
	  \makebox[\linewidth][c]{
	\subfigure[Asymmetric gross error]{\includegraphics[width=0.6\textwidth, page = 3]{R2_figs_simulation/setting_2_cov_x_dep1_max_depth_2_n_3000_p_400}}
		\subfigure[Skewed and heavy-tailed distribution]{\includegraphics[width=0.6\textwidth, page = 4]{R2_figs_simulation/setting_2_cov_x_dep1_max_depth_2_n_3000_p_400}}
		}
 \caption{Boxplots of RMSEs on the test sets for 100 independent runs using data generated from Setting 2 ($g = g_2$, $S = S_1$, $n = 3000$, $p = 400$).  Panel (a) corresponds to clean training sets with Gaussian errors ($D_0$);  panel (b) corresponds to symmetric gross error contaminated ($D_1$) data; panel (c) corresponds to asymmetric gross error contaminated ($D_2$) data; and panel (d) corresponds to skewed distributed ($D_3$) and heavy-tailed distributed ($D_4$) data. In panels (b) and (c), the yellow and blue boxplots correspond to 10\% and 20\% outliers, respectively.  In panel (d), the yellow bloxplots correspond to data with log-normaly distributed errors ($D_3$) and the blue boxplots correspond to data with heavy-tailed distributed errors ($D_4$).}
  \label{figure:setting2}
\end{figure}

\begin{figure}[tp]
  \centering
    \makebox[\linewidth][c]{
	\subfigure[No outliers]{\includegraphics[width=0.6\textwidth, page = 1]{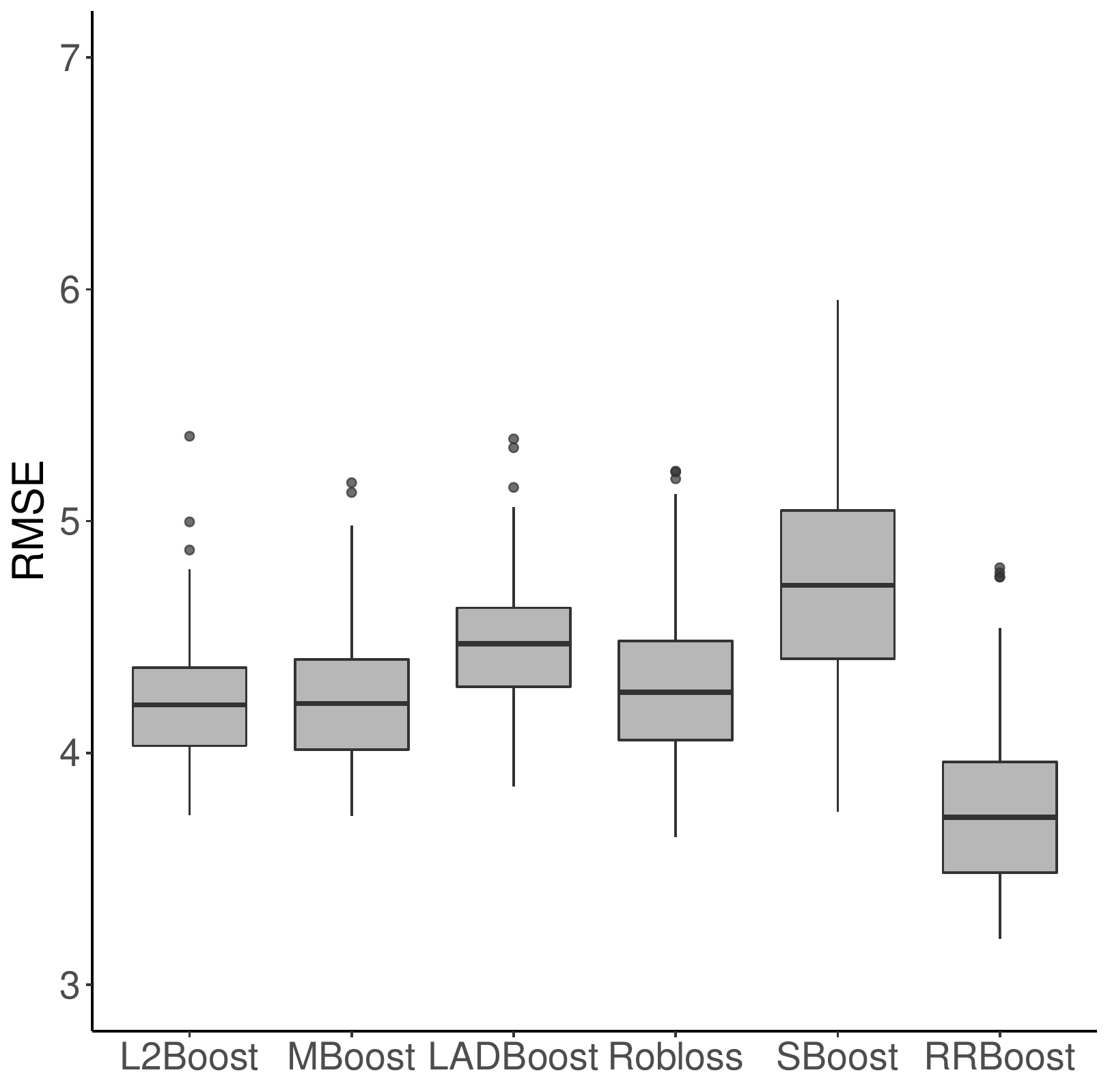}}
	\subfigure[Symmetric gross error]{\includegraphics[width=0.6\textwidth, page = 2]{R2_figs_simulation/setting_3_cov_x_dep2_max_depth_1_n_300_p_400}}
	}
	  \makebox[\linewidth][c]{
	\subfigure[Asymmetric gross error]{\includegraphics[width=0.6\textwidth, page = 3]{R2_figs_simulation/setting_3_cov_x_dep2_max_depth_1_n_300_p_400}}
		\subfigure[Skewed and heavy-tailed distribution]{\includegraphics[width=0.6\textwidth, page = 4]{R2_figs_simulation/setting_3_cov_x_dep2_max_depth_1_n_300_p_400}}
		}
 \caption{Boxplots of RMSEs on the test sets for 100 independent runs using data generated from Setting 3 ($g = g_3$, $S = S_2$, $n = 300$, $p = 400$).  Panel (a) corresponds to clean training sets with Gaussian errors ($D_0$);  panel (b) corresponds to symmetric gross error contaminated ($D_1$) data; panel (c) corresponds to asymmetric gross error contaminated ($D_2$) data; and panel (d) corresponds to skewed distributed ($D_3$) and heavy-tailed distributed ($D_4$) data. In panels (b) and (c), the yellow and blue boxplots correspond to 10\% and 20\% outliers, respectively.  In panel (d), the yellow bloxplots correspond to data with log-normaly distributed errors ($D_3$) and the blue boxplots correspond to data with heavy-tailed distributed errors ($D_4$).}
 \label{figure:setting3}
\end{figure}

\Cref{figure:setting1,figure:setting2,figure:setting3} show the boxplots of the test root-mean-square 
error (RMSE) at early stopping times 
for the 100 independent runs of the experiment, for each of the three settings above (the summary statistics 
are displayed in \Cref{table:setting1,table:setting2,table:setting3}).  

\begin{table}
\hspace{-4ex}
\begin{tabular}{lcccccc}
  \hline
& L2Boost & MBoost & LADBoost & Robloss & SBoost & RRBoost \\ 
  \hline
  $D_0$           & 1.00 (0.04) & 1.01 (0.04) & 1.08 (0.05) & 1.02 (0.04) & 1.61 (0.12) & 1.01 (0.06) \\ 
  $D_1(10\%)$ & 2.03 (0.19) & 1.61 (0.32) & 1.18 (0.07) & 1.15 (0.06) & 1.61 (0.12) & 1.05 (0.08) \\ 
  $D_1(20\%)$ & 2.23 (0.25) & 2.14 (0.18) & 1.29 (0.09) & 1.39 (0.13) & 1.52 (0.14) & 1.09 (0.06) \\ 
  $D_2(10\%)$ & 2.46 (0.19) & 1.86 (0.45) & 1.18 (0.07) & 1.20 (0.11) & 1.58 (0.13) & 1.05 (0.08) \\ 
  $D_2(20\%)$ & 3.53 (0.29) & 3.49 (0.31) & 1.45 (0.16) & 1.72 (0.31) & 1.50 (0.12) & 1.09 (0.07) \\ 
  $D_3$            & 1.29 (0.12) & 1.12 (0.06) & 1.15 (0.06) & 1.10 (0.05) & 1.62 (0.12) & 1.12 (0.10) \\ 
  $D_4$            & 15.3 (37.0) & 1.54 (0.14) & 1.33 (0.09) & 1.37 (0.10) & 1.62 (0.12) & 1.31 (0.09) \\ 
   \hline
\end{tabular}
\caption{Summary statistics of RMSEs on the test sets for 
Setting 1 ($g = g_1$, $S = S_0$, $n = 300$, $p = 10$). Numbers are averages and standard deviations calculated from 100 independent runs of the experiment.}
\label{table:setting1}
\end{table}

\begin{table}
\hspace{-4ex}
\begin{tabular}{lcccccc}
  \hline
 & L2Boost & MBoost & LADBoost & Robloss & SBoost & RRBoost \\ 
  \hline
  $D_0$            & 0.90 (0.03) & 0.91 (0.03) & 0.96 (0.05) & 0.93 (0.03) & 1.11 (0.15) & 0.83 (0.02) \\ 
  $D_1 (10\%)$ & 1.30 (0.07) & 1.13 (0.15) & 0.98 (0.04) & 0.96 (0.03) & 1.01 (0.10) & 0.83 (0.02) \\ 
  $D_1 (20\%)$ & 1.39 (0.07) & 1.40 (0.09) & 1.02 (0.05) & 1.01 (0.04) & 0.93 (0.09) & 0.82 (0.03) \\ 
  $D_2 (10\%)$ & 1.93 (0.07) & 1.39 (0.29) & 0.99 (0.04) & 0.97 (0.03) & 0.98 (0.08) & 0.82 (0.02) \\ 
  $D_2 (20\%)$ & 3.16 (0.08) & 3.18 (0.16) & 1.09 (0.05) & 1.10 (0.04) & 0.86 (0.04) & 0.82 (0.02) \\ 
  $D_3$            & 1.09 (0.06) & 0.96 (0.03) & 1.02 (0.04) & 0.98 (0.03) & 1.23 (0.14) & 0.89 (0.03) \\ 
  $D_4$            & 115 (488)   & 1.10 (0.05) & 1.05 (0.05) & 1.01 (0.04) & 0.89 (0.07) & 0.84 (0.02) \\ 
   \hline
\end{tabular}
\caption{Summary statistics of RMSEs on the test sets for 
Setting 2 ($g = g_2$, $S = S_1$, $n = 3000$, $p = 400$). 
Numbers are averages and standard deviations calculated 
from 100 independent runs of the experiment.}
\label{table:setting2}
\end{table}

\begin{table}
\hspace{-4ex}
\begin{tabular}{lcccccc}
  \hline
  & L2Boost & MBoost & LADBoost & Robloss & SBoost & RRBoost \\ 
  \hline
  $D_0$             & 4.22 (0.28) & 4.23 (0.30) & 4.47 (0.31) & 4.31 (0.36) & 4.77 (0.46) & 3.77 (0.36) \\ 
  $D_1 (10\%)$ & 6.87 (0.92) & 5.71 (0.86) & 4.85 (0.41) & 4.75 (0.41) & 4.49 (0.39) & 3.74 (0.28) \\ 
  $D_1 (20\%)$ & 7.79 (1.33) & 7.90 (1.46) & 5.11 (0.38) & 5.27 (0.38) & 4.16 (0.37) & 3.71 (0.27) \\ 
  $D_2 (10\%)$ & 8.34 (0.74) & 6.71 (1.31) & 4.90 (0.37) & 4.88 (0.38) & 4.48 (0.46) & 3.71 (0.26) \\ 
  $D_2 (20\%)$ & 13.0 (1.31) & 13.2 (1.41) & 5.65 (0.44) & 6.09 (0.7) & 4.16 (0.42) & 3.69 (0.33) \\ 
  $D_3$            & 5.07 (0.47) & 4.54 (0.32) & 4.74 (0.35) & 4.57 (0.37) & 4.73 (0.53) & 3.93 (0.31) \\ 
  $D_4$            & 81.8 (242) & 5.47 (0.36) & 5.11 (0.37) & 5.15 (0.34) & 4.59 (0.55) & 4.07 (0.38) \\ 
   \hline
\end{tabular}
\caption{Summary statistics of RMSEs on the test sets for 
Setting 3 ($g = g_3$, $S = S_2$, $n = 300$, $p = 400$). 
Numbers are averages and standard deviations calculated 
from 100 independent runs of the experiment.}
\label{table:setting3}
\end{table}

We find that when there are no outliers in the training and validation sets
\texttt{L2Boost}, \texttt{MBoost}, \texttt{Robloss} and 
\texttt{RRBoost} performed equally well. As expected the lack of efficiency of 
\texttt{SBoost} and \texttt{LADBoost} were clearly reflected in their increased
prediction errors. Note that in Settings 2 and 3, \texttt{RRBoost} 
outperforms \texttt{L2Boost}, which we believe is likely due to 
the fact that \texttt{RRBoost} requires more iterations to converge than \texttt{L2Boost}, 
and thus may be resulting in a better approximation to the optimization path.  

The performance of \texttt{L2Boost}, \texttt{MBoost}, \texttt{LADBoost}, and \texttt{Robloss} seriously deteriorated 
when outliers were present in the data (both in the symmetric and asymmetric cases), 
while \texttt{RRBoost} achieved the lowest average test RMSE among all methods, with a remarkably stable 
performance. The worst contamination setting for 
\texttt{MBoost}, \texttt{LADBoost}, and \texttt{Robloss} was $D_2$ with 20\% outliers. 
It is interesting to note that for some contaminated settings 
the performance of \texttt{MBoost} was very similar to that of 
\texttt{L2Boost}. A possible explanation 
is that \texttt{MBoost} selects the tuning constant for the loss function as the 
90\% quantile of the residuals. 
It is not clear how to address this issue without assuming that the proportion of outliers is known. 

Skewed errors ($D_3$) affected all predictors, but the damage was generally the least among all the 
contamination settings. Not surprisingly, heavy-tailed symmetric errors ($D_4$) 
produced the largest prediction error, with \texttt{L2Boost} the most negatively affected, and 
\texttt{RRBoost} achieving the best performance in all three cases.

We also report the fractions of variables recovered (see Section \ref{var:imp}) using the 
variable importance measure in Section \ref{robust-imp}. The results 
can be found in Tables \ref{imp:setting1}, \ref{imp:setting2} and \ref{imp:setting3}.  
We note that for Settings 1 and 2 \texttt{RRBoost} an \texttt{LADBoost} 
recovered almost all variables for clean and contaminated data, whereas the other boosting methods 
(especially \texttt{L2Boost} and \texttt{MBoost}) recovered fewer variables when trained with contaminated data compared to clean data, with the exception of $D_3$ (which also least affected the prediction errors).   
In Setting 3 \texttt{RRBoost} recovered over 99\% of the variables for
all contaminations, except for $D_4$ where it identified 96\% of them. 
This setting was more challenging for the other methods, 
especially for \texttt{L2Boost} and \texttt{MBoost}. 
In particular, \texttt{L2Boost} and \texttt{MBoost} recovered less than $30\%$ 
of variables under $D_1(20\%)$ and $D_2(20\%)$, and \texttt{L2Boost} rarely selected the true variables
when atypical observations were present. 

Analyzing the complete set of results of our numerical experiments (reported in the Appendix) we can make the 
following general observations: 
\begin{itemize}
\item Error distributions: The observations from the three
cases discussed above 
extend to the rest of our experiments. Specifically, asymmetric gross errors 
in $D_2(20\%)$ affected \texttt{MBoost},  \texttt{LADBoost},  and \texttt{Robloss}
the most, while \texttt{L2Boost} was most sensitive to heavy-tailed errors ($D_4$). 
Skewed errors in $D_3$ only had a  mild effect on  \texttt{L2Boost},  \texttt{MBoost},  \texttt{LADBoost},  and \texttt{Robloss} compared to the other cases. 
The performance of \texttt{RRBoost} was very stable across different types of contamination, 
whereas the performance of the other boosting methods
significantly worsened when the percentage contamination increases from 10\% to 20\% for $D_1$ and $D_2$.   
\item Sample sizes: As expected, larger training samples improved the prediction error of all boosting methods. 
\item Correlation structures of the features: it is not surprising that 
the effect of the correlation structure of the
explanatory variables depended on the regression function. 
For $g_1$, all boosting methods tended to perform better with correlated features ($S_1$ and $S_2$) 
than in the independent case ($S_0$). 
For $g_2$ and $g_3$ the best scenario was 
$S_0$, and in most cases the second best was $S_2$. 
\item Depths of the tree learners: Although  
the optimal tree depth varied for different scenarios, we
observed  that for $g_1$ and $g_3$, trees with depth $d = 1$ worked the best,
while for $g_2$  it was $d = 2$. 
The deepest trees considered here ($d=3$) usually resulted in the worst performances. 
\item Variable selection results:  we observed similar patterns for the fraction of variables recovered
 for different regression functions and correlation structures.   
 We analyzed the results of different methods for the optimal tree depth at each setting 
 ($d = 1$ for $g_1$ and $g_3$, and $d = 1$ or $2$ for $g_2$, see Appendix), 
 treating each sample size setting separately.   
 Since the skewed error distribution $D_3$ did not affect visibly the variable recovery rate 
 in general (except for a few cases where it mildly affected $\texttt{L2Boost}$) we 
 do not include it in the discussion below. 
\begin{itemize}
\item When $n = 300$, \texttt{RRBoost} had the best variable recovery rate in almost all tables.  
\texttt{RRBoost} outperformed \texttt{L2Boost} and \texttt{MBoost}
by a significant amount for different types of contaminations. The latter were 
negatively affected by $D_1$ and $D_2$ (more severely by 20\% of outliers). 
\texttt{LADBoost} and \texttt{Robloss} produced similar results to those of  
\texttt{RRBoost} under $D_1(10\%)$ and $D_2(10\%)$ contaminations. 

When $p = 400$ the differences among methods were more prominent.
Table \ref{imp:setting3} is representative of the 
situations where the contrast between \texttt{RRBoost} 
and the other methods was particularly notable. 
\texttt{L2Boost} recovered very few variables under $D_4$, which could be expected given
its poor predictive performance.
\item  When $n = 3000$, \texttt{RRBoost}, \texttt{LADBoost} and \texttt{Robloss}  
performed similarly, recovering almost all variables for clean and contaminated settings.   
\texttt{L2Boost} and  \texttt{MBoost} were negatively affected by $D_1$ and $D_2$ (the case
$g = g_1$, $S = S_0$ was an exception were all methods worked well, 
probably due to the combination of an additive function 
and independent features).  
\texttt{L2Boost} was again severely affected by $D_4$ (recovering only 
$20\% \sim 30\%$ features on average). As $p$ increases from $p = 10$ to $p = 400$, we 
observed a decrease in variable recovery rates for  \texttt{L2Boost} and  \texttt{MBoost}. 
In  particular, with $D_1(20\%)$ and $D_2(20\%)$ contaminations their average recovery rates 
usually dropped from $70\% \sim 90\%$ to $40\%  \sim 60\%$. 
\end{itemize}
\end{itemize}

Overall, \texttt{RRBoost} resulted in more accurate predictions for all data generation schemes considered
in our experiment, and the associated variable importance measure can be used as a reliable variable selection method.

\begin{table}
\centering
\begin{tabular}{lccccc}
  \hline
 & L2Boost & MBoost & LADBoost & Robloss & RRBoost \\ 
  \hline
  $D_0$            & 1.00 (0.00) & 1.00 (0.00) & 1.00 (0.00) & 1.00 (0.00) & 1.00 (0.00) \\ 
  $D_1 (10\%)$     & 0.48 (0.27) & 0.77 (0.30) & 1.00 (0.00) & 1.00 (0.02) & 1.00 (0.02) \\ 
  $D_1 (20\%)$     & 0.39 (0.27) & 0.27 (0.23) & 1.00 (0.02) & 0.97 (0.07) & 1.00 (0.00) \\ 
  $D_2 (10\%)$     & 0.50 (0.30) & 0.85 (0.21) & 1.00 (0.02) & 0.99 (0.03) & 1.00 (0.00) \\ 
  $D_2 (20\%)$     & 0.31 (0.22) & 0.45 (0.25) & 0.99 (0.04) & 0.91 (0.16) & 1.00 (0.00) \\ 
  $D_3$            & 0.99 (0.04) & 1.00 (0.02) & 1.00 (0.00) & 1.00 (0.00) & 1.00 (0.00) \\ 
  $D_4$            & 0.36 (0.27) & 0.92 (0.14) & 0.98 (0.06) & 0.97 (0.07) & 0.98 (0.06) \\ 
   \hline
\end{tabular}
\caption{Summary statistics of the fractions of variables recovered for Setting 1 ($g = g_1$, $S = S_0$, $n = 300$, $p = 10$). Numbers are averages and standard deviations calculated from 100 independent runs of the experiment.
}
\label{imp:setting1}
\end{table}

\begin{table}
\centering
\begin{tabular}{lccccc}
  \hline
 & L2Boost & MBoost & LADBoost & Robloss & RRBoost \\ 
  \hline
  $D_0$        & 1.00 (0.00) & 1.00 (0.00) & 1.00 (0.00) & 1.00 (0.00) & 1.00 (0.00) \\ 
  $D_1 (10\%)$ & 0.74 (0.12) & 0.87 (0.14) & 1.00 (0.00) & 1.00 (0.00) & 1.00 (0.00) \\ 
  $D_1 (20\%)$ & 0.66 (0.13) & 0.66 (0.14) & 0.97 (0.09) & 0.98 (0.07) & 1.00 (0.00) \\ 
  $D_2 (10\%)$ & 0.68 (0.12) & 0.78 (0.10) & 1.00 (0.00) & 1.00 (0.00) & 1.00 (0.00) \\ 
  $D_2 (20\%)$ & 0.53 (0.11) & 0.57 (0.11) & 0.98 (0.07) & 0.91 (0.12) & 1.00 (0.00) \\ 
  $D_3$        & 0.82 (0.12) & 1.00 (0.00) & 0.99 (0.04) & 1.00 (0.00) & 1.00 (0.00) \\ 
  $D_4$        & 0.03 (0.08) & 0.88 (0.13) & 0.97 (0.08) & 0.96 (0.09) & 1.00 (0.00) \\ 
   \hline
\end{tabular}
\caption{Summary statistics of the fractions of variables recovered for Setting 2 ($g = g_2$, $S = S_1$, $n = 3000$, $p = 400$). Numbers are averages and standard deviations calculated from 100 independent runs of the experiment.
}
\label{imp:setting2}
\end{table}

\begin{table}
\centering
\begin{tabular}{lccccc}
  \hline
  & L2Boost & MBoost & LADBoost & Robloss & RRBoost \\ 
  \hline
  $D_0$        & 0.96 (0.08) & 0.97 (0.07) & 0.91 (0.12) & 0.95 (0.10) & 0.99 (0.04) \\ 
  $D_1 (10\%)$ & 0.33 (0.10) & 0.55 (0.22) & 0.80 (0.16) & 0.83 (0.17) & 0.99 (0.03) \\ 
  $D_1 (20\%)$ & 0.23 (0.11) & 0.24 (0.12) & 0.67 (0.15) & 0.64 (0.17) & 1.00 (0.03) \\ 
  $D_2 (10\%)$ & 0.33 (0.09) & 0.51 (0.19) & 0.79 (0.15) & 0.83 (0.15) & 1.00 (0.00) \\ 
  $D_2 (20\%)$ & 0.28 (0.10) & 0.29 (0.10) & 0.65 (0.15) & 0.57 (0.14) & 0.99 (0.04) \\ 
  $D_3$        & 0.68 (0.20) & 0.89 (0.14) & 0.87 (0.15) & 0.92 (0.13) & 0.99 (0.04) \\ 
  $D_4$        & 0.05 (0.11) & 0.53 (0.13) & 0.68 (0.17) & 0.66 (0.16) & 0.96 (0.09) \\ 
   \hline
\end{tabular}
\caption{Summary statistics of the fractions of variables recovered for Setting 3 ($g = g_3, S = S_2, n = 300, p = 400$). Numbers are averages and standard deviations calculated from 100 independent runs of the experiment.}
\label{imp:setting3}
\end{table}

\section{Empirical studies} \label{sec:emp}

We evaluated the performance of our proposed algorithm on benchmark data sets,
where we might find very different
regression functions and 
correlation structures from the ones used in our
simulation experiments.   
 We considered 5 data sets 
 (see Table \ref{dat-chari} for details)
 where previous analyses found
potential outliers. 
 \begin{table}[tp] 
\centering
\begin{tabular}{llllll}
  \hline
Data &  Size $(n)$ & Dimension $(p)$ & Source \\
  \hline
airfoil & 1503 & 5 & UCI \citep{brooks1989airfoil} \\
abalone & 4177 & 7 & UCI \citep{nash1994population}\\
wage & 3000 & 8 & \texttt{ISLR} R package \citep{james2013islr} \\
nir & 103 & 94 & \texttt{pls} R package \citep{wehrens2007pls} \\
glass & 180 & 486 & Serneels et al.  \citep{serneels2005partial} \\ \hline
\end{tabular}
\caption{Summary of the characteristics of the data sets used in the experiments.}
\label{dat-chari}
\end{table}

Comparing different predictors on non-synthetic data sets 
is typically complicated by the fact that generally there is no 
fixed test set on which to evaluate their performances. 
For our experiments we randomly split 
each data set into a  training set (composed of 60\% of the data), a validation
set (20\% of the data) and a test set with the remaining 20\%. We ran the
boosting algorithms on the training sets, and used the validation sets to
select an early stopping time.  To avoid the effect of potential 
outliers in the test set, prediction performance was evaluated
with a trimmed root-mean-square error (TRMSE) as follows. Let 
$\tilde{\mathcal{I}}_{\mbox{test}}$
be the observations in the test set with prediction errors deviating from their median by less than 
3 times of their MAD:
$$
\tilde{\mathcal{I}}_{\mbox{test}} = \{i: i \in \mathcal{I}_{\mbox{test}} \ \text{and} \  
\left| \hat{F}_{T_{\text{stop}}}(\mathbf{x}_i)
- y_i - \hat{\mu}_r 
\right| < 3 \hat{\sigma}_r \} \, ,
$$
where $\hat{\mu}_r  = 
\text{median}(\hat{F}_{T_{\text{stop}}}(\mathbf{x}_i)- y_i)$ and  
$\hat{\sigma}_r  = \text{MAD}(  
\hat{F}_{T_{\text{stop}}}(\mathbf{x}_i)
- y_i, i \in \mathcal{I}_{\mbox{test}})$. 
The TRMSE is 
$$
\mbox{TRMSE} \, = \ 
\sqrt{\frac{1}{|\tilde{\mathcal{I}}_{\mbox{test}}|} \, 
\sum_{i \in \tilde{\mathcal{I}}_{\mbox{test}}} \, \left( 
\hat{F}_{T_{\text{stop}}}(\mathbf{x}_i)
- y_i \right)^2 } \, .
$$
In order to reduce the natural variability introduced by the random partitions,
we repeated this experiment 50 times with different random splits.  

%

We used the same base learners and number of iterations as in Section \ref{sec:simu},
and show results for the tree depth that produced the best results.  Where similar
performances were observed for base learners of different depths, we show the
results for the simpler one (corresponding to a smaller depth).  We picked $d  =
2$ for the ``airfoil'' data, and $d = 1$ for the rest of data sets. Results for
all depths can be found in the Appendix. 
We initialized \texttt{RRBoost} as in  Section \ref{set-up}, using
\texttt{LADTrees} with depth 0, 1, 2, 3 and 4. The minimum number of
observations per node was set at 10, 20 or 30 for the ``airfoil'', ``abalone'',
and ``wage'' data sets; and at 5, 10 or 15 for the ``nir'' and ``glass'' data sets
(on account of their smaller sample sizes). 

The results are summarized in Table \ref{t:real.clean}. 
We note that \texttt{RRBoost} resulted in the smallest
TRMSE for the ``airfoil'' and
``abalone'' data sets (in the latter case 
\texttt{LADBoost} is a close second). 
\texttt{RRBoost} also performed better than the alternatives 
with the ``nir'' data set, but the magnitudes of 
the differences are less notable considering the 
Monte Carlo uncertainty reflected in the standard errors. 
Finally, we note that all predictors
performed similarly on the ``wage'' and ``glass'' data sets. 
\begin{table}[htp]
\begin{tabular}{lccccc}
  \hline 
Data &  L2Boost & MBoost & LADBoost & Robloss & RRBoost \\ 
  \hline
airfoil  &  1.72 (0.13) & 1.67 (0.11) & 2.05 (0.16) & 1.67 (0.12) & 1.58 (0.12) \\ 
abalone & 1.78 (0.06) & 1.73 (0.05) & 1.66 (0.06) & 1.70 (0.05) & 1.64 (0.05) \\  
wage   & 24.7 (0.77) & 24.0 (0.79) & 23.9 (0.72) & 23.9 (0.71) & 23.9 (0.78) \\  
nir    & 3.89 (1.00) & 3.95 (1.08) & 4.42 (1.21) & 3.98 (0.93) & 3.74 (1.01)  \\  
glass   & 0.10 (0.04) & 0.11 (0.04) & 0.12 (0.02) & 0.13 (0.03) & 0.10 (0.02) \\  
   \hline
\end{tabular}
\caption{Average and standard deviation of 
trimmed root-mean-square errors (TRMSEs) computed on 50 random training / validation / test data splits.} 
\label{t:real.clean}
\end{table}

\subsection{Added atypical observations}

To further explore the behaviour of these predictors on 
non-synthetic data (where we do not control the shape of
the regression function, or the correlation structure 
among the available features), we performed another
experiment by randomly adding outliers to the 
training and validations sets obtained before.  
Thus, this study can be considered as an extension of our experiments
reported in Section \ref{sec:simu}, where now both the 
regression function and variable correlation structure remain
unknown. 

For each data set, new response variables were
generated as follows: 
$$
y_i = y^{\ast}_i + C \epsilon_i \, ,
$$ 
where $y^{\ast}_i$ denote the original responses, $\epsilon_i$ are random
errors, and $C > 0$ controls the desired signal-to-noise ratio (SNR) as in Section
\ref{SNR}. 
We used $\text{SNR} = 10$ for the high dimensional ``glass'' data  and
$\text{SNR} = 6$ for the other data sets.  For the distribution of errors
($\epsilon_i$) we followed the settings with $D_1(20\%)$, $D_2(20\%)$ and $D_4$ error distributions in Section \ref{set-up}, which were shown to significantly affect the performance of predictors in most simulated cases  (see Section \ref{sec:results}). 

The results are shown in Table \ref{t:real.synthetic}. 
We see that the performance of \texttt{RRBoost} was notably stable, and, for
each data set, its results were closest to those for the original data set
(Table \ref{t:real.clean}) across  different contamination settings.  
Both \texttt{L2Boost} and \texttt{MBoost} were seriously affected by
contaminations (the worst being $D_2(20\%)$).  
The performance of \texttt{Robloss}  was
relatively good, but clearly worse than that of 
\texttt{LADBoost} for all data sets except for the ``glass'' data set. 
The deterioration with respect to the
original data was most pronounced for the ``airfoil'' and ``nir'' data sets.  It
is interesting to note that, as it was the case for the original ``wage'' and
``glass'' data sets, \texttt{LADBoost}, \texttt{Robloss} and \texttt{RRBoost}
produced similar results on their contaminated versions.  However, Table
\ref{t:real.synthetic} shows that \texttt{RRBoost} resulted in best predictions
across contaminations and data sets, and also suffered the least performance
degradation when compared with its behaviour on the original data sets
(see Table \ref{t:real.clean}).  We provided  top variables ranked using the robust variable importance (see \Cref{robust-imp})  in the Appendix.  There is no clear consensus among the rankings obtained with the different boosting methods, but we often observed overlaps in top ranked variables produced by \texttt{RRBoost} using the original data and data with added contamination.

\begin{table}
\begin{centering}
\begin{tabular}{llccccc}
  \hline 
	Data & Error & L2Boost & MBoost & LADBoost & Robloss & RRBoost \\ 
  \hline
airfoil  & $D_1 (20\%)$ & 6.49 (0.66) & 6.24 (0.50) & 3.21 (0.28) & 3.47 (0.31) & 2.56 (0.20) \\ 
         & $D_2 (20\%)$ & 12.7 (0.53) & 12.5 (0.68) & 3.68 (0.27) & 4.14 (0.34) & 2.56 (0.20) \\ 
         & $D_4$        & 9.04 (4.32) & 4.15 (0.33) & 3.34 (0.25) & 3.49 (0.24) & 3.06 (0.25)\\   \hline 
abalone  & $D_1 (20\%)$ & 2.29 (0.20) & 2.26 (0.18) & 1.77 (0.06) & 1.81 (0.05) & 1.74 (0.06) \\ 
         & $D_2 (20\%)$ & 5.91 (0.17) & 5.69 (0.14) & 1.98 (0.07) & 2.19 (0.07) & 1.74 (0.06) \\ 
         & $D_4$        & 4.43 (3.12) & 1.93 (0.06) & 1.78 (0.07) & 1.83 (0.06) & 1.75 (0.06)  \\  \hline
wage     & $D_1 (20\%)$ & 29.9 (1.92) & 29.3 (1.63) & 24.4 (0.87) & 24.5 (0.83) & 24.1 (0.84) \\ 
         & $D_2 (20\%)$ & 77.6 (2.41) & 75.2 (2.96) & 26.9 (1.03) & 29.1 (1.23) & 24.1 (0.82) \\ 
         & $D_4$        & 46.4 (16.1) & 25.6 (0.87) & 24.6 (0.82) & 24.8 (0.92) & 24.3 (0.80)  \\  \hline 
nir      & $D_1 (20\%)$ & 13.9 (4.74) & 12.7 (3.60) & 5.58 (1.43) & 5.96 (1.32) & 4.71 (1.20) \\ 
         & $D_2 (20\%)$ & 18.3 (3.23) & 16.1 (3.45) & 6.59 (1.64) & 7.83 (2.54) & 4.98 (1.28) \\ 
         & $D_4$        & 10.3 (3.11) & 6.95 (1.54) & 5.48 (1.63) & 6.06 (1.40) & 5.36 (1.32) \\  \hline 
glass    & $D_1 (20\%)$ & 0.43 (0.46) & 0.33 (0.30) & 0.28 (0.12) & 0.23 (0.13) & 0.22 (0.06) \\ 
         & $D_2 (20\%)$ & 1.16 (0.48) & 1.12 (0.61) & 0.30 (0.15) & 0.22 (0.17) & 0.22 (0.07) \\ 
         & $D_4$        & 0.46 (0.50) & 0.21 (0.16) & 0.28 (0.13) & 0.24 (0.15) & 0.28 (0.14) \\ 
   \hline
\end{tabular}
\end{centering}
\caption{Average and standard deviation of trimmed root-mean-square errors
	(TRMSEs) computed on 50 random training / validation / test data splits with
	added outliers following the settings in Section \ref{set-up}.}
\label{t:real.synthetic}
\end{table}


\section{Conclusion} \label{sec:conclusion}

In this paper we proposed a new robust boosting algorithm (\texttt{RRBoost})
based on best practices for robust regression estimation.  Our approach uses
bounded loss functions, which are known to provide strong protection against
outliers, and do not require that the possible proportion of outliers in the
data be known or estimated beforehand.  Extensive numerical experiments showed
that our \texttt{RRBoost} algorithm can be much more robust than existing
alternatives in the literature, while behaving very similarly to the standard
\texttt{L2Boost} when the data do not contain atypical observations.  In
particular, the second stage of \texttt{RRBoost} significantly improves the
prediction accuracy of the initial \texttt{SBoost} fit without affecting the
overall robustness of the resulting predictor. Using a permutation-based
variable importance ranking we also showed that in our experiments
\texttt{RRBoost} generally achieved the highest variable selection accuracy.

\section{Acknowledgements.}

The authors wish to thank the Associate Editor and two
anonymous referees for valuable comments which led to  
an improved version of the original paper.  This research was partially supported by    
the Natural Sciences and Engineering Research Council of Canada [Discovery Grant RGPIN-2016-04288].

\bibliographystyle{apalike} 
\bibliography{R2_mybibfile}

\end{document}


\title{Appendix for ``Robust Boosting for Regression Problems''}
\author{Xiaomeng Ju and Mat\'{i}as Salibi\'{a}n-Barrera }

\maketitle

\let\thefootnote\relax\footnotetext{Contact: xiaomeng.ju@stat.ubc.ca (Xiaomeng Ju);  Department of Statistics,  2207 Main Mall, University of British Columbia, Vancouver, BC,  V6T1Z4, Canada. 
}

\section*{Abstract}
This supplementary document reports all the results of the numerical
experiments described in ``Robust Boosting for Regression Problems''.

\section{Simulation}
\subsection{Function $g_1$}
\begin{table}[H]
\centering
\begin{tabular}{lcccccc}
  \hline
 & L2Boost & MBoost & LADBoost & Robloss & SBoost & RRBoost \\ 
  \hline
$D_0$ & 1.00 (0.04) & 1.01 (0.04) & 1.08 (0.05) & 1.02 (0.04) & 1.61 (0.12) & 1.01 (0.06) \\ 
  $D_1 (10\%)$ & 2.03 (0.19) & 1.61 (0.32) & 1.18 (0.07) & 1.15 (0.06) & 1.61 (0.12) & 1.05 (0.08) \\ 
  $D_1 (20\%)$ & 2.23 (0.25) & 2.14 (0.18) & 1.29 (0.09) & 1.39 (0.13) & 1.52 (0.14) & 1.09 (0.06) \\ 
  $D_2 (10\%)$ & 2.46 (0.19) & 1.86 (0.45) & 1.18 (0.07) & 1.20 (0.11) & 1.58 (0.13) & 1.05 (0.08) \\ 
  $D_2 (20\%)$ & 3.53 (0.29) & 3.49 (0.31) & 1.45 (0.16) & 1.72 (0.31) & 1.50 (0.12) & 1.09 (0.07) \\ 
  $D_3$ & 1.29 (0.12) & 1.12 (0.06) & 1.15 (0.06) & 1.10 (0.05) & 1.62 (0.12) & 1.12 (0.10) \\ 
  $D_4$ & 15.31 (36.96) & 1.54 (0.14) & 1.33 (0.09) & 1.37 (0.10) & 1.62 (0.12) & 1.31 (0.09) \\ 
   \hline
\end{tabular}
\caption{Summary statistics of RMSEs on the test sets by L2Boost, MBoost, LADBoost, Robloss, SBoost, and RRBoost applied with tree learners of $d$ = 1 for clean ($D_0$), symmetric gross error contaminated ($D_1$), asymmetric gross error contaminated ($D_2$),  skewed distributed ($D_3$),  and heavy-tailed distributed  ($D_4$) data generated from $g$ = $g_1$ S = $S_0$ $n$ = 300 $p$ = 10, displayed in the form of: mean (SD) calculated from 100 independent runs of the experiment.} 
\end{table}
\begin{table}[H]
\centering
\begin{tabular}{lccccc}
  \hline
 & L2Boost & MBoost & LADBoost & Robloss & RRBoost \\ 
  \hline
$D_0$ & 1.00 (0.00) & 1.00 (0.00) & 1.00 (0.00) & 1.00 (0.00) & 1.00 (0.00) \\ 
  $D_1 (10\%)$ & 0.48 (0.27) & 0.77 (0.30) & 1.00 (0.00) & 1.00 (0.02) & 1.00 (0.02) \\ 
  $D_1 (20\%)$ & 0.39 (0.27) & 0.27 (0.23) & 1.00 (0.02) & 0.97 (0.07) & 1.00 (0.00) \\ 
  $D_2 (10\%)$ & 0.50 (0.30) & 0.85 (0.21) & 1.00 (0.02) & 0.99 (0.03) & 1.00 (0.00) \\ 
  $D_2 (20\%)$ & 0.31 (0.22) & 0.45 (0.25) & 0.99 (0.04) & 0.91 (0.16) & 1.00 (0.00) \\ 
  $D_3$ & 0.99 (0.04) & 1.00 (0.02) & 1.00 (0.00) & 1.00 (0.00) & 1.00 (0.00) \\ 
  $D_4$ & 0.36 (0.27) & 0.92 (0.14) & 0.98 (0.06) & 0.97 (0.07) & 0.98 (0.06) \\ 
   \hline
\end{tabular}
\caption{Fractions of variables recovered by L2Boost, MBoost, LADBoost, Robloss, and RRBoost applied with tree learners of $d$ = 1 for clean ($D_0$), symmetric gross error contaminated ($D_1$), asymmetric gross error contaminated ($D_2$),  skewed distributed ($D_3$),  and heavy-tailed distributed  ($D_4$) data generated from $g$ = $g_1$ S = $S_0$ $n$ = 300 $p$ = 10, displayed in the form of: mean (SD) calculated from 100 independent runs of the experiment.} 
\end{table}
\begin{table}[H]
\centering
\begin{tabular}{lcccccc}
  \hline
 & L2Boost & MBoost & LADBoost & Robloss & SBoost & RRBoost \\ 
  \hline
$D_0$ & 1.22 (0.07) & 1.23 (0.06) & 1.23 (0.06) & 1.23 (0.06) & 1.66 (0.11) & 1.20 (0.05) \\ 
  $D_1 (10\%)$ & 2.19 (0.23) & 1.84 (0.31) & 1.35 (0.09) & 1.41 (0.10) & 1.64 (0.13) & 1.24 (0.08) \\ 
  $D_1 (20\%)$ & 2.51 (0.28) & 2.48 (0.26) & 1.51 (0.13) & 1.60 (0.15) & 1.62 (0.14) & 1.31 (0.09) \\ 
  $D_2 (10\%)$ & 2.59 (0.25) & 2.07 (0.44) & 1.38 (0.09) & 1.42 (0.12) & 1.62 (0.12) & 1.23 (0.07) \\ 
  $D_2 (20\%)$ & 3.68 (0.27) & 3.71 (0.32) & 1.68 (0.16) & 1.87 (0.18) & 1.60 (0.14) & 1.31 (0.08) \\ 
  $D_3$ & 1.50 (0.13) & 1.34 (0.08) & 1.32 (0.08) & 1.33 (0.08) & 1.68 (0.11) & 1.29 (0.07) \\ 
  $D_4$ & 15.02 (30.29) & 1.69 (0.13) & 1.51 (0.11) & 1.54 (0.12) & 1.66 (0.14) & 1.44 (0.09) \\ 
   \hline
\end{tabular}
\caption{Summary statistics of RMSEs on the test sets by L2Boost, MBoost, LADBoost, Robloss, SBoost, and RRBoost applied with tree learners of $d$ = 2 for clean ($D_0$), symmetric gross error contaminated ($D_1$), asymmetric gross error contaminated ($D_2$),  skewed distributed ($D_3$),  and heavy-tailed distributed  ($D_4$) data generated from $g$ = $g_1$ S = $S_0$ $n$ = 300 $p$ = 10, displayed in the form of: mean (SD) calculated from 100 independent runs of the experiment.} 
\end{table}
\begin{table}[H]
\centering
\begin{tabular}{lccccc}
  \hline
 & L2Boost & MBoost & LADBoost & Robloss & RRBoost \\ 
  \hline
$D_0$ & 1.00 (0.02) & 1.00 (0.00) & 1.00 (0.02) & 1.00 (0.00) & 1.00 (0.00) \\ 
  $D_1 (10\%)$ & 0.49 (0.21) & 0.73 (0.26) & 0.99 (0.03) & 0.99 (0.04) & 1.00 (0.00) \\ 
  $D_1 (20\%)$ & 0.39 (0.20) & 0.33 (0.19) & 0.95 (0.10) & 0.90 (0.14) & 0.99 (0.04) \\ 
  $D_2 (10\%)$ & 0.47 (0.21) & 0.72 (0.24) & 0.99 (0.04) & 0.97 (0.08) & 1.00 (0.03) \\ 
  $D_2 (20\%)$ & 0.40 (0.19) & 0.48 (0.21) & 0.91 (0.14) & 0.83 (0.19) & 0.99 (0.04) \\ 
  $D_3$ & 0.94 (0.13) & 0.99 (0.04) & 1.00 (0.02) & 0.99 (0.03) & 1.00 (0.03) \\ 
  $D_4$ & 0.38 (0.24) & 0.84 (0.17) & 0.94 (0.10) & 0.92 (0.12) & 0.95 (0.09) \\ 
   \hline
\end{tabular}
\caption{Fractions of variables recovered by L2Boost, MBoost, LADBoost, Robloss, and RRBoost applied with tree learners of $d$ = 2 for clean ($D_0$), symmetric gross error contaminated ($D_1$), asymmetric gross error contaminated ($D_2$),  skewed distributed ($D_3$),  and heavy-tailed distributed  ($D_4$) data generated from $g$ = $g_1$ S = $S_0$ $n$ = 300 $p$ = 10, displayed in the form of: mean (SD) calculated from 100 independent runs of the experiment.} 
\end{table}
\begin{table}[H]
\centering
\begin{tabular}{lcccccc}
  \hline
 & L2Boost & MBoost & LADBoost & Robloss & SBoost & RRBoost \\ 
  \hline
$D_0$ & 1.34 (0.07) & 1.36 (0.09) & 1.35 (0.08) & 1.34 (0.09) & 1.68 (0.11) & 1.31 (0.06) \\ 
  $D_1 (10\%)$ & 2.44 (0.30) & 2.03 (0.37) & 1.50 (0.10) & 1.54 (0.11) & 1.68 (0.13) & 1.36 (0.07) \\ 
  $D_1 (20\%)$ & 2.88 (0.37) & 2.87 (0.34) & 1.66 (0.13) & 1.74 (0.14) & 1.66 (0.12) & 1.41 (0.09) \\ 
  $D_2 (10\%)$ & 2.76 (0.30) & 2.26 (0.50) & 1.53 (0.10) & 1.60 (0.13) & 1.68 (0.12) & 1.35 (0.08) \\ 
  $D_2 (20\%)$ & 3.86 (0.31) & 3.93 (0.32) & 1.79 (0.15) & 1.95 (0.16) & 1.66 (0.16) & 1.42 (0.14) \\ 
  $D_3$ & 1.60 (0.14) & 1.47 (0.09) & 1.43 (0.09) & 1.45 (0.09) & 1.70 (0.11) & 1.39 (0.09) \\ 
  $D_4$ & 15.65 (30.32) & 1.81 (0.13) & 1.63 (0.12) & 1.68 (0.12) & 1.72 (0.12) & 1.54 (0.10) \\ 
   \hline
\end{tabular}
\caption{Summary statistics of RMSEs on the test sets by L2Boost, MBoost, LADBoost, Robloss, SBoost, and RRBoost applied with tree learners of $d$ = 3 for clean ($D_0$), symmetric gross error contaminated ($D_1$), asymmetric gross error contaminated ($D_2$),  skewed distributed ($D_3$),  and heavy-tailed distributed  ($D_4$) data generated from $g$ = $g_1$ S = $S_0$ $n$ = 300 $p$ = 10, displayed in the form of: mean (SD) calculated from 100 independent runs of the experiment.} 
\end{table}
\begin{table}[H]
\centering
\begin{tabular}{lccccc}
  \hline
 & L2Boost & MBoost & LADBoost & Robloss & RRBoost \\ 
  \hline
$D_0$ & 0.99 (0.03) & 0.99 (0.03) & 0.99 (0.03) & 1.00 (0.02) & 0.99 (0.04) \\ 
  $D_1 (10\%)$ & 0.56 (0.17) & 0.72 (0.22) & 0.94 (0.10) & 0.97 (0.09) & 1.00 (0.03) \\ 
  $D_1 (20\%)$ & 0.47 (0.18) & 0.45 (0.18) & 0.90 (0.13) & 0.83 (0.17) & 0.98 (0.06) \\ 
  $D_2 (10\%)$ & 0.56 (0.19) & 0.72 (0.21) & 0.95 (0.09) & 0.95 (0.11) & 0.99 (0.04) \\ 
  $D_2 (20\%)$ & 0.47 (0.19) & 0.52 (0.21) & 0.87 (0.14) & 0.76 (0.18) & 0.98 (0.06) \\ 
  $D_3$ & 0.92 (0.15) & 0.97 (0.07) & 0.97 (0.07) & 0.97 (0.07) & 0.98 (0.06) \\ 
  $D_4$ & 0.48 (0.20) & 0.76 (0.20) & 0.88 (0.13) & 0.85 (0.18) & 0.91 (0.12) \\ 
   \hline
\end{tabular}
\caption{Fractions of variables recovered by L2Boost, MBoost, LADBoost, Robloss, and RRBoost applied with tree learners of $d$ = 3 for clean ($D_0$), symmetric gross error contaminated ($D_1$), asymmetric gross error contaminated ($D_2$),  skewed distributed ($D_3$),  and heavy-tailed distributed  ($D_4$) data generated from $g$ = $g_1$ S = $S_0$ $n$ = 300 $p$ = 10, displayed in the form of: mean (SD) calculated from 100 independent runs of the experiment.} 
\end{table}
\begin{table}[H]
\centering
\begin{tabular}{lcccccc}
  \hline
 & L2Boost & MBoost & LADBoost & Robloss & SBoost & RRBoost \\ 
  \hline
$D_0$ & 0.76 (0.02) & 0.76 (0.02) & 0.77 (0.02) & 0.76 (0.02) & 1.07 (0.13) & 0.75 (0.02) \\ 
  $D_1 (10\%)$ & 1.27 (0.07) & 1.04 (0.22) & 0.79 (0.02) & 0.79 (0.04) & 0.99 (0.12) & 0.77 (0.18) \\ 
  $D_1 (20\%)$ & 1.46 (0.10) & 1.46 (0.09) & 0.81 (0.02) & 0.83 (0.05) & 0.89 (0.08) & 0.76 (0.02) \\ 
  $D_2 (10\%)$ & 1.88 (0.07) & 1.19 (0.32) & 0.80 (0.02) & 0.81 (0.04) & 0.98 (0.11) & 0.77 (0.17) \\ 
  $D_2 (20\%)$ & 3.14 (0.08) & 3.09 (0.10) & 0.90 (0.10) & 1.00 (0.14) & 0.88 (0.06) & 0.76 (0.03) \\ 
  $D_3$ & 0.88 (0.08) & 0.80 (0.02) & 0.86 (0.02) & 0.83 (0.02) & 1.12 (0.09) & 0.81 (0.02) \\ 
  $D_4$ & 10.82 (30.46) & 0.95 (0.03) & 0.82 (0.02) & 0.84 (0.02) & 0.93 (0.05) & 0.79 (0.02) \\ 
   \hline
\end{tabular}
\caption{Summary statistics of RMSEs on the test sets by L2Boost, MBoost, LADBoost, Robloss, SBoost, and RRBoost applied with tree learners of $d$ = 1 for clean ($D_0$), symmetric gross error contaminated ($D_1$), asymmetric gross error contaminated ($D_2$),  skewed distributed ($D_3$),  and heavy-tailed distributed  ($D_4$) data generated from $g$ = $g_1$ S = $S_0$ $n$ = 3000 $p$ = 10, displayed in the form of: mean (SD) calculated from 100 independent runs of the experiment.} 
\end{table}
\begin{table}[H]
\centering
\begin{tabular}{lccccc}
  \hline
 & L2Boost & MBoost & LADBoost & Robloss & RRBoost \\ 
  \hline
$D_0$ & 1.00 (0.00) & 1.00 (0.00) & 1.00 (0.00) & 1.00 (0.00) & 1.00 (0.00) \\ 
  $D_1 (10\%)$ & 1.00 (0.00) & 1.00 (0.02) & 1.00 (0.00) & 1.00 (0.00) & 1.00 (0.00) \\ 
  $D_1 (20\%)$ & 0.99 (0.06) & 1.00 (0.02) & 1.00 (0.00) & 1.00 (0.02) & 1.00 (0.00) \\ 
  $D_2 (10\%)$ & 1.00 (0.00) & 0.99 (0.04) & 1.00 (0.00) & 1.00 (0.00) & 1.00 (0.00) \\ 
  $D_2 (20\%)$ & 0.99 (0.06) & 0.94 (0.16) & 0.99 (0.04) & 0.98 (0.06) & 1.00 (0.00) \\ 
  $D_3$ & 1.00 (0.00) & 1.00 (0.00) & 1.00 (0.00) & 1.00 (0.00) & 1.00 (0.00) \\ 
  $D_4$ & 0.29 (0.31) & 1.00 (0.00) & 1.00 (0.00) & 1.00 (0.00) & 1.00 (0.00) \\ 
   \hline
\end{tabular}
\caption{Fractions of variables recovered by L2Boost, MBoost, LADBoost, Robloss, and RRBoost applied with tree learners of $d$ = 1 for clean ($D_0$), symmetric gross error contaminated ($D_1$), asymmetric gross error contaminated ($D_2$),  skewed distributed ($D_3$),  and heavy-tailed distributed  ($D_4$) data generated from $g$ = $g_1$ S = $S_0$ $n$ = 3000 $p$ = 10, displayed in the form of: mean (SD) calculated from 100 independent runs of the experiment.} 
\end{table}
\begin{table}[H]
\centering
\begin{tabular}{lcccccc}
  \hline
 & L2Boost & MBoost & LADBoost & Robloss & SBoost & RRBoost \\ 
  \hline
$D_0$ & 0.85 (0.02) & 0.85 (0.02) & 0.88 (0.03) & 0.86 (0.03) & 0.96 (0.09) & 0.80 (0.02) \\ 
  $D_1 (10\%)$ & 1.42 (0.09) & 1.15 (0.20) & 0.91 (0.03) & 0.90 (0.03) & 0.90 (0.06) & 0.80 (0.02) \\ 
  $D_1 (20\%)$ & 1.62 (0.09) & 1.63 (0.09) & 0.96 (0.04) & 0.96 (0.04) & 0.86 (0.04) & 0.80 (0.02) \\ 
  $D_2 (10\%)$ & 1.99 (0.08) & 1.35 (0.33) & 0.92 (0.03) & 0.91 (0.03) & 0.90 (0.06) & 0.80 (0.02) \\ 
  $D_2 (20\%)$ & 3.21 (0.09) & 3.14 (0.11) & 1.01 (0.04) & 1.07 (0.05) & 0.84 (0.04) & 0.79 (0.02) \\ 
  $D_3$ & 1.00 (0.05) & 0.90 (0.03) & 0.94 (0.03) & 0.91 (0.03) & 1.09 (0.09) & 0.89 (0.02) \\ 
  $D_4$ & 67.81 (347.07) & 1.08 (0.04) & 0.97 (0.04) & 0.98 (0.04) & 0.87 (0.05) & 0.83 (0.02) \\ 
   \hline
\end{tabular}
\caption{Summary statistics of RMSEs on the test sets by L2Boost, MBoost, LADBoost, Robloss, SBoost, and RRBoost applied with tree learners of $d$ = 2 for clean ($D_0$), symmetric gross error contaminated ($D_1$), asymmetric gross error contaminated ($D_2$),  skewed distributed ($D_3$),  and heavy-tailed distributed  ($D_4$) data generated from $g$ = $g_1$ S = $S_0$ $n$ = 3000 $p$ = 10, displayed in the form of: mean (SD) calculated from 100 independent runs of the experiment.} 
\end{table}
\begin{table}[H]
\centering
\begin{tabular}{lccccc}
  \hline
 & L2Boost & MBoost & LADBoost & Robloss & RRBoost \\ 
  \hline
$D_0$ & 1.00 (0.00) & 1.00 (0.00) & 1.00 (0.00) & 1.00 (0.00) & 1.00 (0.00) \\ 
  $D_1 (10\%)$ & 0.99 (0.03) & 1.00 (0.00) & 1.00 (0.00) & 1.00 (0.00) & 1.00 (0.00) \\ 
  $D_1 (20\%)$ & 0.90 (0.18) & 0.93 (0.12) & 1.00 (0.00) & 1.00 (0.00) & 1.00 (0.00) \\ 
  $D_2 (10\%)$ & 0.98 (0.07) & 1.00 (0.02) & 1.00 (0.00) & 1.00 (0.00) & 1.00 (0.00) \\ 
  $D_2 (20\%)$ & 0.90 (0.19) & 0.90 (0.15) & 1.00 (0.00) & 1.00 (0.00) & 1.00 (0.00) \\ 
  $D_3$ & 1.00 (0.02) & 1.00 (0.00) & 1.00 (0.00) & 1.00 (0.00) & 1.00 (0.00) \\ 
  $D_4$ & 0.27 (0.19) & 1.00 (0.00) & 1.00 (0.00) & 1.00 (0.00) & 1.00 (0.00) \\ 
   \hline
\end{tabular}
\caption{Fractions of variables recovered by L2Boost, MBoost, LADBoost, Robloss, and RRBoost applied with tree learners of $d$ = 2 for clean ($D_0$), symmetric gross error contaminated ($D_1$), asymmetric gross error contaminated ($D_2$),  skewed distributed ($D_3$),  and heavy-tailed distributed  ($D_4$) data generated from $g$ = $g_1$ S = $S_0$ $n$ = 3000 $p$ = 10, displayed in the form of: mean (SD) calculated from 100 independent runs of the experiment.} 
\end{table}
\begin{table}[H]
\centering
\begin{tabular}{lcccccc}
  \hline
 & L2Boost & MBoost & LADBoost & Robloss & SBoost & RRBoost \\ 
  \hline
$D_0$ & 0.90 (0.03) & 0.90 (0.03) & 0.96 (0.04) & 0.93 (0.04) & 1.04 (0.07) & 0.84 (0.02) \\ 
  $D_1 (10\%)$ & 1.56 (0.09) & 1.26 (0.21) & 1.01 (0.04) & 0.98 (0.04) & 0.98 (0.06) & 0.84 (0.02) \\ 
  $D_1 (20\%)$ & 1.77 (0.09) & 1.78 (0.10) & 1.07 (0.05) & 1.07 (0.05) & 0.92 (0.04) & 0.84 (0.02) \\ 
  $D_2 (10\%)$ & 2.09 (0.08) & 1.46 (0.31) & 1.00 (0.04) & 0.99 (0.04) & 0.98 (0.06) & 0.84 (0.02) \\ 
  $D_2 (20\%)$ & 3.28 (0.09) & 3.21 (0.10) & 1.13 (0.05) & 1.19 (0.05) & 0.92 (0.05) & 0.84 (0.04) \\ 
  $D_3$ & 1.10 (0.07) & 0.97 (0.03) & 1.02 (0.03) & 0.98 (0.04) & 1.13 (0.09) & 0.93 (0.03) \\ 
  $D_4$ & 69.59 (350.61) & 1.17 (0.05) & 1.08 (0.04) & 1.07 (0.04) & 0.94 (0.04) & 0.87 (0.03) \\ 
   \hline
\end{tabular}
\caption{Summary statistics of RMSEs on the test sets by L2Boost, MBoost, LADBoost, Robloss, SBoost, and RRBoost applied with tree learners of $d$ = 3 for clean ($D_0$), symmetric gross error contaminated ($D_1$), asymmetric gross error contaminated ($D_2$),  skewed distributed ($D_3$),  and heavy-tailed distributed  ($D_4$) data generated from $g$ = $g_1$ S = $S_0$ $n$ = 3000 $p$ = 10, displayed in the form of: mean (SD) calculated from 100 independent runs of the experiment.} 
\end{table}
\begin{table}[H]
\centering
\begin{tabular}{lccccc}
  \hline
 & L2Boost & MBoost & LADBoost & Robloss & RRBoost \\ 
  \hline
$D_0$ & 1.00 (0.00) & 1.00 (0.00) & 1.00 (0.00) & 1.00 (0.00) & 1.00 (0.00) \\ 
  $D_1 (10\%)$ & 0.96 (0.09) & 1.00 (0.00) & 1.00 (0.00) & 1.00 (0.00) & 1.00 (0.00) \\ 
  $D_1 (20\%)$ & 0.78 (0.18) & 0.82 (0.16) & 1.00 (0.00) & 1.00 (0.00) & 1.00 (0.00) \\ 
  $D_2 (10\%)$ & 0.92 (0.15) & 1.00 (0.00) & 1.00 (0.00) & 1.00 (0.00) & 1.00 (0.00) \\ 
  $D_2 (20\%)$ & 0.84 (0.18) & 0.92 (0.11) & 1.00 (0.00) & 1.00 (0.00) & 1.00 (0.02) \\ 
  $D_3$ & 1.00 (0.00) & 1.00 (0.00) & 1.00 (0.00) & 1.00 (0.00) & 1.00 (0.00) \\ 
  $D_4$ & 0.44 (0.19) & 1.00 (0.00) & 1.00 (0.00) & 1.00 (0.00) & 1.00 (0.00) \\ 
   \hline
\end{tabular}
\caption{Fractions of variables recovered by L2Boost, MBoost, LADBoost, Robloss, and RRBoost applied with tree learners of $d$ = 3 for clean ($D_0$), symmetric gross error contaminated ($D_1$), asymmetric gross error contaminated ($D_2$),  skewed distributed ($D_3$),  and heavy-tailed distributed  ($D_4$) data generated from $g$ = $g_1$ S = $S_0$ $n$ = 3000 $p$ = 10, displayed in the form of: mean (SD) calculated from 100 independent runs of the experiment.} 
\end{table}
\begin{table}[H]
\centering
\begin{tabular}{lcccccc}
  \hline
 & L2Boost & MBoost & LADBoost & Robloss & SBoost & RRBoost \\ 
  \hline
$D_0$ & 1.29 (0.07) & 1.28 (0.06) & 1.44 (0.11) & 1.31 (0.07) & 1.69 (0.13) & 1.28 (0.08) \\ 
  $D_1 (10\%)$ & 2.00 (0.16) & 1.73 (0.23) & 1.59 (0.12) & 1.46 (0.13) & 1.70 (0.12) & 1.31 (0.09) \\ 
  $D_1 (20\%)$ & 2.23 (0.18) & 2.23 (0.13) & 1.73 (0.12) & 1.68 (0.14) & 1.70 (0.14) & 1.41 (0.14) \\ 
  $D_2 (10\%)$ & 2.30 (0.19) & 1.93 (0.33) & 1.57 (0.15) & 1.48 (0.14) & 1.71 (0.11) & 1.32 (0.10) \\ 
  $D_2 (20\%)$ & 3.07 (0.21) & 3.08 (0.24) & 1.80 (0.13) & 1.87 (0.17) & 1.69 (0.14) & 1.41 (0.14) \\ 
  $D_3$ & 1.44 (0.13) & 1.34 (0.08) & 1.49 (0.13) & 1.35 (0.08) & 1.69 (0.13) & 1.35 (0.11) \\ 
  $D_4$ & 10.81 (15.55) & 1.63 (0.15) & 1.67 (0.12) & 1.55 (0.14) & 1.74 (0.13) & 1.52 (0.16) \\ 
   \hline
\end{tabular}
\caption{Summary statistics of RMSEs on the test sets by L2Boost, MBoost, LADBoost, Robloss, SBoost, and RRBoost applied with tree learners of $d$ = 1 for clean ($D_0$), symmetric gross error contaminated ($D_1$), asymmetric gross error contaminated ($D_2$),  skewed distributed ($D_3$),  and heavy-tailed distributed  ($D_4$) data generated from $g$ = $g_1$ S = $S_0$ $n$ = 300 $p$ = 400, displayed in the form of: mean (SD) calculated from 100 independent runs of the experiment.} 
\end{table}
\begin{table}[H]
\centering
\begin{tabular}{lccccc}
  \hline
 & L2Boost & MBoost & LADBoost & Robloss & RRBoost \\ 
  \hline
$D_0$ & 1.00 (0.02) & 1.00 (0.00) & 0.94 (0.13) & 1.00 (0.03) & 0.99 (0.03) \\ 
  $D_1 (10\%)$ & 0.10 (0.13) & 0.44 (0.33) & 0.74 (0.25) & 0.86 (0.19) & 0.99 (0.05) \\ 
  $D_1 (20\%)$ & 0.05 (0.10) & 0.04 (0.09) & 0.49 (0.25) & 0.50 (0.29) & 0.94 (0.11) \\ 
  $D_2 (10\%)$ & 0.09 (0.13) & 0.43 (0.30) & 0.79 (0.24) & 0.90 (0.17) & 0.98 (0.06) \\ 
  $D_2 (20\%)$ & 0.06 (0.09) & 0.06 (0.11) & 0.46 (0.29) & 0.40 (0.24) & 0.94 (0.13) \\ 
  $D_3$ & 0.87 (0.19) & 0.97 (0.08) & 0.90 (0.15) & 0.99 (0.05) & 0.97 (0.08) \\ 
  $D_4$ & 0.01 (0.03) & 0.55 (0.25) & 0.60 (0.24) & 0.73 (0.23) & 0.82 (0.21) \\ 
   \hline
\end{tabular}
\caption{Fractions of variables recovered by L2Boost, MBoost, LADBoost, Robloss, and RRBoost applied with tree learners of $d$ = 1 for clean ($D_0$), symmetric gross error contaminated ($D_1$), asymmetric gross error contaminated ($D_2$),  skewed distributed ($D_3$),  and heavy-tailed distributed  ($D_4$) data generated from $g$ = $g_1$ S = $S_0$ $n$ = 300 $p$ = 400, displayed in the form of: mean (SD) calculated from 100 independent runs of the experiment.} 
\end{table}
\begin{table}[H]
\centering
\begin{tabular}{lcccccc}
  \hline
 & L2Boost & MBoost & LADBoost & Robloss & SBoost & RRBoost \\ 
  \hline
$D_0$ & 1.39 (0.12) & 1.38 (0.12) & 1.59 (0.14) & 1.44 (0.12) & 1.73 (0.11) & 1.40 (0.12) \\ 
  $D_1 (10\%)$ & 2.28 (0.22) & 1.98 (0.27) & 1.74 (0.11) & 1.68 (0.14) & 1.75 (0.12) & 1.46 (0.14) \\ 
  $D_1 (20\%)$ & 2.66 (0.24) & 2.68 (0.20) & 1.84 (0.12) & 1.87 (0.13) & 1.77 (0.13) & 1.57 (0.15) \\ 
  $D_2 (10\%)$ & 2.55 (0.26) & 2.17 (0.39) & 1.75 (0.13) & 1.68 (0.16) & 1.75 (0.11) & 1.46 (0.15) \\ 
  $D_2 (20\%)$ & 3.38 (0.29) & 3.38 (0.27) & 1.92 (0.13) & 2.02 (0.18) & 1.78 (0.13) & 1.59 (0.17) \\ 
  $D_3$ & 1.59 (0.14) & 1.49 (0.11) & 1.65 (0.12) & 1.53 (0.13) & 1.75 (0.11) & 1.48 (0.11) \\ 
  $D_4$ & 9.56 (11.66) & 1.78 (0.16) & 1.79 (0.13) & 1.74 (0.13) & 1.80 (0.12) & 1.68 (0.15) \\ 
   \hline
\end{tabular}
\caption{Summary statistics of RMSEs on the test sets by L2Boost, MBoost, LADBoost, Robloss, SBoost, and RRBoost applied with tree learners of $d$ = 2 for clean ($D_0$), symmetric gross error contaminated ($D_1$), asymmetric gross error contaminated ($D_2$),  skewed distributed ($D_3$),  and heavy-tailed distributed  ($D_4$) data generated from $g$ = $g_1$ S = $S_0$ $n$ = 300 $p$ = 400, displayed in the form of: mean (SD) calculated from 100 independent runs of the experiment.} 
\end{table}
\begin{table}[H]
\centering
\begin{tabular}{lccccc}
  \hline
 & L2Boost & MBoost & LADBoost & Robloss & RRBoost \\ 
  \hline
$D_0$ & 0.96 (0.08) & 0.97 (0.08) & 0.86 (0.17) & 0.95 (0.10) & 0.96 (0.08) \\ 
  $D_1 (10\%)$ & 0.11 (0.12) & 0.31 (0.22) & 0.57 (0.24) & 0.66 (0.22) & 0.92 (0.11) \\ 
  $D_1 (20\%)$ & 0.06 (0.10) & 0.05 (0.10) & 0.38 (0.22) & 0.37 (0.20) & 0.86 (0.18) \\ 
  $D_2 (10\%)$ & 0.11 (0.13) & 0.32 (0.20) & 0.60 (0.26) & 0.66 (0.25) & 0.91 (0.14) \\ 
  $D_2 (20\%)$ & 0.07 (0.11) & 0.07 (0.12) & 0.39 (0.25) & 0.34 (0.18) & 0.79 (0.19) \\ 
  $D_3$ & 0.74 (0.24) & 0.89 (0.15) & 0.75 (0.22) & 0.86 (0.16) & 0.89 (0.14) \\ 
  $D_4$ & 0.03 (0.07) & 0.39 (0.20) & 0.44 (0.22) & 0.51 (0.22) & 0.64 (0.22) \\ 
   \hline
\end{tabular}
\caption{Fractions of variables recovered by L2Boost, MBoost, LADBoost, Robloss, and RRBoost applied with tree learners of $d$ = 2 for clean ($D_0$), symmetric gross error contaminated ($D_1$), asymmetric gross error contaminated ($D_2$),  skewed distributed ($D_3$),  and heavy-tailed distributed  ($D_4$) data generated from $g$ = $g_1$ S = $S_0$ $n$ = 300 $p$ = 400, displayed in the form of: mean (SD) calculated from 100 independent runs of the experiment.} 
\end{table}
\begin{table}[H]
\centering
\begin{tabular}{lcccccc}
  \hline
 & L2Boost & MBoost & LADBoost & Robloss & SBoost & RRBoost \\ 
  \hline
$D_0$ & 1.54 (0.14) & 1.58 (0.13) & 1.71 (0.13) & 1.60 (0.13) & 1.73 (0.11) & 1.53 (0.13) \\ 
  $D_1 (10\%)$ & 2.55 (0.31) & 2.24 (0.41) & 1.84 (0.13) & 1.82 (0.15) & 1.79 (0.11) & 1.61 (0.14) \\ 
  $D_1 (20\%)$ & 3.13 (0.36) & 3.13 (0.35) & 1.93 (0.10) & 2.04 (0.18) & 1.80 (0.11) & 1.66 (0.13) \\ 
  $D_2 (10\%)$ & 2.76 (0.30) & 2.42 (0.49) & 1.86 (0.12) & 1.86 (0.17) & 1.77 (0.12) & 1.60 (0.15) \\ 
  $D_2 (20\%)$ & 3.70 (0.33) & 3.69 (0.31) & 2.02 (0.16) & 2.20 (0.24) & 1.81 (0.17) & 1.72 (0.19) \\ 
  $D_3$ & 1.69 (0.14) & 1.66 (0.15) & 1.76 (0.12) & 1.68 (0.13) & 1.76 (0.12) & 1.61 (0.14) \\ 
  $D_4$ & 10.03 (12.13) & 1.92 (0.22) & 1.90 (0.14) & 1.89 (0.17) & 1.83 (0.12) & 1.76 (0.14) \\ 
   \hline
\end{tabular}
\caption{Summary statistics of RMSEs on the test sets by L2Boost, MBoost, LADBoost, Robloss, SBoost, and RRBoost applied with tree learners of $d$ = 3 for clean ($D_0$), symmetric gross error contaminated ($D_1$), asymmetric gross error contaminated ($D_2$),  skewed distributed ($D_3$),  and heavy-tailed distributed  ($D_4$) data generated from $g$ = $g_1$ S = $S_0$ $n$ = 300 $p$ = 400, displayed in the form of: mean (SD) calculated from 100 independent runs of the experiment.} 
\end{table}
\begin{table}[H]
\centering
\begin{tabular}{lccccc}
  \hline
 & L2Boost & MBoost & LADBoost & Robloss & RRBoost \\ 
  \hline
$D_0$ & 0.85 (0.17) & 0.84 (0.17) & 0.69 (0.22) & 0.82 (0.17) & 0.86 (0.15) \\ 
  $D_1 (10\%)$ & 0.14 (0.13) & 0.28 (0.16) & 0.46 (0.23) & 0.56 (0.21) & 0.79 (0.18) \\ 
  $D_1 (20\%)$ & 0.07 (0.10) & 0.05 (0.11) & 0.32 (0.17) & 0.36 (0.19) & 0.73 (0.18) \\ 
  $D_2 (10\%)$ & 0.14 (0.14) & 0.30 (0.17) & 0.45 (0.22) & 0.55 (0.21) & 0.80 (0.19) \\ 
  $D_2 (20\%)$ & 0.09 (0.13) & 0.08 (0.14) & 0.34 (0.20) & 0.33 (0.17) & 0.67 (0.22) \\ 
  $D_3$ & 0.63 (0.21) & 0.69 (0.20) & 0.63 (0.23) & 0.69 (0.20) & 0.78 (0.20) \\ 
  $D_4$ & 0.05 (0.11) & 0.39 (0.16) & 0.43 (0.22) & 0.44 (0.18) & 0.53 (0.22) \\ 
   \hline
\end{tabular}
\caption{Fractions of variables recovered by L2Boost, MBoost, LADBoost, Robloss, and RRBoost applied with tree learners of $d$ = 3 for clean ($D_0$), symmetric gross error contaminated ($D_1$), asymmetric gross error contaminated ($D_2$),  skewed distributed ($D_3$),  and heavy-tailed distributed  ($D_4$) data generated from $g$ = $g_1$ S = $S_0$ $n$ = 300 $p$ = 400, displayed in the form of: mean (SD) calculated from 100 independent runs of the experiment.} 
\end{table}
\begin{table}[H]
\centering
\begin{tabular}{lcccccc}
  \hline
 & L2Boost & MBoost & LADBoost & Robloss & SBoost & RRBoost \\ 
  \hline
$D_0$ & 0.70 (0.02) & 0.70 (0.02) & 0.75 (0.02) & 0.71 (0.02) & 1.28 (0.16) & 0.67 (0.02) \\ 
  $D_1 (10\%)$ & 1.24 (0.06) & 1.01 (0.18) & 0.78 (0.03) & 0.75 (0.02) & 1.16 (0.17) & 0.70 (0.21) \\ 
  $D_1 (20\%)$ & 1.38 (0.07) & 1.40 (0.07) & 0.83 (0.03) & 0.83 (0.04) & 1.01 (0.18) & 0.68 (0.02) \\ 
  $D_2 (10\%)$ & 1.65 (0.07) & 1.16 (0.26) & 0.79 (0.02) & 0.77 (0.03) & 1.19 (0.18) & 0.67 (0.02) \\ 
  $D_2 (20\%)$ & 2.57 (0.07) & 2.52 (0.08) & 0.90 (0.05) & 0.95 (0.07) & 0.98 (0.15) & 0.71 (0.21) \\ 
  $D_3$ & 0.90 (0.06) & 0.76 (0.02) & 0.79 (0.02) & 0.76 (0.02) & 1.26 (0.14) & 0.71 (0.02) \\ 
  $D_4$ & 29.76 (132.37) & 0.96 (0.04) & 0.86 (0.03) & 0.84 (0.03) & 1.04 (0.13) & 0.76 (0.03) \\ 
   \hline
\end{tabular}
\caption{Summary statistics of RMSEs on the test sets by L2Boost, MBoost, LADBoost, Robloss, SBoost, and RRBoost applied with tree learners of $d$ = 1 for clean ($D_0$), symmetric gross error contaminated ($D_1$), asymmetric gross error contaminated ($D_2$),  skewed distributed ($D_3$),  and heavy-tailed distributed  ($D_4$) data generated from $g$ = $g_1$ S = $S_0$ $n$ = 3000 $p$ = 400, displayed in the form of: mean (SD) calculated from 100 independent runs of the experiment.} 
\end{table}
\begin{table}[H]
\centering
\begin{tabular}{lccccc}
  \hline
 & L2Boost & MBoost & LADBoost & Robloss & RRBoost \\ 
  \hline
$D_0$ & 1.00 (0.00) & 1.00 (0.00) & 1.00 (0.00) & 1.00 (0.00) & 1.00 (0.00) \\ 
  $D_1 (10\%)$ & 1.00 (0.00) & 1.00 (0.00) & 1.00 (0.00) & 1.00 (0.00) & 1.00 (0.00) \\ 
  $D_1 (20\%)$ & 0.93 (0.10) & 0.92 (0.12) & 1.00 (0.00) & 1.00 (0.00) & 1.00 (0.00) \\ 
  $D_2 (10\%)$ & 1.00 (0.03) & 1.00 (0.02) & 1.00 (0.00) & 1.00 (0.00) & 1.00 (0.00) \\ 
  $D_2 (20\%)$ & 0.97 (0.07) & 0.93 (0.12) & 1.00 (0.02) & 1.00 (0.02) & 1.00 (0.00) \\ 
  $D_3$ & 1.00 (0.00) & 1.00 (0.00) & 1.00 (0.00) & 1.00 (0.00) & 1.00 (0.00) \\ 
  $D_4$ & 0.02 (0.07) & 1.00 (0.00) & 1.00 (0.00) & 1.00 (0.00) & 1.00 (0.00) \\ 
   \hline
\end{tabular}
\caption{Fractions of variables recovered by L2Boost, MBoost, LADBoost, Robloss, and RRBoost applied with tree learners of $d$ = 1 for clean ($D_0$), symmetric gross error contaminated ($D_1$), asymmetric gross error contaminated ($D_2$),  skewed distributed ($D_3$),  and heavy-tailed distributed  ($D_4$) data generated from $g$ = $g_1$ S = $S_0$ $n$ = 3000 $p$ = 400, displayed in the form of: mean (SD) calculated from 100 independent runs of the experiment.} 
\end{table}
\begin{table}[H]
\centering
\begin{tabular}{lcccccc}
  \hline
 & L2Boost & MBoost & LADBoost & Robloss & SBoost & RRBoost \\ 
  \hline
$D_0$ & 0.77 (0.03) & 0.78 (0.03) & 0.86 (0.04) & 0.81 (0.04) & 1.19 (0.16) & 0.72 (0.02) \\ 
  $D_1 (10\%)$ & 1.35 (0.10) & 1.09 (0.15) & 0.92 (0.05) & 0.87 (0.05) & 1.09 (0.16) & 0.72 (0.03) \\ 
  $D_1 (20\%)$ & 1.54 (0.11) & 1.58 (0.10) & 0.96 (0.05) & 0.94 (0.06) & 0.95 (0.12) & 0.71 (0.02) \\ 
  $D_2 (10\%)$ & 1.74 (0.10) & 1.23 (0.23) & 0.91 (0.05) & 0.87 (0.05) & 1.06 (0.13) & 0.72 (0.03) \\ 
  $D_2 (20\%)$ & 2.65 (0.08) & 2.58 (0.11) & 0.99 (0.06) & 1.02 (0.05) & 0.93 (0.10) & 0.71 (0.03) \\ 
  $D_3$ & 0.96 (0.07) & 0.82 (0.04) & 0.90 (0.03) & 0.85 (0.04) & 1.18 (0.14) & 0.76 (0.03) \\ 
  $D_4$ & 26.78 (87.01) & 1.01 (0.05) & 0.97 (0.05) & 0.93 (0.05) & 1.01 (0.10) & 0.78 (0.03) \\ 
   \hline
\end{tabular}
\caption{Summary statistics of RMSEs on the test sets by L2Boost, MBoost, LADBoost, Robloss, SBoost, and RRBoost applied with tree learners of $d$ = 2 for clean ($D_0$), symmetric gross error contaminated ($D_1$), asymmetric gross error contaminated ($D_2$),  skewed distributed ($D_3$),  and heavy-tailed distributed  ($D_4$) data generated from $g$ = $g_1$ S = $S_0$ $n$ = 3000 $p$ = 400, displayed in the form of: mean (SD) calculated from 100 independent runs of the experiment.} 
\end{table}
\begin{table}[H]
\centering
\begin{tabular}{lccccc}
  \hline
 & L2Boost & MBoost & LADBoost & Robloss & RRBoost \\ 
  \hline
$D_0$ & 1.00 (0.00) & 1.00 (0.00) & 1.00 (0.00) & 1.00 (0.00) & 1.00 (0.00) \\ 
  $D_1 (10\%)$ & 0.99 (0.05) & 1.00 (0.02) & 1.00 (0.00) & 1.00 (0.00) & 1.00 (0.00) \\ 
  $D_1 (20\%)$ & 0.78 (0.22) & 0.79 (0.20) & 1.00 (0.00) & 1.00 (0.00) & 1.00 (0.00) \\ 
  $D_2 (10\%)$ & 0.96 (0.09) & 1.00 (0.02) & 1.00 (0.00) & 1.00 (0.00) & 1.00 (0.00) \\ 
  $D_2 (20\%)$ & 0.83 (0.15) & 0.82 (0.14) & 1.00 (0.00) & 1.00 (0.00) & 1.00 (0.00) \\ 
  $D_3$ & 1.00 (0.00) & 1.00 (0.00) & 1.00 (0.00) & 1.00 (0.00) & 1.00 (0.00) \\ 
  $D_4$ & 0.02 (0.07) & 1.00 (0.00) & 1.00 (0.00) & 1.00 (0.00) & 1.00 (0.00) \\ 
   \hline
\end{tabular}
\caption{Fractions of variables recovered by L2Boost, MBoost, LADBoost, Robloss, and RRBoost applied with tree learners of $d$ = 2 for clean ($D_0$), symmetric gross error contaminated ($D_1$), asymmetric gross error contaminated ($D_2$),  skewed distributed ($D_3$),  and heavy-tailed distributed  ($D_4$) data generated from $g$ = $g_1$ S = $S_0$ $n$ = 3000 $p$ = 400, displayed in the form of: mean (SD) calculated from 100 independent runs of the experiment.} 
\end{table}
\begin{table}[H]
\centering
\begin{tabular}{lcccccc}
  \hline
 & L2Boost & MBoost & LADBoost & Robloss & SBoost & RRBoost \\ 
  \hline
$D_0$ & 0.83 (0.04) & 0.85 (0.04) & 0.95 (0.05) & 0.89 (0.05) & 1.22 (0.12) & 0.78 (0.03) \\ 
  $D_1 (10\%)$ & 1.54 (0.10) & 1.22 (0.15) & 1.02 (0.05) & 0.97 (0.06) & 1.14 (0.11) & 0.77 (0.03) \\ 
  $D_1 (20\%)$ & 1.74 (0.10) & 1.74 (0.11) & 1.09 (0.06) & 1.08 (0.07) & 1.03 (0.09) & 0.78 (0.03) \\ 
  $D_2 (10\%)$ & 1.91 (0.10) & 1.36 (0.19) & 1.02 (0.05) & 0.95 (0.06) & 1.15 (0.12) & 0.78 (0.02) \\ 
  $D_2 (20\%)$ & 2.76 (0.10) & 2.72 (0.12) & 1.16 (0.07) & 1.20 (0.08) & 1.05 (0.08) & 0.78 (0.05) \\ 
  $D_3$ & 1.04 (0.06) & 0.90 (0.05) & 1.00 (0.05) & 0.93 (0.05) & 1.23 (0.14) & 0.83 (0.03) \\ 
  $D_4$ & 49.12 (208.32) & 1.13 (0.06) & 1.08 (0.05) & 1.06 (0.06) & 1.12 (0.09) & 0.87 (0.04) \\ 
   \hline
\end{tabular}
\caption{Summary statistics of RMSEs on the test sets by L2Boost, MBoost, LADBoost, Robloss, SBoost, and RRBoost applied with tree learners of $d$ = 3 for clean ($D_0$), symmetric gross error contaminated ($D_1$), asymmetric gross error contaminated ($D_2$),  skewed distributed ($D_3$),  and heavy-tailed distributed  ($D_4$) data generated from $g$ = $g_1$ S = $S_0$ $n$ = 3000 $p$ = 400, displayed in the form of: mean (SD) calculated from 100 independent runs of the experiment.} 
\end{table}
\begin{table}[H]
\centering
\begin{tabular}{lccccc}
  \hline
 & L2Boost & MBoost & LADBoost & Robloss & RRBoost \\ 
  \hline
$D_0$ & 1.00 (0.00) & 1.00 (0.00) & 1.00 (0.00) & 1.00 (0.00) & 1.00 (0.00) \\ 
  $D_1 (10\%)$ & 0.81 (0.22) & 1.00 (0.00) & 1.00 (0.00) & 1.00 (0.00) & 1.00 (0.00) \\ 
  $D_1 (20\%)$ & 0.61 (0.22) & 0.60 (0.21) & 1.00 (0.00) & 1.00 (0.00) & 1.00 (0.00) \\ 
  $D_2 (10\%)$ & 0.77 (0.23) & 1.00 (0.00) & 1.00 (0.00) & 1.00 (0.00) & 1.00 (0.00) \\ 
  $D_2 (20\%)$ & 0.62 (0.22) & 0.78 (0.13) & 1.00 (0.00) & 1.00 (0.00) & 1.00 (0.03) \\ 
  $D_3$ & 1.00 (0.00) & 1.00 (0.00) & 1.00 (0.00) & 1.00 (0.00) & 1.00 (0.00) \\ 
  $D_4$ & 0.02 (0.07) & 1.00 (0.00) & 1.00 (0.00) & 1.00 (0.00) & 1.00 (0.00) \\ 
   \hline
\end{tabular}
\caption{Fractions of variables recovered by L2Boost, MBoost, LADBoost, Robloss, and RRBoost applied with tree learners of $d$ = 3 for clean ($D_0$), symmetric gross error contaminated ($D_1$), asymmetric gross error contaminated ($D_2$),  skewed distributed ($D_3$),  and heavy-tailed distributed  ($D_4$) data generated from $g$ = $g_1$ S = $S_0$ $n$ = 3000 $p$ = 400, displayed in the form of: mean (SD) calculated from 100 independent runs of the experiment.} 
\end{table}
\begin{table}[H]
\centering
\begin{tabular}{lcccccc}
  \hline
 & L2Boost & MBoost & LADBoost & Robloss & SBoost & RRBoost \\ 
  \hline
$D_0$ & 0.91 (0.04) & 0.92 (0.04) & 1.00 (0.05) & 0.93 (0.04) & 1.53 (0.17) & 0.91 (0.04) \\ 
  $D_1 (10\%)$ & 1.74 (0.19) & 1.42 (0.29) & 1.07 (0.07) & 1.03 (0.06) & 1.47 (0.17) & 0.97 (0.16) \\ 
  $D_1 (20\%)$ & 1.97 (0.31) & 1.87 (0.24) & 1.19 (0.11) & 1.25 (0.14) & 1.43 (0.17) & 0.98 (0.05) \\ 
  $D_2 (10\%)$ & 2.17 (0.20) & 1.63 (0.40) & 1.09 (0.08) & 1.10 (0.17) & 1.50 (0.18) & 0.94 (0.06) \\ 
  $D_2 (20\%)$ & 3.21 (0.23) & 3.20 (0.27) & 1.36 (0.25) & 1.54 (0.28) & 1.46 (0.17) & 0.99 (0.08) \\ 
  $D_3$ & 1.19 (0.13) & 1.02 (0.05) & 1.06 (0.06) & 1.00 (0.05) & 1.51 (0.15) & 1.01 (0.06) \\ 
  $D_4$ & 29.38 (183.94) & 1.37 (0.12) & 1.21 (0.08) & 1.22 (0.10) & 1.44 (0.17) & 1.17 (0.11) \\ 
   \hline
\end{tabular}
\caption{Summary statistics of RMSEs on the test sets by L2Boost, MBoost, LADBoost, Robloss, SBoost, and RRBoost applied with tree learners of $d$ = 1 for clean ($D_0$), symmetric gross error contaminated ($D_1$), asymmetric gross error contaminated ($D_2$),  skewed distributed ($D_3$),  and heavy-tailed distributed  ($D_4$) data generated from $g$ = $g_1$ S = $S_1$ $n$ = 300 $p$ = 10, displayed in the form of: mean (SD) calculated from 100 independent runs of the experiment.} 
\end{table}
\begin{table}[H]
\centering
\begin{tabular}{lccccc}
  \hline
 & L2Boost & MBoost & LADBoost & Robloss & RRBoost \\ 
  \hline
$D_0$ & 1.00 (0.02) & 1.00 (0.03) & 0.97 (0.08) & 1.00 (0.03) & 1.00 (0.02) \\ 
  $D_1 (10\%)$ & 0.40 (0.25) & 0.64 (0.33) & 0.91 (0.11) & 0.95 (0.09) & 0.99 (0.05) \\ 
  $D_1 (20\%)$ & 0.32 (0.23) & 0.21 (0.15) & 0.86 (0.14) & 0.84 (0.16) & 0.98 (0.06) \\ 
  $D_2 (10\%)$ & 0.41 (0.23) & 0.70 (0.25) & 0.91 (0.13) & 0.95 (0.10) & 0.98 (0.06) \\ 
  $D_2 (20\%)$ & 0.31 (0.19) & 0.39 (0.18) & 0.82 (0.15) & 0.75 (0.20) & 0.97 (0.07) \\ 
  $D_3$ & 0.89 (0.15) & 0.96 (0.08) & 0.96 (0.09) & 0.98 (0.06) & 0.99 (0.05) \\ 
  $D_4$ & 0.32 (0.28) & 0.73 (0.22) & 0.84 (0.15) & 0.84 (0.14) & 0.88 (0.13) \\ 
   \hline
\end{tabular}
\caption{Fractions of variables recovered by L2Boost, MBoost, LADBoost, Robloss, and RRBoost applied with tree learners of $d$ = 1 for clean ($D_0$), symmetric gross error contaminated ($D_1$), asymmetric gross error contaminated ($D_2$),  skewed distributed ($D_3$),  and heavy-tailed distributed  ($D_4$) data generated from $g$ = $g_1$ S = $S_1$ $n$ = 300 $p$ = 10, displayed in the form of: mean (SD) calculated from 100 independent runs of the experiment.} 
\end{table}
\begin{table}[H]
\centering
\begin{tabular}{lcccccc}
  \hline
 & L2Boost & MBoost & LADBoost & Robloss & SBoost & RRBoost \\ 
  \hline
$D_0$ & 1.04 (0.06) & 1.05 (0.06) & 1.04 (0.06) & 1.05 (0.06) & 1.46 (0.14) & 1.05 (0.07) \\ 
  $D_1 (10\%)$ & 1.90 (0.24) & 1.56 (0.30) & 1.14 (0.07) & 1.18 (0.09) & 1.44 (0.16) & 1.06 (0.06) \\ 
  $D_1 (20\%)$ & 2.19 (0.30) & 2.19 (0.30) & 1.31 (0.12) & 1.39 (0.17) & 1.40 (0.17) & 1.10 (0.07) \\ 
  $D_2 (10\%)$ & 2.29 (0.25) & 1.79 (0.43) & 1.16 (0.08) & 1.20 (0.10) & 1.43 (0.17) & 1.07 (0.07) \\ 
  $D_2 (20\%)$ & 3.35 (0.28) & 3.44 (0.35) & 1.43 (0.18) & 1.61 (0.26) & 1.43 (0.16) & 1.12 (0.08) \\ 
  $D_3$ & 1.28 (0.15) & 1.16 (0.08) & 1.11 (0.07) & 1.13 (0.07) & 1.51 (0.17) & 1.12 (0.07) \\ 
  $D_4$ & 35.17 (184.99) & 1.46 (0.14) & 1.30 (0.11) & 1.33 (0.11) & 1.47 (0.18) & 1.25 (0.11) \\ 
   \hline
\end{tabular}
\caption{Summary statistics of RMSEs on the test sets by L2Boost, MBoost, LADBoost, Robloss, SBoost, and RRBoost applied with tree learners of $d$ = 2 for clean ($D_0$), symmetric gross error contaminated ($D_1$), asymmetric gross error contaminated ($D_2$),  skewed distributed ($D_3$),  and heavy-tailed distributed  ($D_4$) data generated from $g$ = $g_1$ S = $S_1$ $n$ = 300 $p$ = 10, displayed in the form of: mean (SD) calculated from 100 independent runs of the experiment.} 
\end{table}
\begin{table}[H]
\centering
\begin{tabular}{lccccc}
  \hline
 & L2Boost & MBoost & LADBoost & Robloss & RRBoost \\ 
  \hline
$D_0$ & 0.94 (0.10) & 0.95 (0.09) & 0.93 (0.11) & 0.95 (0.10) & 0.95 (0.09) \\ 
  $D_1 (10\%)$ & 0.39 (0.22) & 0.57 (0.26) & 0.88 (0.12) & 0.82 (0.18) & 0.94 (0.11) \\ 
  $D_1 (20\%)$ & 0.33 (0.20) & 0.28 (0.17) & 0.79 (0.15) & 0.71 (0.19) & 0.92 (0.12) \\ 
  $D_2 (10\%)$ & 0.41 (0.19) & 0.58 (0.20) & 0.87 (0.14) & 0.83 (0.18) & 0.95 (0.09) \\ 
  $D_2 (20\%)$ & 0.34 (0.20) & 0.44 (0.22) & 0.75 (0.17) & 0.63 (0.18) & 0.90 (0.14) \\ 
  $D_3$ & 0.69 (0.20) & 0.85 (0.16) & 0.91 (0.13) & 0.89 (0.13) & 0.94 (0.10) \\ 
  $D_4$ & 0.30 (0.19) & 0.55 (0.23) & 0.77 (0.19) & 0.72 (0.17) & 0.78 (0.17) \\ 
   \hline
\end{tabular}
\caption{Fractions of variables recovered by L2Boost, MBoost, LADBoost, Robloss, and RRBoost applied with tree learners of $d$ = 2 for clean ($D_0$), symmetric gross error contaminated ($D_1$), asymmetric gross error contaminated ($D_2$),  skewed distributed ($D_3$),  and heavy-tailed distributed  ($D_4$) data generated from $g$ = $g_1$ S = $S_1$ $n$ = 300 $p$ = 10, displayed in the form of: mean (SD) calculated from 100 independent runs of the experiment.} 
\end{table}
\begin{table}[H]
\centering
\begin{tabular}{lcccccc}
  \hline
 & L2Boost & MBoost & LADBoost & Robloss & SBoost & RRBoost \\ 
  \hline
$D_0$ & 1.11 (0.07) & 1.14 (0.06) & 1.12 (0.06) & 1.13 (0.07) & 1.51 (0.16) & 1.10 (0.08) \\ 
  $D_1 (10\%)$ & 2.04 (0.28) & 1.69 (0.37) & 1.20 (0.09) & 1.26 (0.10) & 1.43 (0.15) & 1.12 (0.06) \\ 
  $D_1 (20\%)$ & 2.55 (0.40) & 2.53 (0.38) & 1.37 (0.13) & 1.46 (0.16) & 1.42 (0.16) & 1.16 (0.07) \\ 
  $D_2 (10\%)$ & 2.41 (0.29) & 1.89 (0.48) & 1.22 (0.08) & 1.29 (0.10) & 1.44 (0.14) & 1.12 (0.07) \\ 
  $D_2 (20\%)$ & 3.53 (0.35) & 3.60 (0.34) & 1.47 (0.17) & 1.65 (0.20) & 1.45 (0.17) & 1.17 (0.07) \\ 
  $D_3$ & 1.35 (0.14) & 1.24 (0.09) & 1.18 (0.07) & 1.21 (0.08) & 1.49 (0.16) & 1.18 (0.09) \\ 
  $D_4$ & 35.22 (184.62) & 1.49 (0.14) & 1.36 (0.11) & 1.39 (0.13) & 1.45 (0.15) & 1.30 (0.10) \\ 
   \hline
\end{tabular}
\caption{Summary statistics of RMSEs on the test sets by L2Boost, MBoost, LADBoost, Robloss, SBoost, and RRBoost applied with tree learners of $d$ = 3 for clean ($D_0$), symmetric gross error contaminated ($D_1$), asymmetric gross error contaminated ($D_2$),  skewed distributed ($D_3$),  and heavy-tailed distributed  ($D_4$) data generated from $g$ = $g_1$ S = $S_1$ $n$ = 300 $p$ = 10, displayed in the form of: mean (SD) calculated from 100 independent runs of the experiment.} 
\end{table}
\begin{table}[H]
\centering
\begin{tabular}{lccccc}
  \hline
 & L2Boost & MBoost & LADBoost & Robloss & RRBoost \\ 
  \hline
$D_0$ & 0.90 (0.13) & 0.90 (0.13) & 0.92 (0.11) & 0.90 (0.13) & 0.93 (0.11) \\ 
  $D_1 (10\%)$ & 0.47 (0.20) & 0.53 (0.22) & 0.85 (0.14) & 0.84 (0.16) & 0.88 (0.14) \\ 
  $D_1 (20\%)$ & 0.43 (0.18) & 0.40 (0.18) & 0.77 (0.14) & 0.66 (0.18) & 0.86 (0.14) \\ 
  $D_2 (10\%)$ & 0.46 (0.18) & 0.54 (0.21) & 0.83 (0.15) & 0.80 (0.16) & 0.89 (0.12) \\ 
  $D_2 (20\%)$ & 0.40 (0.19) & 0.45 (0.18) & 0.72 (0.16) & 0.61 (0.18) & 0.87 (0.13) \\ 
  $D_3$ & 0.69 (0.22) & 0.83 (0.16) & 0.85 (0.13) & 0.84 (0.15) & 0.87 (0.12) \\ 
  $D_4$ & 0.47 (0.20) & 0.51 (0.21) & 0.72 (0.16) & 0.66 (0.23) & 0.79 (0.15) \\ 
   \hline
\end{tabular}
\caption{Fractions of variables recovered by L2Boost, MBoost, LADBoost, Robloss, and RRBoost applied with tree learners of $d$ = 3 for clean ($D_0$), symmetric gross error contaminated ($D_1$), asymmetric gross error contaminated ($D_2$),  skewed distributed ($D_3$),  and heavy-tailed distributed  ($D_4$) data generated from $g$ = $g_1$ S = $S_1$ $n$ = 300 $p$ = 10, displayed in the form of: mean (SD) calculated from 100 independent runs of the experiment.} 
\end{table}
\begin{table}[H]
\centering
\begin{tabular}{lcccccc}
  \hline
 & L2Boost & MBoost & LADBoost & Robloss & SBoost & RRBoost \\ 
  \hline
$D_0$ & 0.71 (0.02) & 0.71 (0.02) & 0.73 (0.02) & 0.71 (0.02) & 1.14 (0.17) & 0.70 (0.02) \\ 
  $D_1 (10\%)$ & 1.16 (0.06) & 0.97 (0.21) & 0.74 (0.02) & 0.73 (0.02) & 1.08 (0.22) & 0.73 (0.22) \\ 
  $D_1 (20\%)$ & 1.32 (0.08) & 1.32 (0.07) & 0.76 (0.02) & 0.77 (0.05) & 0.98 (0.19) & 0.72 (0.18) \\ 
  $D_2 (10\%)$ & 1.74 (0.06) & 1.13 (0.28) & 0.75 (0.02) & 0.75 (0.04) & 1.06 (0.20) & 0.74 (0.26) \\ 
  $D_2 (20\%)$ & 2.93 (0.08) & 2.88 (0.09) & 0.85 (0.09) & 0.89 (0.10) & 0.95 (0.19) & 0.70 (0.02) \\ 
  $D_3$ & 0.80 (0.04) & 0.75 (0.02) & 0.81 (0.02) & 0.78 (0.02) & 1.18 (0.21) & 0.76 (0.02) \\ 
  $D_4$ & 9.27 (52.18) & 0.87 (0.03) & 0.77 (0.02) & 0.78 (0.02) & 0.95 (0.13) & 0.75 (0.02) \\ 
   \hline
\end{tabular}
\caption{Summary statistics of RMSEs on the test sets by L2Boost, MBoost, LADBoost, Robloss, SBoost, and RRBoost applied with tree learners of $d$ = 1 for clean ($D_0$), symmetric gross error contaminated ($D_1$), asymmetric gross error contaminated ($D_2$),  skewed distributed ($D_3$),  and heavy-tailed distributed  ($D_4$) data generated from $g$ = $g_1$ S = $S_1$ $n$ = 3000 $p$ = 10, displayed in the form of: mean (SD) calculated from 100 independent runs of the experiment.} 
\end{table}
\begin{table}[H]
\centering
\begin{tabular}{lccccc}
  \hline
 & L2Boost & MBoost & LADBoost & Robloss & RRBoost \\ 
  \hline
$D_0$ & 1.00 (0.00) & 1.00 (0.00) & 1.00 (0.00) & 1.00 (0.00) & 1.00 (0.00) \\ 
  $D_1 (10\%)$ & 0.95 (0.11) & 1.00 (0.02) & 1.00 (0.00) & 1.00 (0.00) & 1.00 (0.00) \\ 
  $D_1 (20\%)$ & 0.74 (0.21) & 0.75 (0.20) & 1.00 (0.00) & 1.00 (0.00) & 1.00 (0.00) \\ 
  $D_2 (10\%)$ & 0.95 (0.11) & 0.99 (0.04) & 1.00 (0.00) & 1.00 (0.02) & 1.00 (0.00) \\ 
  $D_2 (20\%)$ & 0.79 (0.19) & 0.67 (0.25) & 0.99 (0.04) & 0.99 (0.04) & 1.00 (0.00) \\ 
  $D_3$ & 1.00 (0.00) & 1.00 (0.00) & 1.00 (0.00) & 1.00 (0.00) & 1.00 (0.00) \\ 
  $D_4$ & 0.26 (0.24) & 1.00 (0.00) & 1.00 (0.00) & 1.00 (0.00) & 1.00 (0.00) \\ 
   \hline
\end{tabular}
\caption{Fractions of variables recovered by L2Boost, MBoost, LADBoost, Robloss, and RRBoost applied with tree learners of $d$ = 1 for clean ($D_0$), symmetric gross error contaminated ($D_1$), asymmetric gross error contaminated ($D_2$),  skewed distributed ($D_3$),  and heavy-tailed distributed  ($D_4$) data generated from $g$ = $g_1$ S = $S_1$ $n$ = 3000 $p$ = 10, displayed in the form of: mean (SD) calculated from 100 independent runs of the experiment.} 
\end{table}
\begin{table}[H]
\centering
\begin{tabular}{lcccccc}
  \hline
 & L2Boost & MBoost & LADBoost & Robloss & SBoost & RRBoost \\ 
  \hline
$D_0$ & 0.77 (0.02) & 0.77 (0.02) & 0.79 (0.02) & 0.78 (0.02) & 0.94 (0.15) & 0.98 (0.85) \\ 
  $D_1 (10\%)$ & 1.22 (0.07) & 1.05 (0.17) & 0.82 (0.02) & 0.81 (0.02) & 0.93 (0.18) & 0.98 (0.69) \\ 
  $D_1 (20\%)$ & 1.39 (0.12) & 1.39 (0.10) & 0.85 (0.03) & 0.86 (0.03) & 0.97 (0.22) & 1.20 (1.00) \\ 
  $D_2 (10\%)$ & 1.79 (0.06) & 1.27 (0.31) & 0.83 (0.02) & 0.82 (0.03) & 0.95 (0.21) & 1.14 (0.96) \\ 
  $D_2 (20\%)$ & 2.96 (0.10) & 2.89 (0.11) & 0.90 (0.03) & 0.96 (0.05) & 0.90 (0.19) & 1.00 (0.73) \\ 
  $D_3$ & 0.91 (0.04) & 0.81 (0.02) & 0.85 (0.02) & 0.83 (0.02) & 1.15 (0.23) & 1.07 (0.71) \\ 
  $D_4$ & 38.89 (177.81) & 0.97 (0.03) & 0.86 (0.03) & 0.88 (0.03) & 0.93 (0.18) & 0.80 (0.21) \\ 
   \hline
\end{tabular}
\caption{Summary statistics of RMSEs on the test sets by L2Boost, MBoost, LADBoost, Robloss, SBoost, and RRBoost applied with tree learners of $d$ = 2 for clean ($D_0$), symmetric gross error contaminated ($D_1$), asymmetric gross error contaminated ($D_2$),  skewed distributed ($D_3$),  and heavy-tailed distributed  ($D_4$) data generated from $g$ = $g_1$ S = $S_1$ $n$ = 3000 $p$ = 10, displayed in the form of: mean (SD) calculated from 100 independent runs of the experiment.} 
\end{table}
\begin{table}[H]
\centering
\begin{tabular}{lccccc}
  \hline
 & L2Boost & MBoost & LADBoost & Robloss & RRBoost \\ 
  \hline
$D_0$ & 1.00 (0.00) & 1.00 (0.00) & 1.00 (0.00) & 1.00 (0.00) & 0.99 (0.05) \\ 
  $D_1 (10\%)$ & 0.69 (0.21) & 0.88 (0.18) & 1.00 (0.00) & 1.00 (0.00) & 0.99 (0.04) \\ 
  $D_1 (20\%)$ & 0.56 (0.19) & 0.60 (0.18) & 1.00 (0.00) & 1.00 (0.02) & 1.00 (0.03) \\ 
  $D_2 (10\%)$ & 0.60 (0.20) & 0.91 (0.14) & 1.00 (0.00) & 1.00 (0.00) & 0.99 (0.04) \\ 
  $D_2 (20\%)$ & 0.50 (0.17) & 0.52 (0.19) & 1.00 (0.00) & 1.00 (0.00) & 1.00 (0.02) \\ 
  $D_3$ & 0.97 (0.09) & 1.00 (0.00) & 1.00 (0.00) & 1.00 (0.00) & 0.99 (0.04) \\ 
  $D_4$ & 0.27 (0.17) & 0.98 (0.07) & 1.00 (0.00) & 1.00 (0.02) & 1.00 (0.02) \\ 
   \hline
\end{tabular}
\caption{Fractions of variables recovered by L2Boost, MBoost, LADBoost, Robloss, and RRBoost applied with tree learners of $d$ = 2 for clean ($D_0$), symmetric gross error contaminated ($D_1$), asymmetric gross error contaminated ($D_2$),  skewed distributed ($D_3$),  and heavy-tailed distributed  ($D_4$) data generated from $g$ = $g_1$ S = $S_1$ $n$ = 3000 $p$ = 10, displayed in the form of: mean (SD) calculated from 100 independent runs of the experiment.} 
\end{table}
\begin{table}[H]
\centering
\begin{tabular}{lcccccc}
  \hline
 & L2Boost & MBoost & LADBoost & Robloss & SBoost & RRBoost \\ 
  \hline
$D_0$ & 0.81 (0.02) & 0.81 (0.02) & 0.83 (0.02) & 0.81 (0.03) & 0.96 (0.12) & 0.80 (0.07) \\ 
  $D_1 (10\%)$ & 1.31 (0.07) & 1.12 (0.17) & 0.87 (0.03) & 0.86 (0.03) & 0.89 (0.08) & 0.79 (0.12) \\ 
  $D_1 (20\%)$ & 1.48 (0.11) & 1.48 (0.10) & 0.92 (0.04) & 0.93 (0.04) & 0.84 (0.07) & 0.77 (0.03) \\ 
  $D_2 (10\%)$ & 1.86 (0.07) & 1.33 (0.31) & 0.88 (0.03) & 0.87 (0.02) & 0.91 (0.10) & 0.79 (0.06) \\ 
  $D_2 (20\%)$ & 3.00 (0.09) & 2.92 (0.13) & 0.97 (0.04) & 1.03 (0.04) & 0.86 (0.07) & 0.77 (0.03) \\ 
  $D_3$ & 0.96 (0.06) & 0.86 (0.03) & 0.89 (0.03) & 0.87 (0.02) & 1.09 (0.16) & 0.94 (0.13) \\ 
  $D_4$ & 38.29 (153.70) & 1.02 (0.04) & 0.93 (0.03) & 0.94 (0.04) & 0.89 (0.10) & 0.81 (0.04) \\ 
   \hline
\end{tabular}
\caption{Summary statistics of RMSEs on the test sets by L2Boost, MBoost, LADBoost, Robloss, SBoost, and RRBoost applied with tree learners of $d$ = 3 for clean ($D_0$), symmetric gross error contaminated ($D_1$), asymmetric gross error contaminated ($D_2$),  skewed distributed ($D_3$),  and heavy-tailed distributed  ($D_4$) data generated from $g$ = $g_1$ S = $S_1$ $n$ = 3000 $p$ = 10, displayed in the form of: mean (SD) calculated from 100 independent runs of the experiment.} 
\end{table}
\begin{table}[H]
\centering
\begin{tabular}{lccccc}
  \hline
 & L2Boost & MBoost & LADBoost & Robloss & RRBoost \\ 
  \hline
$D_0$ & 1.00 (0.00) & 1.00 (0.00) & 1.00 (0.00) & 1.00 (0.00) & 1.00 (0.00) \\ 
  $D_1 (10\%)$ & 0.66 (0.21) & 0.86 (0.18) & 1.00 (0.00) & 1.00 (0.00) & 1.00 (0.03) \\ 
  $D_1 (20\%)$ & 0.61 (0.17) & 0.58 (0.17) & 1.00 (0.02) & 1.00 (0.00) & 1.00 (0.00) \\ 
  $D_2 (10\%)$ & 0.57 (0.19) & 0.90 (0.15) & 1.00 (0.00) & 1.00 (0.00) & 1.00 (0.00) \\ 
  $D_2 (20\%)$ & 0.52 (0.17) & 0.59 (0.19) & 1.00 (0.02) & 1.00 (0.03) & 1.00 (0.00) \\ 
  $D_3$ & 0.96 (0.09) & 1.00 (0.00) & 1.00 (0.00) & 1.00 (0.00) & 1.00 (0.00) \\ 
  $D_4$ & 0.45 (0.18) & 0.96 (0.09) & 1.00 (0.02) & 0.99 (0.03) & 1.00 (0.00) \\ 
   \hline
\end{tabular}
\caption{Fractions of variables recovered by L2Boost, MBoost, LADBoost, Robloss, and RRBoost applied with tree learners of $d$ = 3 for clean ($D_0$), symmetric gross error contaminated ($D_1$), asymmetric gross error contaminated ($D_2$),  skewed distributed ($D_3$),  and heavy-tailed distributed  ($D_4$) data generated from $g$ = $g_1$ S = $S_1$ $n$ = 3000 $p$ = 10, displayed in the form of: mean (SD) calculated from 100 independent runs of the experiment.} 
\end{table}
\begin{table}[H]
\centering
\begin{tabular}{lcccccc}
  \hline
 & L2Boost & MBoost & LADBoost & Robloss & SBoost & RRBoost \\ 
  \hline
$D_0$ & 1.21 (0.08) & 1.22 (0.09) & 1.37 (0.11) & 1.25 (0.08) & 1.39 (0.19) & 1.14 (0.08) \\ 
  $D_1 (10\%)$ & 1.65 (0.22) & 1.49 (0.16) & 1.47 (0.13) & 1.35 (0.13) & 1.44 (0.16) & 1.18 (0.10) \\ 
  $D_1 (20\%)$ & 1.90 (0.27) & 1.91 (0.25) & 1.57 (0.11) & 1.48 (0.15) & 1.52 (0.17) & 1.27 (0.14) \\ 
  $D_2 (10\%)$ & 2.00 (0.21) & 1.69 (0.26) & 1.46 (0.12) & 1.37 (0.10) & 1.49 (0.17) & 1.18 (0.10) \\ 
  $D_2 (20\%)$ & 2.74 (0.21) & 2.82 (0.22) & 1.58 (0.12) & 1.65 (0.22) & 1.55 (0.14) & 1.25 (0.11) \\ 
  $D_3$ & 1.33 (0.13) & 1.29 (0.10) & 1.39 (0.11) & 1.29 (0.11) & 1.43 (0.19) & 1.20 (0.09) \\ 
  $D_4$ & 15.13 (39.95) & 1.42 (0.12) & 1.52 (0.13) & 1.43 (0.12) & 1.50 (0.16) & 1.34 (0.11) \\ 
   \hline
\end{tabular}
\caption{Summary statistics of RMSEs on the test sets by L2Boost, MBoost, LADBoost, Robloss, SBoost, and RRBoost applied with tree learners of $d$ = 1 for clean ($D_0$), symmetric gross error contaminated ($D_1$), asymmetric gross error contaminated ($D_2$),  skewed distributed ($D_3$),  and heavy-tailed distributed  ($D_4$) data generated from $g$ = $g_1$ S = $S_1$ $n$ = 300 $p$ = 400, displayed in the form of: mean (SD) calculated from 100 independent runs of the experiment.} 
\end{table}
\begin{table}[H]
\centering
\begin{tabular}{lccccc}
  \hline
 & L2Boost & MBoost & LADBoost & Robloss & RRBoost \\ 
  \hline
$D_0$ & 0.56 (0.15) & 0.55 (0.16) & 0.51 (0.15) & 0.52 (0.14) & 0.61 (0.18) \\ 
  $D_1 (10\%)$ & 0.16 (0.08) & 0.27 (0.15) & 0.42 (0.16) & 0.43 (0.14) & 0.56 (0.15) \\ 
  $D_1 (20\%)$ & 0.10 (0.10) & 0.09 (0.10) & 0.31 (0.16) & 0.31 (0.16) & 0.52 (0.14) \\ 
  $D_2 (10\%)$ & 0.15 (0.11) & 0.27 (0.14) & 0.44 (0.15) & 0.44 (0.13) & 0.55 (0.16) \\ 
  $D_2 (20\%)$ & 0.11 (0.10) & 0.11 (0.11) & 0.35 (0.15) & 0.29 (0.13) & 0.52 (0.15) \\ 
  $D_3$ & 0.42 (0.15) & 0.49 (0.15) & 0.47 (0.13) & 0.48 (0.15) & 0.55 (0.15) \\ 
  $D_4$ & 0.03 (0.07) & 0.32 (0.14) & 0.34 (0.16) & 0.38 (0.13) & 0.46 (0.11) \\ 
   \hline
\end{tabular}
\caption{Fractions of variables recovered by L2Boost, MBoost, LADBoost, Robloss, and RRBoost applied with tree learners of $d$ = 1 for clean ($D_0$), symmetric gross error contaminated ($D_1$), asymmetric gross error contaminated ($D_2$),  skewed distributed ($D_3$),  and heavy-tailed distributed  ($D_4$) data generated from $g$ = $g_1$ S = $S_1$ $n$ = 300 $p$ = 400, displayed in the form of: mean (SD) calculated from 100 independent runs of the experiment.} 
\end{table}
\begin{table}[H]
\centering
\begin{tabular}{lcccccc}
  \hline
 & L2Boost & MBoost & LADBoost & Robloss & SBoost & RRBoost \\ 
  \hline
$D_0$ & 1.16 (0.11) & 1.17 (0.10) & 1.37 (0.15) & 1.24 (0.13) & 1.46 (0.16) & 1.14 (0.08) \\ 
  $D_1 (10\%)$ & 1.89 (0.26) & 1.65 (0.21) & 1.51 (0.17) & 1.40 (0.17) & 1.50 (0.16) & 1.19 (0.11) \\ 
  $D_1 (20\%)$ & 2.28 (0.31) & 2.32 (0.30) & 1.64 (0.13) & 1.61 (0.17) & 1.54 (0.16) & 1.28 (0.17) \\ 
  $D_2 (10\%)$ & 2.19 (0.25) & 1.87 (0.31) & 1.49 (0.16) & 1.42 (0.17) & 1.50 (0.16) & 1.18 (0.13) \\ 
  $D_2 (20\%)$ & 3.01 (0.27) & 3.09 (0.27) & 1.65 (0.16) & 1.75 (0.22) & 1.54 (0.14) & 1.26 (0.15) \\ 
  $D_3$ & 1.37 (0.16) & 1.25 (0.14) & 1.41 (0.15) & 1.28 (0.14) & 1.47 (0.16) & 1.20 (0.11) \\ 
  $D_4$ & 13.66 (35.05) & 1.52 (0.17) & 1.59 (0.14) & 1.49 (0.16) & 1.53 (0.15) & 1.37 (0.13) \\ 
   \hline
\end{tabular}
\caption{Summary statistics of RMSEs on the test sets by L2Boost, MBoost, LADBoost, Robloss, SBoost, and RRBoost applied with tree learners of $d$ = 2 for clean ($D_0$), symmetric gross error contaminated ($D_1$), asymmetric gross error contaminated ($D_2$),  skewed distributed ($D_3$),  and heavy-tailed distributed  ($D_4$) data generated from $g$ = $g_1$ S = $S_1$ $n$ = 300 $p$ = 400, displayed in the form of: mean (SD) calculated from 100 independent runs of the experiment.} 
\end{table}
\begin{table}[H]
\centering
\begin{tabular}{lccccc}
  \hline
 & L2Boost & MBoost & LADBoost & Robloss & RRBoost \\ 
  \hline
$D_0$ & 0.52 (0.15) & 0.51 (0.15) & 0.49 (0.15) & 0.53 (0.15) & 0.56 (0.17) \\ 
  $D_1 (10\%)$ & 0.18 (0.09) & 0.24 (0.10) & 0.39 (0.17) & 0.42 (0.14) & 0.52 (0.14) \\ 
  $D_1 (20\%)$ & 0.12 (0.11) & 0.12 (0.11) & 0.29 (0.15) & 0.30 (0.13) & 0.50 (0.13) \\ 
  $D_2 (10\%)$ & 0.17 (0.11) & 0.26 (0.13) & 0.43 (0.17) & 0.41 (0.15) & 0.50 (0.13) \\ 
  $D_2 (20\%)$ & 0.12 (0.11) & 0.11 (0.12) & 0.34 (0.15) & 0.29 (0.14) & 0.51 (0.14) \\ 
  $D_3$ & 0.39 (0.13) & 0.46 (0.13) & 0.45 (0.14) & 0.50 (0.13) & 0.51 (0.13) \\ 
  $D_4$ & 0.06 (0.09) & 0.29 (0.13) & 0.30 (0.16) & 0.37 (0.15) & 0.43 (0.13) \\ 
   \hline
\end{tabular}
\caption{Fractions of variables recovered by L2Boost, MBoost, LADBoost, Robloss, and RRBoost applied with tree learners of $d$ = 2 for clean ($D_0$), symmetric gross error contaminated ($D_1$), asymmetric gross error contaminated ($D_2$),  skewed distributed ($D_3$),  and heavy-tailed distributed  ($D_4$) data generated from $g$ = $g_1$ S = $S_1$ $n$ = 300 $p$ = 400, displayed in the form of: mean (SD) calculated from 100 independent runs of the experiment.} 
\end{table}
\begin{table}[H]
\centering
\begin{tabular}{lcccccc}
  \hline
 & L2Boost & MBoost & LADBoost & Robloss & SBoost & RRBoost \\ 
  \hline
$D_0$ & 1.24 (0.11) & 1.25 (0.12) & 1.41 (0.16) & 1.27 (0.14) & 1.44 (0.15) & 1.20 (0.10) \\ 
  $D_1 (10\%)$ & 2.15 (0.27) & 1.91 (0.30) & 1.56 (0.17) & 1.52 (0.19) & 1.47 (0.16) & 1.26 (0.12) \\ 
  $D_1 (20\%)$ & 2.65 (0.38) & 2.72 (0.36) & 1.69 (0.15) & 1.71 (0.22) & 1.56 (0.15) & 1.34 (0.15) \\ 
  $D_2 (10\%)$ & 2.36 (0.27) & 2.08 (0.40) & 1.53 (0.17) & 1.49 (0.18) & 1.49 (0.14) & 1.25 (0.11) \\ 
  $D_2 (20\%)$ & 3.28 (0.34) & 3.34 (0.32) & 1.70 (0.19) & 1.90 (0.29) & 1.53 (0.15) & 1.33 (0.13) \\ 
  $D_3$ & 1.40 (0.16) & 1.33 (0.15) & 1.46 (0.16) & 1.34 (0.16) & 1.45 (0.17) & 1.26 (0.10) \\ 
  $D_4$ & 14.59 (37.01) & 1.63 (0.22) & 1.63 (0.17) & 1.57 (0.20) & 1.53 (0.16) & 1.41 (0.14) \\ 
   \hline
\end{tabular}
\caption{Summary statistics of RMSEs on the test sets by L2Boost, MBoost, LADBoost, Robloss, SBoost, and RRBoost applied with tree learners of $d$ = 3 for clean ($D_0$), symmetric gross error contaminated ($D_1$), asymmetric gross error contaminated ($D_2$),  skewed distributed ($D_3$),  and heavy-tailed distributed  ($D_4$) data generated from $g$ = $g_1$ S = $S_1$ $n$ = 300 $p$ = 400, displayed in the form of: mean (SD) calculated from 100 independent runs of the experiment.} 
\end{table}
\begin{table}[H]
\centering
\begin{tabular}{lccccc}
  \hline
 & L2Boost & MBoost & LADBoost & Robloss & RRBoost \\ 
  \hline
$D_0$ & 0.45 (0.14) & 0.50 (0.14) & 0.50 (0.16) & 0.51 (0.15) & 0.51 (0.14) \\ 
  $D_1 (10\%)$ & 0.19 (0.08) & 0.25 (0.12) & 0.40 (0.18) & 0.39 (0.16) & 0.49 (0.12) \\ 
  $D_1 (20\%)$ & 0.13 (0.12) & 0.13 (0.11) & 0.28 (0.16) & 0.29 (0.14) & 0.50 (0.13) \\ 
  $D_2 (10\%)$ & 0.18 (0.11) & 0.25 (0.12) & 0.41 (0.17) & 0.40 (0.16) & 0.51 (0.13) \\ 
  $D_2 (20\%)$ & 0.13 (0.11) & 0.12 (0.12) & 0.34 (0.16) & 0.30 (0.13) & 0.49 (0.13) \\ 
  $D_3$ & 0.39 (0.14) & 0.43 (0.16) & 0.45 (0.14) & 0.48 (0.14) & 0.51 (0.13) \\ 
  $D_4$ & 0.10 (0.12) & 0.30 (0.13) & 0.33 (0.17) & 0.36 (0.15) & 0.42 (0.15) \\ 
   \hline
\end{tabular}
\caption{Fractions of variables recovered by L2Boost, MBoost, LADBoost, Robloss, and RRBoost applied with tree learners of $d$ = 3 for clean ($D_0$), symmetric gross error contaminated ($D_1$), asymmetric gross error contaminated ($D_2$),  skewed distributed ($D_3$),  and heavy-tailed distributed  ($D_4$) data generated from $g$ = $g_1$ S = $S_1$ $n$ = 300 $p$ = 400, displayed in the form of: mean (SD) calculated from 100 independent runs of the experiment.} 
\end{table}
\begin{table}[H]
\centering
\begin{tabular}{lcccccc}
  \hline
 & L2Boost & MBoost & LADBoost & Robloss & SBoost & RRBoost \\ 
  \hline
$D_0$ & 0.64 (0.02) & 0.65 (0.02) & 0.69 (0.02) & 0.65 (0.02) & 1.42 (0.23) & 0.64 (0.02) \\ 
  $D_1 (10\%)$ & 1.17 (0.05) & 0.97 (0.18) & 0.73 (0.02) & 0.70 (0.03) & 1.35 (0.21) & 0.66 (0.12) \\ 
  $D_1 (20\%)$ & 1.30 (0.08) & 1.31 (0.08) & 0.76 (0.03) & 0.77 (0.04) & 1.15 (0.22) & 0.66 (0.03) \\ 
  $D_2 (10\%)$ & 1.56 (0.05) & 1.10 (0.23) & 0.74 (0.03) & 0.72 (0.03) & 1.25 (0.21) & 0.65 (0.08) \\ 
  $D_2 (20\%)$ & 2.37 (0.06) & 2.31 (0.07) & 0.84 (0.04) & 0.90 (0.09) & 1.09 (0.23) & 0.67 (0.17) \\ 
  $D_3$ & 0.82 (0.05) & 0.70 (0.02) & 0.74 (0.02) & 0.70 (0.02) & 1.41 (0.21) & 0.68 (0.02) \\ 
  $D_4$ & 37.12 (219.79) & 0.90 (0.04) & 0.79 (0.03) & 0.78 (0.03) & 1.17 (0.19) & 0.75 (0.03) \\ 
   \hline
\end{tabular}
\caption{Summary statistics of RMSEs on the test sets by L2Boost, MBoost, LADBoost, Robloss, SBoost, and RRBoost applied with tree learners of $d$ = 1 for clean ($D_0$), symmetric gross error contaminated ($D_1$), asymmetric gross error contaminated ($D_2$),  skewed distributed ($D_3$),  and heavy-tailed distributed  ($D_4$) data generated from $g$ = $g_1$ S = $S_1$ $n$ = 3000 $p$ = 400, displayed in the form of: mean (SD) calculated from 100 independent runs of the experiment.} 
\end{table}
\begin{table}[H]
\centering
\begin{tabular}{lccccc}
  \hline
 & L2Boost & MBoost & LADBoost & Robloss & RRBoost \\ 
  \hline
$D_0$ & 1.00 (0.00) & 1.00 (0.00) & 1.00 (0.00) & 1.00 (0.00) & 1.00 (0.00) \\ 
  $D_1 (10\%)$ & 0.55 (0.15) & 0.89 (0.17) & 1.00 (0.00) & 1.00 (0.00) & 1.00 (0.02) \\ 
  $D_1 (20\%)$ & 0.45 (0.12) & 0.44 (0.12) & 1.00 (0.00) & 1.00 (0.03) & 1.00 (0.00) \\ 
  $D_2 (10\%)$ & 0.57 (0.15) & 0.90 (0.15) & 1.00 (0.00) & 1.00 (0.00) & 1.00 (0.00) \\ 
  $D_2 (20\%)$ & 0.47 (0.12) & 0.48 (0.13) & 1.00 (0.02) & 0.98 (0.07) & 1.00 (0.00) \\ 
  $D_3$ & 0.99 (0.05) & 1.00 (0.00) & 1.00 (0.00) & 1.00 (0.00) & 1.00 (0.00) \\ 
  $D_4$ & 0.02 (0.08) & 0.98 (0.05) & 1.00 (0.00) & 1.00 (0.00) & 1.00 (0.00) \\ 
   \hline
\end{tabular}
\caption{Fractions of variables recovered by L2Boost, MBoost, LADBoost, Robloss, and RRBoost applied with tree learners of $d$ = 1 for clean ($D_0$), symmetric gross error contaminated ($D_1$), asymmetric gross error contaminated ($D_2$),  skewed distributed ($D_3$),  and heavy-tailed distributed  ($D_4$) data generated from $g$ = $g_1$ S = $S_1$ $n$ = 3000 $p$ = 400, displayed in the form of: mean (SD) calculated from 100 independent runs of the experiment.} 
\end{table}
\begin{table}[H]
\centering
\begin{tabular}{lcccccc}
  \hline
 & L2Boost & MBoost & LADBoost & Robloss & SBoost & RRBoost \\ 
  \hline
$D_0$ & 0.70 (0.02) & 0.70 (0.03) & 0.74 (0.03) & 0.70 (0.02) & 1.20 (0.23) & 0.69 (0.10) \\ 
  $D_1 (10\%)$ & 1.17 (0.08) & 0.96 (0.14) & 0.78 (0.03) & 0.76 (0.03) & 1.09 (0.18) & 0.88 (0.81) \\ 
  $D_1 (20\%)$ & 1.36 (0.12) & 1.36 (0.11) & 0.82 (0.04) & 0.83 (0.04) & 0.99 (0.18) & 0.74 (0.37) \\ 
  $D_2 (10\%)$ & 1.56 (0.07) & 1.10 (0.21) & 0.78 (0.04) & 0.75 (0.03) & 1.06 (0.18) & 0.73 (0.33) \\ 
  $D_2 (20\%)$ & 2.40 (0.07) & 2.32 (0.10) & 0.87 (0.04) & 0.89 (0.04) & 0.93 (0.18) & 0.70 (0.22) \\ 
  $D_3$ & 0.85 (0.06) & 0.73 (0.02) & 0.78 (0.03) & 0.74 (0.03) & 1.29 (0.20) & 0.87 (0.59) \\ 
  $D_4$ & 58.57 (284.03) & 0.88 (0.04) & 0.83 (0.04) & 0.82 (0.04) & 1.04 (0.18) & 0.74 (0.05) \\ 
   \hline
\end{tabular}
\caption{Summary statistics of RMSEs on the test sets by L2Boost, MBoost, LADBoost, Robloss, SBoost, and RRBoost applied with tree learners of $d$ = 2 for clean ($D_0$), symmetric gross error contaminated ($D_1$), asymmetric gross error contaminated ($D_2$),  skewed distributed ($D_3$),  and heavy-tailed distributed  ($D_4$) data generated from $g$ = $g_1$ S = $S_1$ $n$ = 3000 $p$ = 400, displayed in the form of: mean (SD) calculated from 100 independent runs of the experiment.} 
\end{table}
\begin{table}[H]
\centering
\begin{tabular}{lccccc}
  \hline
 & L2Boost & MBoost & LADBoost & Robloss & RRBoost \\ 
  \hline
$D_0$ & 1.00 (0.00) & 1.00 (0.00) & 1.00 (0.00) & 1.00 (0.00) & 1.00 (0.02) \\ 
  $D_1 (10\%)$ & 0.50 (0.13) & 0.81 (0.23) & 1.00 (0.00) & 1.00 (0.00) & 1.00 (0.03) \\ 
  $D_1 (20\%)$ & 0.42 (0.14) & 0.40 (0.14) & 1.00 (0.02) & 0.99 (0.04) & 0.99 (0.04) \\ 
  $D_2 (10\%)$ & 0.45 (0.10) & 0.81 (0.21) & 1.00 (0.00) & 1.00 (0.00) & 1.00 (0.03) \\ 
  $D_2 (20\%)$ & 0.39 (0.11) & 0.40 (0.11) & 1.00 (0.03) & 0.99 (0.04) & 0.99 (0.04) \\ 
  $D_3$ & 0.97 (0.08) & 1.00 (0.00) & 1.00 (0.00) & 1.00 (0.00) & 1.00 (0.02) \\ 
  $D_4$ & 0.02 (0.06) & 0.97 (0.08) & 1.00 (0.03) & 1.00 (0.02) & 1.00 (0.00) \\ 
   \hline
\end{tabular}
\caption{Fractions of variables recovered by L2Boost, MBoost, LADBoost, Robloss, and RRBoost applied with tree learners of $d$ = 2 for clean ($D_0$), symmetric gross error contaminated ($D_1$), asymmetric gross error contaminated ($D_2$),  skewed distributed ($D_3$),  and heavy-tailed distributed  ($D_4$) data generated from $g$ = $g_1$ S = $S_1$ $n$ = 3000 $p$ = 400, displayed in the form of: mean (SD) calculated from 100 independent runs of the experiment.} 
\end{table}
\begin{table}[H]
\centering
\begin{tabular}{lcccccc}
  \hline
 & L2Boost & MBoost & LADBoost & Robloss & SBoost & RRBoost \\ 
  \hline
$D_0$ & 0.72 (0.03) & 0.73 (0.03) & 0.79 (0.03) & 0.75 (0.03) & 1.13 (0.19) & 0.73 (0.09) \\ 
  $D_1 (10\%)$ & 1.26 (0.06) & 1.06 (0.12) & 0.86 (0.04) & 0.83 (0.04) & 1.04 (0.17) & 0.73 (0.08) \\ 
  $D_1 (20\%)$ & 1.43 (0.11) & 1.43 (0.13) & 0.90 (0.04) & 0.93 (0.05) & 0.92 (0.12) & 0.71 (0.07) \\ 
  $D_2 (10\%)$ & 1.63 (0.07) & 1.19 (0.20) & 0.84 (0.03) & 0.82 (0.04) & 1.03 (0.12) & 0.71 (0.04) \\ 
  $D_2 (20\%)$ & 2.44 (0.07) & 2.41 (0.10) & 0.96 (0.05) & 1.00 (0.05) & 0.94 (0.10) & 0.70 (0.03) \\ 
  $D_3$ & 0.89 (0.06) & 0.78 (0.03) & 0.83 (0.03) & 0.79 (0.03) & 1.14 (0.20) & 0.79 (0.11) \\ 
  $D_4$ & 56.42 (274.38) & 0.97 (0.05) & 0.91 (0.04) & 0.90 (0.05) & 1.04 (0.15) & 0.79 (0.05) \\ 
   \hline
\end{tabular}
\caption{Summary statistics of RMSEs on the test sets by L2Boost, MBoost, LADBoost, Robloss, SBoost, and RRBoost applied with tree learners of $d$ = 3 for clean ($D_0$), symmetric gross error contaminated ($D_1$), asymmetric gross error contaminated ($D_2$),  skewed distributed ($D_3$),  and heavy-tailed distributed  ($D_4$) data generated from $g$ = $g_1$ S = $S_1$ $n$ = 3000 $p$ = 400, displayed in the form of: mean (SD) calculated from 100 independent runs of the experiment.} 
\end{table}
\begin{table}[H]
\centering
\begin{tabular}{lccccc}
  \hline
 & L2Boost & MBoost & LADBoost & Robloss & RRBoost \\ 
  \hline
$D_0$ & 1.00 (0.00) & 1.00 (0.00) & 1.00 (0.00) & 1.00 (0.00) & 1.00 (0.00) \\ 
  $D_1 (10\%)$ & 0.43 (0.14) & 0.74 (0.21) & 1.00 (0.00) & 1.00 (0.02) & 1.00 (0.00) \\ 
  $D_1 (20\%)$ & 0.38 (0.15) & 0.39 (0.14) & 0.99 (0.05) & 0.97 (0.08) & 1.00 (0.00) \\ 
  $D_2 (10\%)$ & 0.38 (0.12) & 0.71 (0.22) & 1.00 (0.00) & 1.00 (0.00) & 1.00 (0.02) \\ 
  $D_2 (20\%)$ & 0.32 (0.11) & 0.40 (0.11) & 0.96 (0.08) & 0.96 (0.08) & 1.00 (0.00) \\ 
  $D_3$ & 0.91 (0.13) & 1.00 (0.00) & 1.00 (0.00) & 1.00 (0.00) & 1.00 (0.00) \\ 
  $D_4$ & 0.03 (0.08) & 0.84 (0.20) & 0.98 (0.05) & 0.99 (0.04) & 1.00 (0.00) \\ 
   \hline
\end{tabular}
\caption{Fractions of variables recovered by L2Boost, MBoost, LADBoost, Robloss, and RRBoost applied with tree learners of $d$ = 3 for clean ($D_0$), symmetric gross error contaminated ($D_1$), asymmetric gross error contaminated ($D_2$),  skewed distributed ($D_3$),  and heavy-tailed distributed  ($D_4$) data generated from $g$ = $g_1$ S = $S_1$ $n$ = 3000 $p$ = 400, displayed in the form of: mean (SD) calculated from 100 independent runs of the experiment.} 
\end{table}
\begin{table}[H]
\centering
\begin{tabular}{lcccccc}
  \hline
 & L2Boost & MBoost & LADBoost & Robloss & SBoost & RRBoost \\ 
  \hline
$D_0$ & 0.91 (0.04) & 0.92 (0.04) & 0.99 (0.05) & 0.92 (0.04) & 1.48 (0.15) & 0.91 (0.03) \\ 
  $D_1 (10\%)$ & 1.78 (0.19) & 1.44 (0.26) & 1.07 (0.06) & 1.07 (0.11) & 1.48 (0.15) & 0.94 (0.05) \\ 
  $D_1 (20\%)$ & 2.00 (0.24) & 1.95 (0.19) & 1.20 (0.10) & 1.29 (0.18) & 1.46 (0.17) & 0.98 (0.08) \\ 
  $D_2 (10\%)$ & 2.16 (0.19) & 1.64 (0.38) & 1.10 (0.08) & 1.09 (0.11) & 1.47 (0.15) & 0.96 (0.18) \\ 
  $D_2 (20\%)$ & 3.19 (0.23) & 3.15 (0.26) & 1.32 (0.15) & 1.51 (0.24) & 1.45 (0.17) & 0.98 (0.06) \\ 
  $D_3$ & 1.20 (0.13) & 1.03 (0.05) & 1.05 (0.05) & 1.01 (0.05) & 1.51 (0.17) & 1.01 (0.05) \\ 
  $D_4$ & 26.42 (103.14) & 1.40 (0.10) & 1.22 (0.10) & 1.24 (0.10) & 1.49 (0.16) & 1.20 (0.08) \\ 
   \hline
\end{tabular}
\caption{Summary statistics of RMSEs on the test sets by L2Boost, MBoost, LADBoost, Robloss, SBoost, and RRBoost applied with tree learners of $d$ = 1 for clean ($D_0$), symmetric gross error contaminated ($D_1$), asymmetric gross error contaminated ($D_2$),  skewed distributed ($D_3$),  and heavy-tailed distributed  ($D_4$) data generated from $g$ = $g_1$ S = $S_2$ $n$ = 300 $p$ = 10, displayed in the form of: mean (SD) calculated from 100 independent runs of the experiment.} 
\end{table}
\begin{table}[H]
\centering
\begin{tabular}{lccccc}
  \hline
 & L2Boost & MBoost & LADBoost & Robloss & RRBoost \\ 
  \hline
$D_0$ & 1.00 (0.00) & 1.00 (0.02) & 1.00 (0.02) & 1.00 (0.02) & 1.00 (0.00) \\ 
  $D_1 (10\%)$ & 0.45 (0.25) & 0.71 (0.31) & 0.98 (0.05) & 0.99 (0.05) & 1.00 (0.03) \\ 
  $D_1 (20\%)$ & 0.29 (0.25) & 0.25 (0.24) & 0.94 (0.11) & 0.90 (0.17) & 1.00 (0.03) \\ 
  $D_2 (10\%)$ & 0.44 (0.28) & 0.76 (0.24) & 0.98 (0.07) & 0.98 (0.06) & 1.00 (0.02) \\ 
  $D_2 (20\%)$ & 0.40 (0.25) & 0.43 (0.24) & 0.88 (0.16) & 0.81 (0.20) & 0.99 (0.03) \\ 
  $D_3$ & 0.93 (0.13) & 0.99 (0.04) & 0.99 (0.03) & 1.00 (0.02) & 1.00 (0.02) \\ 
  $D_4$ & 0.32 (0.31) & 0.72 (0.23) & 0.91 (0.14) & 0.90 (0.15) & 0.93 (0.12) \\ 
   \hline
\end{tabular}
\caption{Fractions of variables recovered by L2Boost, MBoost, LADBoost, Robloss, and RRBoost applied with tree learners of $d$ = 1 for clean ($D_0$), symmetric gross error contaminated ($D_1$), asymmetric gross error contaminated ($D_2$),  skewed distributed ($D_3$),  and heavy-tailed distributed  ($D_4$) data generated from $g$ = $g_1$ S = $S_2$ $n$ = 300 $p$ = 10, displayed in the form of: mean (SD) calculated from 100 independent runs of the experiment.} 
\end{table}
\begin{table}[H]
\centering
\begin{tabular}{lcccccc}
  \hline
 & L2Boost & MBoost & LADBoost & Robloss & SBoost & RRBoost \\ 
  \hline
$D_0$ & 1.07 (0.06) & 1.06 (0.06) & 1.07 (0.05) & 1.07 (0.05) & 1.47 (0.15) & 1.03 (0.05) \\ 
  $D_1 (10\%)$ & 1.88 (0.22) & 1.59 (0.30) & 1.18 (0.08) & 1.23 (0.10) & 1.46 (0.16) & 1.08 (0.06) \\ 
  $D_1 (20\%)$ & 2.31 (0.35) & 2.27 (0.27) & 1.35 (0.15) & 1.41 (0.15) & 1.46 (0.18) & 1.12 (0.06) \\ 
  $D_2 (10\%)$ & 2.26 (0.23) & 1.76 (0.41) & 1.19 (0.08) & 1.22 (0.10) & 1.44 (0.16) & 1.08 (0.06) \\ 
  $D_2 (20\%)$ & 3.29 (0.27) & 3.33 (0.30) & 1.42 (0.14) & 1.57 (0.17) & 1.46 (0.15) & 1.13 (0.08) \\ 
  $D_3$ & 1.32 (0.13) & 1.16 (0.06) & 1.14 (0.06) & 1.14 (0.06) & 1.51 (0.17) & 1.13 (0.09) \\ 
  $D_4$ & 40.84 (263.02) & 1.47 (0.14) & 1.34 (0.12) & 1.38 (0.12) & 1.46 (0.16) & 1.27 (0.10) \\ 
   \hline
\end{tabular}
\caption{Summary statistics of RMSEs on the test sets by L2Boost, MBoost, LADBoost, Robloss, SBoost, and RRBoost applied with tree learners of $d$ = 2 for clean ($D_0$), symmetric gross error contaminated ($D_1$), asymmetric gross error contaminated ($D_2$),  skewed distributed ($D_3$),  and heavy-tailed distributed  ($D_4$) data generated from $g$ = $g_1$ S = $S_2$ $n$ = 300 $p$ = 10, displayed in the form of: mean (SD) calculated from 100 independent runs of the experiment.} 
\end{table}
\begin{table}[H]
\centering
\begin{tabular}{lccccc}
  \hline
 & L2Boost & MBoost & LADBoost & Robloss & RRBoost \\ 
  \hline
$D_0$ & 0.97 (0.09) & 0.97 (0.08) & 0.99 (0.05) & 0.97 (0.09) & 0.99 (0.05) \\ 
  $D_1 (10\%)$ & 0.42 (0.20) & 0.64 (0.27) & 0.93 (0.11) & 0.86 (0.16) & 0.98 (0.06) \\ 
  $D_1 (20\%)$ & 0.35 (0.19) & 0.30 (0.17) & 0.83 (0.16) & 0.73 (0.21) & 0.96 (0.08) \\ 
  $D_2 (10\%)$ & 0.41 (0.19) & 0.64 (0.21) & 0.93 (0.14) & 0.87 (0.18) & 0.98 (0.08) \\ 
  $D_2 (20\%)$ & 0.33 (0.16) & 0.43 (0.19) & 0.79 (0.18) & 0.61 (0.22) & 0.95 (0.10) \\ 
  $D_3$ & 0.74 (0.22) & 0.92 (0.13) & 0.95 (0.10) & 0.93 (0.12) & 0.98 (0.07) \\ 
  $D_4$ & 0.28 (0.21) & 0.60 (0.24) & 0.82 (0.18) & 0.74 (0.21) & 0.82 (0.13) \\ 
   \hline
\end{tabular}
\caption{Fractions of variables recovered by L2Boost, MBoost, LADBoost, Robloss, and RRBoost applied with tree learners of $d$ = 2 for clean ($D_0$), symmetric gross error contaminated ($D_1$), asymmetric gross error contaminated ($D_2$),  skewed distributed ($D_3$),  and heavy-tailed distributed  ($D_4$) data generated from $g$ = $g_1$ S = $S_2$ $n$ = 300 $p$ = 10, displayed in the form of: mean (SD) calculated from 100 independent runs of the experiment.} 
\end{table}
\begin{table}[H]
\centering
\begin{tabular}{lcccccc}
  \hline
 & L2Boost & MBoost & LADBoost & Robloss & SBoost & RRBoost \\ 
  \hline
$D_0$ & 1.13 (0.07) & 1.13 (0.07) & 1.16 (0.07) & 1.14 (0.08) & 1.46 (0.16) & 1.11 (0.07) \\ 
  $D_1 (10\%)$ & 2.08 (0.27) & 1.69 (0.37) & 1.28 (0.09) & 1.30 (0.11) & 1.49 (0.15) & 1.14 (0.07) \\ 
  $D_1 (20\%)$ & 2.60 (0.36) & 2.62 (0.32) & 1.45 (0.13) & 1.49 (0.16) & 1.45 (0.16) & 1.18 (0.07) \\ 
  $D_2 (10\%)$ & 2.38 (0.32) & 1.88 (0.47) & 1.28 (0.08) & 1.31 (0.11) & 1.45 (0.14) & 1.14 (0.07) \\ 
  $D_2 (20\%)$ & 3.45 (0.31) & 3.53 (0.32) & 1.48 (0.14) & 1.63 (0.16) & 1.46 (0.16) & 1.19 (0.07) \\ 
  $D_3$ & 1.38 (0.15) & 1.23 (0.09) & 1.22 (0.08) & 1.22 (0.09) & 1.50 (0.14) & 1.18 (0.08) \\ 
  $D_4$ & 42.19 (263.24) & 1.52 (0.13) & 1.40 (0.12) & 1.43 (0.13) & 1.50 (0.16) & 1.31 (0.09) \\ 
   \hline
\end{tabular}
\caption{Summary statistics of RMSEs on the test sets by L2Boost, MBoost, LADBoost, Robloss, SBoost, and RRBoost applied with tree learners of $d$ = 3 for clean ($D_0$), symmetric gross error contaminated ($D_1$), asymmetric gross error contaminated ($D_2$),  skewed distributed ($D_3$),  and heavy-tailed distributed  ($D_4$) data generated from $g$ = $g_1$ S = $S_2$ $n$ = 300 $p$ = 10, displayed in the form of: mean (SD) calculated from 100 independent runs of the experiment.} 
\end{table}
\begin{table}[H]
\centering
\begin{tabular}{lccccc}
  \hline
 & L2Boost & MBoost & LADBoost & Robloss & RRBoost \\ 
  \hline
$D_0$ & 0.93 (0.13) & 0.94 (0.10) & 0.95 (0.09) & 0.93 (0.12) & 0.97 (0.09) \\ 
  $D_1 (10\%)$ & 0.46 (0.19) & 0.58 (0.22) & 0.86 (0.15) & 0.82 (0.17) & 0.96 (0.09) \\ 
  $D_1 (20\%)$ & 0.50 (0.23) & 0.47 (0.21) & 0.80 (0.17) & 0.67 (0.22) & 0.90 (0.14) \\ 
  $D_2 (10\%)$ & 0.48 (0.19) & 0.57 (0.24) & 0.84 (0.15) & 0.77 (0.19) & 0.94 (0.10) \\ 
  $D_2 (20\%)$ & 0.46 (0.18) & 0.46 (0.19) & 0.70 (0.18) & 0.59 (0.23) & 0.91 (0.12) \\ 
  $D_3$ & 0.70 (0.22) & 0.83 (0.19) & 0.92 (0.12) & 0.86 (0.17) & 0.91 (0.13) \\ 
  $D_4$ & 0.39 (0.22) & 0.52 (0.22) & 0.78 (0.17) & 0.66 (0.24) & 0.78 (0.15) \\ 
   \hline
\end{tabular}
\caption{Fractions of variables recovered by L2Boost, MBoost, LADBoost, Robloss, and RRBoost applied with tree learners of $d$ = 3 for clean ($D_0$), symmetric gross error contaminated ($D_1$), asymmetric gross error contaminated ($D_2$),  skewed distributed ($D_3$),  and heavy-tailed distributed  ($D_4$) data generated from $g$ = $g_1$ S = $S_2$ $n$ = 300 $p$ = 10, displayed in the form of: mean (SD) calculated from 100 independent runs of the experiment.} 
\end{table}
\begin{table}[H]
\centering
\begin{tabular}{lcccccc}
  \hline
 & L2Boost & MBoost & LADBoost & Robloss & SBoost & RRBoost \\ 
  \hline
$D_0$ & 0.70 (0.02) & 0.70 (0.02) & 0.71 (0.02) & 0.70 (0.02) & 1.14 (0.18) & 0.69 (0.01) \\ 
  $D_1 (10\%)$ & 1.16 (0.06) & 0.96 (0.20) & 0.72 (0.02) & 0.72 (0.03) & 1.06 (0.20) & 0.69 (0.02) \\ 
  $D_1 (20\%)$ & 1.34 (0.09) & 1.33 (0.07) & 0.75 (0.04) & 0.76 (0.07) & 0.94 (0.17) & 0.69 (0.01) \\ 
  $D_2 (10\%)$ & 1.71 (0.06) & 1.14 (0.29) & 0.74 (0.03) & 0.74 (0.04) & 1.07 (0.20) & 0.69 (0.01) \\ 
  $D_2 (20\%)$ & 2.88 (0.07) & 2.83 (0.09) & 0.82 (0.09) & 0.91 (0.14) & 0.90 (0.16) & 0.69 (0.02) \\ 
  $D_3$ & 0.79 (0.04) & 0.74 (0.02) & 0.79 (0.02) & 0.76 (0.02) & 1.22 (0.21) & 0.75 (0.02) \\ 
  $D_4$ & 8.75 (19.47) & 0.86 (0.03) & 0.76 (0.02) & 0.77 (0.02) & 0.93 (0.14) & 0.73 (0.02) \\ 
   \hline
\end{tabular}
\caption{Summary statistics of RMSEs on the test sets by L2Boost, MBoost, LADBoost, Robloss, SBoost, and RRBoost applied with tree learners of $d$ = 1 for clean ($D_0$), symmetric gross error contaminated ($D_1$), asymmetric gross error contaminated ($D_2$),  skewed distributed ($D_3$),  and heavy-tailed distributed  ($D_4$) data generated from $g$ = $g_1$ S = $S_2$ $n$ = 3000 $p$ = 10, displayed in the form of: mean (SD) calculated from 100 independent runs of the experiment.} 
\end{table}
\begin{table}[H]
\centering
\begin{tabular}{lccccc}
  \hline
 & L2Boost & MBoost & LADBoost & Robloss & RRBoost \\ 
  \hline
$D_0$ & 1.00 (0.00) & 1.00 (0.00) & 1.00 (0.00) & 1.00 (0.00) & 1.00 (0.00) \\ 
  $D_1 (10\%)$ & 0.96 (0.10) & 0.99 (0.03) & 1.00 (0.00) & 1.00 (0.00) & 1.00 (0.00) \\ 
  $D_1 (20\%)$ & 0.77 (0.23) & 0.79 (0.17) & 1.00 (0.00) & 1.00 (0.00) & 1.00 (0.00) \\ 
  $D_2 (10\%)$ & 0.98 (0.06) & 0.99 (0.05) & 1.00 (0.00) & 1.00 (0.00) & 1.00 (0.00) \\ 
  $D_2 (20\%)$ & 0.83 (0.19) & 0.81 (0.21) & 1.00 (0.02) & 0.99 (0.04) & 1.00 (0.00) \\ 
  $D_3$ & 1.00 (0.02) & 1.00 (0.00) & 1.00 (0.00) & 1.00 (0.00) & 1.00 (0.00) \\ 
  $D_4$ & 0.25 (0.28) & 1.00 (0.00) & 1.00 (0.00) & 1.00 (0.00) & 1.00 (0.00) \\ 
   \hline
\end{tabular}
\caption{Fractions of variables recovered by L2Boost, MBoost, LADBoost, Robloss, and RRBoost applied with tree learners of $d$ = 1 for clean ($D_0$), symmetric gross error contaminated ($D_1$), asymmetric gross error contaminated ($D_2$),  skewed distributed ($D_3$),  and heavy-tailed distributed  ($D_4$) data generated from $g$ = $g_1$ S = $S_2$ $n$ = 3000 $p$ = 10, displayed in the form of: mean (SD) calculated from 100 independent runs of the experiment.} 
\end{table}
\begin{table}[H]
\centering
\begin{tabular}{lcccccc}
  \hline
 & L2Boost & MBoost & LADBoost & Robloss & SBoost & RRBoost \\ 
  \hline
$D_0$ & 0.76 (0.02) & 0.76 (0.02) & 0.78 (0.02) & 0.77 (0.02) & 0.98 (0.15) & 1.24 (1.16) \\ 
  $D_1 (10\%)$ & 1.24 (0.07) & 1.05 (0.17) & 0.81 (0.02) & 0.80 (0.02) & 0.90 (0.15) & 1.00 (0.70) \\ 
  $D_1 (20\%)$ & 1.41 (0.11) & 1.43 (0.11) & 0.85 (0.03) & 0.86 (0.03) & 0.92 (0.19) & 1.10 (0.80) \\ 
  $D_2 (10\%)$ & 1.79 (0.07) & 1.26 (0.31) & 0.82 (0.03) & 0.82 (0.02) & 0.91 (0.14) & 1.18 (1.00) \\ 
  $D_2 (20\%)$ & 2.92 (0.08) & 2.84 (0.08) & 0.90 (0.03) & 0.95 (0.04) & 0.84 (0.14) & 0.79 (0.29) \\ 
  $D_3$ & 0.90 (0.04) & 0.81 (0.02) & 0.84 (0.02) & 0.82 (0.02) & 1.10 (0.17) & 1.04 (0.72) \\ 
  $D_4$ & 24.37 (72.10) & 0.95 (0.04) & 0.86 (0.02) & 0.87 (0.03) & 0.86 (0.14) & 0.76 (0.04) \\ 
   \hline
\end{tabular}
\caption{Summary statistics of RMSEs on the test sets by L2Boost, MBoost, LADBoost, Robloss, SBoost, and RRBoost applied with tree learners of $d$ = 2 for clean ($D_0$), symmetric gross error contaminated ($D_1$), asymmetric gross error contaminated ($D_2$),  skewed distributed ($D_3$),  and heavy-tailed distributed  ($D_4$) data generated from $g$ = $g_1$ S = $S_2$ $n$ = 3000 $p$ = 10, displayed in the form of: mean (SD) calculated from 100 independent runs of the experiment.} 
\end{table}
\begin{table}[H]
\centering
\begin{tabular}{lccccc}
  \hline
 & L2Boost & MBoost & LADBoost & Robloss & RRBoost \\ 
  \hline
$D_0$ & 1.00 (0.00) & 1.00 (0.00) & 1.00 (0.00) & 1.00 (0.00) & 1.00 (0.03) \\ 
  $D_1 (10\%)$ & 0.76 (0.20) & 0.92 (0.14) & 1.00 (0.00) & 1.00 (0.00) & 0.99 (0.05) \\ 
  $D_1 (20\%)$ & 0.66 (0.22) & 0.64 (0.21) & 1.00 (0.00) & 1.00 (0.00) & 0.99 (0.04) \\ 
  $D_2 (10\%)$ & 0.71 (0.21) & 0.95 (0.11) & 1.00 (0.00) & 1.00 (0.00) & 1.00 (0.03) \\ 
  $D_2 (20\%)$ & 0.61 (0.19) & 0.61 (0.20) & 1.00 (0.00) & 1.00 (0.00) & 1.00 (0.00) \\ 
  $D_3$ & 1.00 (0.00) & 1.00 (0.00) & 1.00 (0.00) & 1.00 (0.00) & 1.00 (0.00) \\ 
  $D_4$ & 0.27 (0.20) & 1.00 (0.00) & 1.00 (0.00) & 1.00 (0.00) & 1.00 (0.00) \\ 
   \hline
\end{tabular}
\caption{Fractions of variables recovered by L2Boost, MBoost, LADBoost, Robloss, and RRBoost applied with tree learners of $d$ = 2 for clean ($D_0$), symmetric gross error contaminated ($D_1$), asymmetric gross error contaminated ($D_2$),  skewed distributed ($D_3$),  and heavy-tailed distributed  ($D_4$) data generated from $g$ = $g_1$ S = $S_2$ $n$ = 3000 $p$ = 10, displayed in the form of: mean (SD) calculated from 100 independent runs of the experiment.} 
\end{table}
\begin{table}[H]
\centering
\begin{tabular}{lcccccc}
  \hline
 & L2Boost & MBoost & LADBoost & Robloss & SBoost & RRBoost \\ 
  \hline
$D_0$ & 0.80 (0.02) & 0.79 (0.02) & 0.82 (0.02) & 0.80 (0.02) & 0.96 (0.11) & 0.77 (0.06) \\ 
  $D_1 (10\%)$ & 1.33 (0.07) & 1.12 (0.18) & 0.86 (0.02) & 0.85 (0.03) & 0.90 (0.08) & 0.77 (0.04) \\ 
  $D_1 (20\%)$ & 1.51 (0.09) & 1.51 (0.10) & 0.91 (0.03) & 0.91 (0.04) & 0.83 (0.06) & 0.75 (0.02) \\ 
  $D_2 (10\%)$ & 1.85 (0.07) & 1.31 (0.29) & 0.87 (0.03) & 0.86 (0.03) & 0.88 (0.07) & 0.76 (0.03) \\ 
  $D_2 (20\%)$ & 2.96 (0.08) & 2.90 (0.10) & 0.97 (0.04) & 1.02 (0.04) & 0.83 (0.06) & 0.75 (0.02) \\ 
  $D_3$ & 0.95 (0.05) & 0.84 (0.03) & 0.88 (0.03) & 0.85 (0.02) & 1.04 (0.11) & 0.88 (0.11) \\ 
  $D_4$ & 31.68 (114.74) & 1.01 (0.04) & 0.91 (0.04) & 0.91 (0.04) & 0.86 (0.08) & 0.80 (0.03) \\ 
   \hline
\end{tabular}
\caption{Summary statistics of RMSEs on the test sets by L2Boost, MBoost, LADBoost, Robloss, SBoost, and RRBoost applied with tree learners of $d$ = 3 for clean ($D_0$), symmetric gross error contaminated ($D_1$), asymmetric gross error contaminated ($D_2$),  skewed distributed ($D_3$),  and heavy-tailed distributed  ($D_4$) data generated from $g$ = $g_1$ S = $S_2$ $n$ = 3000 $p$ = 10, displayed in the form of: mean (SD) calculated from 100 independent runs of the experiment.} 
\end{table}
\begin{table}[H]
\centering
\begin{tabular}{lccccc}
  \hline
 & L2Boost & MBoost & LADBoost & Robloss & RRBoost \\ 
  \hline
$D_0$ & 1.00 (0.00) & 1.00 (0.00) & 1.00 (0.00) & 1.00 (0.00) & 1.00 (0.02) \\ 
  $D_1 (10\%)$ & 0.70 (0.21) & 0.91 (0.14) & 1.00 (0.00) & 1.00 (0.00) & 1.00 (0.00) \\ 
  $D_1 (20\%)$ & 0.60 (0.20) & 0.60 (0.20) & 1.00 (0.00) & 1.00 (0.00) & 1.00 (0.00) \\ 
  $D_2 (10\%)$ & 0.62 (0.21) & 0.93 (0.12) & 1.00 (0.00) & 1.00 (0.00) & 1.00 (0.00) \\ 
  $D_2 (20\%)$ & 0.55 (0.21) & 0.66 (0.16) & 1.00 (0.00) & 1.00 (0.00) & 1.00 (0.00) \\ 
  $D_3$ & 0.99 (0.05) & 1.00 (0.00) & 1.00 (0.00) & 1.00 (0.00) & 0.99 (0.04) \\ 
  $D_4$ & 0.41 (0.20) & 0.99 (0.06) & 1.00 (0.00) & 1.00 (0.00) & 1.00 (0.00) \\ 
   \hline
\end{tabular}
\caption{Fractions of variables recovered by L2Boost, MBoost, LADBoost, Robloss, and RRBoost applied with tree learners of $d$ = 3 for clean ($D_0$), symmetric gross error contaminated ($D_1$), asymmetric gross error contaminated ($D_2$),  skewed distributed ($D_3$),  and heavy-tailed distributed  ($D_4$) data generated from $g$ = $g_1$ S = $S_2$ $n$ = 3000 $p$ = 10, displayed in the form of: mean (SD) calculated from 100 independent runs of the experiment.} 
\end{table}
\begin{table}[H]
\centering
\begin{tabular}{lcccccc}
  \hline
 & L2Boost & MBoost & LADBoost & Robloss & SBoost & RRBoost \\ 
  \hline
$D_0$ & 1.22 (0.09) & 1.21 (0.09) & 1.36 (0.10) & 1.22 (0.09) & 1.45 (0.17) & 1.16 (0.08) \\ 
  $D_1 (10\%)$ & 1.75 (0.20) & 1.54 (0.20) & 1.48 (0.12) & 1.35 (0.13) & 1.48 (0.16) & 1.18 (0.10) \\ 
  $D_1 (20\%)$ & 1.98 (0.19) & 1.99 (0.18) & 1.61 (0.12) & 1.53 (0.14) & 1.52 (0.17) & 1.26 (0.14) \\ 
  $D_2 (10\%)$ & 2.01 (0.17) & 1.67 (0.24) & 1.47 (0.12) & 1.35 (0.11) & 1.48 (0.17) & 1.18 (0.08) \\ 
  $D_2 (20\%)$ & 2.75 (0.22) & 2.80 (0.23) & 1.63 (0.17) & 1.63 (0.15) & 1.57 (0.13) & 1.25 (0.11) \\ 
  $D_3$ & 1.37 (0.12) & 1.28 (0.10) & 1.40 (0.11) & 1.28 (0.10) & 1.48 (0.17) & 1.20 (0.09) \\ 
  $D_4$ & 14.60 (32.81) & 1.45 (0.11) & 1.55 (0.13) & 1.44 (0.13) & 1.51 (0.17) & 1.37 (0.14) \\ 
   \hline
\end{tabular}
\caption{Summary statistics of RMSEs on the test sets by L2Boost, MBoost, LADBoost, Robloss, SBoost, and RRBoost applied with tree learners of $d$ = 1 for clean ($D_0$), symmetric gross error contaminated ($D_1$), asymmetric gross error contaminated ($D_2$),  skewed distributed ($D_3$),  and heavy-tailed distributed  ($D_4$) data generated from $g$ = $g_1$ S = $S_2$ $n$ = 300 $p$ = 400, displayed in the form of: mean (SD) calculated from 100 independent runs of the experiment.} 
\end{table}
\begin{table}[H]
\centering
\begin{tabular}{lccccc}
  \hline
 & L2Boost & MBoost & LADBoost & Robloss & RRBoost \\ 
  \hline
$D_0$ & 0.63 (0.17) & 0.62 (0.16) & 0.55 (0.15) & 0.61 (0.17) & 0.67 (0.19) \\ 
  $D_1 (10\%)$ & 0.12 (0.10) & 0.26 (0.16) & 0.47 (0.19) & 0.49 (0.16) & 0.65 (0.17) \\ 
  $D_1 (20\%)$ & 0.06 (0.09) & 0.06 (0.09) & 0.29 (0.18) & 0.32 (0.19) & 0.57 (0.16) \\ 
  $D_2 (10\%)$ & 0.12 (0.12) & 0.29 (0.15) & 0.47 (0.18) & 0.51 (0.18) & 0.63 (0.17) \\ 
  $D_2 (20\%)$ & 0.08 (0.10) & 0.06 (0.09) & 0.35 (0.19) & 0.31 (0.17) & 0.56 (0.14) \\ 
  $D_3$ & 0.47 (0.20) & 0.56 (0.17) & 0.52 (0.16) & 0.55 (0.15) & 0.59 (0.17) \\ 
  $D_4$ & 0.01 (0.06) & 0.31 (0.17) & 0.35 (0.18) & 0.39 (0.19) & 0.49 (0.17) \\ 
   \hline
\end{tabular}
\caption{Fractions of variables recovered by L2Boost, MBoost, LADBoost, Robloss, and RRBoost applied with tree learners of $d$ = 1 for clean ($D_0$), symmetric gross error contaminated ($D_1$), asymmetric gross error contaminated ($D_2$),  skewed distributed ($D_3$),  and heavy-tailed distributed  ($D_4$) data generated from $g$ = $g_1$ S = $S_2$ $n$ = 300 $p$ = 400, displayed in the form of: mean (SD) calculated from 100 independent runs of the experiment.} 
\end{table}
\begin{table}[H]
\centering
\begin{tabular}{lcccccc}
  \hline
 & L2Boost & MBoost & LADBoost & Robloss & SBoost & RRBoost \\ 
  \hline
$D_0$ & 1.21 (0.13) & 1.21 (0.12) & 1.41 (0.14) & 1.23 (0.12) & 1.47 (0.15) & 1.17 (0.09) \\ 
  $D_1 (10\%)$ & 1.99 (0.23) & 1.68 (0.25) & 1.54 (0.15) & 1.41 (0.18) & 1.53 (0.16) & 1.21 (0.09) \\ 
  $D_1 (20\%)$ & 2.40 (0.25) & 2.38 (0.24) & 1.67 (0.14) & 1.63 (0.19) & 1.53 (0.16) & 1.28 (0.13) \\ 
  $D_2 (10\%)$ & 2.20 (0.23) & 1.83 (0.32) & 1.53 (0.15) & 1.41 (0.14) & 1.50 (0.14) & 1.21 (0.09) \\ 
  $D_2 (20\%)$ & 2.99 (0.28) & 3.05 (0.26) & 1.67 (0.15) & 1.73 (0.20) & 1.58 (0.16) & 1.30 (0.16) \\ 
  $D_3$ & 1.40 (0.13) & 1.29 (0.15) & 1.43 (0.15) & 1.30 (0.13) & 1.51 (0.16) & 1.21 (0.09) \\ 
  $D_4$ & 13.21 (33.32) & 1.51 (0.13) & 1.63 (0.13) & 1.48 (0.17) & 1.54 (0.17) & 1.39 (0.14) \\ 
   \hline
\end{tabular}
\caption{Summary statistics of RMSEs on the test sets by L2Boost, MBoost, LADBoost, Robloss, SBoost, and RRBoost applied with tree learners of $d$ = 2 for clean ($D_0$), symmetric gross error contaminated ($D_1$), asymmetric gross error contaminated ($D_2$),  skewed distributed ($D_3$),  and heavy-tailed distributed  ($D_4$) data generated from $g$ = $g_1$ S = $S_2$ $n$ = 300 $p$ = 400, displayed in the form of: mean (SD) calculated from 100 independent runs of the experiment.} 
\end{table}
\begin{table}[H]
\centering
\begin{tabular}{lccccc}
  \hline
 & L2Boost & MBoost & LADBoost & Robloss & RRBoost \\ 
  \hline
$D_0$ & 0.57 (0.19) & 0.58 (0.17) & 0.51 (0.15) & 0.56 (0.16) & 0.61 (0.15) \\ 
  $D_1 (10\%)$ & 0.13 (0.11) & 0.25 (0.15) & 0.41 (0.20) & 0.44 (0.18) & 0.59 (0.15) \\ 
  $D_1 (20\%)$ & 0.07 (0.10) & 0.06 (0.10) & 0.28 (0.17) & 0.30 (0.17) & 0.52 (0.15) \\ 
  $D_2 (10\%)$ & 0.14 (0.11) & 0.26 (0.14) & 0.43 (0.18) & 0.47 (0.18) & 0.59 (0.15) \\ 
  $D_2 (20\%)$ & 0.09 (0.11) & 0.08 (0.12) & 0.34 (0.17) & 0.27 (0.13) & 0.55 (0.14) \\ 
  $D_3$ & 0.43 (0.17) & 0.55 (0.17) & 0.51 (0.17) & 0.50 (0.15) & 0.57 (0.18) \\ 
  $D_4$ & 0.05 (0.09) & 0.30 (0.14) & 0.31 (0.18) & 0.37 (0.16) & 0.44 (0.16) \\ 
   \hline
\end{tabular}
\caption{Fractions of variables recovered by L2Boost, MBoost, LADBoost, Robloss, and RRBoost applied with tree learners of $d$ = 2 for clean ($D_0$), symmetric gross error contaminated ($D_1$), asymmetric gross error contaminated ($D_2$),  skewed distributed ($D_3$),  and heavy-tailed distributed  ($D_4$) data generated from $g$ = $g_1$ S = $S_2$ $n$ = 300 $p$ = 400, displayed in the form of: mean (SD) calculated from 100 independent runs of the experiment.} 
\end{table}
\begin{table}[H]
\centering
\begin{tabular}{lcccccc}
  \hline
 & L2Boost & MBoost & LADBoost & Robloss & SBoost & RRBoost \\ 
  \hline
$D_0$ & 1.24 (0.10) & 1.26 (0.11) & 1.43 (0.14) & 1.27 (0.09) & 1.46 (0.16) & 1.24 (0.09) \\ 
  $D_1 (10\%)$ & 2.20 (0.26) & 1.84 (0.31) & 1.58 (0.17) & 1.45 (0.18) & 1.49 (0.16) & 1.29 (0.12) \\ 
  $D_1 (20\%)$ & 2.79 (0.34) & 2.78 (0.30) & 1.73 (0.16) & 1.74 (0.26) & 1.56 (0.17) & 1.36 (0.13) \\ 
  $D_2 (10\%)$ & 2.38 (0.24) & 2.00 (0.39) & 1.56 (0.16) & 1.46 (0.19) & 1.50 (0.14) & 1.26 (0.11) \\ 
  $D_2 (20\%)$ & 3.23 (0.32) & 3.31 (0.33) & 1.73 (0.19) & 1.82 (0.29) & 1.56 (0.17) & 1.37 (0.15) \\ 
  $D_3$ & 1.40 (0.14) & 1.33 (0.13) & 1.50 (0.17) & 1.35 (0.12) & 1.48 (0.17) & 1.29 (0.11) \\ 
  $D_4$ & 13.56 (31.49) & 1.59 (0.20) & 1.68 (0.17) & 1.51 (0.18) & 1.54 (0.18) & 1.45 (0.15) \\ 
   \hline
\end{tabular}
\caption{Summary statistics of RMSEs on the test sets by L2Boost, MBoost, LADBoost, Robloss, SBoost, and RRBoost applied with tree learners of $d$ = 3 for clean ($D_0$), symmetric gross error contaminated ($D_1$), asymmetric gross error contaminated ($D_2$),  skewed distributed ($D_3$),  and heavy-tailed distributed  ($D_4$) data generated from $g$ = $g_1$ S = $S_2$ $n$ = 300 $p$ = 400, displayed in the form of: mean (SD) calculated from 100 independent runs of the experiment.} 
\end{table}
\begin{table}[H]
\centering
\begin{tabular}{lccccc}
  \hline
 & L2Boost & MBoost & LADBoost & Robloss & RRBoost \\ 
  \hline
$D_0$ & 0.48 (0.16) & 0.51 (0.17) & 0.51 (0.18) & 0.54 (0.16) & 0.57 (0.16) \\ 
  $D_1 (10\%)$ & 0.15 (0.12) & 0.25 (0.14) & 0.37 (0.19) & 0.42 (0.19) & 0.57 (0.15) \\ 
  $D_1 (20\%)$ & 0.08 (0.12) & 0.06 (0.11) & 0.28 (0.18) & 0.30 (0.15) & 0.52 (0.16) \\ 
  $D_2 (10\%)$ & 0.16 (0.12) & 0.26 (0.11) & 0.41 (0.19) & 0.43 (0.19) & 0.57 (0.16) \\ 
  $D_2 (20\%)$ & 0.10 (0.13) & 0.08 (0.12) & 0.30 (0.17) & 0.27 (0.13) & 0.53 (0.15) \\ 
  $D_3$ & 0.40 (0.17) & 0.46 (0.19) & 0.45 (0.17) & 0.48 (0.17) & 0.54 (0.14) \\ 
  $D_4$ & 0.09 (0.13) & 0.30 (0.12) & 0.30 (0.17) & 0.35 (0.16) & 0.42 (0.17) \\ 
   \hline
\end{tabular}
\caption{Fractions of variables recovered by L2Boost, MBoost, LADBoost, Robloss, and RRBoost applied with tree learners of $d$ = 3 for clean ($D_0$), symmetric gross error contaminated ($D_1$), asymmetric gross error contaminated ($D_2$),  skewed distributed ($D_3$),  and heavy-tailed distributed  ($D_4$) data generated from $g$ = $g_1$ S = $S_2$ $n$ = 300 $p$ = 400, displayed in the form of: mean (SD) calculated from 100 independent runs of the experiment.} 
\end{table}
\begin{table}[H]
\centering
\begin{tabular}{lcccccc}
  \hline
 & L2Boost & MBoost & LADBoost & Robloss & SBoost & RRBoost \\ 
  \hline
$D_0$ & 0.64 (0.02) & 0.65 (0.02) & 0.69 (0.02) & 0.65 (0.02) & 1.37 (0.23) & 0.63 (0.02) \\ 
  $D_1 (10\%)$ & 1.18 (0.06) & 0.94 (0.17) & 0.72 (0.02) & 0.69 (0.02) & 1.34 (0.22) & 0.65 (0.15) \\ 
  $D_1 (20\%)$ & 1.28 (0.08) & 1.30 (0.08) & 0.75 (0.03) & 0.76 (0.03) & 1.14 (0.24) & 0.64 (0.02) \\ 
  $D_2 (10\%)$ & 1.54 (0.06) & 1.09 (0.25) & 0.73 (0.03) & 0.70 (0.02) & 1.28 (0.23) & 0.65 (0.10) \\ 
  $D_2 (20\%)$ & 2.34 (0.06) & 2.30 (0.08) & 0.82 (0.05) & 0.87 (0.08) & 1.06 (0.23) & 0.64 (0.02) \\ 
  $D_3$ & 0.83 (0.07) & 0.69 (0.02) & 0.73 (0.02) & 0.70 (0.02) & 1.44 (0.23) & 0.67 (0.03) \\ 
  $D_4$ & 59.20 (288.14) & 0.88 (0.04) & 0.78 (0.03) & 0.77 (0.03) & 1.16 (0.19) & 0.74 (0.03) \\ 
   \hline
\end{tabular}
\caption{Summary statistics of RMSEs on the test sets by L2Boost, MBoost, LADBoost, Robloss, SBoost, and RRBoost applied with tree learners of $d$ = 1 for clean ($D_0$), symmetric gross error contaminated ($D_1$), asymmetric gross error contaminated ($D_2$),  skewed distributed ($D_3$),  and heavy-tailed distributed  ($D_4$) data generated from $g$ = $g_1$ S = $S_2$ $n$ = 3000 $p$ = 400, displayed in the form of: mean (SD) calculated from 100 independent runs of the experiment.} 
\end{table}
\begin{table}[H]
\centering
\begin{tabular}{lccccc}
  \hline
 & L2Boost & MBoost & LADBoost & Robloss & RRBoost \\ 
  \hline
$D_0$ & 1.00 (0.00) & 1.00 (0.00) & 1.00 (0.00) & 1.00 (0.00) & 1.00 (0.00) \\ 
  $D_1 (10\%)$ & 0.61 (0.15) & 0.91 (0.17) & 1.00 (0.00) & 1.00 (0.00) & 1.00 (0.00) \\ 
  $D_1 (20\%)$ & 0.57 (0.16) & 0.57 (0.16) & 1.00 (0.00) & 1.00 (0.00) & 1.00 (0.00) \\ 
  $D_2 (10\%)$ & 0.58 (0.16) & 0.92 (0.12) & 1.00 (0.00) & 1.00 (0.00) & 1.00 (0.00) \\ 
  $D_2 (20\%)$ & 0.57 (0.17) & 0.57 (0.16) & 1.00 (0.00) & 0.98 (0.06) & 1.00 (0.00) \\ 
  $D_3$ & 1.00 (0.03) & 1.00 (0.00) & 1.00 (0.00) & 1.00 (0.00) & 1.00 (0.00) \\ 
  $D_4$ & 0.03 (0.08) & 1.00 (0.02) & 1.00 (0.00) & 1.00 (0.00) & 1.00 (0.00) \\ 
   \hline
\end{tabular}
\caption{Fractions of variables recovered by L2Boost, MBoost, LADBoost, Robloss, and RRBoost applied with tree learners of $d$ = 1 for clean ($D_0$), symmetric gross error contaminated ($D_1$), asymmetric gross error contaminated ($D_2$),  skewed distributed ($D_3$),  and heavy-tailed distributed  ($D_4$) data generated from $g$ = $g_1$ S = $S_2$ $n$ = 3000 $p$ = 400, displayed in the form of: mean (SD) calculated from 100 independent runs of the experiment.} 
\end{table}
\begin{table}[H]
\centering
\begin{tabular}{lcccccc}
  \hline
 & L2Boost & MBoost & LADBoost & Robloss & SBoost & RRBoost \\ 
  \hline
$D_0$ & 0.69 (0.02) & 0.69 (0.02) & 0.75 (0.03) & 0.70 (0.03) & 1.19 (0.18) & 0.66 (0.07) \\ 
  $D_1 (10\%)$ & 1.20 (0.08) & 0.98 (0.14) & 0.79 (0.03) & 0.76 (0.03) & 1.08 (0.14) & 0.71 (0.38) \\ 
  $D_1 (20\%)$ & 1.36 (0.11) & 1.39 (0.12) & 0.84 (0.03) & 0.84 (0.04) & 1.01 (0.17) & 0.76 (0.54) \\ 
  $D_2 (10\%)$ & 1.57 (0.09) & 1.11 (0.22) & 0.79 (0.03) & 0.75 (0.03) & 1.05 (0.14) & 0.68 (0.29) \\ 
  $D_2 (20\%)$ & 2.40 (0.07) & 2.31 (0.09) & 0.87 (0.05) & 0.89 (0.04) & 0.94 (0.15) & 0.66 (0.15) \\ 
  $D_3$ & 0.84 (0.05) & 0.73 (0.03) & 0.79 (0.03) & 0.74 (0.03) & 1.21 (0.20) & 0.70 (0.05) \\ 
  $D_4$ & 55.24 (260.59) & 0.90 (0.05) & 0.85 (0.04) & 0.82 (0.04) & 1.07 (0.16) & 0.73 (0.05) \\ 
   \hline
\end{tabular}
\caption{Summary statistics of RMSEs on the test sets by L2Boost, MBoost, LADBoost, Robloss, SBoost, and RRBoost applied with tree learners of $d$ = 2 for clean ($D_0$), symmetric gross error contaminated ($D_1$), asymmetric gross error contaminated ($D_2$),  skewed distributed ($D_3$),  and heavy-tailed distributed  ($D_4$) data generated from $g$ = $g_1$ S = $S_2$ $n$ = 3000 $p$ = 400, displayed in the form of: mean (SD) calculated from 100 independent runs of the experiment.} 
\end{table}
\begin{table}[H]
\centering
\begin{tabular}{lccccc}
  \hline
 & L2Boost & MBoost & LADBoost & Robloss & RRBoost \\ 
  \hline
$D_0$ & 1.00 (0.00) & 1.00 (0.00) & 1.00 (0.00) & 1.00 (0.00) & 1.00 (0.02) \\ 
  $D_1 (10\%)$ & 0.53 (0.15) & 0.86 (0.21) & 1.00 (0.00) & 1.00 (0.00) & 1.00 (0.03) \\ 
  $D_1 (20\%)$ & 0.45 (0.18) & 0.45 (0.16) & 1.00 (0.00) & 1.00 (0.00) & 1.00 (0.02) \\ 
  $D_2 (10\%)$ & 0.50 (0.15) & 0.85 (0.19) & 1.00 (0.00) & 1.00 (0.00) & 1.00 (0.00) \\ 
  $D_2 (20\%)$ & 0.46 (0.16) & 0.45 (0.16) & 1.00 (0.00) & 1.00 (0.00) & 1.00 (0.03) \\ 
  $D_3$ & 0.98 (0.11) & 1.00 (0.00) & 1.00 (0.00) & 1.00 (0.00) & 0.99 (0.05) \\ 
  $D_4$ & 0.02 (0.07) & 0.99 (0.04) & 1.00 (0.00) & 1.00 (0.00) & 1.00 (0.00) \\ 
   \hline
\end{tabular}
\caption{Fractions of variables recovered by L2Boost, MBoost, LADBoost, Robloss, and RRBoost applied with tree learners of $d$ = 2 for clean ($D_0$), symmetric gross error contaminated ($D_1$), asymmetric gross error contaminated ($D_2$),  skewed distributed ($D_3$),  and heavy-tailed distributed  ($D_4$) data generated from $g$ = $g_1$ S = $S_2$ $n$ = 3000 $p$ = 400, displayed in the form of: mean (SD) calculated from 100 independent runs of the experiment.} 
\end{table}
\begin{table}[H]
\centering
\begin{tabular}{lcccccc}
  \hline
 & L2Boost & MBoost & LADBoost & Robloss & SBoost & RRBoost \\ 
  \hline
$D_0$ & 0.72 (0.03) & 0.71 (0.03) & 0.80 (0.03) & 0.74 (0.03) & 1.13 (0.17) & 0.70 (0.06) \\ 
  $D_1 (10\%)$ & 1.28 (0.07) & 1.06 (0.13) & 0.85 (0.03) & 0.82 (0.03) & 1.03 (0.11) & 0.69 (0.03) \\ 
  $D_1 (20\%)$ & 1.44 (0.11) & 1.45 (0.12) & 0.92 (0.04) & 0.91 (0.05) & 0.96 (0.10) & 0.69 (0.03) \\ 
  $D_2 (10\%)$ & 1.64 (0.09) & 1.19 (0.19) & 0.86 (0.04) & 0.81 (0.03) & 1.04 (0.13) & 0.70 (0.03) \\ 
  $D_2 (20\%)$ & 2.45 (0.08) & 2.39 (0.12) & 0.98 (0.05) & 0.99 (0.05) & 0.95 (0.11) & 0.69 (0.02) \\ 
  $D_3$ & 0.90 (0.06) & 0.76 (0.03) & 0.84 (0.04) & 0.78 (0.03) & 1.12 (0.14) & 0.74 (0.06) \\ 
  $D_4$ & 56.23 (257.88) & 0.97 (0.04) & 0.91 (0.04) & 0.88 (0.04) & 1.04 (0.12) & 0.78 (0.03) \\ 
   \hline
\end{tabular}
\caption{Summary statistics of RMSEs on the test sets by L2Boost, MBoost, LADBoost, Robloss, SBoost, and RRBoost applied with tree learners of $d$ = 3 for clean ($D_0$), symmetric gross error contaminated ($D_1$), asymmetric gross error contaminated ($D_2$),  skewed distributed ($D_3$),  and heavy-tailed distributed  ($D_4$) data generated from $g$ = $g_1$ S = $S_2$ $n$ = 3000 $p$ = 400, displayed in the form of: mean (SD) calculated from 100 independent runs of the experiment.} 
\end{table}
\begin{table}[H]
\centering
\begin{tabular}{lccccc}
  \hline
 & L2Boost & MBoost & LADBoost & Robloss & RRBoost \\ 
  \hline
$D_0$ & 1.00 (0.00) & 1.00 (0.00) & 1.00 (0.00) & 1.00 (0.00) & 1.00 (0.00) \\ 
  $D_1 (10\%)$ & 0.42 (0.17) & 0.77 (0.23) & 1.00 (0.00) & 1.00 (0.00) & 1.00 (0.00) \\ 
  $D_1 (20\%)$ & 0.34 (0.17) & 0.36 (0.17) & 1.00 (0.00) & 1.00 (0.02) & 1.00 (0.00) \\ 
  $D_2 (10\%)$ & 0.37 (0.17) & 0.75 (0.21) & 1.00 (0.00) & 1.00 (0.00) & 1.00 (0.00) \\ 
  $D_2 (20\%)$ & 0.34 (0.15) & 0.49 (0.12) & 1.00 (0.00) & 0.99 (0.05) & 1.00 (0.00) \\ 
  $D_3$ & 0.94 (0.14) & 1.00 (0.00) & 1.00 (0.00) & 1.00 (0.00) & 1.00 (0.00) \\ 
  $D_4$ & 0.04 (0.09) & 0.92 (0.14) & 1.00 (0.04) & 1.00 (0.02) & 1.00 (0.00) \\ 
   \hline
\end{tabular}
\caption{Fractions of variables recovered by L2Boost, MBoost, LADBoost, Robloss, and RRBoost applied with tree learners of $d$ = 3 for clean ($D_0$), symmetric gross error contaminated ($D_1$), asymmetric gross error contaminated ($D_2$),  skewed distributed ($D_3$),  and heavy-tailed distributed  ($D_4$) data generated from $g$ = $g_1$ S = $S_2$ $n$ = 3000 $p$ = 400, displayed in the form of: mean (SD) calculated from 100 independent runs of the experiment.} 
\end{table}

\subsection{Function $g_2$}
\begin{table}[H]
\centering
\begin{tabular}{lcccccc}
  \hline
 & L2Boost & MBoost & LADBoost & Robloss & SBoost & RRBoost \\ 
  \hline
$D_0$ & 1.25 (0.04) & 1.25 (0.04) & 1.25 (0.05) & 1.24 (0.04) & 1.23 (0.12) & 1.17 (0.06) \\ 
  $D_1 (10\%)$ & 1.73 (0.22) & 1.48 (0.18) & 1.27 (0.05) & 1.29 (0.06) & 1.26 (0.11) & 1.20 (0.08) \\ 
  $D_1 (20\%)$ & 1.99 (0.33) & 1.91 (0.22) & 1.30 (0.05) & 1.33 (0.06) & 1.29 (0.10) & 1.22 (0.06) \\ 
  $D_2 (10\%)$ & 2.16 (0.22) & 1.70 (0.36) & 1.27 (0.05) & 1.30 (0.06) & 1.26 (0.11) & 1.20 (0.06) \\ 
  $D_2 (20\%)$ & 3.33 (0.30) & 3.32 (0.30) & 1.37 (0.09) & 1.47 (0.13) & 1.30 (0.10) & 1.23 (0.08) \\ 
  $D_3$ & 1.34 (0.09) & 1.29 (0.05) & 1.28 (0.05) & 1.28 (0.05) & 1.32 (0.12) & 1.22 (0.06) \\ 
  $D_4$ & 15.09 (36.44) & 1.38 (0.08) & 1.32 (0.06) & 1.35 (0.07) & 1.32 (0.09) & 1.28 (0.07) \\ 
   \hline
\end{tabular}
\caption{Summary statistics of RMSEs on the test sets by L2Boost, MBoost, LADBoost, Robloss, SBoost, and RRBoost applied with tree learners of $d$ = 1 for clean ($D_0$), symmetric gross error contaminated ($D_1$), asymmetric gross error contaminated ($D_2$),  skewed distributed ($D_3$),  and heavy-tailed distributed  ($D_4$) data generated from $g$ = $g_2$ S = $S_0$ $n$ = 300 $p$ = 10, displayed in the form of: mean (SD) calculated from 100 independent runs of the experiment.} 
\end{table}
\begin{table}[H]
\centering
\begin{tabular}{lccccc}
  \hline
 & L2Boost & MBoost & LADBoost & Robloss & RRBoost \\ 
  \hline
$D_0$ & 0.80 (0.10) & 0.82 (0.11) & 0.80 (0.11) & 0.80 (0.10) & 0.80 (0.10) \\ 
  $D_1 (10\%)$ & 0.46 (0.20) & 0.61 (0.20) & 0.80 (0.11) & 0.78 (0.10) & 0.79 (0.09) \\ 
  $D_1 (20\%)$ & 0.42 (0.21) & 0.34 (0.19) & 0.79 (0.09) & 0.76 (0.11) & 0.81 (0.11) \\ 
  $D_2 (10\%)$ & 0.55 (0.19) & 0.69 (0.13) & 0.78 (0.09) & 0.78 (0.09) & 0.79 (0.09) \\ 
  $D_2 (20\%)$ & 0.45 (0.21) & 0.53 (0.20) & 0.75 (0.10) & 0.73 (0.12) & 0.79 (0.09) \\ 
  $D_3$ & 0.78 (0.12) & 0.80 (0.10) & 0.78 (0.09) & 0.78 (0.09) & 0.80 (0.10) \\ 
  $D_4$ & 0.34 (0.24) & 0.69 (0.12) & 0.73 (0.10) & 0.72 (0.13) & 0.78 (0.08) \\ 
   \hline
\end{tabular}
\caption{Fractions of variables recovered by L2Boost, MBoost, LADBoost, Robloss, and RRBoost applied with tree learners of $d$ = 1 for clean ($D_0$), symmetric gross error contaminated ($D_1$), asymmetric gross error contaminated ($D_2$),  skewed distributed ($D_3$),  and heavy-tailed distributed  ($D_4$) data generated from $g$ = $g_2$ S = $S_0$ $n$ = 300 $p$ = 10, displayed in the form of: mean (SD) calculated from 100 independent runs of the experiment.} 
\end{table}
\begin{table}[H]
\centering
\begin{tabular}{lcccccc}
  \hline
 & L2Boost & MBoost & LADBoost & Robloss & SBoost & RRBoost \\ 
  \hline
$D_0$ & 1.12 (0.06) & 1.12 (0.06) & 1.10 (0.06) & 1.10 (0.06) & 1.29 (0.11) & 1.09 (0.06) \\ 
  $D_1 (10\%)$ & 1.87 (0.27) & 1.52 (0.29) & 1.16 (0.07) & 1.19 (0.08) & 1.27 (0.10) & 1.09 (0.07) \\ 
  $D_1 (20\%)$ & 2.29 (0.37) & 2.28 (0.30) & 1.25 (0.08) & 1.29 (0.10) & 1.25 (0.09) & 1.11 (0.05) \\ 
  $D_2 (10\%)$ & 2.33 (0.24) & 1.77 (0.46) & 1.18 (0.07) & 1.22 (0.09) & 1.25 (0.09) & 1.11 (0.07) \\ 
  $D_2 (20\%)$ & 3.48 (0.33) & 3.53 (0.30) & 1.34 (0.11) & 1.49 (0.18) & 1.24 (0.07) & 1.13 (0.06) \\ 
  $D_3$ & 1.29 (0.10) & 1.19 (0.07) & 1.15 (0.06) & 1.17 (0.07) & 1.32 (0.11) & 1.14 (0.06) \\ 
  $D_4$ & 15.11 (30.02) & 1.40 (0.12) & 1.28 (0.10) & 1.31 (0.11) & 1.29 (0.11) & 1.20 (0.09) \\ 
   \hline
\end{tabular}
\caption{Summary statistics of RMSEs on the test sets by L2Boost, MBoost, LADBoost, Robloss, SBoost, and RRBoost applied with tree learners of $d$ = 2 for clean ($D_0$), symmetric gross error contaminated ($D_1$), asymmetric gross error contaminated ($D_2$),  skewed distributed ($D_3$),  and heavy-tailed distributed  ($D_4$) data generated from $g$ = $g_2$ S = $S_0$ $n$ = 300 $p$ = 10, displayed in the form of: mean (SD) calculated from 100 independent runs of the experiment.} 
\end{table}
\begin{table}[H]
\centering
\begin{tabular}{lccccc}
  \hline
 & L2Boost & MBoost & LADBoost & Robloss & RRBoost \\ 
  \hline
$D_0$ & 0.85 (0.12) & 0.84 (0.12) & 0.85 (0.12) & 0.85 (0.12) & 0.86 (0.12) \\ 
  $D_1 (10\%)$ & 0.56 (0.18) & 0.71 (0.15) & 0.81 (0.11) & 0.80 (0.10) & 0.84 (0.12) \\ 
  $D_1 (20\%)$ & 0.51 (0.20) & 0.48 (0.18) & 0.81 (0.11) & 0.76 (0.07) & 0.84 (0.12) \\ 
  $D_2 (10\%)$ & 0.61 (0.16) & 0.72 (0.15) & 0.80 (0.10) & 0.79 (0.10) & 0.84 (0.12) \\ 
  $D_2 (20\%)$ & 0.54 (0.20) & 0.58 (0.20) & 0.78 (0.09) & 0.77 (0.12) & 0.86 (0.12) \\ 
  $D_3$ & 0.76 (0.09) & 0.81 (0.11) & 0.84 (0.12) & 0.83 (0.12) & 0.84 (0.12) \\ 
  $D_4$ & 0.38 (0.20) & 0.72 (0.13) & 0.80 (0.11) & 0.74 (0.12) & 0.79 (0.09) \\ 
   \hline
\end{tabular}
\caption{Fractions of variables recovered by L2Boost, MBoost, LADBoost, Robloss, and RRBoost applied with tree learners of $d$ = 2 for clean ($D_0$), symmetric gross error contaminated ($D_1$), asymmetric gross error contaminated ($D_2$),  skewed distributed ($D_3$),  and heavy-tailed distributed  ($D_4$) data generated from $g$ = $g_2$ S = $S_0$ $n$ = 300 $p$ = 10, displayed in the form of: mean (SD) calculated from 100 independent runs of the experiment.} 
\end{table}
\begin{table}[H]
\centering
\begin{tabular}{lcccccc}
  \hline
 & L2Boost & MBoost & LADBoost & Robloss & SBoost & RRBoost \\ 
  \hline
$D_0$ & 1.12 (0.06) & 1.11 (0.05) & 1.14 (0.05) & 1.12 (0.05) & 1.28 (0.11) & 1.13 (0.05) \\ 
  $D_1 (10\%)$ & 2.13 (0.26) & 1.64 (0.37) & 1.21 (0.06) & 1.22 (0.07) & 1.30 (0.11) & 1.14 (0.06) \\ 
  $D_1 (20\%)$ & 2.75 (0.37) & 2.77 (0.36) & 1.29 (0.08) & 1.36 (0.10) & 1.28 (0.08) & 1.15 (0.06) \\ 
  $D_2 (10\%)$ & 2.56 (0.24) & 1.92 (0.58) & 1.23 (0.07) & 1.22 (0.07) & 1.28 (0.10) & 1.14 (0.07) \\ 
  $D_2 (20\%)$ & 3.74 (0.29) & 3.80 (0.28) & 1.40 (0.10) & 1.54 (0.14) & 1.27 (0.09) & 1.17 (0.06) \\ 
  $D_3$ & 1.28 (0.12) & 1.19 (0.06) & 1.19 (0.06) & 1.18 (0.05) & 1.32 (0.11) & 1.18 (0.06) \\ 
  $D_4$ & 15.55 (30.11) & 1.45 (0.14) & 1.33 (0.08) & 1.32 (0.09) & 1.31 (0.10) & 1.27 (0.08) \\ 
   \hline
\end{tabular}
\caption{Summary statistics of RMSEs on the test sets by L2Boost, MBoost, LADBoost, Robloss, SBoost, and RRBoost applied with tree learners of $d$ = 3 for clean ($D_0$), symmetric gross error contaminated ($D_1$), asymmetric gross error contaminated ($D_2$),  skewed distributed ($D_3$),  and heavy-tailed distributed  ($D_4$) data generated from $g$ = $g_2$ S = $S_0$ $n$ = 300 $p$ = 10, displayed in the form of: mean (SD) calculated from 100 independent runs of the experiment.} 
\end{table}
\begin{table}[H]
\centering
\begin{tabular}{lccccc}
  \hline
 & L2Boost & MBoost & LADBoost & Robloss & RRBoost \\ 
  \hline
$D_0$ & 0.86 (0.12) & 0.86 (0.12) & 0.86 (0.12) & 0.86 (0.12) & 0.88 (0.13) \\ 
  $D_1 (10\%)$ & 0.67 (0.17) & 0.74 (0.14) & 0.82 (0.11) & 0.78 (0.09) & 0.86 (0.12) \\ 
  $D_1 (20\%)$ & 0.60 (0.18) & 0.57 (0.18) & 0.80 (0.10) & 0.78 (0.09) & 0.85 (0.12) \\ 
  $D_2 (10\%)$ & 0.67 (0.15) & 0.74 (0.12) & 0.82 (0.12) & 0.82 (0.11) & 0.86 (0.12) \\ 
  $D_2 (20\%)$ & 0.63 (0.19) & 0.61 (0.20) & 0.80 (0.10) & 0.78 (0.08) & 0.84 (0.12) \\ 
  $D_3$ & 0.80 (0.10) & 0.82 (0.11) & 0.84 (0.12) & 0.84 (0.12) & 0.84 (0.12) \\ 
  $D_4$ & 0.53 (0.21) & 0.76 (0.09) & 0.78 (0.10) & 0.78 (0.08) & 0.80 (0.10) \\ 
   \hline
\end{tabular}
\caption{Fractions of variables recovered by L2Boost, MBoost, LADBoost, Robloss, and RRBoost applied with tree learners of $d$ = 3 for clean ($D_0$), symmetric gross error contaminated ($D_1$), asymmetric gross error contaminated ($D_2$),  skewed distributed ($D_3$),  and heavy-tailed distributed  ($D_4$) data generated from $g$ = $g_2$ S = $S_0$ $n$ = 300 $p$ = 10, displayed in the form of: mean (SD) calculated from 100 independent runs of the experiment.} 
\end{table}
\begin{table}[H]
\centering
\begin{tabular}{lcccccc}
  \hline
 & L2Boost & MBoost & LADBoost & Robloss & SBoost & RRBoost \\ 
  \hline
$D_0$ & 1.10 (0.02) & 1.10 (0.02) & 1.10 (0.02) & 1.10 (0.02) & 1.00 (0.03) & 0.97 (0.03) \\ 
  $D_1 (10\%)$ & 1.29 (0.05) & 1.20 (0.07) & 1.12 (0.03) & 1.12 (0.03) & 1.00 (0.03) & 0.98 (0.03) \\ 
  $D_1 (20\%)$ & 1.36 (0.08) & 1.35 (0.06) & 1.13 (0.03) & 1.14 (0.03) & 1.01 (0.03) & 1.00 (0.03) \\ 
  $D_2 (10\%)$ & 1.88 (0.06) & 1.42 (0.24) & 1.13 (0.02) & 1.13 (0.03) & 1.00 (0.03) & 0.98 (0.03) \\ 
  $D_2 (20\%)$ & 3.06 (0.08) & 2.99 (0.10) & 1.21 (0.06) & 1.26 (0.08) & 1.00 (0.03) & 0.99 (0.03) \\ 
  $D_3$ & 1.14 (0.03) & 1.12 (0.02) & 1.14 (0.02) & 1.13 (0.02) & 1.12 (0.04) & 1.03 (0.03) \\ 
  $D_4$ & 10.52 (30.16) & 1.17 (0.03) & 1.13 (0.03) & 1.14 (0.03) & 1.01 (0.03) & 1.00 (0.03) \\ 
   \hline
\end{tabular}
\caption{Summary statistics of RMSEs on the test sets by L2Boost, MBoost, LADBoost, Robloss, SBoost, and RRBoost applied with tree learners of $d$ = 1 for clean ($D_0$), symmetric gross error contaminated ($D_1$), asymmetric gross error contaminated ($D_2$),  skewed distributed ($D_3$),  and heavy-tailed distributed  ($D_4$) data generated from $g$ = $g_2$ S = $S_0$ $n$ = 3000 $p$ = 10, displayed in the form of: mean (SD) calculated from 100 independent runs of the experiment.} 
\end{table}
\begin{table}[H]
\centering
\begin{tabular}{lccccc}
  \hline
 & L2Boost & MBoost & LADBoost & Robloss & RRBoost \\ 
  \hline
$D_0$ & 0.98 (0.07) & 0.97 (0.08) & 0.98 (0.08) & 0.98 (0.08) & 0.98 (0.06) \\ 
  $D_1 (10\%)$ & 0.77 (0.09) & 0.87 (0.13) & 0.96 (0.09) & 0.96 (0.09) & 0.98 (0.08) \\ 
  $D_1 (20\%)$ & 0.72 (0.10) & 0.72 (0.08) & 0.95 (0.10) & 0.94 (0.11) & 0.98 (0.08) \\ 
  $D_2 (10\%)$ & 0.77 (0.08) & 0.86 (0.13) & 0.95 (0.10) & 0.94 (0.10) & 0.98 (0.07) \\ 
  $D_2 (20\%)$ & 0.74 (0.09) & 0.63 (0.16) & 0.91 (0.12) & 0.87 (0.13) & 0.95 (0.10) \\ 
  $D_3$ & 0.95 (0.10) & 0.95 (0.10) & 0.95 (0.10) & 0.97 (0.08) & 0.97 (0.08) \\ 
  $D_4$ & 0.22 (0.25) & 0.88 (0.13) & 0.90 (0.12) & 0.91 (0.12) & 0.92 (0.12) \\ 
   \hline
\end{tabular}
\caption{Fractions of variables recovered by L2Boost, MBoost, LADBoost, Robloss, and RRBoost applied with tree learners of $d$ = 1 for clean ($D_0$), symmetric gross error contaminated ($D_1$), asymmetric gross error contaminated ($D_2$),  skewed distributed ($D_3$),  and heavy-tailed distributed  ($D_4$) data generated from $g$ = $g_2$ S = $S_0$ $n$ = 3000 $p$ = 10, displayed in the form of: mean (SD) calculated from 100 independent runs of the experiment.} 
\end{table}
\begin{table}[H]
\centering
\begin{tabular}{lcccccc}
  \hline
 & L2Boost & MBoost & LADBoost & Robloss & SBoost & RRBoost \\ 
  \hline
$D_0$ & 0.87 (0.02) & 0.87 (0.03) & 0.88 (0.03) & 0.87 (0.03) & 0.86 (0.03) & 0.80 (0.02) \\ 
  $D_1 (10\%)$ & 1.23 (0.06) & 1.07 (0.14) & 0.90 (0.02) & 0.90 (0.03) & 0.85 (0.03) & 0.81 (0.02) \\ 
  $D_1 (20\%)$ & 1.35 (0.09) & 1.36 (0.08) & 0.93 (0.03) & 0.94 (0.03) & 0.84 (0.02) & 0.81 (0.02) \\ 
  $D_2 (10\%)$ & 1.85 (0.07) & 1.30 (0.31) & 0.91 (0.03) & 0.92 (0.03) & 0.85 (0.02) & 0.81 (0.02) \\ 
  $D_2 (20\%)$ & 3.04 (0.08) & 2.98 (0.10) & 1.00 (0.04) & 1.04 (0.04) & 0.83 (0.02) & 0.81 (0.02) \\ 
  $D_3$ & 0.98 (0.05) & 0.92 (0.03) & 0.94 (0.02) & 0.92 (0.02) & 0.99 (0.03) & 0.90 (0.02) \\ 
  $D_4$ & 67.17 (341.35) & 1.03 (0.04) & 0.94 (0.03) & 0.95 (0.04) & 0.84 (0.02) & 0.83 (0.02) \\ 
   \hline
\end{tabular}
\caption{Summary statistics of RMSEs on the test sets by L2Boost, MBoost, LADBoost, Robloss, SBoost, and RRBoost applied with tree learners of $d$ = 2 for clean ($D_0$), symmetric gross error contaminated ($D_1$), asymmetric gross error contaminated ($D_2$),  skewed distributed ($D_3$),  and heavy-tailed distributed  ($D_4$) data generated from $g$ = $g_2$ S = $S_0$ $n$ = 3000 $p$ = 10, displayed in the form of: mean (SD) calculated from 100 independent runs of the experiment.} 
\end{table}
\begin{table}[H]
\centering
\begin{tabular}{lccccc}
  \hline
 & L2Boost & MBoost & LADBoost & Robloss & RRBoost \\ 
  \hline
$D_0$ & 1.00 (0.00) & 1.00 (0.00) & 1.00 (0.00) & 1.00 (0.00) & 1.00 (0.00) \\ 
  $D_1 (10\%)$ & 0.77 (0.07) & 0.90 (0.13) & 1.00 (0.00) & 1.00 (0.02) & 1.00 (0.00) \\ 
  $D_1 (20\%)$ & 0.74 (0.11) & 0.74 (0.11) & 1.00 (0.00) & 1.00 (0.02) & 1.00 (0.00) \\ 
  $D_2 (10\%)$ & 0.76 (0.07) & 0.92 (0.12) & 1.00 (0.00) & 1.00 (0.00) & 1.00 (0.00) \\ 
  $D_2 (20\%)$ & 0.73 (0.09) & 0.76 (0.09) & 0.99 (0.04) & 1.00 (0.04) & 1.00 (0.00) \\ 
  $D_3$ & 0.96 (0.09) & 1.00 (0.00) & 1.00 (0.00) & 1.00 (0.00) & 1.00 (0.00) \\ 
  $D_4$ & 0.27 (0.22) & 0.92 (0.12) & 0.98 (0.06) & 0.99 (0.05) & 1.00 (0.00) \\ 
   \hline
\end{tabular}
\caption{Fractions of variables recovered by L2Boost, MBoost, LADBoost, Robloss, and RRBoost applied with tree learners of $d$ = 2 for clean ($D_0$), symmetric gross error contaminated ($D_1$), asymmetric gross error contaminated ($D_2$),  skewed distributed ($D_3$),  and heavy-tailed distributed  ($D_4$) data generated from $g$ = $g_2$ S = $S_0$ $n$ = 3000 $p$ = 10, displayed in the form of: mean (SD) calculated from 100 independent runs of the experiment.} 
\end{table}
\begin{table}[H]
\centering
\begin{tabular}{lcccccc}
  \hline
 & L2Boost & MBoost & LADBoost & Robloss & SBoost & RRBoost \\ 
  \hline
$D_0$ & 0.89 (0.03) & 0.89 (0.03) & 0.90 (0.03) & 0.89 (0.03) & 0.92 (0.05) & 0.81 (0.02) \\ 
  $D_1 (10\%)$ & 1.22 (0.07) & 1.08 (0.10) & 0.93 (0.03) & 0.93 (0.03) & 0.88 (0.04) & 0.80 (0.02) \\ 
  $D_1 (20\%)$ & 1.37 (0.08) & 1.39 (0.10) & 0.96 (0.03) & 0.98 (0.04) & 0.85 (0.03) & 0.80 (0.02) \\ 
  $D_2 (10\%)$ & 1.84 (0.07) & 1.34 (0.31) & 0.94 (0.03) & 0.94 (0.03) & 0.88 (0.04) & 0.81 (0.02) \\ 
  $D_2 (20\%)$ & 3.04 (0.09) & 3.07 (0.11) & 1.03 (0.04) & 1.07 (0.05) & 0.84 (0.03) & 0.80 (0.02) \\ 
  $D_3$ & 1.02 (0.05) & 0.93 (0.03) & 0.96 (0.03) & 0.94 (0.03) & 1.03 (0.05) & 0.89 (0.02) \\ 
  $D_4$ & 68.95 (345.12) & 1.06 (0.04) & 0.98 (0.03) & 0.98 (0.03) & 0.84 (0.03) & 0.82 (0.02) \\ 
   \hline
\end{tabular}
\caption{Summary statistics of RMSEs on the test sets by L2Boost, MBoost, LADBoost, Robloss, SBoost, and RRBoost applied with tree learners of $d$ = 3 for clean ($D_0$), symmetric gross error contaminated ($D_1$), asymmetric gross error contaminated ($D_2$),  skewed distributed ($D_3$),  and heavy-tailed distributed  ($D_4$) data generated from $g$ = $g_2$ S = $S_0$ $n$ = 3000 $p$ = 10, displayed in the form of: mean (SD) calculated from 100 independent runs of the experiment.} 
\end{table}
\begin{table}[H]
\centering
\begin{tabular}{lccccc}
  \hline
 & L2Boost & MBoost & LADBoost & Robloss & RRBoost \\ 
  \hline
$D_0$ & 1.00 (0.00) & 1.00 (0.00) & 1.00 (0.00) & 1.00 (0.00) & 1.00 (0.00) \\ 
  $D_1 (10\%)$ & 0.78 (0.08) & 0.90 (0.12) & 1.00 (0.00) & 1.00 (0.00) & 1.00 (0.00) \\ 
  $D_1 (20\%)$ & 0.78 (0.09) & 0.79 (0.10) & 1.00 (0.00) & 0.99 (0.04) & 1.00 (0.00) \\ 
  $D_2 (10\%)$ & 0.77 (0.07) & 0.91 (0.12) & 1.00 (0.00) & 1.00 (0.00) & 1.00 (0.00) \\ 
  $D_2 (20\%)$ & 0.78 (0.08) & 0.80 (0.11) & 1.00 (0.04) & 0.99 (0.05) & 1.00 (0.00) \\ 
  $D_3$ & 0.96 (0.09) & 1.00 (0.00) & 1.00 (0.00) & 1.00 (0.00) & 1.00 (0.00) \\ 
  $D_4$ & 0.46 (0.23) & 0.95 (0.10) & 0.98 (0.06) & 0.99 (0.04) & 1.00 (0.00) \\ 
   \hline
\end{tabular}
\caption{Fractions of variables recovered by L2Boost, MBoost, LADBoost, Robloss, and RRBoost applied with tree learners of $d$ = 3 for clean ($D_0$), symmetric gross error contaminated ($D_1$), asymmetric gross error contaminated ($D_2$),  skewed distributed ($D_3$),  and heavy-tailed distributed  ($D_4$) data generated from $g$ = $g_2$ S = $S_0$ $n$ = 3000 $p$ = 10, displayed in the form of: mean (SD) calculated from 100 independent runs of the experiment.} 
\end{table}
\begin{table}[H]
\centering
\begin{tabular}{lcccccc}
  \hline
 & L2Boost & MBoost & LADBoost & Robloss & SBoost & RRBoost \\ 
  \hline
$D_0$ & 1.20 (0.04) & 1.20 (0.05) & 1.22 (0.05) & 1.20 (0.05) & 1.22 (0.14) & 1.13 (0.07) \\ 
  $D_1 (10\%)$ & 1.60 (0.20) & 1.40 (0.16) & 1.25 (0.07) & 1.25 (0.06) & 1.29 (0.15) & 1.17 (0.07) \\ 
  $D_1 (20\%)$ & 1.90 (0.30) & 1.91 (0.29) & 1.29 (0.08) & 1.29 (0.07) & 1.34 (0.12) & 1.19 (0.07) \\ 
  $D_2 (10\%)$ & 1.98 (0.22) & 1.58 (0.26) & 1.24 (0.06) & 1.26 (0.05) & 1.30 (0.11) & 1.17 (0.06) \\ 
  $D_2 (20\%)$ & 2.85 (0.26) & 2.86 (0.28) & 1.33 (0.08) & 1.44 (0.12) & 1.33 (0.10) & 1.19 (0.07) \\ 
  $D_3$ & 1.27 (0.07) & 1.23 (0.05) & 1.24 (0.06) & 1.23 (0.05) & 1.29 (0.13) & 1.18 (0.07) \\ 
  $D_4$ & 10.57 (14.99) & 1.32 (0.07) & 1.30 (0.07) & 1.28 (0.06) & 1.33 (0.12) & 1.24 (0.09) \\ 
   \hline
\end{tabular}
\caption{Summary statistics of RMSEs on the test sets by L2Boost, MBoost, LADBoost, Robloss, SBoost, and RRBoost applied with tree learners of $d$ = 1 for clean ($D_0$), symmetric gross error contaminated ($D_1$), asymmetric gross error contaminated ($D_2$),  skewed distributed ($D_3$),  and heavy-tailed distributed  ($D_4$) data generated from $g$ = $g_2$ S = $S_0$ $n$ = 300 $p$ = 400, displayed in the form of: mean (SD) calculated from 100 independent runs of the experiment.} 
\end{table}
\begin{table}[H]
\centering
\begin{tabular}{lccccc}
  \hline
 & L2Boost & MBoost & LADBoost & Robloss & RRBoost \\ 
  \hline
$D_0$ & 0.75 (0.05) & 0.74 (0.06) & 0.73 (0.09) & 0.75 (0.05) & 0.76 (0.05) \\ 
  $D_1 (10\%)$ & 0.39 (0.16) & 0.56 (0.18) & 0.71 (0.10) & 0.72 (0.09) & 0.74 (0.04) \\ 
  $D_1 (20\%)$ & 0.22 (0.15) & 0.20 (0.15) & 0.68 (0.12) & 0.67 (0.12) & 0.75 (0.06) \\ 
  $D_2 (10\%)$ & 0.38 (0.16) & 0.59 (0.15) & 0.72 (0.10) & 0.74 (0.07) & 0.75 (0.04) \\ 
  $D_2 (20\%)$ & 0.26 (0.19) & 0.28 (0.20) & 0.67 (0.12) & 0.64 (0.13) & 0.74 (0.04) \\ 
  $D_3$ & 0.71 (0.09) & 0.74 (0.09) & 0.74 (0.06) & 0.74 (0.07) & 0.75 (0.00) \\ 
  $D_4$ & 0.02 (0.06) & 0.63 (0.13) & 0.64 (0.12) & 0.66 (0.13) & 0.72 (0.10) \\ 
   \hline
\end{tabular}
\caption{Fractions of variables recovered by L2Boost, MBoost, LADBoost, Robloss, and RRBoost applied with tree learners of $d$ = 1 for clean ($D_0$), symmetric gross error contaminated ($D_1$), asymmetric gross error contaminated ($D_2$),  skewed distributed ($D_3$),  and heavy-tailed distributed  ($D_4$) data generated from $g$ = $g_2$ S = $S_0$ $n$ = 300 $p$ = 400, displayed in the form of: mean (SD) calculated from 100 independent runs of the experiment.} 
\end{table}
\begin{table}[H]
\centering
\begin{tabular}{lcccccc}
  \hline
 & L2Boost & MBoost & LADBoost & Robloss & SBoost & RRBoost \\ 
  \hline
$D_0$ & 1.11 (0.07) & 1.10 (0.07) & 1.16 (0.08) & 1.10 (0.07) & 1.31 (0.13) & 1.07 (0.07) \\ 
  $D_1 (10\%)$ & 1.97 (0.26) & 1.54 (0.30) & 1.23 (0.10) & 1.20 (0.10) & 1.32 (0.12) & 1.08 (0.07) \\ 
  $D_1 (20\%)$ & 2.38 (0.31) & 2.41 (0.32) & 1.31 (0.10) & 1.37 (0.13) & 1.31 (0.11) & 1.10 (0.07) \\ 
  $D_2 (10\%)$ & 2.29 (0.26) & 1.74 (0.38) & 1.22 (0.09) & 1.21 (0.09) & 1.30 (0.10) & 1.09 (0.08) \\ 
  $D_2 (20\%)$ & 3.16 (0.28) & 3.21 (0.30) & 1.40 (0.11) & 1.54 (0.18) & 1.28 (0.10) & 1.12 (0.08) \\ 
  $D_3$ & 1.24 (0.13) & 1.15 (0.08) & 1.19 (0.08) & 1.14 (0.08) & 1.32 (0.12) & 1.13 (0.08) \\ 
  $D_4$ & 9.65 (11.66) & 1.37 (0.14) & 1.34 (0.10) & 1.29 (0.10) & 1.35 (0.12) & 1.22 (0.09) \\ 
   \hline
\end{tabular}
\caption{Summary statistics of RMSEs on the test sets by L2Boost, MBoost, LADBoost, Robloss, SBoost, and RRBoost applied with tree learners of $d$ = 2 for clean ($D_0$), symmetric gross error contaminated ($D_1$), asymmetric gross error contaminated ($D_2$),  skewed distributed ($D_3$),  and heavy-tailed distributed  ($D_4$) data generated from $g$ = $g_2$ S = $S_0$ $n$ = 300 $p$ = 400, displayed in the form of: mean (SD) calculated from 100 independent runs of the experiment.} 
\end{table}
\begin{table}[H]
\centering
\begin{tabular}{lccccc}
  \hline
 & L2Boost & MBoost & LADBoost & Robloss & RRBoost \\ 
  \hline
$D_0$ & 0.75 (0.02) & 0.75 (0.02) & 0.74 (0.04) & 0.76 (0.05) & 0.75 (0.02) \\ 
  $D_1 (10\%)$ & 0.37 (0.18) & 0.59 (0.17) & 0.72 (0.09) & 0.73 (0.08) & 0.76 (0.04) \\ 
  $D_1 (20\%)$ & 0.23 (0.18) & 0.22 (0.18) & 0.67 (0.12) & 0.64 (0.13) & 0.76 (0.04) \\ 
  $D_2 (10\%)$ & 0.39 (0.17) & 0.61 (0.14) & 0.74 (0.07) & 0.72 (0.09) & 0.76 (0.05) \\ 
  $D_2 (20\%)$ & 0.26 (0.18) & 0.27 (0.19) & 0.69 (0.11) & 0.64 (0.13) & 0.75 (0.02) \\ 
  $D_3$ & 0.71 (0.10) & 0.74 (0.06) & 0.75 (0.02) & 0.75 (0.04) & 0.76 (0.04) \\ 
  $D_4$ & 0.06 (0.12) & 0.63 (0.13) & 0.67 (0.13) & 0.68 (0.11) & 0.74 (0.06) \\ 
   \hline
\end{tabular}
\caption{Fractions of variables recovered by L2Boost, MBoost, LADBoost, Robloss, and RRBoost applied with tree learners of $d$ = 2 for clean ($D_0$), symmetric gross error contaminated ($D_1$), asymmetric gross error contaminated ($D_2$),  skewed distributed ($D_3$),  and heavy-tailed distributed  ($D_4$) data generated from $g$ = $g_2$ S = $S_0$ $n$ = 300 $p$ = 400, displayed in the form of: mean (SD) calculated from 100 independent runs of the experiment.} 
\end{table}
\begin{table}[H]
\centering
\begin{tabular}{lcccccc}
  \hline
 & L2Boost & MBoost & LADBoost & Robloss & SBoost & RRBoost \\ 
  \hline
$D_0$ & 1.08 (0.06) & 1.07 (0.06) & 1.23 (0.08) & 1.11 (0.06) & 1.26 (0.13) & 1.14 (0.07) \\ 
  $D_1 (10\%)$ & 2.31 (0.27) & 1.79 (0.42) & 1.34 (0.08) & 1.28 (0.11) & 1.29 (0.12) & 1.19 (0.08) \\ 
  $D_1 (20\%)$ & 2.94 (0.32) & 2.98 (0.33) & 1.41 (0.08) & 1.51 (0.15) & 1.32 (0.10) & 1.20 (0.08) \\ 
  $D_2 (10\%)$ & 2.60 (0.28) & 2.00 (0.49) & 1.33 (0.09) & 1.29 (0.11) & 1.29 (0.10) & 1.18 (0.07) \\ 
  $D_2 (20\%)$ & 3.51 (0.30) & 3.56 (0.34) & 1.54 (0.11) & 1.74 (0.21) & 1.33 (0.08) & 1.25 (0.08) \\ 
  $D_3$ & 1.25 (0.13) & 1.15 (0.08) & 1.26 (0.07) & 1.17 (0.07) & 1.29 (0.11) & 1.20 (0.07) \\ 
  $D_4$ & 9.80 (11.83) & 1.51 (0.15) & 1.42 (0.09) & 1.38 (0.12) & 1.36 (0.11) & 1.32 (0.10) \\ 
   \hline
\end{tabular}
\caption{Summary statistics of RMSEs on the test sets by L2Boost, MBoost, LADBoost, Robloss, SBoost, and RRBoost applied with tree learners of $d$ = 3 for clean ($D_0$), symmetric gross error contaminated ($D_1$), asymmetric gross error contaminated ($D_2$),  skewed distributed ($D_3$),  and heavy-tailed distributed  ($D_4$) data generated from $g$ = $g_2$ S = $S_0$ $n$ = 300 $p$ = 400, displayed in the form of: mean (SD) calculated from 100 independent runs of the experiment.} 
\end{table}
\begin{table}[H]
\centering
\begin{tabular}{lccccc}
  \hline
 & L2Boost & MBoost & LADBoost & Robloss & RRBoost \\ 
  \hline
$D_0$ & 0.76 (0.05) & 0.76 (0.04) & 0.76 (0.06) & 0.77 (0.07) & 0.76 (0.05) \\ 
  $D_1 (10\%)$ & 0.41 (0.20) & 0.63 (0.18) & 0.75 (0.04) & 0.74 (0.06) & 0.76 (0.04) \\ 
  $D_1 (20\%)$ & 0.24 (0.18) & 0.24 (0.18) & 0.72 (0.08) & 0.69 (0.11) & 0.76 (0.05) \\ 
  $D_2 (10\%)$ & 0.44 (0.17) & 0.64 (0.17) & 0.75 (0.04) & 0.75 (0.04) & 0.76 (0.05) \\ 
  $D_2 (20\%)$ & 0.28 (0.17) & 0.26 (0.17) & 0.73 (0.08) & 0.69 (0.13) & 0.75 (0.00) \\ 
  $D_3$ & 0.75 (0.02) & 0.76 (0.04) & 0.76 (0.05) & 0.75 (0.02) & 0.75 (0.02) \\ 
  $D_4$ & 0.15 (0.18) & 0.70 (0.11) & 0.70 (0.10) & 0.72 (0.09) & 0.73 (0.08) \\ 
   \hline
\end{tabular}
\caption{Fractions of variables recovered by L2Boost, MBoost, LADBoost, Robloss, and RRBoost applied with tree learners of $d$ = 3 for clean ($D_0$), symmetric gross error contaminated ($D_1$), asymmetric gross error contaminated ($D_2$),  skewed distributed ($D_3$),  and heavy-tailed distributed  ($D_4$) data generated from $g$ = $g_2$ S = $S_0$ $n$ = 300 $p$ = 400, displayed in the form of: mean (SD) calculated from 100 independent runs of the experiment.} 
\end{table}
\begin{table}[H]
\centering
\begin{tabular}{lcccccc}
  \hline
 & L2Boost & MBoost & LADBoost & Robloss & SBoost & RRBoost \\ 
  \hline
$D_0$ & 1.03 (0.02) & 1.03 (0.03) & 1.04 (0.03) & 1.03 (0.03) & 0.92 (0.03) & 0.88 (0.02) \\ 
  $D_1 (10\%)$ & 1.20 (0.04) & 1.13 (0.06) & 1.05 (0.02) & 1.05 (0.03) & 0.93 (0.03) & 0.89 (0.02) \\ 
  $D_1 (20\%)$ & 1.25 (0.05) & 1.25 (0.05) & 1.06 (0.03) & 1.07 (0.03) & 0.93 (0.03) & 0.90 (0.03) \\ 
  $D_2 (10\%)$ & 1.61 (0.04) & 1.30 (0.16) & 1.07 (0.02) & 1.07 (0.03) & 0.92 (0.03) & 0.89 (0.03) \\ 
  $D_2 (20\%)$ & 2.47 (0.06) & 2.41 (0.07) & 1.14 (0.03) & 1.20 (0.05) & 0.95 (0.04) & 0.92 (0.03) \\ 
  $D_3$ & 1.10 (0.04) & 1.05 (0.03) & 1.06 (0.03) & 1.05 (0.03) & 0.99 (0.03) & 0.91 (0.02) \\ 
  $D_4$ & 29.29 (130.19) & 1.10 (0.03) & 1.08 (0.02) & 1.08 (0.03) & 0.94 (0.03) & 0.92 (0.03) \\ 
   \hline
\end{tabular}
\caption{Summary statistics of RMSEs on the test sets by L2Boost, MBoost, LADBoost, Robloss, SBoost, and RRBoost applied with tree learners of $d$ = 1 for clean ($D_0$), symmetric gross error contaminated ($D_1$), asymmetric gross error contaminated ($D_2$),  skewed distributed ($D_3$),  and heavy-tailed distributed  ($D_4$) data generated from $g$ = $g_2$ S = $S_0$ $n$ = 3000 $p$ = 400, displayed in the form of: mean (SD) calculated from 100 independent runs of the experiment.} 
\end{table}
\begin{table}[H]
\centering
\begin{tabular}{lccccc}
  \hline
 & L2Boost & MBoost & LADBoost & Robloss & RRBoost \\ 
  \hline
$D_0$ & 0.95 (0.10) & 0.96 (0.09) & 0.95 (0.10) & 0.96 (0.09) & 0.95 (0.10) \\ 
  $D_1 (10\%)$ & 0.75 (0.00) & 0.79 (0.09) & 0.89 (0.13) & 0.90 (0.12) & 0.95 (0.10) \\ 
  $D_1 (20\%)$ & 0.72 (0.08) & 0.72 (0.08) & 0.85 (0.12) & 0.82 (0.11) & 0.94 (0.11) \\ 
  $D_2 (10\%)$ & 0.75 (0.00) & 0.78 (0.08) & 0.83 (0.12) & 0.87 (0.13) & 0.94 (0.11) \\ 
  $D_2 (20\%)$ & 0.73 (0.06) & 0.66 (0.12) & 0.77 (0.07) & 0.78 (0.08) & 0.92 (0.12) \\ 
  $D_3$ & 0.79 (0.09) & 0.91 (0.12) & 0.89 (0.13) & 0.90 (0.12) & 0.94 (0.11) \\ 
  $D_4$ & 0.04 (0.12) & 0.79 (0.09) & 0.83 (0.12) & 0.85 (0.12) & 0.85 (0.12) \\ 
   \hline
\end{tabular}
\caption{Fractions of variables recovered by L2Boost, MBoost, LADBoost, Robloss, and RRBoost applied with tree learners of $d$ = 1 for clean ($D_0$), symmetric gross error contaminated ($D_1$), asymmetric gross error contaminated ($D_2$),  skewed distributed ($D_3$),  and heavy-tailed distributed  ($D_4$) data generated from $g$ = $g_2$ S = $S_0$ $n$ = 3000 $p$ = 400, displayed in the form of: mean (SD) calculated from 100 independent runs of the experiment.} 
\end{table}
\begin{table}[H]
\centering
\begin{tabular}{lcccccc}
  \hline
 & L2Boost & MBoost & LADBoost & Robloss & SBoost & RRBoost \\ 
  \hline
$D_0$ & 0.79 (0.03) & 0.79 (0.03) & 0.82 (0.03) & 0.80 (0.03) & 1.02 (0.07) & 0.73 (0.02) \\ 
  $D_1 (10\%)$ & 1.13 (0.06) & 0.99 (0.12) & 0.84 (0.04) & 0.82 (0.03) & 0.92 (0.07) & 0.72 (0.03) \\ 
  $D_1 (20\%)$ & 1.24 (0.08) & 1.24 (0.08) & 0.86 (0.04) & 0.87 (0.04) & 0.82 (0.05) & 0.71 (0.02) \\ 
  $D_2 (10\%)$ & 1.56 (0.06) & 1.17 (0.22) & 0.87 (0.04) & 0.85 (0.04) & 0.90 (0.06) & 0.72 (0.02) \\ 
  $D_2 (20\%)$ & 2.46 (0.07) & 2.39 (0.10) & 0.96 (0.04) & 0.99 (0.05) & 0.81 (0.05) & 0.72 (0.02) \\ 
  $D_3$ & 0.92 (0.06) & 0.82 (0.03) & 0.86 (0.03) & 0.84 (0.03) & 1.01 (0.07) & 0.75 (0.03) \\ 
  $D_4$ & 28.91 (92.28) & 0.93 (0.04) & 0.88 (0.03) & 0.88 (0.04) & 0.84 (0.04) & 0.75 (0.03) \\ 
   \hline
\end{tabular}
\caption{Summary statistics of RMSEs on the test sets by L2Boost, MBoost, LADBoost, Robloss, SBoost, and RRBoost applied with tree learners of $d$ = 2 for clean ($D_0$), symmetric gross error contaminated ($D_1$), asymmetric gross error contaminated ($D_2$),  skewed distributed ($D_3$),  and heavy-tailed distributed  ($D_4$) data generated from $g$ = $g_2$ S = $S_0$ $n$ = 3000 $p$ = 400, displayed in the form of: mean (SD) calculated from 100 independent runs of the experiment.} 
\end{table}
\begin{table}[H]
\centering
\begin{tabular}{lccccc}
  \hline
 & L2Boost & MBoost & LADBoost & Robloss & RRBoost \\ 
  \hline
$D_0$ & 1.00 (0.03) & 1.00 (0.03) & 0.98 (0.07) & 0.99 (0.04) & 1.00 (0.00) \\ 
  $D_1 (10\%)$ & 0.74 (0.04) & 0.84 (0.12) & 0.95 (0.10) & 0.98 (0.07) & 1.00 (0.00) \\ 
  $D_1 (20\%)$ & 0.72 (0.09) & 0.72 (0.08) & 0.91 (0.12) & 0.89 (0.12) & 1.00 (0.00) \\ 
  $D_2 (10\%)$ & 0.75 (0.03) & 0.81 (0.11) & 0.93 (0.11) & 0.97 (0.08) & 1.00 (0.00) \\ 
  $D_2 (20\%)$ & 0.74 (0.06) & 0.75 (0.03) & 0.84 (0.12) & 0.85 (0.12) & 1.00 (0.00) \\ 
  $D_3$ & 0.83 (0.12) & 0.99 (0.04) & 0.96 (0.09) & 0.98 (0.06) & 1.00 (0.00) \\ 
  $D_4$ & 0.02 (0.10) & 0.82 (0.11) & 0.87 (0.13) & 0.91 (0.12) & 0.98 (0.07) \\ 
   \hline
\end{tabular}
\caption{Fractions of variables recovered by L2Boost, MBoost, LADBoost, Robloss, and RRBoost applied with tree learners of $d$ = 2 for clean ($D_0$), symmetric gross error contaminated ($D_1$), asymmetric gross error contaminated ($D_2$),  skewed distributed ($D_3$),  and heavy-tailed distributed  ($D_4$) data generated from $g$ = $g_2$ S = $S_0$ $n$ = 3000 $p$ = 400, displayed in the form of: mean (SD) calculated from 100 independent runs of the experiment.} 
\end{table}
\begin{table}[H]
\centering
\begin{tabular}{lcccccc}
  \hline
 & L2Boost & MBoost & LADBoost & Robloss & SBoost & RRBoost \\ 
  \hline
$D_0$ & 0.82 (0.04) & 0.81 (0.03) & 0.84 (0.03) & 0.83 (0.03) & 1.03 (0.08) & 0.72 (0.02) \\ 
  $D_1 (10\%)$ & 1.14 (0.06) & 1.02 (0.08) & 0.87 (0.03) & 0.86 (0.04) & 0.95 (0.07) & 0.72 (0.02) \\ 
  $D_1 (20\%)$ & 1.31 (0.10) & 1.34 (0.10) & 0.91 (0.03) & 0.92 (0.04) & 0.88 (0.07) & 0.71 (0.03) \\ 
  $D_2 (10\%)$ & 1.56 (0.08) & 1.23 (0.21) & 0.89 (0.03) & 0.86 (0.04) & 0.93 (0.07) & 0.72 (0.02) \\ 
  $D_2 (20\%)$ & 2.48 (0.08) & 2.56 (0.12) & 0.99 (0.04) & 1.02 (0.04) & 0.82 (0.05) & 0.71 (0.02) \\ 
  $D_3$ & 0.95 (0.05) & 0.86 (0.03) & 0.89 (0.03) & 0.86 (0.04) & 1.02 (0.08) & 0.76 (0.02) \\ 
  $D_4$ & 48.98 (204.87) & 0.96 (0.04) & 0.92 (0.04) & 0.91 (0.04) & 0.89 (0.05) & 0.76 (0.03) \\ 
   \hline
\end{tabular}
\caption{Summary statistics of RMSEs on the test sets by L2Boost, MBoost, LADBoost, Robloss, SBoost, and RRBoost applied with tree learners of $d$ = 3 for clean ($D_0$), symmetric gross error contaminated ($D_1$), asymmetric gross error contaminated ($D_2$),  skewed distributed ($D_3$),  and heavy-tailed distributed  ($D_4$) data generated from $g$ = $g_2$ S = $S_0$ $n$ = 3000 $p$ = 400, displayed in the form of: mean (SD) calculated from 100 independent runs of the experiment.} 
\end{table}
\begin{table}[H]
\centering
\begin{tabular}{lccccc}
  \hline
 & L2Boost & MBoost & LADBoost & Robloss & RRBoost \\ 
  \hline
$D_0$ & 1.00 (0.00) & 1.00 (0.00) & 0.99 (0.04) & 1.00 (0.03) & 1.00 (0.00) \\ 
  $D_1 (10\%)$ & 0.75 (0.00) & 0.84 (0.12) & 0.98 (0.07) & 0.99 (0.04) & 1.00 (0.00) \\ 
  $D_1 (20\%)$ & 0.75 (0.00) & 0.75 (0.00) & 0.95 (0.10) & 0.94 (0.11) & 1.00 (0.00) \\ 
  $D_2 (10\%)$ & 0.75 (0.00) & 0.82 (0.11) & 0.98 (0.06) & 0.99 (0.04) & 1.00 (0.00) \\ 
  $D_2 (20\%)$ & 0.75 (0.00) & 0.74 (0.04) & 0.92 (0.12) & 0.92 (0.12) & 1.00 (0.00) \\ 
  $D_3$ & 0.87 (0.13) & 0.99 (0.04) & 0.97 (0.08) & 0.99 (0.05) & 1.00 (0.00) \\ 
  $D_4$ & 0.03 (0.09) & 0.88 (0.13) & 0.91 (0.12) & 0.94 (0.11) & 0.99 (0.04) \\ 
   \hline
\end{tabular}
\caption{Fractions of variables recovered by L2Boost, MBoost, LADBoost, Robloss, and RRBoost applied with tree learners of $d$ = 3 for clean ($D_0$), symmetric gross error contaminated ($D_1$), asymmetric gross error contaminated ($D_2$),  skewed distributed ($D_3$),  and heavy-tailed distributed  ($D_4$) data generated from $g$ = $g_2$ S = $S_0$ $n$ = 3000 $p$ = 400, displayed in the form of: mean (SD) calculated from 100 independent runs of the experiment.} 
\end{table}
\begin{table}[H]
\centering
\begin{tabular}{lcccccc}
  \hline
 & L2Boost & MBoost & LADBoost & Robloss & SBoost & RRBoost \\ 
  \hline
$D_0$ & 1.18 (0.04) & 1.19 (0.04) & 1.21 (0.04) & 1.19 (0.04) & 1.36 (0.11) & 1.18 (0.05) \\ 
  $D_1 (10\%)$ & 1.97 (0.20) & 1.59 (0.31) & 1.25 (0.05) & 1.25 (0.06) & 1.33 (0.10) & 1.20 (0.05) \\ 
  $D_1 (20\%)$ & 2.26 (0.45) & 2.11 (0.31) & 1.31 (0.06) & 1.36 (0.10) & 1.31 (0.08) & 1.21 (0.05) \\ 
  $D_2 (10\%)$ & 2.67 (0.26) & 1.95 (0.61) & 1.27 (0.05) & 1.28 (0.06) & 1.32 (0.09) & 1.20 (0.05) \\ 
  $D_2 (20\%)$ & 4.27 (0.32) & 4.30 (0.39) & 1.44 (0.25) & 1.64 (0.37) & 1.32 (0.10) & 1.21 (0.04) \\ 
  $D_3$ & 1.41 (0.13) & 1.27 (0.06) & 1.29 (0.05) & 1.27 (0.05) & 1.42 (0.09) & 1.29 (0.06) \\ 
  $D_4$ & 41.28 (258.62) & 1.56 (0.13) & 1.37 (0.08) & 1.40 (0.09) & 1.34 (0.10) & 1.30 (0.08) \\ 
   \hline
\end{tabular}
\caption{Summary statistics of RMSEs on the test sets by L2Boost, MBoost, LADBoost, Robloss, SBoost, and RRBoost applied with tree learners of $d$ = 1 for clean ($D_0$), symmetric gross error contaminated ($D_1$), asymmetric gross error contaminated ($D_2$),  skewed distributed ($D_3$),  and heavy-tailed distributed  ($D_4$) data generated from $g$ = $g_2$ S = $S_1$ $n$ = 300 $p$ = 10, displayed in the form of: mean (SD) calculated from 100 independent runs of the experiment.} 
\end{table}
\begin{table}[H]
\centering
\begin{tabular}{lccccc}
  \hline
 & L2Boost & MBoost & LADBoost & Robloss & RRBoost \\ 
  \hline
$D_0$ & 0.90 (0.13) & 0.88 (0.13) & 0.88 (0.13) & 0.88 (0.13) & 0.90 (0.12) \\ 
  $D_1 (10\%)$ & 0.38 (0.18) & 0.61 (0.26) & 0.87 (0.13) & 0.86 (0.13) & 0.88 (0.13) \\ 
  $D_1 (20\%)$ & 0.31 (0.15) & 0.25 (0.06) & 0.84 (0.14) & 0.82 (0.14) & 0.87 (0.13) \\ 
  $D_2 (10\%)$ & 0.42 (0.19) & 0.73 (0.20) & 0.87 (0.14) & 0.87 (0.13) & 0.88 (0.13) \\ 
  $D_2 (20\%)$ & 0.35 (0.16) & 0.44 (0.17) & 0.82 (0.15) & 0.80 (0.17) & 0.87 (0.13) \\ 
  $D_3$ & 0.78 (0.18) & 0.86 (0.14) & 0.89 (0.13) & 0.87 (0.13) & 0.88 (0.13) \\ 
  $D_4$ & 0.29 (0.24) & 0.69 (0.19) & 0.82 (0.15) & 0.78 (0.15) & 0.84 (0.12) \\ 
   \hline
\end{tabular}
\caption{Fractions of variables recovered by L2Boost, MBoost, LADBoost, Robloss, and RRBoost applied with tree learners of $d$ = 1 for clean ($D_0$), symmetric gross error contaminated ($D_1$), asymmetric gross error contaminated ($D_2$),  skewed distributed ($D_3$),  and heavy-tailed distributed  ($D_4$) data generated from $g$ = $g_2$ S = $S_1$ $n$ = 300 $p$ = 10, displayed in the form of: mean (SD) calculated from 100 independent runs of the experiment.} 
\end{table}
\begin{table}[H]
\centering
\begin{tabular}{lcccccc}
  \hline
 & L2Boost & MBoost & LADBoost & Robloss & SBoost & RRBoost \\ 
  \hline
$D_0$ & 1.26 (0.05) & 1.27 (0.07) & 1.31 (0.07) & 1.28 (0.06) & 1.40 (0.13) & 1.26 (0.06) \\ 
  $D_1 (10\%)$ & 2.22 (0.30) & 1.74 (0.38) & 1.37 (0.07) & 1.35 (0.08) & 1.39 (0.11) & 1.27 (0.05) \\ 
  $D_1 (20\%)$ & 2.78 (0.43) & 2.73 (0.37) & 1.44 (0.09) & 1.48 (0.10) & 1.37 (0.09) & 1.28 (0.06) \\ 
  $D_2 (10\%)$ & 2.84 (0.32) & 2.15 (0.69) & 1.38 (0.08) & 1.37 (0.07) & 1.39 (0.09) & 1.27 (0.05) \\ 
  $D_2 (20\%)$ & 4.52 (0.40) & 4.66 (0.43) & 1.53 (0.10) & 1.66 (0.16) & 1.37 (0.09) & 1.27 (0.06) \\ 
  $D_3$ & 1.50 (0.14) & 1.36 (0.07) & 1.39 (0.06) & 1.35 (0.06) & 1.51 (0.18) & 1.35 (0.06) \\ 
  $D_4$ & 47.56 (259.10) & 1.61 (0.16) & 1.47 (0.11) & 1.47 (0.10) & 1.42 (0.10) & 1.38 (0.08) \\ 
   \hline
\end{tabular}
\caption{Summary statistics of RMSEs on the test sets by L2Boost, MBoost, LADBoost, Robloss, SBoost, and RRBoost applied with tree learners of $d$ = 2 for clean ($D_0$), symmetric gross error contaminated ($D_1$), asymmetric gross error contaminated ($D_2$),  skewed distributed ($D_3$),  and heavy-tailed distributed  ($D_4$) data generated from $g$ = $g_2$ S = $S_1$ $n$ = 300 $p$ = 10, displayed in the form of: mean (SD) calculated from 100 independent runs of the experiment.} 
\end{table}
\begin{table}[H]
\centering
\begin{tabular}{lccccc}
  \hline
 & L2Boost & MBoost & LADBoost & Robloss & RRBoost \\ 
  \hline
$D_0$ & 0.88 (0.13) & 0.88 (0.14) & 0.87 (0.13) & 0.87 (0.13) & 0.88 (0.13) \\ 
  $D_1 (10\%)$ & 0.46 (0.15) & 0.69 (0.22) & 0.88 (0.13) & 0.85 (0.14) & 0.91 (0.12) \\ 
  $D_1 (20\%)$ & 0.40 (0.17) & 0.38 (0.15) & 0.82 (0.13) & 0.79 (0.17) & 0.87 (0.13) \\ 
  $D_2 (10\%)$ & 0.48 (0.18) & 0.68 (0.19) & 0.86 (0.13) & 0.84 (0.13) & 0.87 (0.13) \\ 
  $D_2 (20\%)$ & 0.36 (0.16) & 0.43 (0.19) & 0.84 (0.13) & 0.73 (0.15) & 0.87 (0.13) \\ 
  $D_3$ & 0.73 (0.16) & 0.83 (0.15) & 0.84 (0.12) & 0.84 (0.13) & 0.90 (0.13) \\ 
  $D_4$ & 0.34 (0.20) & 0.70 (0.17) & 0.81 (0.14) & 0.78 (0.16) & 0.84 (0.13) \\ 
   \hline
\end{tabular}
\caption{Fractions of variables recovered by L2Boost, MBoost, LADBoost, Robloss, and RRBoost applied with tree learners of $d$ = 2 for clean ($D_0$), symmetric gross error contaminated ($D_1$), asymmetric gross error contaminated ($D_2$),  skewed distributed ($D_3$),  and heavy-tailed distributed  ($D_4$) data generated from $g$ = $g_2$ S = $S_1$ $n$ = 300 $p$ = 10, displayed in the form of: mean (SD) calculated from 100 independent runs of the experiment.} 
\end{table}
\begin{table}[H]
\centering
\begin{tabular}{lcccccc}
  \hline
 & L2Boost & MBoost & LADBoost & Robloss & SBoost & RRBoost \\ 
  \hline
$D_0$ & 1.28 (0.06) & 1.29 (0.06) & 1.35 (0.06) & 1.30 (0.06) & 1.39 (0.14) & 1.29 (0.06) \\ 
  $D_1 (10\%)$ & 2.62 (0.32) & 1.99 (0.52) & 1.44 (0.09) & 1.42 (0.09) & 1.39 (0.14) & 1.29 (0.06) \\ 
  $D_1 (20\%)$ & 3.37 (0.44) & 3.39 (0.44) & 1.55 (0.11) & 1.63 (0.12) & 1.40 (0.12) & 1.31 (0.06) \\ 
  $D_2 (10\%)$ & 3.17 (0.36) & 2.36 (0.85) & 1.42 (0.09) & 1.42 (0.08) & 1.41 (0.12) & 1.30 (0.06) \\ 
  $D_2 (20\%)$ & 4.82 (0.41) & 4.95 (0.45) & 1.61 (0.11) & 1.80 (0.18) & 1.40 (0.08) & 1.32 (0.07) \\ 
  $D_3$ & 1.52 (0.15) & 1.37 (0.07) & 1.42 (0.07) & 1.37 (0.07) & 1.46 (0.13) & 1.37 (0.06) \\ 
  $D_4$ & 49.87 (259.36) & 1.69 (0.18) & 1.57 (0.11) & 1.57 (0.11) & 1.45 (0.15) & 1.41 (0.09) \\ 
   \hline
\end{tabular}
\caption{Summary statistics of RMSEs on the test sets by L2Boost, MBoost, LADBoost, Robloss, SBoost, and RRBoost applied with tree learners of $d$ = 3 for clean ($D_0$), symmetric gross error contaminated ($D_1$), asymmetric gross error contaminated ($D_2$),  skewed distributed ($D_3$),  and heavy-tailed distributed  ($D_4$) data generated from $g$ = $g_2$ S = $S_1$ $n$ = 300 $p$ = 10, displayed in the form of: mean (SD) calculated from 100 independent runs of the experiment.} 
\end{table}
\begin{table}[H]
\centering
\begin{tabular}{lccccc}
  \hline
 & L2Boost & MBoost & LADBoost & Robloss & RRBoost \\ 
  \hline
$D_0$ & 0.87 (0.14) & 0.87 (0.14) & 0.88 (0.13) & 0.87 (0.13) & 0.89 (0.12) \\ 
  $D_1 (10\%)$ & 0.59 (0.17) & 0.72 (0.22) & 0.88 (0.13) & 0.86 (0.13) & 0.88 (0.13) \\ 
  $D_1 (20\%)$ & 0.53 (0.18) & 0.50 (0.18) & 0.84 (0.13) & 0.80 (0.14) & 0.87 (0.13) \\ 
  $D_2 (10\%)$ & 0.59 (0.17) & 0.71 (0.20) & 0.87 (0.13) & 0.85 (0.14) & 0.87 (0.13) \\ 
  $D_2 (20\%)$ & 0.48 (0.18) & 0.50 (0.20) & 0.84 (0.14) & 0.75 (0.16) & 0.86 (0.12) \\ 
  $D_3$ & 0.79 (0.14) & 0.83 (0.14) & 0.88 (0.14) & 0.86 (0.12) & 0.90 (0.12) \\ 
  $D_4$ & 0.48 (0.22) & 0.75 (0.15) & 0.82 (0.14) & 0.75 (0.14) & 0.84 (0.14) \\ 
   \hline
\end{tabular}
\caption{Fractions of variables recovered by L2Boost, MBoost, LADBoost, Robloss, and RRBoost applied with tree learners of $d$ = 3 for clean ($D_0$), symmetric gross error contaminated ($D_1$), asymmetric gross error contaminated ($D_2$),  skewed distributed ($D_3$),  and heavy-tailed distributed  ($D_4$) data generated from $g$ = $g_2$ S = $S_1$ $n$ = 300 $p$ = 10, displayed in the form of: mean (SD) calculated from 100 independent runs of the experiment.} 
\end{table}
\begin{table}[H]
\centering
\begin{tabular}{lcccccc}
  \hline
 & L2Boost & MBoost & LADBoost & Robloss & SBoost & RRBoost \\ 
  \hline
$D_0$ & 1.02 (0.02) & 1.02 (0.02) & 1.03 (0.02) & 1.02 (0.02) & 1.02 (0.03) & 1.00 (0.02) \\ 
  $D_1 (10\%)$ & 1.39 (0.07) & 1.24 (0.17) & 1.05 (0.02) & 1.04 (0.02) & 1.02 (0.03) & 1.01 (0.02) \\ 
  $D_1 (20\%)$ & 1.54 (0.11) & 1.55 (0.09) & 1.06 (0.03) & 1.07 (0.06) & 1.01 (0.02) & 1.00 (0.02) \\ 
  $D_2 (10\%)$ & 2.30 (0.08) & 1.55 (0.41) & 1.06 (0.03) & 1.06 (0.04) & 1.02 (0.02) & 1.01 (0.02) \\ 
  $D_2 (20\%)$ & 3.98 (0.11) & 3.92 (0.14) & 1.14 (0.08) & 1.19 (0.09) & 1.01 (0.02) & 1.01 (0.04) \\ 
  $D_3$ & 1.10 (0.03) & 1.07 (0.02) & 1.15 (0.02) & 1.11 (0.02) & 1.26 (0.03) & 1.13 (0.03) \\ 
  $D_4$ & 13.20 (73.33) & 1.16 (0.04) & 1.07 (0.03) & 1.09 (0.03) & 1.01 (0.02) & 1.01 (0.02) \\ 
   \hline
\end{tabular}
\caption{Summary statistics of RMSEs on the test sets by L2Boost, MBoost, LADBoost, Robloss, SBoost, and RRBoost applied with tree learners of $d$ = 1 for clean ($D_0$), symmetric gross error contaminated ($D_1$), asymmetric gross error contaminated ($D_2$),  skewed distributed ($D_3$),  and heavy-tailed distributed  ($D_4$) data generated from $g$ = $g_2$ S = $S_1$ $n$ = 3000 $p$ = 10, displayed in the form of: mean (SD) calculated from 100 independent runs of the experiment.} 
\end{table}
\begin{table}[H]
\centering
\begin{tabular}{lccccc}
  \hline
 & L2Boost & MBoost & LADBoost & Robloss & RRBoost \\ 
  \hline
$D_0$ & 1.00 (0.00) & 1.00 (0.00) & 1.00 (0.00) & 1.00 (0.00) & 1.00 (0.00) \\ 
  $D_1 (10\%)$ & 0.81 (0.16) & 0.90 (0.14) & 1.00 (0.02) & 1.00 (0.02) & 1.00 (0.02) \\ 
  $D_1 (20\%)$ & 0.73 (0.17) & 0.75 (0.16) & 0.99 (0.05) & 1.00 (0.02) & 1.00 (0.00) \\ 
  $D_2 (10\%)$ & 0.82 (0.16) & 0.88 (0.13) & 1.00 (0.02) & 0.99 (0.05) & 1.00 (0.02) \\ 
  $D_2 (20\%)$ & 0.76 (0.17) & 0.60 (0.17) & 0.99 (0.05) & 0.98 (0.06) & 1.00 (0.02) \\ 
  $D_3$ & 0.97 (0.08) & 1.00 (0.00) & 1.00 (0.00) & 1.00 (0.00) & 1.00 (0.00) \\ 
  $D_4$ & 0.25 (0.23) & 0.95 (0.10) & 0.98 (0.06) & 0.97 (0.08) & 1.00 (0.02) \\ 
   \hline
\end{tabular}
\caption{Fractions of variables recovered by L2Boost, MBoost, LADBoost, Robloss, and RRBoost applied with tree learners of $d$ = 1 for clean ($D_0$), symmetric gross error contaminated ($D_1$), asymmetric gross error contaminated ($D_2$),  skewed distributed ($D_3$),  and heavy-tailed distributed  ($D_4$) data generated from $g$ = $g_2$ S = $S_1$ $n$ = 3000 $p$ = 10, displayed in the form of: mean (SD) calculated from 100 independent runs of the experiment.} 
\end{table}
\begin{table}[H]
\centering
\begin{tabular}{lcccccc}
  \hline
 & L2Boost & MBoost & LADBoost & Robloss & SBoost & RRBoost \\ 
  \hline
$D_0$ & 1.05 (0.03) & 1.05 (0.03) & 1.07 (0.03) & 1.06 (0.02) & 1.03 (0.03) & 1.00 (0.02) \\ 
  $D_1 (10\%)$ & 1.46 (0.08) & 1.29 (0.16) & 1.09 (0.03) & 1.09 (0.03) & 1.02 (0.02) & 1.00 (0.02) \\ 
  $D_1 (20\%)$ & 1.56 (0.09) & 1.57 (0.09) & 1.12 (0.04) & 1.12 (0.03) & 1.00 (0.02) & 0.99 (0.02) \\ 
  $D_2 (10\%)$ & 2.35 (0.08) & 1.67 (0.45) & 1.11 (0.03) & 1.11 (0.03) & 1.01 (0.02) & 0.99 (0.02) \\ 
  $D_2 (20\%)$ & 4.00 (0.11) & 3.96 (0.16) & 1.20 (0.05) & 1.23 (0.04) & 1.00 (0.03) & 0.99 (0.02) \\ 
  $D_3$ & 1.19 (0.05) & 1.10 (0.03) & 1.18 (0.03) & 1.14 (0.03) & 1.33 (0.09) & 1.14 (0.03) \\ 
  $D_4$ & 54.71 (249.72) & 1.23 (0.05) & 1.14 (0.04) & 1.15 (0.04) & 1.00 (0.02) & 1.00 (0.02) \\ 
   \hline
\end{tabular}
\caption{Summary statistics of RMSEs on the test sets by L2Boost, MBoost, LADBoost, Robloss, SBoost, and RRBoost applied with tree learners of $d$ = 2 for clean ($D_0$), symmetric gross error contaminated ($D_1$), asymmetric gross error contaminated ($D_2$),  skewed distributed ($D_3$),  and heavy-tailed distributed  ($D_4$) data generated from $g$ = $g_2$ S = $S_1$ $n$ = 3000 $p$ = 10, displayed in the form of: mean (SD) calculated from 100 independent runs of the experiment.} 
\end{table}
\begin{table}[H]
\centering
\begin{tabular}{lccccc}
  \hline
 & L2Boost & MBoost & LADBoost & Robloss & RRBoost \\ 
  \hline
$D_0$ & 1.00 (0.02) & 1.00 (0.02) & 1.00 (0.00) & 1.00 (0.00) & 1.00 (0.00) \\ 
  $D_1 (10\%)$ & 0.76 (0.14) & 0.86 (0.14) & 1.00 (0.04) & 1.00 (0.02) & 1.00 (0.00) \\ 
  $D_1 (20\%)$ & 0.68 (0.16) & 0.68 (0.16) & 1.00 (0.00) & 0.98 (0.06) & 1.00 (0.00) \\ 
  $D_2 (10\%)$ & 0.66 (0.14) & 0.81 (0.19) & 0.99 (0.05) & 0.99 (0.05) & 1.00 (0.00) \\ 
  $D_2 (20\%)$ & 0.52 (0.13) & 0.61 (0.19) & 0.95 (0.10) & 0.95 (0.10) & 1.00 (0.00) \\ 
  $D_3$ & 0.89 (0.12) & 0.99 (0.05) & 1.00 (0.02) & 1.00 (0.04) & 1.00 (0.00) \\ 
  $D_4$ & 0.35 (0.20) & 0.90 (0.12) & 0.98 (0.07) & 0.98 (0.08) & 1.00 (0.02) \\ 
   \hline
\end{tabular}
\caption{Fractions of variables recovered by L2Boost, MBoost, LADBoost, Robloss, and RRBoost applied with tree learners of $d$ = 2 for clean ($D_0$), symmetric gross error contaminated ($D_1$), asymmetric gross error contaminated ($D_2$),  skewed distributed ($D_3$),  and heavy-tailed distributed  ($D_4$) data generated from $g$ = $g_2$ S = $S_1$ $n$ = 3000 $p$ = 10, displayed in the form of: mean (SD) calculated from 100 independent runs of the experiment.} 
\end{table}
\begin{table}[H]
\centering
\begin{tabular}{lcccccc}
  \hline
 & L2Boost & MBoost & LADBoost & Robloss & SBoost & RRBoost \\ 
  \hline
$D_0$ & 1.06 (0.03) & 1.06 (0.03) & 1.11 (0.03) & 1.08 (0.03) & 1.12 (0.14) & 1.04 (0.30) \\ 
  $D_1 (10\%)$ & 1.49 (0.09) & 1.30 (0.15) & 1.13 (0.03) & 1.11 (0.03) & 1.12 (0.17) & 1.01 (0.03) \\ 
  $D_1 (20\%)$ & 1.67 (0.11) & 1.70 (0.12) & 1.17 (0.04) & 1.15 (0.04) & 1.06 (0.11) & 1.01 (0.03) \\ 
  $D_2 (10\%)$ & 2.38 (0.10) & 1.72 (0.46) & 1.14 (0.03) & 1.12 (0.03) & 1.12 (0.14) & 1.01 (0.03) \\ 
  $D_2 (20\%)$ & 4.03 (0.12) & 4.07 (0.16) & 1.23 (0.04) & 1.26 (0.04) & 1.06 (0.06) & 1.00 (0.02) \\ 
  $D_3$ & 1.21 (0.06) & 1.12 (0.03) & 1.21 (0.03) & 1.15 (0.03) & 1.33 (0.16) & 1.28 (0.86) \\ 
  $D_4$ & 53.48 (215.75) & 1.26 (0.04) & 1.18 (0.04) & 1.17 (0.04) & 1.11 (0.14) & 1.03 (0.03) \\ 
   \hline
\end{tabular}
\caption{Summary statistics of RMSEs on the test sets by L2Boost, MBoost, LADBoost, Robloss, SBoost, and RRBoost applied with tree learners of $d$ = 3 for clean ($D_0$), symmetric gross error contaminated ($D_1$), asymmetric gross error contaminated ($D_2$),  skewed distributed ($D_3$),  and heavy-tailed distributed  ($D_4$) data generated from $g$ = $g_2$ S = $S_1$ $n$ = 3000 $p$ = 10, displayed in the form of: mean (SD) calculated from 100 independent runs of the experiment.} 
\end{table}
\begin{table}[H]
\centering
\begin{tabular}{lccccc}
  \hline
 & L2Boost & MBoost & LADBoost & Robloss & RRBoost \\ 
  \hline
$D_0$ & 1.00 (0.00) & 1.00 (0.02) & 1.00 (0.02) & 1.00 (0.02) & 1.00 (0.00) \\ 
  $D_1 (10\%)$ & 0.80 (0.13) & 0.88 (0.16) & 0.98 (0.06) & 0.99 (0.05) & 1.00 (0.00) \\ 
  $D_1 (20\%)$ & 0.76 (0.15) & 0.77 (0.15) & 0.99 (0.04) & 0.98 (0.07) & 1.00 (0.00) \\ 
  $D_2 (10\%)$ & 0.66 (0.16) & 0.80 (0.18) & 0.98 (0.06) & 0.99 (0.05) & 1.00 (0.02) \\ 
  $D_2 (20\%)$ & 0.55 (0.17) & 0.59 (0.19) & 0.96 (0.09) & 0.93 (0.11) & 1.00 (0.00) \\ 
  $D_3$ & 0.88 (0.13) & 0.99 (0.05) & 1.00 (0.02) & 1.00 (0.02) & 0.99 (0.06) \\ 
  $D_4$ & 0.51 (0.21) & 0.89 (0.12) & 0.96 (0.10) & 0.97 (0.08) & 0.99 (0.05) \\ 
   \hline
\end{tabular}
\caption{Fractions of variables recovered by L2Boost, MBoost, LADBoost, Robloss, and RRBoost applied with tree learners of $d$ = 3 for clean ($D_0$), symmetric gross error contaminated ($D_1$), asymmetric gross error contaminated ($D_2$),  skewed distributed ($D_3$),  and heavy-tailed distributed  ($D_4$) data generated from $g$ = $g_2$ S = $S_1$ $n$ = 3000 $p$ = 10, displayed in the form of: mean (SD) calculated from 100 independent runs of the experiment.} 
\end{table}
\begin{table}[H]
\centering
\begin{tabular}{lcccccc}
  \hline
 & L2Boost & MBoost & LADBoost & Robloss & SBoost & RRBoost \\ 
  \hline
$D_0$ & 1.14 (0.06) & 1.14 (0.07) & 1.21 (0.08) & 1.15 (0.08) & 1.45 (0.21) & 1.10 (0.06) \\ 
  $D_1 (10\%)$ & 1.76 (0.13) & 1.52 (0.22) & 1.26 (0.09) & 1.24 (0.10) & 1.38 (0.16) & 1.13 (0.06) \\ 
  $D_1 (20\%)$ & 1.95 (0.31) & 1.95 (0.28) & 1.33 (0.11) & 1.39 (0.16) & 1.33 (0.12) & 1.15 (0.07) \\ 
  $D_2 (10\%)$ & 2.28 (0.18) & 1.86 (0.46) & 1.30 (0.11) & 1.28 (0.12) & 1.37 (0.13) & 1.13 (0.06) \\ 
  $D_2 (20\%)$ & 3.43 (0.27) & 3.58 (0.37) & 1.50 (0.13) & 1.60 (0.20) & 1.32 (0.13) & 1.16 (0.06) \\ 
  $D_3$ & 1.38 (0.15) & 1.21 (0.07) & 1.26 (0.08) & 1.21 (0.07) & 1.44 (0.18) & 1.15 (0.06) \\ 
  $D_4$ & 19.61 (49.07) & 1.50 (0.13) & 1.39 (0.12) & 1.39 (0.12) & 1.36 (0.15) & 1.22 (0.09) \\ 
   \hline
\end{tabular}
\caption{Summary statistics of RMSEs on the test sets by L2Boost, MBoost, LADBoost, Robloss, SBoost, and RRBoost applied with tree learners of $d$ = 1 for clean ($D_0$), symmetric gross error contaminated ($D_1$), asymmetric gross error contaminated ($D_2$),  skewed distributed ($D_3$),  and heavy-tailed distributed  ($D_4$) data generated from $g$ = $g_2$ S = $S_1$ $n$ = 300 $p$ = 400, displayed in the form of: mean (SD) calculated from 100 independent runs of the experiment.} 
\end{table}
\begin{table}[H]
\centering
\begin{tabular}{lccccc}
  \hline
 & L2Boost & MBoost & LADBoost & Robloss & RRBoost \\ 
  \hline
$D_0$ & 0.81 (0.14) & 0.80 (0.16) & 0.80 (0.14) & 0.78 (0.16) & 0.83 (0.14) \\ 
  $D_1 (10\%)$ & 0.27 (0.06) & 0.48 (0.24) & 0.74 (0.17) & 0.72 (0.15) & 0.80 (0.14) \\ 
  $D_1 (20\%)$ & 0.24 (0.04) & 0.24 (0.04) & 0.67 (0.15) & 0.62 (0.14) & 0.80 (0.12) \\ 
  $D_2 (10\%)$ & 0.26 (0.06) & 0.50 (0.18) & 0.74 (0.18) & 0.72 (0.17) & 0.80 (0.13) \\ 
  $D_2 (20\%)$ & 0.25 (0.04) & 0.27 (0.06) & 0.64 (0.16) & 0.58 (0.16) & 0.79 (0.14) \\ 
  $D_3$ & 0.63 (0.19) & 0.76 (0.16) & 0.75 (0.16) & 0.76 (0.16) & 0.82 (0.12) \\ 
  $D_4$ & 0.06 (0.10) & 0.54 (0.16) & 0.64 (0.15) & 0.64 (0.17) & 0.74 (0.13) \\ 
   \hline
\end{tabular}
\caption{Fractions of variables recovered by L2Boost, MBoost, LADBoost, Robloss, and RRBoost applied with tree learners of $d$ = 1 for clean ($D_0$), symmetric gross error contaminated ($D_1$), asymmetric gross error contaminated ($D_2$),  skewed distributed ($D_3$),  and heavy-tailed distributed  ($D_4$) data generated from $g$ = $g_2$ S = $S_1$ $n$ = 300 $p$ = 400, displayed in the form of: mean (SD) calculated from 100 independent runs of the experiment.} 
\end{table}
\begin{table}[H]
\centering
\begin{tabular}{lcccccc}
  \hline
 & L2Boost & MBoost & LADBoost & Robloss & SBoost & RRBoost \\ 
  \hline
$D_0$ & 1.18 (0.05) & 1.19 (0.06) & 1.31 (0.11) & 1.21 (0.07) & 1.36 (0.24) & 1.17 (0.06) \\ 
  $D_1 (10\%)$ & 2.34 (0.24) & 1.77 (0.40) & 1.39 (0.14) & 1.32 (0.12) & 1.38 (0.18) & 1.21 (0.08) \\ 
  $D_1 (20\%)$ & 2.81 (0.26) & 2.82 (0.29) & 1.56 (0.16) & 1.58 (0.20) & 1.37 (0.15) & 1.23 (0.08) \\ 
  $D_2 (10\%)$ & 2.79 (0.29) & 2.12 (0.62) & 1.36 (0.11) & 1.28 (0.11) & 1.37 (0.13) & 1.20 (0.07) \\ 
  $D_2 (20\%)$ & 3.98 (0.35) & 4.06 (0.35) & 1.54 (0.16) & 1.72 (0.22) & 1.33 (0.10) & 1.23 (0.09) \\ 
  $D_3$ & 1.39 (0.20) & 1.24 (0.07) & 1.34 (0.09) & 1.24 (0.08) & 1.40 (0.21) & 1.23 (0.07) \\ 
  $D_4$ & 19.79 (54.17) & 1.57 (0.18) & 1.52 (0.14) & 1.45 (0.16) & 1.40 (0.16) & 1.34 (0.10) \\ 
   \hline
\end{tabular}
\caption{Summary statistics of RMSEs on the test sets by L2Boost, MBoost, LADBoost, Robloss, SBoost, and RRBoost applied with tree learners of $d$ = 2 for clean ($D_0$), symmetric gross error contaminated ($D_1$), asymmetric gross error contaminated ($D_2$),  skewed distributed ($D_3$),  and heavy-tailed distributed  ($D_4$) data generated from $g$ = $g_2$ S = $S_1$ $n$ = 300 $p$ = 400, displayed in the form of: mean (SD) calculated from 100 independent runs of the experiment.} 
\end{table}
\begin{table}[H]
\centering
\begin{tabular}{lccccc}
  \hline
 & L2Boost & MBoost & LADBoost & Robloss & RRBoost \\ 
  \hline
$D_0$ & 0.76 (0.11) & 0.78 (0.10) & 0.80 (0.13) & 0.80 (0.12) & 0.80 (0.13) \\ 
  $D_1 (10\%)$ & 0.29 (0.11) & 0.50 (0.21) & 0.75 (0.15) & 0.74 (0.14) & 0.80 (0.14) \\ 
  $D_1 (20\%)$ & 0.27 (0.07) & 0.26 (0.08) & 0.70 (0.15) & 0.62 (0.17) & 0.77 (0.13) \\ 
  $D_2 (10\%)$ & 0.29 (0.11) & 0.49 (0.22) & 0.76 (0.15) & 0.77 (0.12) & 0.80 (0.13) \\ 
  $D_2 (20\%)$ & 0.27 (0.07) & 0.28 (0.08) & 0.71 (0.17) & 0.62 (0.18) & 0.78 (0.13) \\ 
  $D_3$ & 0.71 (0.15) & 0.78 (0.13) & 0.81 (0.14) & 0.80 (0.12) & 0.80 (0.13) \\ 
  $D_4$ & 0.12 (0.14) & 0.55 (0.17) & 0.68 (0.15) & 0.68 (0.16) & 0.74 (0.15) \\ 
   \hline
\end{tabular}
\caption{Fractions of variables recovered by L2Boost, MBoost, LADBoost, Robloss, and RRBoost applied with tree learners of $d$ = 2 for clean ($D_0$), symmetric gross error contaminated ($D_1$), asymmetric gross error contaminated ($D_2$),  skewed distributed ($D_3$),  and heavy-tailed distributed  ($D_4$) data generated from $g$ = $g_2$ S = $S_1$ $n$ = 300 $p$ = 400, displayed in the form of: mean (SD) calculated from 100 independent runs of the experiment.} 
\end{table}
\begin{table}[H]
\centering
\begin{tabular}{lcccccc}
  \hline
 & L2Boost & MBoost & LADBoost & Robloss & SBoost & RRBoost \\ 
  \hline
$D_0$ & 1.16 (0.05) & 1.20 (0.06) & 1.45 (0.10) & 1.28 (0.08) & 1.29 (0.18) & 1.21 (0.06) \\ 
  $D_1 (10\%)$ & 2.82 (0.31) & 2.13 (0.53) & 1.58 (0.12) & 1.50 (0.10) & 1.31 (0.11) & 1.25 (0.07) \\ 
  $D_1 (20\%)$ & 3.61 (0.39) & 3.64 (0.40) & 1.79 (0.15) & 1.81 (0.20) & 1.39 (0.15) & 1.29 (0.08) \\ 
  $D_2 (10\%)$ & 3.12 (0.33) & 2.40 (0.64) & 1.52 (0.10) & 1.43 (0.11) & 1.37 (0.15) & 1.27 (0.09) \\ 
  $D_2 (20\%)$ & 4.46 (0.38) & 4.53 (0.40) & 1.75 (0.16) & 2.02 (0.26) & 1.42 (0.11) & 1.33 (0.09) \\ 
  $D_3$ & 1.41 (0.18) & 1.29 (0.07) & 1.47 (0.11) & 1.36 (0.08) & 1.31 (0.14) & 1.26 (0.08) \\ 
  $D_4$ & 19.78 (50.03) & 1.78 (0.21) & 1.74 (0.12) & 1.62 (0.14) & 1.39 (0.13) & 1.37 (0.10) \\ 
   \hline
\end{tabular}
\caption{Summary statistics of RMSEs on the test sets by L2Boost, MBoost, LADBoost, Robloss, SBoost, and RRBoost applied with tree learners of $d$ = 3 for clean ($D_0$), symmetric gross error contaminated ($D_1$), asymmetric gross error contaminated ($D_2$),  skewed distributed ($D_3$),  and heavy-tailed distributed  ($D_4$) data generated from $g$ = $g_2$ S = $S_1$ $n$ = 300 $p$ = 400, displayed in the form of: mean (SD) calculated from 100 independent runs of the experiment.} 
\end{table}
\begin{table}[H]
\centering
\begin{tabular}{lccccc}
  \hline
 & L2Boost & MBoost & LADBoost & Robloss & RRBoost \\ 
  \hline
$D_0$ & 0.80 (0.10) & 0.79 (0.10) & 0.82 (0.12) & 0.81 (0.11) & 0.78 (0.11) \\ 
  $D_1 (10\%)$ & 0.35 (0.13) & 0.54 (0.20) & 0.78 (0.16) & 0.74 (0.15) & 0.79 (0.14) \\ 
  $D_1 (20\%)$ & 0.26 (0.07) & 0.26 (0.09) & 0.66 (0.14) & 0.63 (0.16) & 0.77 (0.12) \\ 
  $D_2 (10\%)$ & 0.34 (0.13) & 0.52 (0.20) & 0.78 (0.16) & 0.77 (0.12) & 0.78 (0.14) \\ 
  $D_2 (20\%)$ & 0.28 (0.09) & 0.27 (0.08) & 0.66 (0.15) & 0.58 (0.15) & 0.75 (0.13) \\ 
  $D_3$ & 0.72 (0.14) & 0.78 (0.12) & 0.79 (0.14) & 0.78 (0.11) & 0.76 (0.11) \\ 
  $D_4$ & 0.22 (0.16) & 0.61 (0.16) & 0.67 (0.16) & 0.68 (0.16) & 0.74 (0.15) \\ 
   \hline
\end{tabular}
\caption{Fractions of variables recovered by L2Boost, MBoost, LADBoost, Robloss, and RRBoost applied with tree learners of $d$ = 3 for clean ($D_0$), symmetric gross error contaminated ($D_1$), asymmetric gross error contaminated ($D_2$),  skewed distributed ($D_3$),  and heavy-tailed distributed  ($D_4$) data generated from $g$ = $g_2$ S = $S_1$ $n$ = 300 $p$ = 400, displayed in the form of: mean (SD) calculated from 100 independent runs of the experiment.} 
\end{table}
\begin{table}[H]
\centering
\begin{tabular}{lcccccc}
  \hline
 & L2Boost & MBoost & LADBoost & Robloss & SBoost & RRBoost \\ 
  \hline
$D_0$ & 0.87 (0.02) & 0.88 (0.02) & 0.90 (0.02) & 0.88 (0.02) & 0.88 (0.03) & 0.83 (0.02) \\ 
  $D_1 (10\%)$ & 1.26 (0.08) & 1.08 (0.15) & 0.92 (0.02) & 0.90 (0.02) & 0.86 (0.02) & 0.82 (0.02) \\ 
  $D_1 (20\%)$ & 1.43 (0.10) & 1.45 (0.08) & 0.95 (0.03) & 0.93 (0.03) & 0.86 (0.02) & 0.83 (0.02) \\ 
  $D_2 (10\%)$ & 1.90 (0.07) & 1.33 (0.31) & 0.92 (0.02) & 0.92 (0.02) & 0.86 (0.02) & 0.82 (0.02) \\ 
  $D_2 (20\%)$ & 3.17 (0.09) & 3.11 (0.11) & 0.99 (0.03) & 1.03 (0.03) & 0.86 (0.02) & 0.83 (0.02) \\ 
  $D_3$ & 1.02 (0.08) & 0.92 (0.02) & 0.96 (0.02) & 0.93 (0.02) & 0.99 (0.02) & 0.90 (0.02) \\ 
  $D_4$ & 52.04 (310.45) & 1.04 (0.04) & 0.96 (0.02) & 0.95 (0.03) & 0.86 (0.02) & 0.84 (0.02) \\ 
   \hline
\end{tabular}
\caption{Summary statistics of RMSEs on the test sets by L2Boost, MBoost, LADBoost, Robloss, SBoost, and RRBoost applied with tree learners of $d$ = 1 for clean ($D_0$), symmetric gross error contaminated ($D_1$), asymmetric gross error contaminated ($D_2$),  skewed distributed ($D_3$),  and heavy-tailed distributed  ($D_4$) data generated from $g$ = $g_2$ S = $S_1$ $n$ = 3000 $p$ = 400, displayed in the form of: mean (SD) calculated from 100 independent runs of the experiment.} 
\end{table}
\begin{table}[H]
\centering
\begin{tabular}{lccccc}
  \hline
 & L2Boost & MBoost & LADBoost & Robloss & RRBoost \\ 
  \hline
$D_0$ & 1.00 (0.00) & 1.00 (0.00) & 1.00 (0.00) & 1.00 (0.00) & 1.00 (0.00) \\ 
  $D_1 (10\%)$ & 0.70 (0.16) & 0.85 (0.17) & 1.00 (0.00) & 1.00 (0.03) & 1.00 (0.00) \\ 
  $D_1 (20\%)$ & 0.56 (0.12) & 0.55 (0.11) & 0.98 (0.07) & 0.99 (0.06) & 1.00 (0.00) \\ 
  $D_2 (10\%)$ & 0.69 (0.15) & 0.85 (0.13) & 0.99 (0.04) & 1.00 (0.03) & 1.00 (0.00) \\ 
  $D_2 (20\%)$ & 0.63 (0.16) & 0.55 (0.11) & 0.99 (0.04) & 0.99 (0.04) & 1.00 (0.00) \\ 
  $D_3$ & 0.91 (0.12) & 1.00 (0.00) & 1.00 (0.00) & 1.00 (0.00) & 1.00 (0.00) \\ 
  $D_4$ & 0.03 (0.09) & 0.92 (0.12) & 0.99 (0.06) & 0.98 (0.08) & 1.00 (0.00) \\ 
   \hline
\end{tabular}
\caption{Fractions of variables recovered by L2Boost, MBoost, LADBoost, Robloss, and RRBoost applied with tree learners of $d$ = 1 for clean ($D_0$), symmetric gross error contaminated ($D_1$), asymmetric gross error contaminated ($D_2$),  skewed distributed ($D_3$),  and heavy-tailed distributed  ($D_4$) data generated from $g$ = $g_2$ S = $S_1$ $n$ = 3000 $p$ = 400, displayed in the form of: mean (SD) calculated from 100 independent runs of the experiment.} 
\end{table}
\begin{table}[H]
\centering
\begin{tabular}{lcccccc}
  \hline
 & L2Boost & MBoost & LADBoost & Robloss & SBoost & RRBoost \\ 
  \hline
$D_0$ & 0.91 (0.03) & 0.91 (0.03) & 0.96 (0.05) & 0.93 (0.03) & 1.13 (0.16) & 0.83 (0.02) \\ 
  $D_1 (10\%)$ & 1.31 (0.07) & 1.14 (0.14) & 0.98 (0.04) & 0.96 (0.03) & 1.01 (0.11) & 0.83 (0.02) \\ 
  $D_1 (20\%)$ & 1.39 (0.08) & 1.41 (0.09) & 1.02 (0.05) & 1.01 (0.04) & 0.93 (0.09) & 0.82 (0.02) \\ 
  $D_2 (10\%)$ & 1.93 (0.07) & 1.38 (0.30) & 0.99 (0.04) & 0.98 (0.03) & 0.98 (0.07) & 0.82 (0.02) \\ 
  $D_2 (20\%)$ & 3.16 (0.08) & 3.16 (0.15) & 1.09 (0.05) & 1.10 (0.04) & 0.87 (0.04) & 0.82 (0.02) \\ 
  $D_3$ & 1.08 (0.06) & 0.96 (0.03) & 1.03 (0.04) & 0.98 (0.03) & 1.23 (0.15) & 0.89 (0.03) \\ 
  $D_4$ & 70.14 (350.39) & 1.10 (0.05) & 1.05 (0.04) & 1.01 (0.03) & 0.90 (0.07) & 0.84 (0.02) \\ 
   \hline
\end{tabular}
\caption{Summary statistics of RMSEs on the test sets by L2Boost, MBoost, LADBoost, Robloss, SBoost, and RRBoost applied with tree learners of $d$ = 2 for clean ($D_0$), symmetric gross error contaminated ($D_1$), asymmetric gross error contaminated ($D_2$),  skewed distributed ($D_3$),  and heavy-tailed distributed  ($D_4$) data generated from $g$ = $g_2$ S = $S_1$ $n$ = 3000 $p$ = 400, displayed in the form of: mean (SD) calculated from 100 independent runs of the experiment.} 
\end{table}
\begin{table}[H]
\centering
\begin{tabular}{lccccc}
  \hline
 & L2Boost & MBoost & LADBoost & Robloss & RRBoost \\ 
  \hline
$D_0$ & 1.00 (0.00) & 1.00 (0.00) & 1.00 (0.00) & 1.00 (0.00) & 1.00 (0.00) \\ 
  $D_1 (10\%)$ & 0.73 (0.12) & 0.86 (0.14) & 1.00 (0.03) & 1.00 (0.03) & 1.00 (0.00) \\ 
  $D_1 (20\%)$ & 0.64 (0.13) & 0.64 (0.14) & 0.97 (0.08) & 0.98 (0.07) & 1.00 (0.00) \\ 
  $D_2 (10\%)$ & 0.66 (0.12) & 0.78 (0.12) & 1.00 (0.03) & 1.00 (0.00) & 1.00 (0.00) \\ 
  $D_2 (20\%)$ & 0.52 (0.11) & 0.57 (0.11) & 0.96 (0.09) & 0.92 (0.12) & 1.00 (0.00) \\ 
  $D_3$ & 0.83 (0.12) & 1.00 (0.00) & 1.00 (0.03) & 1.00 (0.00) & 1.00 (0.00) \\ 
  $D_4$ & 0.03 (0.08) & 0.88 (0.13) & 0.97 (0.08) & 0.96 (0.09) & 1.00 (0.00) \\ 
   \hline
\end{tabular}
\caption{Fractions of variables recovered by L2Boost, MBoost, LADBoost, Robloss, and RRBoost applied with tree learners of $d$ = 2 for clean ($D_0$), symmetric gross error contaminated ($D_1$), asymmetric gross error contaminated ($D_2$),  skewed distributed ($D_3$),  and heavy-tailed distributed  ($D_4$) data generated from $g$ = $g_2$ S = $S_1$ $n$ = 3000 $p$ = 400, displayed in the form of: mean (SD) calculated from 100 independent runs of the experiment.} 
\end{table}
\begin{table}[H]
\centering
\begin{tabular}{lcccccc}
  \hline
 & L2Boost & MBoost & LADBoost & Robloss & SBoost & RRBoost \\ 
  \hline
$D_0$ & 0.91 (0.03) & 0.91 (0.03) & 1.00 (0.04) & 0.94 (0.03) & 1.14 (0.18) & 0.90 (0.21) \\ 
  $D_1 (10\%)$ & 1.35 (0.10) & 1.16 (0.13) & 1.04 (0.04) & 0.99 (0.04) & 1.06 (0.13) & 0.85 (0.03) \\ 
  $D_1 (20\%)$ & 1.61 (0.12) & 1.64 (0.14) & 1.08 (0.04) & 1.04 (0.05) & 1.01 (0.13) & 0.85 (0.03) \\ 
  $D_2 (10\%)$ & 2.01 (0.10) & 1.44 (0.35) & 1.04 (0.04) & 0.98 (0.03) & 1.07 (0.13) & 0.85 (0.04) \\ 
  $D_2 (20\%)$ & 3.24 (0.10) & 3.37 (0.16) & 1.14 (0.05) & 1.14 (0.05) & 1.03 (0.07) & 0.84 (0.03) \\ 
  $D_3$ & 1.10 (0.07) & 0.96 (0.03) & 1.06 (0.04) & 0.98 (0.04) & 1.27 (0.28) & 1.03 (0.21) \\ 
  $D_4$ & 79.72 (385.56) & 1.12 (0.04) & 1.09 (0.05) & 1.04 (0.04) & 1.07 (0.19) & 0.90 (0.04) \\ 
   \hline
\end{tabular}
\caption{Summary statistics of RMSEs on the test sets by L2Boost, MBoost, LADBoost, Robloss, SBoost, and RRBoost applied with tree learners of $d$ = 3 for clean ($D_0$), symmetric gross error contaminated ($D_1$), asymmetric gross error contaminated ($D_2$),  skewed distributed ($D_3$),  and heavy-tailed distributed  ($D_4$) data generated from $g$ = $g_2$ S = $S_1$ $n$ = 3000 $p$ = 400, displayed in the form of: mean (SD) calculated from 100 independent runs of the experiment.} 
\end{table}
\begin{table}[H]
\centering
\begin{tabular}{lccccc}
  \hline
 & L2Boost & MBoost & LADBoost & Robloss & RRBoost \\ 
  \hline
$D_0$ & 1.00 (0.00) & 1.00 (0.00) & 1.00 (0.03) & 1.00 (0.00) & 1.00 (0.03) \\ 
  $D_1 (10\%)$ & 0.74 (0.13) & 0.83 (0.14) & 0.97 (0.08) & 0.99 (0.05) & 1.00 (0.00) \\ 
  $D_1 (20\%)$ & 0.71 (0.14) & 0.70 (0.14) & 0.93 (0.11) & 0.96 (0.09) & 1.00 (0.00) \\ 
  $D_2 (10\%)$ & 0.64 (0.13) & 0.79 (0.15) & 0.96 (0.09) & 1.00 (0.03) & 1.00 (0.00) \\ 
  $D_2 (20\%)$ & 0.52 (0.10) & 0.53 (0.10) & 0.94 (0.11) & 0.91 (0.12) & 1.00 (0.00) \\ 
  $D_3$ & 0.85 (0.12) & 0.99 (0.04) & 0.99 (0.06) & 0.99 (0.04) & 1.00 (0.00) \\ 
  $D_4$ & 0.08 (0.13) & 0.84 (0.12) & 0.90 (0.12) & 0.94 (0.11) & 1.00 (0.03) \\ 
   \hline
\end{tabular}
\caption{Fractions of variables recovered by L2Boost, MBoost, LADBoost, Robloss, and RRBoost applied with tree learners of $d$ = 3 for clean ($D_0$), symmetric gross error contaminated ($D_1$), asymmetric gross error contaminated ($D_2$),  skewed distributed ($D_3$),  and heavy-tailed distributed  ($D_4$) data generated from $g$ = $g_2$ S = $S_1$ $n$ = 3000 $p$ = 400, displayed in the form of: mean (SD) calculated from 100 independent runs of the experiment.} 
\end{table}
\begin{table}[H]
\centering
\begin{tabular}{lcccccc}
  \hline
 & L2Boost & MBoost & LADBoost & Robloss & SBoost & RRBoost \\ 
  \hline
$D_0$ & 1.23 (0.04) & 1.23 (0.04) & 1.24 (0.04) & 1.23 (0.05) & 1.26 (0.09) & 1.19 (0.06) \\ 
  $D_1 (10\%)$ & 1.81 (0.25) & 1.53 (0.24) & 1.28 (0.06) & 1.30 (0.06) & 1.27 (0.09) & 1.21 (0.06) \\ 
  $D_1 (20\%)$ & 2.03 (0.29) & 2.01 (0.28) & 1.33 (0.07) & 1.38 (0.08) & 1.28 (0.07) & 1.23 (0.05) \\ 
  $D_2 (10\%)$ & 2.31 (0.20) & 1.79 (0.41) & 1.29 (0.06) & 1.32 (0.08) & 1.27 (0.08) & 1.21 (0.07) \\ 
  $D_2 (20\%)$ & 3.58 (0.27) & 3.57 (0.32) & 1.42 (0.10) & 1.53 (0.12) & 1.31 (0.07) & 1.25 (0.05) \\ 
  $D_3$ & 1.37 (0.10) & 1.29 (0.06) & 1.29 (0.04) & 1.28 (0.05) & 1.36 (0.10) & 1.26 (0.06) \\ 
  $D_4$ & 28.78 (116.61) & 1.47 (0.10) & 1.35 (0.07) & 1.39 (0.08) & 1.32 (0.07) & 1.29 (0.07) \\ 
   \hline
\end{tabular}
\caption{Summary statistics of RMSEs on the test sets by L2Boost, MBoost, LADBoost, Robloss, SBoost, and RRBoost applied with tree learners of $d$ = 1 for clean ($D_0$), symmetric gross error contaminated ($D_1$), asymmetric gross error contaminated ($D_2$),  skewed distributed ($D_3$),  and heavy-tailed distributed  ($D_4$) data generated from $g$ = $g_2$ S = $S_2$ $n$ = 300 $p$ = 10, displayed in the form of: mean (SD) calculated from 100 independent runs of the experiment.} 
\end{table}
\begin{table}[H]
\centering
\begin{tabular}{lccccc}
  \hline
 & L2Boost & MBoost & LADBoost & Robloss & RRBoost \\ 
  \hline
$D_0$ & 0.88 (0.14) & 0.86 (0.13) & 0.87 (0.13) & 0.87 (0.13) & 0.89 (0.13) \\ 
  $D_1 (10\%)$ & 0.47 (0.18) & 0.64 (0.24) & 0.84 (0.16) & 0.84 (0.15) & 0.89 (0.12) \\ 
  $D_1 (20\%)$ & 0.34 (0.17) & 0.36 (0.15) & 0.78 (0.18) & 0.76 (0.18) & 0.88 (0.13) \\ 
  $D_2 (10\%)$ & 0.47 (0.19) & 0.68 (0.21) & 0.87 (0.16) & 0.83 (0.16) & 0.89 (0.12) \\ 
  $D_2 (20\%)$ & 0.41 (0.18) & 0.49 (0.18) & 0.80 (0.16) & 0.74 (0.16) & 0.90 (0.13) \\ 
  $D_3$ & 0.76 (0.20) & 0.86 (0.14) & 0.84 (0.16) & 0.86 (0.13) & 0.90 (0.13) \\ 
  $D_4$ & 0.28 (0.24) & 0.63 (0.17) & 0.78 (0.17) & 0.73 (0.18) & 0.82 (0.15) \\ 
   \hline
\end{tabular}
\caption{Fractions of variables recovered by L2Boost, MBoost, LADBoost, Robloss, and RRBoost applied with tree learners of $d$ = 1 for clean ($D_0$), symmetric gross error contaminated ($D_1$), asymmetric gross error contaminated ($D_2$),  skewed distributed ($D_3$),  and heavy-tailed distributed  ($D_4$) data generated from $g$ = $g_2$ S = $S_2$ $n$ = 300 $p$ = 10, displayed in the form of: mean (SD) calculated from 100 independent runs of the experiment.} 
\end{table}
\begin{table}[H]
\centering
\begin{tabular}{lcccccc}
  \hline
 & L2Boost & MBoost & LADBoost & Robloss & SBoost & RRBoost \\ 
  \hline
$D_0$ & 1.16 (0.06) & 1.16 (0.06) & 1.16 (0.06) & 1.15 (0.06) & 1.32 (0.11) & 1.14 (0.06) \\ 
  $D_1 (10\%)$ & 1.94 (0.28) & 1.60 (0.33) & 1.22 (0.06) & 1.23 (0.07) & 1.29 (0.09) & 1.16 (0.06) \\ 
  $D_1 (20\%)$ & 2.43 (0.37) & 2.44 (0.31) & 1.31 (0.08) & 1.35 (0.10) & 1.28 (0.08) & 1.18 (0.06) \\ 
  $D_2 (10\%)$ & 2.48 (0.28) & 1.92 (0.53) & 1.25 (0.08) & 1.28 (0.09) & 1.28 (0.09) & 1.16 (0.07) \\ 
  $D_2 (20\%)$ & 3.81 (0.36) & 3.88 (0.35) & 1.41 (0.11) & 1.55 (0.14) & 1.26 (0.07) & 1.18 (0.06) \\ 
  $D_3$ & 1.37 (0.13) & 1.25 (0.07) & 1.22 (0.07) & 1.23 (0.06) & 1.36 (0.09) & 1.21 (0.07) \\ 
  $D_4$ & 49.60 (307.70) & 1.51 (0.11) & 1.34 (0.09) & 1.38 (0.10) & 1.34 (0.09) & 1.28 (0.08) \\ 
   \hline
\end{tabular}
\caption{Summary statistics of RMSEs on the test sets by L2Boost, MBoost, LADBoost, Robloss, SBoost, and RRBoost applied with tree learners of $d$ = 2 for clean ($D_0$), symmetric gross error contaminated ($D_1$), asymmetric gross error contaminated ($D_2$),  skewed distributed ($D_3$),  and heavy-tailed distributed  ($D_4$) data generated from $g$ = $g_2$ S = $S_2$ $n$ = 300 $p$ = 10, displayed in the form of: mean (SD) calculated from 100 independent runs of the experiment.} 
\end{table}
\begin{table}[H]
\centering
\begin{tabular}{lccccc}
  \hline
 & L2Boost & MBoost & LADBoost & Robloss & RRBoost \\ 
  \hline
$D_0$ & 0.93 (0.11) & 0.90 (0.13) & 0.95 (0.10) & 0.91 (0.12) & 0.94 (0.11) \\ 
  $D_1 (10\%)$ & 0.54 (0.15) & 0.71 (0.22) & 0.92 (0.12) & 0.90 (0.13) & 0.92 (0.12) \\ 
  $D_1 (20\%)$ & 0.46 (0.16) & 0.46 (0.17) & 0.90 (0.13) & 0.80 (0.18) & 0.92 (0.12) \\ 
  $D_2 (10\%)$ & 0.55 (0.18) & 0.70 (0.18) & 0.92 (0.12) & 0.87 (0.15) & 0.92 (0.12) \\ 
  $D_2 (20\%)$ & 0.49 (0.18) & 0.54 (0.19) & 0.85 (0.17) & 0.78 (0.17) & 0.91 (0.12) \\ 
  $D_3$ & 0.79 (0.17) & 0.88 (0.13) & 0.92 (0.13) & 0.88 (0.14) & 0.95 (0.10) \\ 
  $D_4$ & 0.35 (0.19) & 0.68 (0.15) & 0.84 (0.16) & 0.78 (0.18) & 0.88 (0.14) \\ 
   \hline
\end{tabular}
\caption{Fractions of variables recovered by L2Boost, MBoost, LADBoost, Robloss, and RRBoost applied with tree learners of $d$ = 2 for clean ($D_0$), symmetric gross error contaminated ($D_1$), asymmetric gross error contaminated ($D_2$),  skewed distributed ($D_3$),  and heavy-tailed distributed  ($D_4$) data generated from $g$ = $g_2$ S = $S_2$ $n$ = 300 $p$ = 10, displayed in the form of: mean (SD) calculated from 100 independent runs of the experiment.} 
\end{table}
\begin{table}[H]
\centering
\begin{tabular}{lcccccc}
  \hline
 & L2Boost & MBoost & LADBoost & Robloss & SBoost & RRBoost \\ 
  \hline
$D_0$ & 1.17 (0.06) & 1.18 (0.06) & 1.23 (0.07) & 1.18 (0.05) & 1.32 (0.13) & 1.18 (0.07) \\ 
  $D_1 (10\%)$ & 2.29 (0.26) & 1.80 (0.43) & 1.30 (0.08) & 1.29 (0.07) & 1.33 (0.11) & 1.21 (0.08) \\ 
  $D_1 (20\%)$ & 2.94 (0.40) & 2.98 (0.34) & 1.40 (0.09) & 1.45 (0.11) & 1.32 (0.11) & 1.23 (0.07) \\ 
  $D_2 (10\%)$ & 2.78 (0.30) & 2.11 (0.66) & 1.31 (0.08) & 1.31 (0.09) & 1.30 (0.08) & 1.20 (0.07) \\ 
  $D_2 (20\%)$ & 4.15 (0.37) & 4.19 (0.37) & 1.49 (0.12) & 1.65 (0.16) & 1.29 (0.08) & 1.22 (0.06) \\ 
  $D_3$ & 1.37 (0.15) & 1.26 (0.07) & 1.29 (0.07) & 1.26 (0.06) & 1.38 (0.12) & 1.27 (0.08) \\ 
  $D_4$ & 50.51 (307.88) & 1.52 (0.13) & 1.41 (0.10) & 1.41 (0.11) & 1.36 (0.11) & 1.32 (0.09) \\ 
   \hline
\end{tabular}
\caption{Summary statistics of RMSEs on the test sets by L2Boost, MBoost, LADBoost, Robloss, SBoost, and RRBoost applied with tree learners of $d$ = 3 for clean ($D_0$), symmetric gross error contaminated ($D_1$), asymmetric gross error contaminated ($D_2$),  skewed distributed ($D_3$),  and heavy-tailed distributed  ($D_4$) data generated from $g$ = $g_2$ S = $S_2$ $n$ = 300 $p$ = 10, displayed in the form of: mean (SD) calculated from 100 independent runs of the experiment.} 
\end{table}
\begin{table}[H]
\centering
\begin{tabular}{lccccc}
  \hline
 & L2Boost & MBoost & LADBoost & Robloss & RRBoost \\ 
  \hline
$D_0$ & 0.90 (0.13) & 0.89 (0.12) & 0.92 (0.12) & 0.90 (0.12) & 0.92 (0.12) \\ 
  $D_1 (10\%)$ & 0.65 (0.18) & 0.76 (0.17) & 0.92 (0.12) & 0.88 (0.14) & 0.91 (0.12) \\ 
  $D_1 (20\%)$ & 0.56 (0.16) & 0.55 (0.18) & 0.88 (0.14) & 0.83 (0.16) & 0.90 (0.13) \\ 
  $D_2 (10\%)$ & 0.62 (0.16) & 0.76 (0.17) & 0.94 (0.11) & 0.89 (0.13) & 0.93 (0.11) \\ 
  $D_2 (20\%)$ & 0.56 (0.18) & 0.54 (0.18) & 0.85 (0.13) & 0.81 (0.15) & 0.94 (0.11) \\ 
  $D_3$ & 0.79 (0.16) & 0.86 (0.14) & 0.94 (0.12) & 0.90 (0.12) & 0.93 (0.11) \\ 
  $D_4$ & 0.46 (0.21) & 0.78 (0.14) & 0.88 (0.13) & 0.84 (0.15) & 0.86 (0.15) \\ 
   \hline
\end{tabular}
\caption{Fractions of variables recovered by L2Boost, MBoost, LADBoost, Robloss, and RRBoost applied with tree learners of $d$ = 3 for clean ($D_0$), symmetric gross error contaminated ($D_1$), asymmetric gross error contaminated ($D_2$),  skewed distributed ($D_3$),  and heavy-tailed distributed  ($D_4$) data generated from $g$ = $g_2$ S = $S_2$ $n$ = 300 $p$ = 10, displayed in the form of: mean (SD) calculated from 100 independent runs of the experiment.} 
\end{table}
\begin{table}[H]
\centering
\begin{tabular}{lcccccc}
  \hline
 & L2Boost & MBoost & LADBoost & Robloss & SBoost & RRBoost \\ 
  \hline
$D_0$ & 1.08 (0.02) & 1.09 (0.02) & 1.09 (0.03) & 1.09 (0.02) & 1.02 (0.03) & 0.99 (0.03) \\ 
  $D_1 (10\%)$ & 1.32 (0.06) & 1.23 (0.11) & 1.11 (0.03) & 1.10 (0.03) & 1.03 (0.04) & 1.00 (0.03) \\ 
  $D_1 (20\%)$ & 1.43 (0.07) & 1.42 (0.06) & 1.12 (0.03) & 1.13 (0.03) & 1.03 (0.03) & 1.01 (0.03) \\ 
  $D_2 (10\%)$ & 2.03 (0.07) & 1.51 (0.31) & 1.12 (0.02) & 1.12 (0.03) & 1.02 (0.03) & 1.00 (0.03) \\ 
  $D_2 (20\%)$ & 3.36 (0.09) & 3.26 (0.12) & 1.20 (0.05) & 1.25 (0.05) & 1.03 (0.04) & 1.01 (0.04) \\ 
  $D_3$ & 1.14 (0.04) & 1.11 (0.03) & 1.14 (0.03) & 1.13 (0.02) & 1.18 (0.05) & 1.06 (0.03) \\ 
  $D_4$ & 6.77 (18.70) & 1.17 (0.03) & 1.13 (0.03) & 1.13 (0.03) & 1.04 (0.03) & 1.03 (0.03) \\ 
   \hline
\end{tabular}
\caption{Summary statistics of RMSEs on the test sets by L2Boost, MBoost, LADBoost, Robloss, SBoost, and RRBoost applied with tree learners of $d$ = 1 for clean ($D_0$), symmetric gross error contaminated ($D_1$), asymmetric gross error contaminated ($D_2$),  skewed distributed ($D_3$),  and heavy-tailed distributed  ($D_4$) data generated from $g$ = $g_2$ S = $S_2$ $n$ = 3000 $p$ = 10, displayed in the form of: mean (SD) calculated from 100 independent runs of the experiment.} 
\end{table}
\begin{table}[H]
\centering
\begin{tabular}{lccccc}
  \hline
 & L2Boost & MBoost & LADBoost & Robloss & RRBoost \\ 
  \hline
$D_0$ & 1.00 (0.02) & 1.00 (0.00) & 1.00 (0.00) & 1.00 (0.02) & 1.00 (0.00) \\ 
  $D_1 (10\%)$ & 0.77 (0.19) & 0.92 (0.13) & 1.00 (0.02) & 1.00 (0.04) & 1.00 (0.00) \\ 
  $D_1 (20\%)$ & 0.69 (0.19) & 0.64 (0.17) & 0.99 (0.05) & 0.99 (0.05) & 1.00 (0.00) \\ 
  $D_2 (10\%)$ & 0.81 (0.15) & 0.88 (0.13) & 1.00 (0.02) & 1.00 (0.04) & 1.00 (0.00) \\ 
  $D_2 (20\%)$ & 0.71 (0.19) & 0.60 (0.18) & 0.97 (0.08) & 0.96 (0.09) & 1.00 (0.02) \\ 
  $D_3$ & 0.98 (0.08) & 1.00 (0.00) & 1.00 (0.04) & 0.99 (0.04) & 1.00 (0.00) \\ 
  $D_4$ & 0.23 (0.26) & 0.95 (0.10) & 0.98 (0.07) & 0.97 (0.08) & 0.98 (0.07) \\ 
   \hline
\end{tabular}
\caption{Fractions of variables recovered by L2Boost, MBoost, LADBoost, Robloss, and RRBoost applied with tree learners of $d$ = 1 for clean ($D_0$), symmetric gross error contaminated ($D_1$), asymmetric gross error contaminated ($D_2$),  skewed distributed ($D_3$),  and heavy-tailed distributed  ($D_4$) data generated from $g$ = $g_2$ S = $S_2$ $n$ = 3000 $p$ = 10, displayed in the form of: mean (SD) calculated from 100 independent runs of the experiment.} 
\end{table}
\begin{table}[H]
\centering
\begin{tabular}{lcccccc}
  \hline
 & L2Boost & MBoost & LADBoost & Robloss & SBoost & RRBoost \\ 
  \hline
$D_0$ & 0.91 (0.02) & 0.92 (0.03) & 0.93 (0.02) & 0.92 (0.03) & 0.89 (0.03) & 0.85 (0.02) \\ 
  $D_1 (10\%)$ & 1.32 (0.07) & 1.15 (0.16) & 0.95 (0.03) & 0.96 (0.03) & 0.88 (0.03) & 0.85 (0.02) \\ 
  $D_1 (20\%)$ & 1.46 (0.08) & 1.47 (0.08) & 0.98 (0.03) & 1.00 (0.03) & 0.87 (0.02) & 0.85 (0.02) \\ 
  $D_2 (10\%)$ & 2.02 (0.08) & 1.46 (0.35) & 0.97 (0.03) & 0.98 (0.03) & 0.88 (0.03) & 0.85 (0.02) \\ 
  $D_2 (20\%)$ & 3.36 (0.09) & 3.32 (0.13) & 1.07 (0.04) & 1.12 (0.05) & 0.88 (0.02) & 0.86 (0.02) \\ 
  $D_3$ & 1.05 (0.05) & 0.97 (0.03) & 1.01 (0.03) & 0.99 (0.03) & 1.06 (0.02) & 0.96 (0.02) \\ 
  $D_4$ & 30.16 (105.96) & 1.09 (0.05) & 1.00 (0.03) & 1.02 (0.04) & 0.87 (0.02) & 0.87 (0.02) \\ 
   \hline
\end{tabular}
\caption{Summary statistics of RMSEs on the test sets by L2Boost, MBoost, LADBoost, Robloss, SBoost, and RRBoost applied with tree learners of $d$ = 2 for clean ($D_0$), symmetric gross error contaminated ($D_1$), asymmetric gross error contaminated ($D_2$),  skewed distributed ($D_3$),  and heavy-tailed distributed  ($D_4$) data generated from $g$ = $g_2$ S = $S_2$ $n$ = 3000 $p$ = 10, displayed in the form of: mean (SD) calculated from 100 independent runs of the experiment.} 
\end{table}
\begin{table}[H]
\centering
\begin{tabular}{lccccc}
  \hline
 & L2Boost & MBoost & LADBoost & Robloss & RRBoost \\ 
  \hline
$D_0$ & 1.00 (0.00) & 1.00 (0.00) & 1.00 (0.00) & 1.00 (0.00) & 1.00 (0.00) \\ 
  $D_1 (10\%)$ & 0.80 (0.18) & 0.92 (0.12) & 1.00 (0.00) & 1.00 (0.02) & 1.00 (0.00) \\ 
  $D_1 (20\%)$ & 0.68 (0.20) & 0.67 (0.20) & 1.00 (0.00) & 1.00 (0.00) & 1.00 (0.00) \\ 
  $D_2 (10\%)$ & 0.78 (0.16) & 0.92 (0.12) & 1.00 (0.00) & 1.00 (0.00) & 1.00 (0.00) \\ 
  $D_2 (20\%)$ & 0.70 (0.17) & 0.78 (0.17) & 1.00 (0.04) & 0.99 (0.05) & 1.00 (0.00) \\ 
  $D_3$ & 0.96 (0.09) & 1.00 (0.00) & 1.00 (0.00) & 1.00 (0.00) & 1.00 (0.00) \\ 
  $D_4$ & 0.31 (0.19) & 0.95 (0.10) & 0.99 (0.05) & 0.99 (0.05) & 1.00 (0.00) \\ 
   \hline
\end{tabular}
\caption{Fractions of variables recovered by L2Boost, MBoost, LADBoost, Robloss, and RRBoost applied with tree learners of $d$ = 2 for clean ($D_0$), symmetric gross error contaminated ($D_1$), asymmetric gross error contaminated ($D_2$),  skewed distributed ($D_3$),  and heavy-tailed distributed  ($D_4$) data generated from $g$ = $g_2$ S = $S_2$ $n$ = 3000 $p$ = 10, displayed in the form of: mean (SD) calculated from 100 independent runs of the experiment.} 
\end{table}
\begin{table}[H]
\centering
\begin{tabular}{lcccccc}
  \hline
 & L2Boost & MBoost & LADBoost & Robloss & SBoost & RRBoost \\ 
  \hline
$D_0$ & 0.94 (0.03) & 0.95 (0.03) & 0.97 (0.03) & 0.96 (0.03) & 1.00 (0.07) & 0.87 (0.02) \\ 
  $D_1 (10\%)$ & 1.30 (0.08) & 1.16 (0.12) & 1.00 (0.03) & 1.00 (0.03) & 0.95 (0.05) & 0.87 (0.02) \\ 
  $D_1 (20\%)$ & 1.46 (0.10) & 1.50 (0.10) & 1.03 (0.03) & 1.04 (0.03) & 0.90 (0.04) & 0.86 (0.02) \\ 
  $D_2 (10\%)$ & 2.01 (0.08) & 1.51 (0.36) & 1.02 (0.03) & 1.02 (0.03) & 0.93 (0.04) & 0.86 (0.02) \\ 
  $D_2 (20\%)$ & 3.36 (0.10) & 3.39 (0.14) & 1.12 (0.04) & 1.16 (0.04) & 0.88 (0.02) & 0.86 (0.02) \\ 
  $D_3$ & 1.08 (0.05) & 1.01 (0.03) & 1.05 (0.03) & 1.02 (0.03) & 1.13 (0.06) & 0.97 (0.03) \\ 
  $D_4$ & 36.35 (127.01) & 1.12 (0.04) & 1.05 (0.04) & 1.07 (0.04) & 0.91 (0.04) & 0.89 (0.02) \\ 
   \hline
\end{tabular}
\caption{Summary statistics of RMSEs on the test sets by L2Boost, MBoost, LADBoost, Robloss, SBoost, and RRBoost applied with tree learners of $d$ = 3 for clean ($D_0$), symmetric gross error contaminated ($D_1$), asymmetric gross error contaminated ($D_2$),  skewed distributed ($D_3$),  and heavy-tailed distributed  ($D_4$) data generated from $g$ = $g_2$ S = $S_2$ $n$ = 3000 $p$ = 10, displayed in the form of: mean (SD) calculated from 100 independent runs of the experiment.} 
\end{table}
\begin{table}[H]
\centering
\begin{tabular}{lccccc}
  \hline
 & L2Boost & MBoost & LADBoost & Robloss & RRBoost \\ 
  \hline
$D_0$ & 1.00 (0.00) & 1.00 (0.00) & 1.00 (0.00) & 1.00 (0.00) & 1.00 (0.00) \\ 
  $D_1 (10\%)$ & 0.80 (0.11) & 0.88 (0.13) & 1.00 (0.00) & 1.00 (0.02) & 1.00 (0.00) \\ 
  $D_1 (20\%)$ & 0.81 (0.15) & 0.81 (0.15) & 1.00 (0.02) & 0.99 (0.04) & 1.00 (0.00) \\ 
  $D_2 (10\%)$ & 0.78 (0.13) & 0.90 (0.13) & 1.00 (0.04) & 1.00 (0.00) & 1.00 (0.00) \\ 
  $D_2 (20\%)$ & 0.72 (0.15) & 0.75 (0.16) & 0.99 (0.05) & 0.98 (0.08) & 1.00 (0.00) \\ 
  $D_3$ & 0.93 (0.11) & 1.00 (0.00) & 1.00 (0.00) & 1.00 (0.00) & 1.00 (0.00) \\ 
  $D_4$ & 0.48 (0.19) & 0.94 (0.10) & 0.98 (0.07) & 0.96 (0.09) & 1.00 (0.00) \\ 
   \hline
\end{tabular}
\caption{Fractions of variables recovered by L2Boost, MBoost, LADBoost, Robloss, and RRBoost applied with tree learners of $d$ = 3 for clean ($D_0$), symmetric gross error contaminated ($D_1$), asymmetric gross error contaminated ($D_2$),  skewed distributed ($D_3$),  and heavy-tailed distributed  ($D_4$) data generated from $g$ = $g_2$ S = $S_2$ $n$ = 3000 $p$ = 10, displayed in the form of: mean (SD) calculated from 100 independent runs of the experiment.} 
\end{table}
\begin{table}[H]
\centering
\begin{tabular}{lcccccc}
  \hline
 & L2Boost & MBoost & LADBoost & Robloss & SBoost & RRBoost \\ 
  \hline
$D_0$ & 1.21 (0.06) & 1.20 (0.06) & 1.21 (0.05) & 1.20 (0.05) & 1.26 (0.16) & 1.14 (0.06) \\ 
  $D_1 (10\%)$ & 1.66 (0.20) & 1.44 (0.17) & 1.27 (0.06) & 1.27 (0.06) & 1.29 (0.11) & 1.16 (0.05) \\ 
  $D_1 (20\%)$ & 1.96 (0.34) & 1.97 (0.35) & 1.31 (0.08) & 1.32 (0.09) & 1.31 (0.09) & 1.20 (0.07) \\ 
  $D_2 (10\%)$ & 2.02 (0.18) & 1.63 (0.28) & 1.27 (0.05) & 1.28 (0.07) & 1.30 (0.12) & 1.17 (0.07) \\ 
  $D_2 (20\%)$ & 3.03 (0.28) & 3.02 (0.32) & 1.41 (0.10) & 1.49 (0.17) & 1.31 (0.08) & 1.21 (0.07) \\ 
  $D_3$ & 1.30 (0.08) & 1.25 (0.05) & 1.26 (0.05) & 1.24 (0.06) & 1.30 (0.15) & 1.19 (0.06) \\ 
  $D_4$ & 18.19 (46.43) & 1.37 (0.09) & 1.30 (0.06) & 1.31 (0.08) & 1.35 (0.12) & 1.25 (0.08) \\ 
   \hline
\end{tabular}
\caption{Summary statistics of RMSEs on the test sets by L2Boost, MBoost, LADBoost, Robloss, SBoost, and RRBoost applied with tree learners of $d$ = 1 for clean ($D_0$), symmetric gross error contaminated ($D_1$), asymmetric gross error contaminated ($D_2$),  skewed distributed ($D_3$),  and heavy-tailed distributed  ($D_4$) data generated from $g$ = $g_2$ S = $S_2$ $n$ = 300 $p$ = 400, displayed in the form of: mean (SD) calculated from 100 independent runs of the experiment.} 
\end{table}
\begin{table}[H]
\centering
\begin{tabular}{lccccc}
  \hline
 & L2Boost & MBoost & LADBoost & Robloss & RRBoost \\ 
  \hline
$D_0$ & 0.72 (0.15) & 0.70 (0.14) & 0.69 (0.16) & 0.70 (0.17) & 0.74 (0.15) \\ 
  $D_1 (10\%)$ & 0.37 (0.13) & 0.50 (0.14) & 0.63 (0.15) & 0.61 (0.13) & 0.72 (0.17) \\ 
  $D_1 (20\%)$ & 0.24 (0.14) & 0.24 (0.15) & 0.60 (0.14) & 0.59 (0.13) & 0.71 (0.16) \\ 
  $D_2 (10\%)$ & 0.36 (0.12) & 0.53 (0.13) & 0.61 (0.14) & 0.60 (0.13) & 0.73 (0.15) \\ 
  $D_2 (20\%)$ & 0.26 (0.15) & 0.28 (0.14) & 0.59 (0.14) & 0.55 (0.10) & 0.70 (0.18) \\ 
  $D_3$ & 0.60 (0.13) & 0.65 (0.16) & 0.64 (0.15) & 0.66 (0.15) & 0.76 (0.17) \\ 
  $D_4$ & 0.04 (0.10) & 0.54 (0.11) & 0.59 (0.13) & 0.57 (0.13) & 0.68 (0.14) \\ 
   \hline
\end{tabular}
\caption{Fractions of variables recovered by L2Boost, MBoost, LADBoost, Robloss, and RRBoost applied with tree learners of $d$ = 1 for clean ($D_0$), symmetric gross error contaminated ($D_1$), asymmetric gross error contaminated ($D_2$),  skewed distributed ($D_3$),  and heavy-tailed distributed  ($D_4$) data generated from $g$ = $g_2$ S = $S_2$ $n$ = 300 $p$ = 400, displayed in the form of: mean (SD) calculated from 100 independent runs of the experiment.} 
\end{table}
\begin{table}[H]
\centering
\begin{tabular}{lcccccc}
  \hline
 & L2Boost & MBoost & LADBoost & Robloss & SBoost & RRBoost \\ 
  \hline
$D_0$ & 1.14 (0.08) & 1.14 (0.08) & 1.19 (0.07) & 1.14 (0.08) & 1.34 (0.18) & 1.12 (0.08) \\ 
  $D_1 (10\%)$ & 2.08 (0.26) & 1.63 (0.31) & 1.32 (0.11) & 1.26 (0.10) & 1.35 (0.13) & 1.14 (0.08) \\ 
  $D_1 (20\%)$ & 2.51 (0.32) & 2.52 (0.33) & 1.43 (0.13) & 1.45 (0.13) & 1.35 (0.11) & 1.18 (0.09) \\ 
  $D_2 (10\%)$ & 2.37 (0.24) & 1.86 (0.39) & 1.28 (0.09) & 1.29 (0.10) & 1.32 (0.11) & 1.14 (0.07) \\ 
  $D_2 (20\%)$ & 3.40 (0.34) & 3.45 (0.34) & 1.50 (0.13) & 1.64 (0.15) & 1.29 (0.09) & 1.17 (0.08) \\ 
  $D_3$ & 1.32 (0.14) & 1.21 (0.09) & 1.25 (0.10) & 1.19 (0.09) & 1.33 (0.15) & 1.19 (0.08) \\ 
  $D_4$ & 16.47 (44.82) & 1.43 (0.10) & 1.42 (0.11) & 1.36 (0.11) & 1.41 (0.15) & 1.31 (0.10) \\ 
   \hline
\end{tabular}
\caption{Summary statistics of RMSEs on the test sets by L2Boost, MBoost, LADBoost, Robloss, SBoost, and RRBoost applied with tree learners of $d$ = 2 for clean ($D_0$), symmetric gross error contaminated ($D_1$), asymmetric gross error contaminated ($D_2$),  skewed distributed ($D_3$),  and heavy-tailed distributed  ($D_4$) data generated from $g$ = $g_2$ S = $S_2$ $n$ = 300 $p$ = 400, displayed in the form of: mean (SD) calculated from 100 independent runs of the experiment.} 
\end{table}
\begin{table}[H]
\centering
\begin{tabular}{lccccc}
  \hline
 & L2Boost & MBoost & LADBoost & Robloss & RRBoost \\ 
  \hline
$D_0$ & 0.78 (0.17) & 0.74 (0.16) & 0.76 (0.16) & 0.76 (0.17) & 0.80 (0.12) \\ 
  $D_1 (10\%)$ & 0.37 (0.13) & 0.56 (0.15) & 0.71 (0.18) & 0.69 (0.16) & 0.81 (0.15) \\ 
  $D_1 (20\%)$ & 0.27 (0.16) & 0.26 (0.15) & 0.65 (0.17) & 0.59 (0.13) & 0.77 (0.15) \\ 
  $D_2 (10\%)$ & 0.37 (0.13) & 0.52 (0.10) & 0.74 (0.18) & 0.68 (0.17) & 0.81 (0.14) \\ 
  $D_2 (20\%)$ & 0.28 (0.15) & 0.30 (0.16) & 0.67 (0.17) & 0.57 (0.12) & 0.80 (0.15) \\ 
  $D_3$ & 0.62 (0.15) & 0.69 (0.16) & 0.74 (0.17) & 0.74 (0.17) & 0.79 (0.16) \\ 
  $D_4$ & 0.11 (0.16) & 0.57 (0.12) & 0.64 (0.16) & 0.63 (0.15) & 0.72 (0.15) \\ 
   \hline
\end{tabular}
\caption{Fractions of variables recovered by L2Boost, MBoost, LADBoost, Robloss, and RRBoost applied with tree learners of $d$ = 2 for clean ($D_0$), symmetric gross error contaminated ($D_1$), asymmetric gross error contaminated ($D_2$),  skewed distributed ($D_3$),  and heavy-tailed distributed  ($D_4$) data generated from $g$ = $g_2$ S = $S_2$ $n$ = 300 $p$ = 400, displayed in the form of: mean (SD) calculated from 100 independent runs of the experiment.} 
\end{table}
\begin{table}[H]
\centering
\begin{tabular}{lcccccc}
  \hline
 & L2Boost & MBoost & LADBoost & Robloss & SBoost & RRBoost \\ 
  \hline
$D_0$ & 1.12 (0.06) & 1.13 (0.07) & 1.31 (0.08) & 1.18 (0.07) & 1.29 (0.21) & 1.18 (0.08) \\ 
  $D_1 (10\%)$ & 2.43 (0.26) & 1.95 (0.41) & 1.46 (0.11) & 1.36 (0.10) & 1.33 (0.15) & 1.23 (0.08) \\ 
  $D_1 (20\%)$ & 3.15 (0.32) & 3.13 (0.33) & 1.59 (0.13) & 1.65 (0.18) & 1.39 (0.13) & 1.26 (0.08) \\ 
  $D_2 (10\%)$ & 2.74 (0.27) & 2.19 (0.52) & 1.41 (0.10) & 1.37 (0.10) & 1.36 (0.17) & 1.24 (0.14) \\ 
  $D_2 (20\%)$ & 3.84 (0.38) & 3.87 (0.36) & 1.67 (0.14) & 1.89 (0.20) & 1.36 (0.10) & 1.28 (0.09) \\ 
  $D_3$ & 1.31 (0.15) & 1.21 (0.09) & 1.38 (0.10) & 1.24 (0.08) & 1.31 (0.16) & 1.26 (0.09) \\ 
  $D_4$ & 16.06 (39.50) & 1.58 (0.15) & 1.57 (0.12) & 1.47 (0.13) & 1.40 (0.13) & 1.37 (0.10) \\ 
   \hline
\end{tabular}
\caption{Summary statistics of RMSEs on the test sets by L2Boost, MBoost, LADBoost, Robloss, SBoost, and RRBoost applied with tree learners of $d$ = 3 for clean ($D_0$), symmetric gross error contaminated ($D_1$), asymmetric gross error contaminated ($D_2$),  skewed distributed ($D_3$),  and heavy-tailed distributed  ($D_4$) data generated from $g$ = $g_2$ S = $S_2$ $n$ = 300 $p$ = 400, displayed in the form of: mean (SD) calculated from 100 independent runs of the experiment.} 
\end{table}
\begin{table}[H]
\centering
\begin{tabular}{lccccc}
  \hline
 & L2Boost & MBoost & LADBoost & Robloss & RRBoost \\ 
  \hline
$D_0$ & 0.80 (0.11) & 0.79 (0.12) & 0.82 (0.13) & 0.80 (0.13) & 0.80 (0.14) \\ 
  $D_1 (10\%)$ & 0.42 (0.15) & 0.58 (0.18) & 0.79 (0.14) & 0.75 (0.14) & 0.77 (0.15) \\ 
  $D_1 (20\%)$ & 0.28 (0.16) & 0.28 (0.16) & 0.68 (0.19) & 0.60 (0.14) & 0.76 (0.15) \\ 
  $D_2 (10\%)$ & 0.39 (0.13) & 0.56 (0.17) & 0.80 (0.14) & 0.77 (0.16) & 0.76 (0.14) \\ 
  $D_2 (20\%)$ & 0.30 (0.15) & 0.30 (0.15) & 0.70 (0.15) & 0.61 (0.13) & 0.78 (0.14) \\ 
  $D_3$ & 0.70 (0.15) & 0.78 (0.11) & 0.80 (0.14) & 0.80 (0.13) & 0.77 (0.18) \\ 
  $D_4$ & 0.22 (0.19) & 0.65 (0.14) & 0.67 (0.15) & 0.66 (0.14) & 0.70 (0.15) \\ 
   \hline
\end{tabular}
\caption{Fractions of variables recovered by L2Boost, MBoost, LADBoost, Robloss, and RRBoost applied with tree learners of $d$ = 3 for clean ($D_0$), symmetric gross error contaminated ($D_1$), asymmetric gross error contaminated ($D_2$),  skewed distributed ($D_3$),  and heavy-tailed distributed  ($D_4$) data generated from $g$ = $g_2$ S = $S_2$ $n$ = 300 $p$ = 400, displayed in the form of: mean (SD) calculated from 100 independent runs of the experiment.} 
\end{table}
\begin{table}[H]
\centering
\begin{tabular}{lcccccc}
  \hline
 & L2Boost & MBoost & LADBoost & Robloss & SBoost & RRBoost \\ 
  \hline
$D_0$ & 1.00 (0.02) & 1.00 (0.02) & 1.02 (0.02) & 1.01 (0.02) & 0.91 (0.03) & 0.89 (0.03) \\ 
  $D_1 (10\%)$ & 1.26 (0.04) & 1.14 (0.09) & 1.03 (0.02) & 1.02 (0.02) & 0.92 (0.04) & 0.89 (0.03) \\ 
  $D_1 (20\%)$ & 1.31 (0.06) & 1.31 (0.05) & 1.05 (0.03) & 1.05 (0.03) & 0.93 (0.04) & 0.91 (0.04) \\ 
  $D_2 (10\%)$ & 1.73 (0.06) & 1.33 (0.21) & 1.04 (0.03) & 1.04 (0.03) & 0.91 (0.02) & 0.89 (0.02) \\ 
  $D_2 (20\%)$ & 2.69 (0.07) & 2.58 (0.07) & 1.11 (0.04) & 1.16 (0.04) & 0.96 (0.07) & 0.93 (0.07) \\ 
  $D_3$ & 1.08 (0.05) & 1.03 (0.02) & 1.04 (0.02) & 1.03 (0.02) & 1.01 (0.05) & 0.93 (0.03) \\ 
  $D_4$ & 69.03 (340.61) & 1.09 (0.03) & 1.05 (0.03) & 1.05 (0.03) & 0.94 (0.05) & 0.92 (0.03) \\ 
   \hline
\end{tabular}
\caption{Summary statistics of RMSEs on the test sets by L2Boost, MBoost, LADBoost, Robloss, SBoost, and RRBoost applied with tree learners of $d$ = 1 for clean ($D_0$), symmetric gross error contaminated ($D_1$), asymmetric gross error contaminated ($D_2$),  skewed distributed ($D_3$),  and heavy-tailed distributed  ($D_4$) data generated from $g$ = $g_2$ S = $S_2$ $n$ = 3000 $p$ = 400, displayed in the form of: mean (SD) calculated from 100 independent runs of the experiment.} 
\end{table}
\begin{table}[H]
\centering
\begin{tabular}{lccccc}
  \hline
 & L2Boost & MBoost & LADBoost & Robloss & RRBoost \\ 
  \hline
$D_0$ & 1.00 (0.00) & 0.99 (0.04) & 0.95 (0.10) & 0.99 (0.04) & 0.99 (0.04) \\ 
  $D_1 (10\%)$ & 0.66 (0.15) & 0.83 (0.15) & 0.94 (0.11) & 0.98 (0.07) & 0.99 (0.04) \\ 
  $D_1 (20\%)$ & 0.59 (0.13) & 0.58 (0.12) & 0.88 (0.13) & 0.92 (0.12) & 1.00 (0.00) \\ 
  $D_2 (10\%)$ & 0.68 (0.14) & 0.81 (0.14) & 0.93 (0.11) & 0.96 (0.09) & 1.00 (0.03) \\ 
  $D_2 (20\%)$ & 0.58 (0.13) & 0.51 (0.05) & 0.89 (0.12) & 0.92 (0.12) & 0.99 (0.04) \\ 
  $D_3$ & 0.87 (0.14) & 0.98 (0.06) & 0.96 (0.09) & 0.98 (0.07) & 1.00 (0.03) \\ 
  $D_4$ & 0.07 (0.16) & 0.82 (0.12) & 0.90 (0.12) & 0.92 (0.12) & 0.98 (0.07) \\ 
   \hline
\end{tabular}
\caption{Fractions of variables recovered by L2Boost, MBoost, LADBoost, Robloss, and RRBoost applied with tree learners of $d$ = 1 for clean ($D_0$), symmetric gross error contaminated ($D_1$), asymmetric gross error contaminated ($D_2$),  skewed distributed ($D_3$),  and heavy-tailed distributed  ($D_4$) data generated from $g$ = $g_2$ S = $S_2$ $n$ = 3000 $p$ = 400, displayed in the form of: mean (SD) calculated from 100 independent runs of the experiment.} 
\end{table}
\begin{table}[H]
\centering
\begin{tabular}{lcccccc}
  \hline
 & L2Boost & MBoost & LADBoost & Robloss & SBoost & RRBoost \\ 
  \hline
$D_0$ & 0.83 (0.04) & 0.85 (0.03) & 0.86 (0.04) & 0.85 (0.03) & 0.95 (0.07) & 0.74 (0.02) \\ 
  $D_1 (10\%)$ & 1.23 (0.07) & 1.05 (0.13) & 0.88 (0.04) & 0.89 (0.04) & 0.86 (0.05) & 0.73 (0.02) \\ 
  $D_1 (20\%)$ & 1.34 (0.07) & 1.36 (0.09) & 0.91 (0.04) & 0.93 (0.04) & 0.80 (0.04) & 0.73 (0.02) \\ 
  $D_2 (10\%)$ & 1.73 (0.07) & 1.26 (0.25) & 0.91 (0.04) & 0.91 (0.04) & 0.84 (0.04) & 0.73 (0.02) \\ 
  $D_2 (20\%)$ & 2.69 (0.07) & 2.67 (0.13) & 1.00 (0.05) & 1.04 (0.05) & 0.80 (0.03) & 0.73 (0.02) \\ 
  $D_3$ & 0.97 (0.06) & 0.88 (0.04) & 0.91 (0.04) & 0.89 (0.04) & 0.96 (0.06) & 0.78 (0.02) \\ 
  $D_4$ & 66.85 (310.13) & 0.98 (0.05) & 0.92 (0.04) & 0.94 (0.04) & 0.80 (0.04) & 0.75 (0.02) \\ 
   \hline
\end{tabular}
\caption{Summary statistics of RMSEs on the test sets by L2Boost, MBoost, LADBoost, Robloss, SBoost, and RRBoost applied with tree learners of $d$ = 2 for clean ($D_0$), symmetric gross error contaminated ($D_1$), asymmetric gross error contaminated ($D_2$),  skewed distributed ($D_3$),  and heavy-tailed distributed  ($D_4$) data generated from $g$ = $g_2$ S = $S_2$ $n$ = 3000 $p$ = 400, displayed in the form of: mean (SD) calculated from 100 independent runs of the experiment.} 
\end{table}
\begin{table}[H]
\centering
\begin{tabular}{lccccc}
  \hline
 & L2Boost & MBoost & LADBoost & Robloss & RRBoost \\ 
  \hline
$D_0$ & 1.00 (0.00) & 1.00 (0.00) & 0.99 (0.05) & 1.00 (0.03) & 1.00 (0.00) \\ 
  $D_1 (10\%)$ & 0.67 (0.17) & 0.85 (0.16) & 0.98 (0.07) & 0.99 (0.05) & 1.00 (0.00) \\ 
  $D_1 (20\%)$ & 0.59 (0.13) & 0.61 (0.15) & 0.96 (0.09) & 0.95 (0.10) & 1.00 (0.00) \\ 
  $D_2 (10\%)$ & 0.66 (0.16) & 0.84 (0.16) & 0.98 (0.07) & 0.98 (0.07) & 1.00 (0.00) \\ 
  $D_2 (20\%)$ & 0.58 (0.15) & 0.66 (0.16) & 0.95 (0.10) & 0.94 (0.11) & 1.00 (0.00) \\ 
  $D_3$ & 0.88 (0.13) & 0.99 (0.04) & 0.99 (0.05) & 0.99 (0.04) & 1.00 (0.00) \\ 
  $D_4$ & 0.04 (0.10) & 0.88 (0.13) & 0.95 (0.10) & 0.93 (0.11) & 1.00 (0.00) \\ 
   \hline
\end{tabular}
\caption{Fractions of variables recovered by L2Boost, MBoost, LADBoost, Robloss, and RRBoost applied with tree learners of $d$ = 2 for clean ($D_0$), symmetric gross error contaminated ($D_1$), asymmetric gross error contaminated ($D_2$),  skewed distributed ($D_3$),  and heavy-tailed distributed  ($D_4$) data generated from $g$ = $g_2$ S = $S_2$ $n$ = 3000 $p$ = 400, displayed in the form of: mean (SD) calculated from 100 independent runs of the experiment.} 
\end{table}
\begin{table}[H]
\centering
\begin{tabular}{lcccccc}
  \hline
 & L2Boost & MBoost & LADBoost & Robloss & SBoost & RRBoost \\ 
  \hline
$D_0$ & 0.85 (0.03) & 0.87 (0.04) & 0.91 (0.03) & 0.90 (0.04) & 1.09 (0.09) & 0.76 (0.02) \\ 
  $D_1 (10\%)$ & 1.19 (0.08) & 1.05 (0.09) & 0.93 (0.04) & 0.93 (0.04) & 1.03 (0.09) & 0.75 (0.03) \\ 
  $D_1 (20\%)$ & 1.39 (0.12) & 1.43 (0.12) & 0.97 (0.04) & 0.98 (0.04) & 0.94 (0.07) & 0.75 (0.02) \\ 
  $D_2 (10\%)$ & 1.69 (0.07) & 1.29 (0.25) & 0.94 (0.03) & 0.94 (0.03) & 0.97 (0.06) & 0.75 (0.02) \\ 
  $D_2 (20\%)$ & 2.71 (0.08) & 2.78 (0.11) & 1.05 (0.04) & 1.07 (0.04) & 0.86 (0.04) & 0.74 (0.02) \\ 
  $D_3$ & 0.99 (0.06) & 0.91 (0.04) & 0.95 (0.03) & 0.94 (0.03) & 1.12 (0.09) & 0.80 (0.03) \\ 
  $D_4$ & 64.84 (282.59) & 1.02 (0.04) & 0.98 (0.04) & 0.98 (0.03) & 0.98 (0.06) & 0.80 (0.03) \\ 
   \hline
\end{tabular}
\caption{Summary statistics of RMSEs on the test sets by L2Boost, MBoost, LADBoost, Robloss, SBoost, and RRBoost applied with tree learners of $d$ = 3 for clean ($D_0$), symmetric gross error contaminated ($D_1$), asymmetric gross error contaminated ($D_2$),  skewed distributed ($D_3$),  and heavy-tailed distributed  ($D_4$) data generated from $g$ = $g_2$ S = $S_2$ $n$ = 3000 $p$ = 400, displayed in the form of: mean (SD) calculated from 100 independent runs of the experiment.} 
\end{table}
\begin{table}[H]
\centering
\begin{tabular}{lccccc}
  \hline
 & L2Boost & MBoost & LADBoost & Robloss & RRBoost \\ 
  \hline
$D_0$ & 0.99 (0.04) & 1.00 (0.03) & 0.99 (0.06) & 1.00 (0.00) & 1.00 (0.00) \\ 
  $D_1 (10\%)$ & 0.76 (0.10) & 0.86 (0.13) & 0.97 (0.09) & 0.96 (0.09) & 1.00 (0.00) \\ 
  $D_1 (20\%)$ & 0.72 (0.11) & 0.72 (0.14) & 0.93 (0.11) & 0.93 (0.11) & 1.00 (0.00) \\ 
  $D_2 (10\%)$ & 0.74 (0.10) & 0.83 (0.13) & 0.98 (0.07) & 0.98 (0.07) & 1.00 (0.00) \\ 
  $D_2 (20\%)$ & 0.66 (0.14) & 0.64 (0.14) & 0.93 (0.11) & 0.92 (0.12) & 1.00 (0.00) \\ 
  $D_3$ & 0.85 (0.12) & 0.98 (0.07) & 0.97 (0.08) & 0.99 (0.04) & 1.00 (0.00) \\ 
  $D_4$ & 0.06 (0.12) & 0.87 (0.13) & 0.94 (0.11) & 0.94 (0.11) & 1.00 (0.00) \\ 
   \hline
\end{tabular}
\caption{Fractions of variables recovered by L2Boost, MBoost, LADBoost, Robloss, and RRBoost applied with tree learners of $d$ = 3 for clean ($D_0$), symmetric gross error contaminated ($D_1$), asymmetric gross error contaminated ($D_2$),  skewed distributed ($D_3$),  and heavy-tailed distributed  ($D_4$) data generated from $g$ = $g_2$ S = $S_2$ $n$ = 3000 $p$ = 400, displayed in the form of: mean (SD) calculated from 100 independent runs of the experiment.} 
\end{table}

\subsection{Function $g_3$}
\begin{table}[H]
\centering
\begin{tabular}{lcccccc}
  \hline
 & L2Boost & MBoost & LADBoost & Robloss & SBoost & RRBoost \\ 
  \hline
$D_0$ & 3.15 (0.11) & 3.16 (0.11) & 3.21 (0.12) & 3.18 (0.12) & 3.58 (0.23) & 3.05 (0.17) \\ 
  $D_1 (10\%)$ & 5.95 (0.80) & 4.55 (0.97) & 3.37 (0.14) & 3.46 (0.20) & 3.47 (0.21) & 3.04 (0.12) \\ 
  $D_1 (20\%)$ & 6.97 (0.99) & 6.75 (0.74) & 3.59 (0.20) & 3.76 (0.34) & 3.31 (0.21) & 3.05 (0.12) \\ 
  $D_2 (10\%)$ & 7.47 (0.77) & 5.33 (1.58) & 3.44 (0.16) & 3.52 (0.25) & 3.45 (0.20) & 3.04 (0.14) \\ 
  $D_2 (20\%)$ & 11.76 (0.95) & 11.52 (1.01) & 3.94 (0.49) & 4.48 (0.91) & 3.32 (0.18) & 3.06 (0.13) \\ 
  $D_3$ & 3.80 (0.31) & 3.44 (0.16) & 3.40 (0.13) & 3.38 (0.12) & 3.57 (0.20) & 3.28 (0.15) \\ 
  $D_4$ & 53.84 (126.87) & 4.17 (0.31) & 3.69 (0.20) & 3.83 (0.22) & 3.49 (0.27) & 3.32 (0.19) \\ 
   \hline
\end{tabular}
\caption{Summary statistics of RMSEs on the test sets by L2Boost, MBoost, LADBoost, Robloss, SBoost, and RRBoost applied with tree learners of $d$ = 1 for clean ($D_0$), symmetric gross error contaminated ($D_1$), asymmetric gross error contaminated ($D_2$),  skewed distributed ($D_3$),  and heavy-tailed distributed  ($D_4$) data generated from $g$ = $g_3$ S = $S_0$ $n$ = 300 $p$ = 10, displayed in the form of: mean (SD) calculated from 100 independent runs of the experiment.} 
\end{table}
\begin{table}[H]
\centering
\begin{tabular}{lccccc}
  \hline
 & L2Boost & MBoost & LADBoost & Robloss & RRBoost \\ 
  \hline
$D_0$ & 1.00 (0.00) & 1.00 (0.00) & 1.00 (0.00) & 1.00 (0.00) & 1.00 (0.00) \\ 
  $D_1 (10\%)$ & 0.60 (0.26) & 0.83 (0.25) & 1.00 (0.00) & 1.00 (0.02) & 1.00 (0.00) \\ 
  $D_1 (20\%)$ & 0.44 (0.25) & 0.36 (0.20) & 1.00 (0.02) & 1.00 (0.03) & 1.00 (0.00) \\ 
  $D_2 (10\%)$ & 0.59 (0.24) & 0.85 (0.21) & 1.00 (0.02) & 1.00 (0.02) & 1.00 (0.00) \\ 
  $D_2 (20\%)$ & 0.46 (0.26) & 0.51 (0.23) & 0.99 (0.06) & 0.98 (0.06) & 1.00 (0.00) \\ 
  $D_3$ & 0.98 (0.07) & 1.00 (0.00) & 1.00 (0.00) & 1.00 (0.00) & 1.00 (0.00) \\ 
  $D_4$ & 0.43 (0.30) & 0.92 (0.12) & 0.98 (0.06) & 0.98 (0.07) & 1.00 (0.03) \\ 
   \hline
\end{tabular}
\caption{Fractions of variables recovered by L2Boost, MBoost, LADBoost, Robloss, and RRBoost applied with tree learners of $d$ = 1 for clean ($D_0$), symmetric gross error contaminated ($D_1$), asymmetric gross error contaminated ($D_2$),  skewed distributed ($D_3$),  and heavy-tailed distributed  ($D_4$) data generated from $g$ = $g_3$ S = $S_0$ $n$ = 300 $p$ = 10, displayed in the form of: mean (SD) calculated from 100 independent runs of the experiment.} 
\end{table}
\begin{table}[H]
\centering
\begin{tabular}{lcccccc}
  \hline
 & L2Boost & MBoost & LADBoost & Robloss & SBoost & RRBoost \\ 
  \hline
$D_0$ & 3.73 (0.26) & 3.74 (0.27) & 3.74 (0.25) & 3.73 (0.25) & 4.09 (0.24) & 3.49 (0.18) \\ 
  $D_1 (10\%)$ & 6.74 (0.77) & 5.44 (1.10) & 4.01 (0.29) & 4.08 (0.31) & 3.95 (0.23) & 3.51 (0.20) \\ 
  $D_1 (20\%)$ & 8.22 (1.15) & 8.18 (0.89) & 4.36 (0.36) & 4.47 (0.35) & 3.81 (0.24) & 3.50 (0.17) \\ 
  $D_2 (10\%)$ & 8.25 (0.80) & 6.12 (1.54) & 4.15 (0.29) & 4.21 (0.31) & 4.12 (0.26) & 3.56 (0.22) \\ 
  $D_2 (20\%)$ & 12.40 (1.05) & 12.52 (1.23) & 4.88 (0.41) & 5.27 (0.57) & 3.92 (0.24) & 3.57 (0.19) \\ 
  $D_3$ & 4.36 (0.41) & 4.05 (0.33) & 3.94 (0.26) & 3.98 (0.31) & 4.12 (0.24) & 3.69 (0.20) \\ 
  $D_4$ & 51.52 (103.88) & 5.02 (0.47) & 4.38 (0.33) & 4.51 (0.39) & 4.09 (0.30) & 3.87 (0.24) \\ 
   \hline
\end{tabular}
\caption{Summary statistics of RMSEs on the test sets by L2Boost, MBoost, LADBoost, Robloss, SBoost, and RRBoost applied with tree learners of $d$ = 2 for clean ($D_0$), symmetric gross error contaminated ($D_1$), asymmetric gross error contaminated ($D_2$),  skewed distributed ($D_3$),  and heavy-tailed distributed  ($D_4$) data generated from $g$ = $g_3$ S = $S_0$ $n$ = 300 $p$ = 10, displayed in the form of: mean (SD) calculated from 100 independent runs of the experiment.} 
\end{table}
\begin{table}[H]
\centering
\begin{tabular}{lccccc}
  \hline
 & L2Boost & MBoost & LADBoost & Robloss & RRBoost \\ 
  \hline
$D_0$ & 1.00 (0.02) & 1.00 (0.02) & 1.00 (0.00) & 1.00 (0.00) & 1.00 (0.00) \\ 
  $D_1 (10\%)$ & 0.56 (0.19) & 0.79 (0.24) & 1.00 (0.02) & 0.99 (0.06) & 1.00 (0.00) \\ 
  $D_1 (20\%)$ & 0.50 (0.19) & 0.43 (0.16) & 0.97 (0.08) & 0.97 (0.08) & 1.00 (0.00) \\ 
  $D_2 (10\%)$ & 0.60 (0.19) & 0.84 (0.20) & 0.99 (0.05) & 0.99 (0.06) & 1.00 (0.00) \\ 
  $D_2 (20\%)$ & 0.46 (0.20) & 0.52 (0.22) & 0.93 (0.10) & 0.93 (0.11) & 1.00 (0.00) \\ 
  $D_3$ & 0.94 (0.11) & 1.00 (0.03) & 0.99 (0.03) & 0.99 (0.03) & 1.00 (0.00) \\ 
  $D_4$ & 0.38 (0.23) & 0.85 (0.16) & 0.96 (0.09) & 0.94 (0.10) & 0.99 (0.05) \\ 
   \hline
\end{tabular}
\caption{Fractions of variables recovered by L2Boost, MBoost, LADBoost, Robloss, and RRBoost applied with tree learners of $d$ = 2 for clean ($D_0$), symmetric gross error contaminated ($D_1$), asymmetric gross error contaminated ($D_2$),  skewed distributed ($D_3$),  and heavy-tailed distributed  ($D_4$) data generated from $g$ = $g_3$ S = $S_0$ $n$ = 300 $p$ = 10, displayed in the form of: mean (SD) calculated from 100 independent runs of the experiment.} 
\end{table}
\begin{table}[H]
\centering
\begin{tabular}{lcccccc}
  \hline
 & L2Boost & MBoost & LADBoost & Robloss & SBoost & RRBoost \\ 
  \hline
$D_0$ & 3.87 (0.22) & 3.87 (0.22) & 4.14 (0.22) & 3.94 (0.23) & 4.35 (0.30) & 3.81 (0.16) \\ 
  $D_1 (10\%)$ & 7.83 (0.80) & 6.16 (1.46) & 4.46 (0.27) & 4.45 (0.29) & 4.28 (0.23) & 3.85 (0.22) \\ 
  $D_1 (20\%)$ & 9.70 (1.12) & 9.84 (0.93) & 4.77 (0.37) & 4.99 (0.38) & 4.19 (0.24) & 3.88 (0.20) \\ 
  $D_2 (10\%)$ & 9.01 (1.01) & 6.90 (1.84) & 4.58 (0.29) & 4.62 (0.32) & 4.38 (0.25) & 3.93 (0.20) \\ 
  $D_2 (20\%)$ & 13.34 (1.04) & 13.41 (1.04) & 5.52 (0.43) & 5.85 (0.57) & 4.38 (0.27) & 3.99 (0.22) \\ 
  $D_3$ & 4.61 (0.49) & 4.22 (0.25) & 4.31 (0.24) & 4.23 (0.24) & 4.40 (0.28) & 4.01 (0.20) \\ 
  $D_4$ & 53.46 (104.37) & 5.36 (0.50) & 4.86 (0.31) & 4.90 (0.33) & 4.46 (0.29) & 4.29 (0.28) \\ 
   \hline
\end{tabular}
\caption{Summary statistics of RMSEs on the test sets by L2Boost, MBoost, LADBoost, Robloss, SBoost, and RRBoost applied with tree learners of $d$ = 3 for clean ($D_0$), symmetric gross error contaminated ($D_1$), asymmetric gross error contaminated ($D_2$),  skewed distributed ($D_3$),  and heavy-tailed distributed  ($D_4$) data generated from $g$ = $g_3$ S = $S_0$ $n$ = 300 $p$ = 10, displayed in the form of: mean (SD) calculated from 100 independent runs of the experiment.} 
\end{table}
\begin{table}[H]
\centering
\begin{tabular}{lccccc}
  \hline
 & L2Boost & MBoost & LADBoost & Robloss & RRBoost \\ 
  \hline
$D_0$ & 1.00 (0.00) & 1.00 (0.00) & 1.00 (0.03) & 1.00 (0.00) & 1.00 (0.03) \\ 
  $D_1 (10\%)$ & 0.67 (0.15) & 0.81 (0.18) & 0.97 (0.08) & 0.97 (0.07) & 1.00 (0.02) \\ 
  $D_1 (20\%)$ & 0.58 (0.17) & 0.56 (0.17) & 0.94 (0.10) & 0.88 (0.13) & 0.99 (0.03) \\ 
  $D_2 (10\%)$ & 0.70 (0.15) & 0.80 (0.15) & 0.97 (0.07) & 0.96 (0.09) & 1.00 (0.02) \\ 
  $D_2 (20\%)$ & 0.55 (0.19) & 0.60 (0.17) & 0.87 (0.13) & 0.85 (0.15) & 1.00 (0.03) \\ 
  $D_3$ & 0.90 (0.14) & 0.99 (0.04) & 0.99 (0.05) & 0.98 (0.06) & 1.00 (0.00) \\ 
  $D_4$ & 0.54 (0.21) & 0.80 (0.14) & 0.90 (0.12) & 0.87 (0.13) & 0.94 (0.10) \\ 
   \hline
\end{tabular}
\caption{Fractions of variables recovered by L2Boost, MBoost, LADBoost, Robloss, and RRBoost applied with tree learners of $d$ = 3 for clean ($D_0$), symmetric gross error contaminated ($D_1$), asymmetric gross error contaminated ($D_2$),  skewed distributed ($D_3$),  and heavy-tailed distributed  ($D_4$) data generated from $g$ = $g_3$ S = $S_0$ $n$ = 300 $p$ = 10, displayed in the form of: mean (SD) calculated from 100 independent runs of the experiment.} 
\end{table}
\begin{table}[H]
\centering
\begin{tabular}{lcccccc}
  \hline
 & L2Boost & MBoost & LADBoost & Robloss & SBoost & RRBoost \\ 
  \hline
$D_0$ & 2.59 (0.06) & 2.59 (0.06) & 2.63 (0.06) & 2.60 (0.06) & 2.95 (0.08) & 2.55 (0.06) \\ 
  $D_1 (10\%)$ & 3.74 (0.19) & 3.27 (0.55) & 2.68 (0.07) & 2.67 (0.11) & 2.87 (0.08) & 2.55 (0.06) \\ 
  $D_1 (20\%)$ & 4.09 (0.29) & 4.06 (0.24) & 2.74 (0.08) & 2.77 (0.09) & 2.77 (0.09) & 2.56 (0.06) \\ 
  $D_2 (10\%)$ & 6.08 (0.21) & 3.99 (1.14) & 2.71 (0.06) & 2.74 (0.17) & 2.87 (0.08) & 2.56 (0.06) \\ 
  $D_2 (20\%)$ & 10.38 (0.26) & 10.42 (0.33) & 2.97 (0.24) & 3.24 (0.46) & 2.77 (0.08) & 2.55 (0.06) \\ 
  $D_3$ & 2.90 (0.15) & 2.74 (0.06) & 2.95 (0.07) & 2.84 (0.07) & 3.33 (0.07) & 2.81 (0.07) \\ 
  $D_4$ & 34.07 (104.03) & 3.13 (0.12) & 2.77 (0.07) & 2.82 (0.08) & 2.82 (0.07) & 2.62 (0.07) \\ 
   \hline
\end{tabular}
\caption{Summary statistics of RMSEs on the test sets by L2Boost, MBoost, LADBoost, Robloss, SBoost, and RRBoost applied with tree learners of $d$ = 1 for clean ($D_0$), symmetric gross error contaminated ($D_1$), asymmetric gross error contaminated ($D_2$),  skewed distributed ($D_3$),  and heavy-tailed distributed  ($D_4$) data generated from $g$ = $g_3$ S = $S_0$ $n$ = 3000 $p$ = 10, displayed in the form of: mean (SD) calculated from 100 independent runs of the experiment.} 
\end{table}
\begin{table}[H]
\centering
\begin{tabular}{lccccc}
  \hline
 & L2Boost & MBoost & LADBoost & Robloss & RRBoost \\ 
  \hline
$D_0$ & 1.00 (0.00) & 1.00 (0.00) & 1.00 (0.00) & 1.00 (0.00) & 1.00 (0.00) \\ 
  $D_1 (10\%)$ & 0.99 (0.05) & 1.00 (0.00) & 1.00 (0.00) & 1.00 (0.00) & 1.00 (0.00) \\ 
  $D_1 (20\%)$ & 0.92 (0.12) & 0.94 (0.11) & 1.00 (0.02) & 1.00 (0.02) & 1.00 (0.00) \\ 
  $D_2 (10\%)$ & 1.00 (0.02) & 0.99 (0.04) & 1.00 (0.00) & 1.00 (0.00) & 1.00 (0.00) \\ 
  $D_2 (20\%)$ & 0.97 (0.09) & 0.71 (0.28) & 0.99 (0.03) & 0.99 (0.05) & 1.00 (0.00) \\ 
  $D_3$ & 1.00 (0.00) & 1.00 (0.00) & 1.00 (0.00) & 1.00 (0.00) & 1.00 (0.00) \\ 
  $D_4$ & 0.27 (0.30) & 1.00 (0.00) & 1.00 (0.00) & 1.00 (0.00) & 1.00 (0.00) \\ 
   \hline
\end{tabular}
\caption{Fractions of variables recovered by L2Boost, MBoost, LADBoost, Robloss, and RRBoost applied with tree learners of $d$ = 1 for clean ($D_0$), symmetric gross error contaminated ($D_1$), asymmetric gross error contaminated ($D_2$),  skewed distributed ($D_3$),  and heavy-tailed distributed  ($D_4$) data generated from $g$ = $g_3$ S = $S_0$ $n$ = 3000 $p$ = 10, displayed in the form of: mean (SD) calculated from 100 independent runs of the experiment.} 
\end{table}
\begin{table}[H]
\centering
\begin{tabular}{lcccccc}
  \hline
 & L2Boost & MBoost & LADBoost & Robloss & SBoost & RRBoost \\ 
  \hline
$D_0$ & 2.83 (0.11) & 2.82 (0.11) & 2.89 (0.11) & 2.84 (0.10) & 2.80 (0.07) & 2.59 (0.06) \\ 
  $D_1 (10\%)$ & 4.27 (0.34) & 3.59 (0.53) & 2.96 (0.11) & 2.94 (0.12) & 2.74 (0.07) & 2.59 (0.06) \\ 
  $D_1 (20\%)$ & 4.75 (0.40) & 4.80 (0.41) & 3.05 (0.12) & 3.11 (0.13) & 2.68 (0.07) & 2.60 (0.06) \\ 
  $D_2 (10\%)$ & 6.41 (0.27) & 4.42 (1.16) & 3.02 (0.11) & 3.01 (0.12) & 2.76 (0.07) & 2.60 (0.06) \\ 
  $D_2 (20\%)$ & 10.65 (0.29) & 10.39 (0.34) & 3.30 (0.11) & 3.46 (0.15) & 2.68 (0.07) & 2.59 (0.07) \\ 
  $D_3$ & 3.23 (0.21) & 2.97 (0.12) & 3.11 (0.09) & 3.04 (0.10) & 3.25 (0.08) & 2.91 (0.07) \\ 
  $D_4$ & 231.62 (1194.21) & 3.45 (0.20) & 3.10 (0.14) & 3.15 (0.17) & 2.68 (0.07) & 2.65 (0.07) \\ 
   \hline
\end{tabular}
\caption{Summary statistics of RMSEs on the test sets by L2Boost, MBoost, LADBoost, Robloss, SBoost, and RRBoost applied with tree learners of $d$ = 2 for clean ($D_0$), symmetric gross error contaminated ($D_1$), asymmetric gross error contaminated ($D_2$),  skewed distributed ($D_3$),  and heavy-tailed distributed  ($D_4$) data generated from $g$ = $g_3$ S = $S_0$ $n$ = 3000 $p$ = 10, displayed in the form of: mean (SD) calculated from 100 independent runs of the experiment.} 
\end{table}
\begin{table}[H]
\centering
\begin{tabular}{lccccc}
  \hline
 & L2Boost & MBoost & LADBoost & Robloss & RRBoost \\ 
  \hline
$D_0$ & 1.00 (0.00) & 1.00 (0.00) & 1.00 (0.00) & 1.00 (0.00) & 1.00 (0.00) \\ 
  $D_1 (10\%)$ & 0.95 (0.10) & 0.99 (0.03) & 1.00 (0.00) & 1.00 (0.00) & 1.00 (0.00) \\ 
  $D_1 (20\%)$ & 0.86 (0.15) & 0.87 (0.13) & 1.00 (0.00) & 1.00 (0.00) & 1.00 (0.00) \\ 
  $D_2 (10\%)$ & 0.98 (0.06) & 1.00 (0.03) & 1.00 (0.00) & 1.00 (0.00) & 1.00 (0.00) \\ 
  $D_2 (20\%)$ & 0.91 (0.14) & 0.88 (0.11) & 1.00 (0.00) & 1.00 (0.00) & 1.00 (0.00) \\ 
  $D_3$ & 1.00 (0.00) & 1.00 (0.00) & 1.00 (0.00) & 1.00 (0.00) & 1.00 (0.00) \\ 
  $D_4$ & 0.30 (0.19) & 1.00 (0.00) & 1.00 (0.00) & 1.00 (0.00) & 1.00 (0.00) \\ 
   \hline
\end{tabular}
\caption{Fractions of variables recovered by L2Boost, MBoost, LADBoost, Robloss, and RRBoost applied with tree learners of $d$ = 2 for clean ($D_0$), symmetric gross error contaminated ($D_1$), asymmetric gross error contaminated ($D_2$),  skewed distributed ($D_3$),  and heavy-tailed distributed  ($D_4$) data generated from $g$ = $g_3$ S = $S_0$ $n$ = 3000 $p$ = 10, displayed in the form of: mean (SD) calculated from 100 independent runs of the experiment.} 
\end{table}
\begin{table}[H]
\centering
\begin{tabular}{lcccccc}
  \hline
 & L2Boost & MBoost & LADBoost & Robloss & SBoost & RRBoost \\ 
  \hline
$D_0$ & 2.91 (0.09) & 2.92 (0.08) & 3.13 (0.10) & 2.97 (0.08) & 3.00 (0.10) & 2.70 (0.07) \\ 
  $D_1 (10\%)$ & 4.51 (0.26) & 3.76 (0.58) & 3.20 (0.09) & 3.07 (0.10) & 2.89 (0.08) & 2.69 (0.06) \\ 
  $D_1 (20\%)$ & 5.04 (0.33) & 5.14 (0.36) & 3.33 (0.12) & 3.22 (0.12) & 2.78 (0.08) & 2.68 (0.07) \\ 
  $D_2 (10\%)$ & 6.53 (0.24) & 4.57 (1.13) & 3.26 (0.10) & 3.14 (0.10) & 2.94 (0.11) & 2.71 (0.07) \\ 
  $D_2 (20\%)$ & 10.75 (0.29) & 10.67 (0.32) & 3.60 (0.11) & 3.70 (0.15) & 2.83 (0.09) & 2.69 (0.07) \\ 
  $D_3$ & 3.34 (0.16) & 3.07 (0.08) & 3.28 (0.10) & 3.15 (0.09) & 3.33 (0.09) & 3.07 (0.08) \\ 
  $D_4$ & 240.77 (1207.00) & 3.57 (0.15) & 3.34 (0.12) & 3.28 (0.12) & 2.79 (0.09) & 2.75 (0.07) \\ 
   \hline
\end{tabular}
\caption{Summary statistics of RMSEs on the test sets by L2Boost, MBoost, LADBoost, Robloss, SBoost, and RRBoost applied with tree learners of $d$ = 3 for clean ($D_0$), symmetric gross error contaminated ($D_1$), asymmetric gross error contaminated ($D_2$),  skewed distributed ($D_3$),  and heavy-tailed distributed  ($D_4$) data generated from $g$ = $g_3$ S = $S_0$ $n$ = 3000 $p$ = 10, displayed in the form of: mean (SD) calculated from 100 independent runs of the experiment.} 
\end{table}
\begin{table}[H]
\centering
\begin{tabular}{lccccc}
  \hline
 & L2Boost & MBoost & LADBoost & Robloss & RRBoost \\ 
  \hline
$D_0$ & 1.00 (0.00) & 1.00 (0.00) & 1.00 (0.00) & 1.00 (0.00) & 1.00 (0.00) \\ 
  $D_1 (10\%)$ & 0.85 (0.15) & 0.98 (0.06) & 1.00 (0.00) & 1.00 (0.00) & 1.00 (0.00) \\ 
  $D_1 (20\%)$ & 0.81 (0.14) & 0.80 (0.15) & 1.00 (0.00) & 1.00 (0.00) & 1.00 (0.00) \\ 
  $D_2 (10\%)$ & 0.89 (0.15) & 0.99 (0.04) & 1.00 (0.00) & 1.00 (0.00) & 1.00 (0.00) \\ 
  $D_2 (20\%)$ & 0.80 (0.14) & 0.93 (0.10) & 1.00 (0.00) & 1.00 (0.00) & 1.00 (0.00) \\ 
  $D_3$ & 1.00 (0.00) & 1.00 (0.00) & 1.00 (0.00) & 1.00 (0.00) & 1.00 (0.00) \\ 
  $D_4$ & 0.49 (0.20) & 1.00 (0.00) & 1.00 (0.00) & 1.00 (0.00) & 1.00 (0.00) \\ 
   \hline
\end{tabular}
\caption{Fractions of variables recovered by L2Boost, MBoost, LADBoost, Robloss, and RRBoost applied with tree learners of $d$ = 3 for clean ($D_0$), symmetric gross error contaminated ($D_1$), asymmetric gross error contaminated ($D_2$),  skewed distributed ($D_3$),  and heavy-tailed distributed  ($D_4$) data generated from $g$ = $g_3$ S = $S_0$ $n$ = 3000 $p$ = 10, displayed in the form of: mean (SD) calculated from 100 independent runs of the experiment.} 
\end{table}
\begin{table}[H]
\centering
\begin{tabular}{lcccccc}
  \hline
 & L2Boost & MBoost & LADBoost & Robloss & SBoost & RRBoost \\ 
  \hline
$D_0$ & 3.20 (0.15) & 3.19 (0.16) & 3.29 (0.19) & 3.20 (0.16) & 4.22 (0.40) & 2.93 (0.19) \\ 
  $D_1 (10\%)$ & 5.82 (0.91) & 4.32 (0.94) & 3.38 (0.19) & 3.36 (0.14) & 3.94 (0.37) & 2.94 (0.24) \\ 
  $D_1 (20\%)$ & 6.94 (0.95) & 6.90 (0.87) & 3.55 (0.24) & 3.60 (0.28) & 3.61 (0.33) & 2.93 (0.21) \\ 
  $D_2 (10\%)$ & 6.99 (0.86) & 4.95 (1.25) & 3.47 (0.21) & 3.42 (0.20) & 3.93 (0.31) & 2.93 (0.21) \\ 
  $D_2 (20\%)$ & 10.15 (0.95) & 10.20 (0.96) & 3.99 (0.37) & 4.21 (0.62) & 3.62 (0.36) & 2.91 (0.22) \\ 
  $D_3$ & 3.58 (0.28) & 3.35 (0.18) & 3.37 (0.20) & 3.30 (0.17) & 4.18 (0.39) & 3.11 (0.23) \\ 
  $D_4$ & 35.29 (51.32) & 3.82 (0.29) & 3.61 (0.28) & 3.59 (0.24) & 4.24 (0.45) & 3.32 (0.26) \\ 
   \hline
\end{tabular}
\caption{Summary statistics of RMSEs on the test sets by L2Boost, MBoost, LADBoost, Robloss, SBoost, and RRBoost applied with tree learners of $d$ = 1 for clean ($D_0$), symmetric gross error contaminated ($D_1$), asymmetric gross error contaminated ($D_2$),  skewed distributed ($D_3$),  and heavy-tailed distributed  ($D_4$) data generated from $g$ = $g_3$ S = $S_0$ $n$ = 300 $p$ = 400, displayed in the form of: mean (SD) calculated from 100 independent runs of the experiment.} 
\end{table}
\begin{table}[H]
\centering
\begin{tabular}{lccccc}
  \hline
 & L2Boost & MBoost & LADBoost & Robloss & RRBoost \\ 
  \hline
$D_0$ & 1.00 (0.02) & 1.00 (0.00) & 1.00 (0.00) & 1.00 (0.00) & 1.00 (0.00) \\ 
  $D_1 (10\%)$ & 0.36 (0.18) & 0.75 (0.25) & 0.99 (0.03) & 1.00 (0.03) & 1.00 (0.00) \\ 
  $D_1 (20\%)$ & 0.16 (0.15) & 0.16 (0.14) & 0.96 (0.09) & 0.93 (0.10) & 1.00 (0.00) \\ 
  $D_2 (10\%)$ & 0.37 (0.19) & 0.73 (0.23) & 0.99 (0.05) & 0.99 (0.05) & 1.00 (0.00) \\ 
  $D_2 (20\%)$ & 0.18 (0.16) & 0.23 (0.19) & 0.91 (0.11) & 0.89 (0.13) & 1.00 (0.00) \\ 
  $D_3$ & 0.92 (0.13) & 0.99 (0.03) & 1.00 (0.00) & 1.00 (0.02) & 1.00 (0.02) \\ 
  $D_4$ & 0.03 (0.09) & 0.83 (0.15) & 0.92 (0.12) & 0.94 (0.09) & 0.96 (0.08) \\ 
   \hline
\end{tabular}
\caption{Fractions of variables recovered by L2Boost, MBoost, LADBoost, Robloss, and RRBoost applied with tree learners of $d$ = 1 for clean ($D_0$), symmetric gross error contaminated ($D_1$), asymmetric gross error contaminated ($D_2$),  skewed distributed ($D_3$),  and heavy-tailed distributed  ($D_4$) data generated from $g$ = $g_3$ S = $S_0$ $n$ = 300 $p$ = 400, displayed in the form of: mean (SD) calculated from 100 independent runs of the experiment.} 
\end{table}
\begin{table}[H]
\centering
\begin{tabular}{lcccccc}
  \hline
 & L2Boost & MBoost & LADBoost & Robloss & SBoost & RRBoost \\ 
  \hline
$D_0$ & 3.70 (0.34) & 3.68 (0.34) & 4.05 (0.30) & 3.74 (0.33) & 4.51 (0.47) & 3.54 (0.24) \\ 
  $D_1 (10\%)$ & 7.09 (0.73) & 5.51 (1.09) & 4.38 (0.33) & 4.16 (0.40) & 4.48 (0.36) & 3.64 (0.24) \\ 
  $D_1 (20\%)$ & 8.63 (0.88) & 8.63 (0.76) & 4.80 (0.40) & 4.89 (0.49) & 4.24 (0.33) & 3.67 (0.29) \\ 
  $D_2 (10\%)$ & 8.10 (0.81) & 6.19 (1.21) & 4.61 (0.35) & 4.47 (0.43) & 4.55 (0.33) & 3.62 (0.27) \\ 
  $D_2 (20\%)$ & 11.26 (0.95) & 11.30 (1.04) & 5.46 (0.47) & 5.83 (0.55) & 4.56 (0.36) & 3.77 (0.28) \\ 
  $D_3$ & 4.37 (0.51) & 3.93 (0.33) & 4.20 (0.32) & 3.91 (0.35) & 4.54 (0.38) & 3.71 (0.24) \\ 
  $D_4$ & 33.04 (40.44) & 4.95 (0.49) & 4.66 (0.41) & 4.55 (0.39) & 4.65 (0.36) & 4.20 (0.33) \\ 
   \hline
\end{tabular}
\caption{Summary statistics of RMSEs on the test sets by L2Boost, MBoost, LADBoost, Robloss, SBoost, and RRBoost applied with tree learners of $d$ = 2 for clean ($D_0$), symmetric gross error contaminated ($D_1$), asymmetric gross error contaminated ($D_2$),  skewed distributed ($D_3$),  and heavy-tailed distributed  ($D_4$) data generated from $g$ = $g_3$ S = $S_0$ $n$ = 300 $p$ = 400, displayed in the form of: mean (SD) calculated from 100 independent runs of the experiment.} 
\end{table}
\begin{table}[H]
\centering
\begin{tabular}{lccccc}
  \hline
 & L2Boost & MBoost & LADBoost & Robloss & RRBoost \\ 
  \hline
$D_0$ & 0.99 (0.05) & 0.99 (0.05) & 0.95 (0.08) & 0.99 (0.05) & 0.99 (0.04) \\ 
  $D_1 (10\%)$ & 0.30 (0.15) & 0.63 (0.24) & 0.88 (0.12) & 0.91 (0.11) & 0.98 (0.05) \\ 
  $D_1 (20\%)$ & 0.16 (0.15) & 0.15 (0.14) & 0.79 (0.15) & 0.75 (0.16) & 0.99 (0.05) \\ 
  $D_2 (10\%)$ & 0.31 (0.17) & 0.62 (0.20) & 0.86 (0.13) & 0.88 (0.13) & 0.99 (0.05) \\ 
  $D_2 (20\%)$ & 0.18 (0.14) & 0.23 (0.16) & 0.70 (0.17) & 0.66 (0.15) & 0.96 (0.08) \\ 
  $D_3$ & 0.86 (0.14) & 0.96 (0.09) & 0.92 (0.10) & 0.95 (0.09) & 0.99 (0.03) \\ 
  $D_4$ & 0.05 (0.10) & 0.70 (0.16) & 0.78 (0.13) & 0.81 (0.14) & 0.89 (0.11) \\ 
   \hline
\end{tabular}
\caption{Fractions of variables recovered by L2Boost, MBoost, LADBoost, Robloss, and RRBoost applied with tree learners of $d$ = 2 for clean ($D_0$), symmetric gross error contaminated ($D_1$), asymmetric gross error contaminated ($D_2$),  skewed distributed ($D_3$),  and heavy-tailed distributed  ($D_4$) data generated from $g$ = $g_3$ S = $S_0$ $n$ = 300 $p$ = 400, displayed in the form of: mean (SD) calculated from 100 independent runs of the experiment.} 
\end{table}
\begin{table}[H]
\centering
\begin{tabular}{lcccccc}
  \hline
 & L2Boost & MBoost & LADBoost & Robloss & SBoost & RRBoost \\ 
  \hline
$D_0$ & 3.77 (0.27) & 3.81 (0.27) & 4.68 (0.29) & 4.05 (0.30) & 4.37 (0.53) & 3.95 (0.24) \\ 
  $D_1 (10\%)$ & 8.40 (0.84) & 6.66 (1.43) & 5.13 (0.34) & 4.81 (0.38) & 4.51 (0.39) & 4.10 (0.25) \\ 
  $D_1 (20\%)$ & 10.42 (1.00) & 10.31 (0.89) & 5.58 (0.35) & 5.70 (0.50) & 4.57 (0.34) & 4.26 (0.28) \\ 
  $D_2 (10\%)$ & 9.06 (0.83) & 7.20 (1.46) & 5.31 (0.37) & 5.08 (0.44) & 4.58 (0.45) & 4.16 (0.27) \\ 
  $D_2 (20\%)$ & 12.46 (1.09) & 12.31 (1.01) & 6.14 (0.43) & 6.68 (0.58) & 4.90 (0.36) & 4.42 (0.31) \\ 
  $D_3$ & 4.56 (0.44) & 4.23 (0.32) & 4.84 (0.30) & 4.34 (0.31) & 4.48 (0.40) & 4.12 (0.27) \\ 
  $D_4$ & 33.99 (40.66) & 5.69 (0.66) & 5.46 (0.38) & 5.14 (0.43) & 4.79 (0.40) & 4.62 (0.33) \\ 
   \hline
\end{tabular}
\caption{Summary statistics of RMSEs on the test sets by L2Boost, MBoost, LADBoost, Robloss, SBoost, and RRBoost applied with tree learners of $d$ = 3 for clean ($D_0$), symmetric gross error contaminated ($D_1$), asymmetric gross error contaminated ($D_2$),  skewed distributed ($D_3$),  and heavy-tailed distributed  ($D_4$) data generated from $g$ = $g_3$ S = $S_0$ $n$ = 300 $p$ = 400, displayed in the form of: mean (SD) calculated from 100 independent runs of the experiment.} 
\end{table}
\begin{table}[H]
\centering
\begin{tabular}{lccccc}
  \hline
 & L2Boost & MBoost & LADBoost & Robloss & RRBoost \\ 
  \hline
$D_0$ & 0.96 (0.10) & 0.95 (0.11) & 0.85 (0.14) & 0.94 (0.12) & 0.95 (0.09) \\ 
  $D_1 (10\%)$ & 0.33 (0.15) & 0.55 (0.16) & 0.74 (0.13) & 0.77 (0.14) & 0.95 (0.09) \\ 
  $D_1 (20\%)$ & 0.17 (0.15) & 0.17 (0.15) & 0.63 (0.13) & 0.63 (0.13) & 0.90 (0.11) \\ 
  $D_2 (10\%)$ & 0.35 (0.15) & 0.52 (0.15) & 0.73 (0.13) & 0.76 (0.15) & 0.94 (0.10) \\ 
  $D_2 (20\%)$ & 0.20 (0.13) & 0.22 (0.14) & 0.60 (0.11) & 0.61 (0.12) & 0.87 (0.11) \\ 
  $D_3$ & 0.79 (0.16) & 0.85 (0.15) & 0.81 (0.13) & 0.84 (0.15) & 0.93 (0.12) \\ 
  $D_4$ & 0.12 (0.15) & 0.62 (0.13) & 0.68 (0.16) & 0.66 (0.11) & 0.81 (0.12) \\ 
   \hline
\end{tabular}
\caption{Fractions of variables recovered by L2Boost, MBoost, LADBoost, Robloss, and RRBoost applied with tree learners of $d$ = 3 for clean ($D_0$), symmetric gross error contaminated ($D_1$), asymmetric gross error contaminated ($D_2$),  skewed distributed ($D_3$),  and heavy-tailed distributed  ($D_4$) data generated from $g$ = $g_3$ S = $S_0$ $n$ = 300 $p$ = 400, displayed in the form of: mean (SD) calculated from 100 independent runs of the experiment.} 
\end{table}
\begin{table}[H]
\centering
\begin{tabular}{lcccccc}
  \hline
 & L2Boost & MBoost & LADBoost & Robloss & SBoost & RRBoost \\ 
  \hline
$D_0$ & 2.27 (0.05) & 2.27 (0.06) & 2.36 (0.07) & 2.29 (0.06) & 2.43 (0.07) & 2.05 (0.05) \\ 
  $D_1 (10\%)$ & 3.32 (0.14) & 2.89 (0.37) & 2.43 (0.07) & 2.36 (0.06) & 2.31 (0.06) & 2.04 (0.05) \\ 
  $D_1 (20\%)$ & 3.54 (0.19) & 3.59 (0.19) & 2.54 (0.08) & 2.52 (0.09) & 2.23 (0.07) & 2.04 (0.05) \\ 
  $D_2 (10\%)$ & 5.00 (0.14) & 3.62 (0.83) & 2.47 (0.07) & 2.42 (0.07) & 2.33 (0.08) & 2.04 (0.05) \\ 
  $D_2 (20\%)$ & 8.26 (0.20) & 8.28 (0.35) & 2.72 (0.10) & 2.88 (0.24) & 2.23 (0.06) & 2.04 (0.05) \\ 
  $D_3$ & 2.82 (0.19) & 2.40 (0.06) & 2.53 (0.07) & 2.44 (0.07) & 2.63 (0.06) & 2.22 (0.05) \\ 
  $D_4$ & 104.44 (460.19) & 2.85 (0.12) & 2.61 (0.08) & 2.60 (0.10) & 2.32 (0.05) & 2.13 (0.06) \\ 
   \hline
\end{tabular}
\caption{Summary statistics of RMSEs on the test sets by L2Boost, MBoost, LADBoost, Robloss, SBoost, and RRBoost applied with tree learners of $d$ = 1 for clean ($D_0$), symmetric gross error contaminated ($D_1$), asymmetric gross error contaminated ($D_2$),  skewed distributed ($D_3$),  and heavy-tailed distributed  ($D_4$) data generated from $g$ = $g_3$ S = $S_0$ $n$ = 3000 $p$ = 400, displayed in the form of: mean (SD) calculated from 100 independent runs of the experiment.} 
\end{table}
\begin{table}[H]
\centering
\begin{tabular}{lccccc}
  \hline
 & L2Boost & MBoost & LADBoost & Robloss & RRBoost \\ 
  \hline
$D_0$ & 1.00 (0.00) & 1.00 (0.00) & 1.00 (0.00) & 1.00 (0.00) & 1.00 (0.00) \\ 
  $D_1 (10\%)$ & 0.99 (0.04) & 1.00 (0.00) & 1.00 (0.00) & 1.00 (0.00) & 1.00 (0.00) \\ 
  $D_1 (20\%)$ & 0.92 (0.11) & 0.92 (0.11) & 1.00 (0.00) & 1.00 (0.00) & 1.00 (0.00) \\ 
  $D_2 (10\%)$ & 1.00 (0.02) & 1.00 (0.03) & 1.00 (0.00) & 1.00 (0.00) & 1.00 (0.00) \\ 
  $D_2 (20\%)$ & 0.95 (0.10) & 0.82 (0.20) & 1.00 (0.00) & 1.00 (0.02) & 1.00 (0.00) \\ 
  $D_3$ & 1.00 (0.00) & 1.00 (0.00) & 1.00 (0.00) & 1.00 (0.00) & 1.00 (0.00) \\ 
  $D_4$ & 0.02 (0.07) & 1.00 (0.00) & 1.00 (0.00) & 1.00 (0.00) & 1.00 (0.00) \\ 
   \hline
\end{tabular}
\caption{Fractions of variables recovered by L2Boost, MBoost, LADBoost, Robloss, and RRBoost applied with tree learners of $d$ = 1 for clean ($D_0$), symmetric gross error contaminated ($D_1$), asymmetric gross error contaminated ($D_2$),  skewed distributed ($D_3$),  and heavy-tailed distributed  ($D_4$) data generated from $g$ = $g_3$ S = $S_0$ $n$ = 3000 $p$ = 400, displayed in the form of: mean (SD) calculated from 100 independent runs of the experiment.} 
\end{table}
\begin{table}[H]
\centering
\begin{tabular}{lcccccc}
  \hline
 & L2Boost & MBoost & LADBoost & Robloss & SBoost & RRBoost \\ 
  \hline
$D_0$ & 2.48 (0.18) & 2.50 (0.17) & 2.62 (0.17) & 2.50 (0.16) & 2.59 (0.10) & 2.15 (0.05) \\ 
  $D_1 (10\%)$ & 3.96 (0.31) & 3.34 (0.55) & 2.75 (0.20) & 2.62 (0.19) & 2.44 (0.09) & 2.13 (0.05) \\ 
  $D_1 (20\%)$ & 4.37 (0.38) & 4.48 (0.37) & 2.86 (0.20) & 2.83 (0.24) & 2.32 (0.09) & 2.12 (0.05) \\ 
  $D_2 (10\%)$ & 5.39 (0.25) & 3.97 (0.79) & 2.79 (0.19) & 2.74 (0.22) & 2.47 (0.09) & 2.15 (0.05) \\ 
  $D_2 (20\%)$ & 8.59 (0.28) & 8.43 (0.37) & 3.15 (0.20) & 3.27 (0.24) & 2.34 (0.09) & 2.14 (0.05) \\ 
  $D_3$ & 3.07 (0.27) & 2.65 (0.17) & 2.76 (0.16) & 2.66 (0.17) & 2.69 (0.07) & 2.37 (0.06) \\ 
  $D_4$ & 100.44 (324.83) & 3.20 (0.24) & 2.90 (0.21) & 2.88 (0.23) & 2.35 (0.07) & 2.23 (0.06) \\ 
   \hline
\end{tabular}
\caption{Summary statistics of RMSEs on the test sets by L2Boost, MBoost, LADBoost, Robloss, SBoost, and RRBoost applied with tree learners of $d$ = 2 for clean ($D_0$), symmetric gross error contaminated ($D_1$), asymmetric gross error contaminated ($D_2$),  skewed distributed ($D_3$),  and heavy-tailed distributed  ($D_4$) data generated from $g$ = $g_3$ S = $S_0$ $n$ = 3000 $p$ = 400, displayed in the form of: mean (SD) calculated from 100 independent runs of the experiment.} 
\end{table}
\begin{table}[H]
\centering
\begin{tabular}{lccccc}
  \hline
 & L2Boost & MBoost & LADBoost & Robloss & RRBoost \\ 
  \hline
$D_0$ & 1.00 (0.00) & 1.00 (0.00) & 1.00 (0.00) & 1.00 (0.00) & 1.00 (0.00) \\ 
  $D_1 (10\%)$ & 0.93 (0.09) & 1.00 (0.02) & 1.00 (0.00) & 1.00 (0.00) & 1.00 (0.00) \\ 
  $D_1 (20\%)$ & 0.84 (0.13) & 0.85 (0.13) & 1.00 (0.00) & 1.00 (0.00) & 1.00 (0.00) \\ 
  $D_2 (10\%)$ & 0.93 (0.10) & 1.00 (0.03) & 1.00 (0.00) & 1.00 (0.00) & 1.00 (0.00) \\ 
  $D_2 (20\%)$ & 0.86 (0.13) & 0.85 (0.12) & 1.00 (0.00) & 1.00 (0.00) & 1.00 (0.00) \\ 
  $D_3$ & 1.00 (0.00) & 1.00 (0.00) & 1.00 (0.00) & 1.00 (0.00) & 1.00 (0.00) \\ 
  $D_4$ & 0.02 (0.07) & 1.00 (0.00) & 1.00 (0.00) & 1.00 (0.00) & 1.00 (0.00) \\ 
   \hline
\end{tabular}
\caption{Fractions of variables recovered by L2Boost, MBoost, LADBoost, Robloss, and RRBoost applied with tree learners of $d$ = 2 for clean ($D_0$), symmetric gross error contaminated ($D_1$), asymmetric gross error contaminated ($D_2$),  skewed distributed ($D_3$),  and heavy-tailed distributed  ($D_4$) data generated from $g$ = $g_3$ S = $S_0$ $n$ = 3000 $p$ = 400, displayed in the form of: mean (SD) calculated from 100 independent runs of the experiment.} 
\end{table}
\begin{table}[H]
\centering
\begin{tabular}{lcccccc}
  \hline
 & L2Boost & MBoost & LADBoost & Robloss & SBoost & RRBoost \\ 
  \hline
$D_0$ & 2.50 (0.09) & 2.53 (0.09) & 2.90 (0.12) & 2.65 (0.10) & 3.06 (0.13) & 2.36 (0.09) \\ 
  $D_1 (10\%)$ & 4.11 (0.22) & 3.44 (0.52) & 3.03 (0.10) & 2.75 (0.10) & 2.84 (0.10) & 2.31 (0.07) \\ 
  $D_1 (20\%)$ & 4.66 (0.36) & 4.77 (0.36) & 3.19 (0.13) & 2.95 (0.12) & 2.64 (0.10) & 2.27 (0.07) \\ 
  $D_2 (10\%)$ & 5.52 (0.23) & 4.12 (0.75) & 3.09 (0.12) & 2.88 (0.13) & 2.95 (0.16) & 2.34 (0.08) \\ 
  $D_2 (20\%)$ & 8.79 (0.30) & 8.96 (0.43) & 3.55 (0.17) & 3.57 (0.21) & 2.84 (0.16) & 2.32 (0.06) \\ 
  $D_3$ & 3.11 (0.20) & 2.70 (0.10) & 3.02 (0.12) & 2.77 (0.09) & 3.07 (0.10) & 2.55 (0.09) \\ 
  $D_4$ & 181.97 (814.12) & 3.22 (0.13) & 3.16 (0.13) & 2.99 (0.10) & 2.78 (0.11) & 2.46 (0.08) \\ 
   \hline
\end{tabular}
\caption{Summary statistics of RMSEs on the test sets by L2Boost, MBoost, LADBoost, Robloss, SBoost, and RRBoost applied with tree learners of $d$ = 3 for clean ($D_0$), symmetric gross error contaminated ($D_1$), asymmetric gross error contaminated ($D_2$),  skewed distributed ($D_3$),  and heavy-tailed distributed  ($D_4$) data generated from $g$ = $g_3$ S = $S_0$ $n$ = 3000 $p$ = 400, displayed in the form of: mean (SD) calculated from 100 independent runs of the experiment.} 
\end{table}
\begin{table}[H]
\centering
\begin{tabular}{lccccc}
  \hline
 & L2Boost & MBoost & LADBoost & Robloss & RRBoost \\ 
  \hline
$D_0$ & 1.00 (0.00) & 1.00 (0.00) & 1.00 (0.00) & 1.00 (0.00) & 1.00 (0.00) \\ 
  $D_1 (10\%)$ & 0.79 (0.17) & 0.99 (0.05) & 1.00 (0.00) & 1.00 (0.00) & 1.00 (0.00) \\ 
  $D_1 (20\%)$ & 0.70 (0.12) & 0.70 (0.12) & 1.00 (0.00) & 1.00 (0.00) & 1.00 (0.00) \\ 
  $D_2 (10\%)$ & 0.75 (0.16) & 0.99 (0.05) & 1.00 (0.00) & 1.00 (0.00) & 1.00 (0.00) \\ 
  $D_2 (20\%)$ & 0.69 (0.11) & 0.83 (0.12) & 1.00 (0.00) & 1.00 (0.00) & 1.00 (0.00) \\ 
  $D_3$ & 1.00 (0.00) & 1.00 (0.00) & 1.00 (0.00) & 1.00 (0.00) & 1.00 (0.00) \\ 
  $D_4$ & 0.03 (0.09) & 1.00 (0.00) & 1.00 (0.00) & 1.00 (0.00) & 1.00 (0.00) \\ 
   \hline
\end{tabular}
\caption{Fractions of variables recovered by L2Boost, MBoost, LADBoost, Robloss, and RRBoost applied with tree learners of $d$ = 3 for clean ($D_0$), symmetric gross error contaminated ($D_1$), asymmetric gross error contaminated ($D_2$),  skewed distributed ($D_3$),  and heavy-tailed distributed  ($D_4$) data generated from $g$ = $g_3$ S = $S_0$ $n$ = 3000 $p$ = 400, displayed in the form of: mean (SD) calculated from 100 independent runs of the experiment.} 
\end{table}
\begin{table}[H]
\centering
\begin{tabular}{lcccccc}
  \hline
 & L2Boost & MBoost & LADBoost & Robloss & SBoost & RRBoost \\ 
  \hline
$D_0$ & 4.93 (0.15) & 4.96 (0.16) & 5.09 (0.19) & 4.99 (0.16) & 5.31 (0.58) & 4.80 (0.22) \\ 
  $D_1 (10\%)$ & 8.83 (1.03) & 7.02 (1.44) & 5.32 (0.24) & 5.29 (0.24) & 5.13 (0.51) & 4.74 (0.19) \\ 
  $D_1 (20\%)$ & 9.97 (1.45) & 9.68 (1.16) & 5.65 (0.34) & 5.84 (0.41) & 4.92 (0.35) & 4.72 (0.17) \\ 
  $D_2 (10\%)$ & 11.75 (0.88) & 8.42 (2.69) & 5.45 (0.26) & 5.46 (0.28) & 5.11 (0.48) & 4.75 (0.21) \\ 
  $D_2 (20\%)$ & 18.90 (1.40) & 19.06 (1.69) & 6.35 (1.10) & 6.88 (1.56) & 4.94 (0.37) & 4.73 (0.17) \\ 
  $D_3$ & 5.98 (0.48) & 5.34 (0.20) & 5.46 (0.19) & 5.35 (0.20) & 5.61 (0.47) & 5.26 (0.24) \\ 
  $D_4$ & 180.72 (1143.86) & 6.84 (0.68) & 5.82 (0.40) & 6.01 (0.43) & 5.02 (0.34) & 4.96 (0.26) \\ 
   \hline
\end{tabular}
\caption{Summary statistics of RMSEs on the test sets by L2Boost, MBoost, LADBoost, Robloss, SBoost, and RRBoost applied with tree learners of $d$ = 1 for clean ($D_0$), symmetric gross error contaminated ($D_1$), asymmetric gross error contaminated ($D_2$),  skewed distributed ($D_3$),  and heavy-tailed distributed  ($D_4$) data generated from $g$ = $g_3$ S = $S_1$ $n$ = 300 $p$ = 10, displayed in the form of: mean (SD) calculated from 100 independent runs of the experiment.} 
\end{table}
\begin{table}[H]
\centering
\begin{tabular}{lccccc}
  \hline
 & L2Boost & MBoost & LADBoost & Robloss & RRBoost \\ 
  \hline
$D_0$ & 0.98 (0.06) & 0.99 (0.04) & 0.98 (0.07) & 0.99 (0.05) & 0.99 (0.04) \\ 
  $D_1 (10\%)$ & 0.36 (0.19) & 0.63 (0.32) & 0.95 (0.09) & 0.95 (0.09) & 0.99 (0.04) \\ 
  $D_1 (20\%)$ & 0.26 (0.14) & 0.23 (0.12) & 0.92 (0.11) & 0.90 (0.14) & 1.00 (0.03) \\ 
  $D_2 (10\%)$ & 0.41 (0.21) & 0.76 (0.25) & 0.93 (0.10) & 0.94 (0.09) & 0.99 (0.03) \\ 
  $D_2 (20\%)$ & 0.29 (0.15) & 0.40 (0.21) & 0.88 (0.12) & 0.87 (0.14) & 0.99 (0.04) \\ 
  $D_3$ & 0.89 (0.15) & 0.96 (0.08) & 0.98 (0.07) & 0.98 (0.07) & 0.99 (0.04) \\ 
  $D_4$ & 0.28 (0.24) & 0.68 (0.20) & 0.88 (0.13) & 0.83 (0.14) & 0.95 (0.09) \\ 
   \hline
\end{tabular}
\caption{Fractions of variables recovered by L2Boost, MBoost, LADBoost, Robloss, and RRBoost applied with tree learners of $d$ = 1 for clean ($D_0$), symmetric gross error contaminated ($D_1$), asymmetric gross error contaminated ($D_2$),  skewed distributed ($D_3$),  and heavy-tailed distributed  ($D_4$) data generated from $g$ = $g_3$ S = $S_1$ $n$ = 300 $p$ = 10, displayed in the form of: mean (SD) calculated from 100 independent runs of the experiment.} 
\end{table}
\begin{table}[H]
\centering
\begin{tabular}{lcccccc}
  \hline
 & L2Boost & MBoost & LADBoost & Robloss & SBoost & RRBoost \\ 
  \hline
$D_0$ & 5.51 (0.24) & 5.53 (0.24) & 5.82 (0.29) & 5.59 (0.27) & 6.19 (0.56) & 5.26 (0.26) \\ 
  $D_1 (10\%)$ & 9.46 (1.40) & 7.67 (1.47) & 6.06 (0.30) & 5.97 (0.34) & 5.96 (0.53) & 5.28 (0.29) \\ 
  $D_1 (20\%)$ & 12.18 (1.90) & 12.28 (1.86) & 6.46 (0.42) & 6.68 (0.51) & 5.52 (0.58) & 5.14 (0.31) \\ 
  $D_2 (10\%)$ & 12.55 (1.51) & 9.32 (2.80) & 6.24 (0.36) & 6.26 (0.35) & 6.25 (0.64) & 5.28 (0.28) \\ 
  $D_2 (20\%)$ & 19.55 (1.86) & 19.93 (2.12) & 7.27 (0.69) & 7.78 (0.73) & 6.04 (0.68) & 5.30 (0.28) \\ 
  $D_3$ & 6.52 (0.56) & 5.94 (0.34) & 6.04 (0.29) & 5.96 (0.28) & 6.55 (0.48) & 5.70 (0.28) \\ 
  $D_4$ & 216.53 (1151.15) & 7.19 (0.54) & 6.60 (0.41) & 6.68 (0.44) & 5.63 (0.52) & 5.40 (0.36) \\ 
   \hline
\end{tabular}
\caption{Summary statistics of RMSEs on the test sets by L2Boost, MBoost, LADBoost, Robloss, SBoost, and RRBoost applied with tree learners of $d$ = 2 for clean ($D_0$), symmetric gross error contaminated ($D_1$), asymmetric gross error contaminated ($D_2$),  skewed distributed ($D_3$),  and heavy-tailed distributed  ($D_4$) data generated from $g$ = $g_3$ S = $S_1$ $n$ = 300 $p$ = 10, displayed in the form of: mean (SD) calculated from 100 independent runs of the experiment.} 
\end{table}
\begin{table}[H]
\centering
\begin{tabular}{lccccc}
  \hline
 & L2Boost & MBoost & LADBoost & Robloss & RRBoost \\ 
  \hline
$D_0$ & 0.97 (0.07) & 0.97 (0.07) & 0.95 (0.09) & 0.97 (0.07) & 0.98 (0.06) \\ 
  $D_1 (10\%)$ & 0.44 (0.16) & 0.62 (0.24) & 0.91 (0.12) & 0.91 (0.11) & 0.97 (0.07) \\ 
  $D_1 (20\%)$ & 0.40 (0.17) & 0.37 (0.16) & 0.88 (0.12) & 0.81 (0.16) & 0.99 (0.06) \\ 
  $D_2 (10\%)$ & 0.46 (0.16) & 0.70 (0.22) & 0.91 (0.12) & 0.90 (0.12) & 0.97 (0.07) \\ 
  $D_2 (20\%)$ & 0.40 (0.18) & 0.43 (0.16) & 0.85 (0.13) & 0.79 (0.15) & 0.97 (0.07) \\ 
  $D_3$ & 0.77 (0.16) & 0.94 (0.09) & 0.94 (0.09) & 0.95 (0.09) & 0.97 (0.07) \\ 
  $D_4$ & 0.34 (0.19) & 0.61 (0.18) & 0.82 (0.15) & 0.76 (0.16) & 0.95 (0.08) \\ 
   \hline
\end{tabular}
\caption{Fractions of variables recovered by L2Boost, MBoost, LADBoost, Robloss, and RRBoost applied with tree learners of $d$ = 2 for clean ($D_0$), symmetric gross error contaminated ($D_1$), asymmetric gross error contaminated ($D_2$),  skewed distributed ($D_3$),  and heavy-tailed distributed  ($D_4$) data generated from $g$ = $g_3$ S = $S_1$ $n$ = 300 $p$ = 10, displayed in the form of: mean (SD) calculated from 100 independent runs of the experiment.} 
\end{table}
\begin{table}[H]
\centering
\begin{tabular}{lcccccc}
  \hline
 & L2Boost & MBoost & LADBoost & Robloss & SBoost & RRBoost \\ 
  \hline
$D_0$ & 5.69 (0.24) & 5.67 (0.23) & 6.11 (0.26) & 5.74 (0.23) & 6.38 (0.60) & 5.61 (0.28) \\ 
  $D_1 (10\%)$ & 11.64 (1.23) & 8.76 (2.33) & 6.38 (0.34) & 6.27 (0.33) & 6.38 (0.62) & 5.60 (0.26) \\ 
  $D_1 (20\%)$ & 15.23 (1.99) & 15.26 (1.88) & 6.93 (0.45) & 7.22 (0.60) & 6.21 (0.53) & 5.57 (0.34) \\ 
  $D_2 (10\%)$ & 14.01 (1.60) & 10.29 (3.31) & 6.68 (0.35) & 6.61 (0.43) & 6.58 (0.82) & 5.65 (0.23) \\ 
  $D_2 (20\%)$ & 21.42 (1.90) & 21.76 (2.38) & 8.22 (0.77) & 9.00 (0.80) & 6.55 (0.70) & 5.74 (0.29) \\ 
  $D_3$ & 6.56 (0.62) & 6.10 (0.33) & 6.42 (0.31) & 6.17 (0.30) & 6.55 (0.59) & 5.99 (0.30) \\ 
  $D_4$ & 215.25 (1148.77) & 7.53 (0.74) & 6.94 (0.43) & 6.83 (0.45) & 6.32 (0.51) & 6.01 (0.38) \\ 
   \hline
\end{tabular}
\caption{Summary statistics of RMSEs on the test sets by L2Boost, MBoost, LADBoost, Robloss, SBoost, and RRBoost applied with tree learners of $d$ = 3 for clean ($D_0$), symmetric gross error contaminated ($D_1$), asymmetric gross error contaminated ($D_2$),  skewed distributed ($D_3$),  and heavy-tailed distributed  ($D_4$) data generated from $g$ = $g_3$ S = $S_1$ $n$ = 300 $p$ = 10, displayed in the form of: mean (SD) calculated from 100 independent runs of the experiment.} 
\end{table}
\begin{table}[H]
\centering
\begin{tabular}{lccccc}
  \hline
 & L2Boost & MBoost & LADBoost & Robloss & RRBoost \\ 
  \hline
$D_0$ & 0.95 (0.10) & 0.97 (0.08) & 0.96 (0.09) & 0.98 (0.06) & 0.96 (0.09) \\ 
  $D_1 (10\%)$ & 0.56 (0.17) & 0.69 (0.22) & 0.93 (0.10) & 0.90 (0.10) & 0.96 (0.08) \\ 
  $D_1 (20\%)$ & 0.52 (0.17) & 0.50 (0.17) & 0.87 (0.12) & 0.79 (0.14) & 0.96 (0.08) \\ 
  $D_2 (10\%)$ & 0.59 (0.16) & 0.74 (0.16) & 0.91 (0.11) & 0.90 (0.12) & 0.95 (0.09) \\ 
  $D_2 (20\%)$ & 0.50 (0.17) & 0.52 (0.18) & 0.81 (0.12) & 0.76 (0.14) & 0.95 (0.09) \\ 
  $D_3$ & 0.83 (0.12) & 0.94 (0.11) & 0.93 (0.10) & 0.94 (0.10) & 0.93 (0.10) \\ 
  $D_4$ & 0.51 (0.19) & 0.70 (0.15) & 0.83 (0.13) & 0.78 (0.15) & 0.90 (0.12) \\ 
   \hline
\end{tabular}
\caption{Fractions of variables recovered by L2Boost, MBoost, LADBoost, Robloss, and RRBoost applied with tree learners of $d$ = 3 for clean ($D_0$), symmetric gross error contaminated ($D_1$), asymmetric gross error contaminated ($D_2$),  skewed distributed ($D_3$),  and heavy-tailed distributed  ($D_4$) data generated from $g$ = $g_3$ S = $S_1$ $n$ = 300 $p$ = 10, displayed in the form of: mean (SD) calculated from 100 independent runs of the experiment.} 
\end{table}
\begin{table}[H]
\centering
\begin{tabular}{lcccccc}
  \hline
 & L2Boost & MBoost & LADBoost & Robloss & SBoost & RRBoost \\ 
  \hline
$D_0$ & 4.28 (0.10) & 4.29 (0.10) & 4.33 (0.10) & 4.30 (0.09) & 4.43 (0.23) & 4.21 (0.09) \\ 
  $D_1 (10\%)$ & 5.96 (0.37) & 5.23 (0.76) & 4.40 (0.11) & 4.38 (0.11) & 4.42 (0.25) & 4.23 (0.09) \\ 
  $D_1 (20\%)$ & 6.67 (0.56) & 6.67 (0.44) & 4.46 (0.11) & 4.49 (0.16) & 4.35 (0.22) & 4.22 (0.10) \\ 
  $D_2 (10\%)$ & 10.03 (0.30) & 6.70 (1.85) & 4.45 (0.09) & 4.48 (0.21) & 4.40 (0.22) & 4.22 (0.08) \\ 
  $D_2 (20\%)$ & 17.57 (0.51) & 17.41 (0.55) & 4.82 (0.30) & 5.11 (0.48) & 4.35 (0.21) & 4.23 (0.09) \\ 
  $D_3$ & 4.66 (0.21) & 4.51 (0.09) & 4.94 (0.10) & 4.73 (0.10) & 5.58 (0.11) & 4.81 (0.11) \\ 
  $D_4$ & 58.52 (329.75) & 4.90 (0.17) & 4.50 (0.11) & 4.56 (0.13) & 4.33 (0.12) & 4.29 (0.11) \\ 
   \hline
\end{tabular}
\caption{Summary statistics of RMSEs on the test sets by L2Boost, MBoost, LADBoost, Robloss, SBoost, and RRBoost applied with tree learners of $d$ = 1 for clean ($D_0$), symmetric gross error contaminated ($D_1$), asymmetric gross error contaminated ($D_2$),  skewed distributed ($D_3$),  and heavy-tailed distributed  ($D_4$) data generated from $g$ = $g_3$ S = $S_1$ $n$ = 3000 $p$ = 10, displayed in the form of: mean (SD) calculated from 100 independent runs of the experiment.} 
\end{table}
\begin{table}[H]
\centering
\begin{tabular}{lccccc}
  \hline
 & L2Boost & MBoost & LADBoost & Robloss & RRBoost \\ 
  \hline
$D_0$ & 1.00 (0.00) & 1.00 (0.00) & 1.00 (0.00) & 1.00 (0.00) & 1.00 (0.00) \\ 
  $D_1 (10\%)$ & 0.84 (0.15) & 0.95 (0.11) & 1.00 (0.00) & 1.00 (0.00) & 1.00 (0.00) \\ 
  $D_1 (20\%)$ & 0.68 (0.19) & 0.71 (0.17) & 1.00 (0.00) & 1.00 (0.00) & 1.00 (0.00) \\ 
  $D_2 (10\%)$ & 0.91 (0.11) & 0.96 (0.08) & 1.00 (0.00) & 1.00 (0.00) & 1.00 (0.00) \\ 
  $D_2 (20\%)$ & 0.78 (0.18) & 0.64 (0.20) & 0.99 (0.04) & 0.99 (0.04) & 1.00 (0.00) \\ 
  $D_3$ & 1.00 (0.00) & 1.00 (0.00) & 1.00 (0.00) & 1.00 (0.00) & 1.00 (0.00) \\ 
  $D_4$ & 0.26 (0.22) & 1.00 (0.00) & 1.00 (0.00) & 1.00 (0.00) & 1.00 (0.00) \\ 
   \hline
\end{tabular}
\caption{Fractions of variables recovered by L2Boost, MBoost, LADBoost, Robloss, and RRBoost applied with tree learners of $d$ = 1 for clean ($D_0$), symmetric gross error contaminated ($D_1$), asymmetric gross error contaminated ($D_2$),  skewed distributed ($D_3$),  and heavy-tailed distributed  ($D_4$) data generated from $g$ = $g_3$ S = $S_1$ $n$ = 3000 $p$ = 10, displayed in the form of: mean (SD) calculated from 100 independent runs of the experiment.} 
\end{table}
\begin{table}[H]
\centering
\begin{tabular}{lcccccc}
  \hline
 & L2Boost & MBoost & LADBoost & Robloss & SBoost & RRBoost \\ 
  \hline
$D_0$ & 4.52 (0.12) & 4.52 (0.11) & 4.66 (0.14) & 4.58 (0.15) & 4.42 (0.17) & 4.27 (0.10) \\ 
  $D_1 (10\%)$ & 6.53 (0.33) & 5.73 (0.86) & 4.76 (0.13) & 4.69 (0.15) & 4.39 (0.15) & 4.28 (0.10) \\ 
  $D_1 (20\%)$ & 7.00 (0.39) & 7.08 (0.38) & 4.89 (0.15) & 4.86 (0.18) & 4.33 (0.11) & 4.27 (0.09) \\ 
  $D_2 (10\%)$ & 10.41 (0.36) & 7.39 (2.06) & 4.85 (0.15) & 4.81 (0.16) & 4.38 (0.13) & 4.27 (0.08) \\ 
  $D_2 (20\%)$ & 17.74 (0.49) & 17.63 (0.69) & 5.23 (0.18) & 5.45 (0.22) & 4.34 (0.13) & 4.28 (0.10) \\ 
  $D_3$ & 5.07 (0.20) & 4.77 (0.14) & 5.14 (0.13) & 4.96 (0.13) & 5.42 (0.14) & 4.85 (0.11) \\ 
  $D_4$ & 230.26 (1099.65) & 5.36 (0.22) & 4.93 (0.17) & 4.95 (0.19) & 4.32 (0.11) & 4.31 (0.11) \\ 
   \hline
\end{tabular}
\caption{Summary statistics of RMSEs on the test sets by L2Boost, MBoost, LADBoost, Robloss, SBoost, and RRBoost applied with tree learners of $d$ = 2 for clean ($D_0$), symmetric gross error contaminated ($D_1$), asymmetric gross error contaminated ($D_2$),  skewed distributed ($D_3$),  and heavy-tailed distributed  ($D_4$) data generated from $g$ = $g_3$ S = $S_1$ $n$ = 3000 $p$ = 10, displayed in the form of: mean (SD) calculated from 100 independent runs of the experiment.} 
\end{table}
\begin{table}[H]
\centering
\begin{tabular}{lccccc}
  \hline
 & L2Boost & MBoost & LADBoost & Robloss & RRBoost \\ 
  \hline
$D_0$ & 1.00 (0.00) & 1.00 (0.00) & 1.00 (0.00) & 1.00 (0.00) & 1.00 (0.00) \\ 
  $D_1 (10\%)$ & 0.74 (0.17) & 0.89 (0.15) & 1.00 (0.00) & 1.00 (0.00) & 1.00 (0.00) \\ 
  $D_1 (20\%)$ & 0.59 (0.15) & 0.56 (0.14) & 1.00 (0.00) & 1.00 (0.00) & 1.00 (0.00) \\ 
  $D_2 (10\%)$ & 0.73 (0.17) & 0.90 (0.15) & 1.00 (0.00) & 1.00 (0.00) & 1.00 (0.00) \\ 
  $D_2 (20\%)$ & 0.54 (0.18) & 0.66 (0.16) & 1.00 (0.02) & 0.99 (0.05) & 1.00 (0.00) \\ 
  $D_3$ & 0.99 (0.04) & 1.00 (0.00) & 1.00 (0.00) & 1.00 (0.00) & 1.00 (0.00) \\ 
  $D_4$ & 0.31 (0.16) & 0.99 (0.04) & 1.00 (0.00) & 1.00 (0.02) & 1.00 (0.00) \\ 
   \hline
\end{tabular}
\caption{Fractions of variables recovered by L2Boost, MBoost, LADBoost, Robloss, and RRBoost applied with tree learners of $d$ = 2 for clean ($D_0$), symmetric gross error contaminated ($D_1$), asymmetric gross error contaminated ($D_2$),  skewed distributed ($D_3$),  and heavy-tailed distributed  ($D_4$) data generated from $g$ = $g_3$ S = $S_1$ $n$ = 3000 $p$ = 10, displayed in the form of: mean (SD) calculated from 100 independent runs of the experiment.} 
\end{table}
\begin{table}[H]
\centering
\begin{tabular}{lcccccc}
  \hline
 & L2Boost & MBoost & LADBoost & Robloss & SBoost & RRBoost \\ 
  \hline
$D_0$ & 4.70 (0.10) & 4.69 (0.12) & 5.02 (0.15) & 4.78 (0.14) & 4.84 (0.32) & 4.41 (0.11) \\ 
  $D_1 (10\%)$ & 6.60 (0.42) & 5.79 (0.69) & 5.13 (0.16) & 4.91 (0.16) & 4.58 (0.20) & 4.39 (0.11) \\ 
  $D_1 (20\%)$ & 7.43 (0.49) & 7.57 (0.49) & 5.27 (0.18) & 5.12 (0.18) & 4.39 (0.12) & 4.34 (0.10) \\ 
  $D_2 (10\%)$ & 10.45 (0.37) & 7.63 (1.97) & 5.21 (0.15) & 5.04 (0.16) & 4.82 (0.47) & 4.41 (0.09) \\ 
  $D_2 (20\%)$ & 17.83 (0.48) & 18.11 (0.59) & 5.70 (0.18) & 5.86 (0.25) & 4.76 (0.57) & 4.41 (0.11) \\ 
  $D_3$ & 5.25 (0.22) & 4.94 (0.13) & 5.38 (0.15) & 5.13 (0.13) & 5.67 (0.19) & 5.08 (0.13) \\ 
  $D_4$ & 231.33 (948.29) & 5.60 (0.21) & 5.29 (0.20) & 5.17 (0.20) & 4.37 (0.13) & 4.37 (0.11) \\ 
   \hline
\end{tabular}
\caption{Summary statistics of RMSEs on the test sets by L2Boost, MBoost, LADBoost, Robloss, SBoost, and RRBoost applied with tree learners of $d$ = 3 for clean ($D_0$), symmetric gross error contaminated ($D_1$), asymmetric gross error contaminated ($D_2$),  skewed distributed ($D_3$),  and heavy-tailed distributed  ($D_4$) data generated from $g$ = $g_3$ S = $S_1$ $n$ = 3000 $p$ = 10, displayed in the form of: mean (SD) calculated from 100 independent runs of the experiment.} 
\end{table}
\begin{table}[H]
\centering
\begin{tabular}{lccccc}
  \hline
 & L2Boost & MBoost & LADBoost & Robloss & RRBoost \\ 
  \hline
$D_0$ & 1.00 (0.00) & 1.00 (0.00) & 1.00 (0.00) & 1.00 (0.00) & 1.00 (0.00) \\ 
  $D_1 (10\%)$ & 0.80 (0.14) & 0.91 (0.13) & 1.00 (0.00) & 1.00 (0.00) & 1.00 (0.00) \\ 
  $D_1 (20\%)$ & 0.76 (0.14) & 0.73 (0.16) & 1.00 (0.00) & 1.00 (0.00) & 1.00 (0.00) \\ 
  $D_2 (10\%)$ & 0.80 (0.14) & 0.89 (0.15) & 1.00 (0.00) & 1.00 (0.00) & 1.00 (0.00) \\ 
  $D_2 (20\%)$ & 0.68 (0.18) & 0.72 (0.18) & 1.00 (0.02) & 1.00 (0.02) & 1.00 (0.00) \\ 
  $D_3$ & 0.99 (0.05) & 1.00 (0.00) & 1.00 (0.00) & 1.00 (0.00) & 1.00 (0.00) \\ 
  $D_4$ & 0.49 (0.19) & 0.98 (0.06) & 1.00 (0.00) & 1.00 (0.00) & 1.00 (0.00) \\ 
   \hline
\end{tabular}
\caption{Fractions of variables recovered by L2Boost, MBoost, LADBoost, Robloss, and RRBoost applied with tree learners of $d$ = 3 for clean ($D_0$), symmetric gross error contaminated ($D_1$), asymmetric gross error contaminated ($D_2$),  skewed distributed ($D_3$),  and heavy-tailed distributed  ($D_4$) data generated from $g$ = $g_3$ S = $S_1$ $n$ = 3000 $p$ = 10, displayed in the form of: mean (SD) calculated from 100 independent runs of the experiment.} 
\end{table}
\begin{table}[H]
\centering
\begin{tabular}{lcccccc}
  \hline
 & L2Boost & MBoost & LADBoost & Robloss & SBoost & RRBoost \\ 
  \hline
$D_0$ & 4.68 (0.31) & 4.72 (0.38) & 5.12 (0.38) & 4.80 (0.34) & 5.78 (0.89) & 4.42 (0.42) \\ 
  $D_1 (10\%)$ & 7.97 (0.84) & 6.84 (1.21) & 5.50 (0.47) & 5.37 (0.55) & 5.10 (0.82) & 4.24 (0.32) \\ 
  $D_1 (20\%)$ & 8.80 (1.31) & 8.73 (1.39) & 6.03 (0.57) & 6.26 (0.71) & 4.59 (0.47) & 4.23 (0.31) \\ 
  $D_2 (10\%)$ & 10.05 (0.73) & 7.81 (1.68) & 5.64 (0.45) & 5.44 (0.49) & 5.18 (0.75) & 4.30 (0.36) \\ 
  $D_2 (20\%)$ & 15.28 (1.04) & 15.77 (1.54) & 6.44 (0.71) & 7.23 (1.10) & 4.58 (0.42) & 4.17 (0.31) \\ 
  $D_3$ & 5.89 (0.71) & 5.15 (0.45) & 5.48 (0.46) & 5.16 (0.40) & 5.60 (0.71) & 4.67 (0.39) \\ 
  $D_4$ & 90.48 (255.45) & 6.53 (0.53) & 6.06 (0.59) & 5.98 (0.64) & 4.99 (0.72) & 4.64 (0.39) \\ 
   \hline
\end{tabular}
\caption{Summary statistics of RMSEs on the test sets by L2Boost, MBoost, LADBoost, Robloss, SBoost, and RRBoost applied with tree learners of $d$ = 1 for clean ($D_0$), symmetric gross error contaminated ($D_1$), asymmetric gross error contaminated ($D_2$),  skewed distributed ($D_3$),  and heavy-tailed distributed  ($D_4$) data generated from $g$ = $g_3$ S = $S_1$ $n$ = 300 $p$ = 400, displayed in the form of: mean (SD) calculated from 100 independent runs of the experiment.} 
\end{table}
\begin{table}[H]
\centering
\begin{tabular}{lccccc}
  \hline
 & L2Boost & MBoost & LADBoost & Robloss & RRBoost \\ 
  \hline
$D_0$ & 0.93 (0.10) & 0.95 (0.09) & 0.86 (0.13) & 0.91 (0.12) & 0.98 (0.06) \\ 
  $D_1 (10\%)$ & 0.22 (0.07) & 0.45 (0.24) & 0.79 (0.17) & 0.80 (0.17) & 0.99 (0.04) \\ 
  $D_1 (20\%)$ & 0.19 (0.04) & 0.19 (0.03) & 0.63 (0.19) & 0.55 (0.19) & 0.99 (0.05) \\ 
  $D_2 (10\%)$ & 0.22 (0.06) & 0.43 (0.21) & 0.76 (0.14) & 0.78 (0.16) & 0.99 (0.03) \\ 
  $D_2 (20\%)$ & 0.21 (0.03) & 0.21 (0.05) & 0.62 (0.17) & 0.54 (0.17) & 1.00 (0.02) \\ 
  $D_3$ & 0.64 (0.20) & 0.84 (0.14) & 0.81 (0.15) & 0.85 (0.14) & 0.98 (0.07) \\ 
  $D_4$ & 0.06 (0.10) & 0.44 (0.13) & 0.63 (0.18) & 0.59 (0.19) & 0.97 (0.07) \\ 
   \hline
\end{tabular}
\caption{Fractions of variables recovered by L2Boost, MBoost, LADBoost, Robloss, and RRBoost applied with tree learners of $d$ = 1 for clean ($D_0$), symmetric gross error contaminated ($D_1$), asymmetric gross error contaminated ($D_2$),  skewed distributed ($D_3$),  and heavy-tailed distributed  ($D_4$) data generated from $g$ = $g_3$ S = $S_1$ $n$ = 300 $p$ = 400, displayed in the form of: mean (SD) calculated from 100 independent runs of the experiment.} 
\end{table}
\begin{table}[H]
\centering
\begin{tabular}{lcccccc}
  \hline
 & L2Boost & MBoost & LADBoost & Robloss & SBoost & RRBoost \\ 
  \hline
$D_0$ & 5.29 (0.30) & 5.25 (0.36) & 5.71 (0.33) & 5.34 (0.36) & 6.39 (0.75) & 5.04 (0.37) \\ 
  $D_1 (10\%)$ & 10.41 (1.14) & 7.90 (1.77) & 6.21 (0.47) & 5.98 (0.46) & 6.23 (0.76) & 4.94 (0.38) \\ 
  $D_1 (20\%)$ & 12.72 (1.23) & 12.77 (1.27) & 6.77 (0.66) & 6.98 (0.78) & 5.89 (0.66) & 4.95 (0.39) \\ 
  $D_2 (10\%)$ & 12.21 (1.08) & 9.18 (2.19) & 6.55 (0.54) & 6.40 (0.60) & 6.75 (0.99) & 5.18 (0.39) \\ 
  $D_2 (20\%)$ & 17.51 (1.36) & 17.74 (1.58) & 8.33 (0.78) & 8.92 (0.99) & 6.74 (0.88) & 5.23 (0.38) \\ 
  $D_3$ & 6.28 (0.79) & 5.72 (0.34) & 6.08 (0.45) & 5.74 (0.39) & 6.58 (0.82) & 5.33 (0.33) \\ 
  $D_4$ & 80.98 (216.36) & 6.72 (0.67) & 6.73 (0.62) & 6.43 (0.60) & 6.25 (0.62) & 5.54 (0.44) \\ 
   \hline
\end{tabular}
\caption{Summary statistics of RMSEs on the test sets by L2Boost, MBoost, LADBoost, Robloss, SBoost, and RRBoost applied with tree learners of $d$ = 2 for clean ($D_0$), symmetric gross error contaminated ($D_1$), asymmetric gross error contaminated ($D_2$),  skewed distributed ($D_3$),  and heavy-tailed distributed  ($D_4$) data generated from $g$ = $g_3$ S = $S_1$ $n$ = 300 $p$ = 400, displayed in the form of: mean (SD) calculated from 100 independent runs of the experiment.} 
\end{table}
\begin{table}[H]
\centering
\begin{tabular}{lccccc}
  \hline
 & L2Boost & MBoost & LADBoost & Robloss & RRBoost \\ 
  \hline
$D_0$ & 0.86 (0.13) & 0.87 (0.12) & 0.86 (0.12) & 0.88 (0.12) & 0.95 (0.09) \\ 
  $D_1 (10\%)$ & 0.25 (0.09) & 0.47 (0.23) & 0.78 (0.15) & 0.77 (0.15) & 0.96 (0.09) \\ 
  $D_1 (20\%)$ & 0.21 (0.07) & 0.20 (0.06) & 0.67 (0.16) & 0.55 (0.18) & 0.97 (0.08) \\ 
  $D_2 (10\%)$ & 0.26 (0.10) & 0.42 (0.18) & 0.79 (0.14) & 0.73 (0.15) & 0.94 (0.09) \\ 
  $D_2 (20\%)$ & 0.24 (0.08) & 0.25 (0.09) & 0.62 (0.18) & 0.49 (0.17) & 0.93 (0.10) \\ 
  $D_3$ & 0.64 (0.16) & 0.75 (0.14) & 0.82 (0.14) & 0.82 (0.14) & 0.92 (0.12) \\ 
  $D_4$ & 0.12 (0.12) & 0.52 (0.14) & 0.66 (0.16) & 0.62 (0.16) & 0.88 (0.14) \\ 
   \hline
\end{tabular}
\caption{Fractions of variables recovered by L2Boost, MBoost, LADBoost, Robloss, and RRBoost applied with tree learners of $d$ = 2 for clean ($D_0$), symmetric gross error contaminated ($D_1$), asymmetric gross error contaminated ($D_2$),  skewed distributed ($D_3$),  and heavy-tailed distributed  ($D_4$) data generated from $g$ = $g_3$ S = $S_1$ $n$ = 300 $p$ = 400, displayed in the form of: mean (SD) calculated from 100 independent runs of the experiment.} 
\end{table}
\begin{table}[H]
\centering
\begin{tabular}{lcccccc}
  \hline
 & L2Boost & MBoost & LADBoost & Robloss & SBoost & RRBoost \\ 
  \hline
$D_0$ & 5.28 (0.24) & 5.31 (0.28) & 6.39 (0.37) & 5.55 (0.31) & 5.89 (0.75) & 5.41 (0.26) \\ 
  $D_1 (10\%)$ & 12.36 (1.35) & 9.31 (2.22) & 7.11 (0.56) & 6.78 (0.51) & 6.16 (0.84) & 5.53 (0.34) \\ 
  $D_1 (20\%)$ & 16.06 (1.72) & 16.07 (1.66) & 7.95 (0.60) & 8.15 (0.74) & 6.33 (0.58) & 5.71 (0.39) \\ 
  $D_2 (10\%)$ & 13.64 (1.59) & 10.36 (2.38) & 7.68 (0.56) & 7.40 (0.62) & 6.27 (0.91) & 5.69 (0.35) \\ 
  $D_2 (20\%)$ & 19.77 (1.79) & 19.69 (1.80) & 9.68 (0.72) & 10.40 (1.08) & 7.01 (1.02) & 5.92 (0.43) \\ 
  $D_3$ & 6.37 (0.75) & 5.92 (0.47) & 6.76 (0.38) & 6.08 (0.47) & 6.03 (0.54) & 5.71 (0.30) \\ 
  $D_4$ & 83.85 (212.03) & 7.74 (0.78) & 7.73 (0.56) & 7.34 (0.61) & 6.38 (0.63) & 6.20 (0.44) \\ 
   \hline
\end{tabular}
\caption{Summary statistics of RMSEs on the test sets by L2Boost, MBoost, LADBoost, Robloss, SBoost, and RRBoost applied with tree learners of $d$ = 3 for clean ($D_0$), symmetric gross error contaminated ($D_1$), asymmetric gross error contaminated ($D_2$),  skewed distributed ($D_3$),  and heavy-tailed distributed  ($D_4$) data generated from $g$ = $g_3$ S = $S_1$ $n$ = 300 $p$ = 400, displayed in the form of: mean (SD) calculated from 100 independent runs of the experiment.} 
\end{table}
\begin{table}[H]
\centering
\begin{tabular}{lccccc}
  \hline
 & L2Boost & MBoost & LADBoost & Robloss & RRBoost \\ 
  \hline
$D_0$ & 0.89 (0.13) & 0.91 (0.11) & 0.86 (0.14) & 0.91 (0.11) & 0.91 (0.12) \\ 
  $D_1 (10\%)$ & 0.31 (0.12) & 0.50 (0.18) & 0.77 (0.16) & 0.73 (0.15) & 0.89 (0.13) \\ 
  $D_1 (20\%)$ & 0.23 (0.08) & 0.23 (0.09) & 0.61 (0.18) & 0.52 (0.15) & 0.85 (0.14) \\ 
  $D_2 (10\%)$ & 0.34 (0.13) & 0.49 (0.16) & 0.75 (0.15) & 0.70 (0.15) & 0.88 (0.13) \\ 
  $D_2 (20\%)$ & 0.26 (0.10) & 0.26 (0.09) & 0.56 (0.18) & 0.48 (0.14) & 0.85 (0.15) \\ 
  $D_3$ & 0.75 (0.14) & 0.80 (0.16) & 0.84 (0.15) & 0.82 (0.15) & 0.85 (0.13) \\ 
  $D_4$ & 0.18 (0.13) & 0.58 (0.15) & 0.62 (0.16) & 0.60 (0.13) & 0.76 (0.15) \\ 
   \hline
\end{tabular}
\caption{Fractions of variables recovered by L2Boost, MBoost, LADBoost, Robloss, and RRBoost applied with tree learners of $d$ = 3 for clean ($D_0$), symmetric gross error contaminated ($D_1$), asymmetric gross error contaminated ($D_2$),  skewed distributed ($D_3$),  and heavy-tailed distributed  ($D_4$) data generated from $g$ = $g_3$ S = $S_1$ $n$ = 300 $p$ = 400, displayed in the form of: mean (SD) calculated from 100 independent runs of the experiment.} 
\end{table}
\begin{table}[H]
\centering
\begin{tabular}{lcccccc}
  \hline
 & L2Boost & MBoost & LADBoost & Robloss & SBoost & RRBoost \\ 
  \hline
$D_0$ & 3.56 (0.08) & 3.57 (0.09) & 3.67 (0.09) & 3.58 (0.09) & 3.55 (0.21) & 3.33 (0.08) \\ 
  $D_1 (10\%)$ & 5.32 (0.37) & 4.51 (0.69) & 3.75 (0.10) & 3.68 (0.11) & 3.49 (0.17) & 3.33 (0.08) \\ 
  $D_1 (20\%)$ & 6.12 (0.52) & 6.32 (0.53) & 3.84 (0.11) & 3.81 (0.11) & 3.46 (0.14) & 3.35 (0.07) \\ 
  $D_2 (10\%)$ & 8.22 (0.33) & 5.64 (1.48) & 3.81 (0.10) & 3.75 (0.11) & 3.50 (0.23) & 3.33 (0.08) \\ 
  $D_2 (20\%)$ & 13.92 (0.43) & 13.87 (0.56) & 4.13 (0.12) & 4.27 (0.14) & 3.41 (0.08) & 3.32 (0.07) \\ 
  $D_3$ & 4.21 (0.26) & 3.75 (0.10) & 4.01 (0.09) & 3.86 (0.09) & 4.31 (0.10) & 3.76 (0.09) \\ 
  $D_4$ & 228.91 (1360.55) & 4.21 (0.18) & 3.91 (0.12) & 3.90 (0.12) & 3.45 (0.09) & 3.41 (0.09) \\ 
   \hline
\end{tabular}
\caption{Summary statistics of RMSEs on the test sets by L2Boost, MBoost, LADBoost, Robloss, SBoost, and RRBoost applied with tree learners of $d$ = 1 for clean ($D_0$), symmetric gross error contaminated ($D_1$), asymmetric gross error contaminated ($D_2$),  skewed distributed ($D_3$),  and heavy-tailed distributed  ($D_4$) data generated from $g$ = $g_3$ S = $S_1$ $n$ = 3000 $p$ = 400, displayed in the form of: mean (SD) calculated from 100 independent runs of the experiment.} 
\end{table}
\begin{table}[H]
\centering
\begin{tabular}{lccccc}
  \hline
 & L2Boost & MBoost & LADBoost & Robloss & RRBoost \\ 
  \hline
$D_0$ & 1.00 (0.00) & 1.00 (0.00) & 1.00 (0.00) & 1.00 (0.00) & 1.00 (0.00) \\ 
  $D_1 (10\%)$ & 0.74 (0.14) & 0.92 (0.12) & 1.00 (0.00) & 1.00 (0.00) & 1.00 (0.00) \\ 
  $D_1 (20\%)$ & 0.56 (0.16) & 0.51 (0.15) & 1.00 (0.00) & 1.00 (0.00) & 1.00 (0.00) \\ 
  $D_2 (10\%)$ & 0.76 (0.15) & 0.92 (0.12) & 1.00 (0.00) & 1.00 (0.00) & 1.00 (0.00) \\ 
  $D_2 (20\%)$ & 0.58 (0.16) & 0.57 (0.17) & 1.00 (0.00) & 1.00 (0.00) & 1.00 (0.00) \\ 
  $D_3$ & 0.99 (0.04) & 1.00 (0.00) & 1.00 (0.00) & 1.00 (0.00) & 1.00 (0.00) \\ 
  $D_4$ & 0.04 (0.10) & 1.00 (0.02) & 1.00 (0.00) & 1.00 (0.00) & 1.00 (0.00) \\ 
   \hline
\end{tabular}
\caption{Fractions of variables recovered by L2Boost, MBoost, LADBoost, Robloss, and RRBoost applied with tree learners of $d$ = 1 for clean ($D_0$), symmetric gross error contaminated ($D_1$), asymmetric gross error contaminated ($D_2$),  skewed distributed ($D_3$),  and heavy-tailed distributed  ($D_4$) data generated from $g$ = $g_3$ S = $S_1$ $n$ = 3000 $p$ = 400, displayed in the form of: mean (SD) calculated from 100 independent runs of the experiment.} 
\end{table}
\begin{table}[H]
\centering
\begin{tabular}{lcccccc}
  \hline
 & L2Boost & MBoost & LADBoost & Robloss & SBoost & RRBoost \\ 
  \hline
$D_0$ & 3.76 (0.13) & 3.73 (0.12) & 4.07 (0.16) & 3.82 (0.16) & 3.64 (0.18) & 3.42 (0.07) \\ 
  $D_1 (10\%)$ & 5.99 (0.31) & 5.00 (0.75) & 4.22 (0.17) & 4.00 (0.18) & 3.54 (0.15) & 3.40 (0.07) \\ 
  $D_1 (20\%)$ & 6.31 (0.39) & 6.35 (0.37) & 4.36 (0.21) & 4.19 (0.20) & 3.53 (0.15) & 3.42 (0.07) \\ 
  $D_2 (10\%)$ & 8.64 (0.30) & 6.06 (1.36) & 4.31 (0.22) & 4.14 (0.20) & 3.54 (0.18) & 3.40 (0.09) \\ 
  $D_2 (20\%)$ & 14.07 (0.36) & 14.16 (0.65) & 4.72 (0.22) & 4.92 (0.30) & 3.50 (0.13) & 3.40 (0.07) \\ 
  $D_3$ & 4.57 (0.44) & 3.99 (0.17) & 4.36 (0.18) & 4.10 (0.16) & 4.56 (0.22) & 3.86 (0.10) \\ 
  $D_4$ & 329.86 (1691.99) & 4.65 (0.22) & 4.41 (0.21) & 4.30 (0.23) & 3.47 (0.08) & 3.46 (0.09) \\ 
   \hline
\end{tabular}
\caption{Summary statistics of RMSEs on the test sets by L2Boost, MBoost, LADBoost, Robloss, SBoost, and RRBoost applied with tree learners of $d$ = 2 for clean ($D_0$), symmetric gross error contaminated ($D_1$), asymmetric gross error contaminated ($D_2$),  skewed distributed ($D_3$),  and heavy-tailed distributed  ($D_4$) data generated from $g$ = $g_3$ S = $S_1$ $n$ = 3000 $p$ = 400, displayed in the form of: mean (SD) calculated from 100 independent runs of the experiment.} 
\end{table}
\begin{table}[H]
\centering
\begin{tabular}{lccccc}
  \hline
 & L2Boost & MBoost & LADBoost & Robloss & RRBoost \\ 
  \hline
$D_0$ & 1.00 (0.00) & 1.00 (0.00) & 1.00 (0.00) & 1.00 (0.00) & 1.00 (0.00) \\ 
  $D_1 (10\%)$ & 0.67 (0.15) & 0.92 (0.12) & 1.00 (0.00) & 1.00 (0.00) & 1.00 (0.00) \\ 
  $D_1 (20\%)$ & 0.57 (0.13) & 0.56 (0.13) & 1.00 (0.00) & 1.00 (0.00) & 1.00 (0.00) \\ 
  $D_2 (10\%)$ & 0.66 (0.16) & 0.91 (0.13) & 1.00 (0.00) & 1.00 (0.00) & 1.00 (0.00) \\ 
  $D_2 (20\%)$ & 0.47 (0.13) & 0.58 (0.13) & 1.00 (0.00) & 1.00 (0.00) & 1.00 (0.00) \\ 
  $D_3$ & 0.97 (0.09) & 1.00 (0.00) & 1.00 (0.00) & 1.00 (0.00) & 1.00 (0.00) \\ 
  $D_4$ & 0.03 (0.08) & 0.99 (0.05) & 1.00 (0.00) & 1.00 (0.00) & 1.00 (0.00) \\ 
   \hline
\end{tabular}
\caption{Fractions of variables recovered by L2Boost, MBoost, LADBoost, Robloss, and RRBoost applied with tree learners of $d$ = 2 for clean ($D_0$), symmetric gross error contaminated ($D_1$), asymmetric gross error contaminated ($D_2$),  skewed distributed ($D_3$),  and heavy-tailed distributed  ($D_4$) data generated from $g$ = $g_3$ S = $S_1$ $n$ = 3000 $p$ = 400, displayed in the form of: mean (SD) calculated from 100 independent runs of the experiment.} 
\end{table}
\begin{table}[H]
\centering
\begin{tabular}{lcccccc}
  \hline
 & L2Boost & MBoost & LADBoost & Robloss & SBoost & RRBoost \\ 
  \hline
$D_0$ & 3.94 (0.14) & 3.95 (0.13) & 4.50 (0.18) & 4.08 (0.14) & 5.20 (0.64) & 3.77 (0.10) \\ 
  $D_1 (10\%)$ & 6.11 (0.50) & 5.07 (0.61) & 4.68 (0.15) & 4.27 (0.19) & 4.89 (0.45) & 3.70 (0.11) \\ 
  $D_1 (20\%)$ & 7.21 (0.47) & 7.35 (0.51) & 4.80 (0.19) & 4.50 (0.18) & 4.19 (0.43) & 3.62 (0.11) \\ 
  $D_2 (10\%)$ & 8.83 (0.42) & 6.37 (1.24) & 4.74 (0.18) & 4.47 (0.19) & 4.77 (0.60) & 3.71 (0.13) \\ 
  $D_2 (20\%)$ & 14.38 (0.44) & 14.81 (0.85) & 5.33 (0.22) & 5.48 (0.34) & 4.50 (0.69) & 3.69 (0.12) \\ 
  $D_3$ & 4.72 (0.29) & 4.20 (0.14) & 4.71 (0.17) & 4.38 (0.18) & 5.25 (0.54) & 4.08 (0.11) \\ 
  $D_4$ & 350.82 (1708.69) & 4.91 (0.17) & 4.84 (0.19) & 4.56 (0.18) & 3.92 (0.33) & 3.68 (0.12) \\ 
   \hline
\end{tabular}
\caption{Summary statistics of RMSEs on the test sets by L2Boost, MBoost, LADBoost, Robloss, SBoost, and RRBoost applied with tree learners of $d$ = 3 for clean ($D_0$), symmetric gross error contaminated ($D_1$), asymmetric gross error contaminated ($D_2$),  skewed distributed ($D_3$),  and heavy-tailed distributed  ($D_4$) data generated from $g$ = $g_3$ S = $S_1$ $n$ = 3000 $p$ = 400, displayed in the form of: mean (SD) calculated from 100 independent runs of the experiment.} 
\end{table}
\begin{table}[H]
\centering
\begin{tabular}{lccccc}
  \hline
 & L2Boost & MBoost & LADBoost & Robloss & RRBoost \\ 
  \hline
$D_0$ & 1.00 (0.00) & 1.00 (0.00) & 1.00 (0.00) & 1.00 (0.00) & 1.00 (0.00) \\ 
  $D_1 (10\%)$ & 0.77 (0.13) & 0.93 (0.12) & 1.00 (0.00) & 1.00 (0.00) & 1.00 (0.00) \\ 
  $D_1 (20\%)$ & 0.65 (0.13) & 0.64 (0.14) & 1.00 (0.00) & 1.00 (0.00) & 1.00 (0.00) \\ 
  $D_2 (10\%)$ & 0.73 (0.15) & 0.91 (0.12) & 1.00 (0.00) & 1.00 (0.00) & 1.00 (0.00) \\ 
  $D_2 (20\%)$ & 0.58 (0.14) & 0.58 (0.14) & 0.99 (0.04) & 1.00 (0.02) & 1.00 (0.00) \\ 
  $D_3$ & 0.97 (0.08) & 1.00 (0.00) & 1.00 (0.00) & 1.00 (0.00) & 1.00 (0.00) \\ 
  $D_4$ & 0.07 (0.11) & 0.98 (0.06) & 1.00 (0.00) & 1.00 (0.02) & 1.00 (0.00) \\ 
   \hline
\end{tabular}
\caption{Fractions of variables recovered by L2Boost, MBoost, LADBoost, Robloss, and RRBoost applied with tree learners of $d$ = 3 for clean ($D_0$), symmetric gross error contaminated ($D_1$), asymmetric gross error contaminated ($D_2$),  skewed distributed ($D_3$),  and heavy-tailed distributed  ($D_4$) data generated from $g$ = $g_3$ S = $S_1$ $n$ = 3000 $p$ = 400, displayed in the form of: mean (SD) calculated from 100 independent runs of the experiment.} 
\end{table}
\begin{table}[H]
\centering
\begin{tabular}{lcccccc}
  \hline
 & L2Boost & MBoost & LADBoost & Robloss & SBoost & RRBoost \\ 
  \hline
$D_0$ & 4.23 (0.15) & 4.24 (0.15) & 4.30 (0.17) & 4.28 (0.16) & 4.51 (0.35) & 4.03 (0.20) \\ 
  $D_1 (10\%)$ & 7.27 (0.94) & 5.91 (1.15) & 4.53 (0.19) & 4.59 (0.36) & 4.38 (0.33) & 4.03 (0.16) \\ 
  $D_1 (20\%)$ & 8.77 (1.23) & 8.48 (1.14) & 4.87 (0.32) & 5.23 (0.74) & 4.28 (0.31) & 4.05 (0.18) \\ 
  $D_2 (10\%)$ & 9.82 (0.93) & 7.22 (2.09) & 4.60 (0.23) & 4.71 (0.39) & 4.41 (0.35) & 4.05 (0.16) \\ 
  $D_2 (20\%)$ & 15.66 (1.19) & 15.52 (1.25) & 5.36 (0.52) & 5.96 (0.73) & 4.29 (0.30) & 4.05 (0.17) \\ 
  $D_3$ & 5.19 (0.50) & 4.59 (0.20) & 4.59 (0.16) & 4.57 (0.20) & 4.69 (0.28) & 4.41 (0.18) \\ 
  $D_4$ & 132.21 (535.13) & 5.81 (0.47) & 5.02 (0.33) & 5.24 (0.37) & 4.32 (0.28) & 4.24 (0.22) \\ 
   \hline
\end{tabular}
\caption{Summary statistics of RMSEs on the test sets by L2Boost, MBoost, LADBoost, Robloss, SBoost, and RRBoost applied with tree learners of $d$ = 1 for clean ($D_0$), symmetric gross error contaminated ($D_1$), asymmetric gross error contaminated ($D_2$),  skewed distributed ($D_3$),  and heavy-tailed distributed  ($D_4$) data generated from $g$ = $g_3$ S = $S_2$ $n$ = 300 $p$ = 10, displayed in the form of: mean (SD) calculated from 100 independent runs of the experiment.} 
\end{table}
\begin{table}[H]
\centering
\begin{tabular}{lccccc}
  \hline
 & L2Boost & MBoost & LADBoost & Robloss & RRBoost \\ 
  \hline
$D_0$ & 1.00 (0.02) & 1.00 (0.00) & 1.00 (0.02) & 1.00 (0.03) & 1.00 (0.00) \\ 
  $D_1 (10\%)$ & 0.41 (0.16) & 0.70 (0.30) & 1.00 (0.03) & 0.99 (0.03) & 1.00 (0.00) \\ 
  $D_1 (20\%)$ & 0.33 (0.16) & 0.31 (0.12) & 0.97 (0.07) & 0.94 (0.10) & 1.00 (0.03) \\ 
  $D_2 (10\%)$ & 0.39 (0.17) & 0.77 (0.27) & 1.00 (0.03) & 0.98 (0.07) & 1.00 (0.00) \\ 
  $D_2 (20\%)$ & 0.35 (0.15) & 0.44 (0.18) & 0.96 (0.09) & 0.89 (0.15) & 0.99 (0.03) \\ 
  $D_3$ & 0.91 (0.16) & 0.98 (0.05) & 0.99 (0.04) & 1.00 (0.03) & 1.00 (0.00) \\ 
  $D_4$ & 0.37 (0.32) & 0.76 (0.20) & 0.94 (0.11) & 0.90 (0.13) & 0.99 (0.04) \\ 
   \hline
\end{tabular}
\caption{Fractions of variables recovered by L2Boost, MBoost, LADBoost, Robloss, and RRBoost applied with tree learners of $d$ = 1 for clean ($D_0$), symmetric gross error contaminated ($D_1$), asymmetric gross error contaminated ($D_2$),  skewed distributed ($D_3$),  and heavy-tailed distributed  ($D_4$) data generated from $g$ = $g_3$ S = $S_2$ $n$ = 300 $p$ = 10, displayed in the form of: mean (SD) calculated from 100 independent runs of the experiment.} 
\end{table}
\begin{table}[H]
\centering
\begin{tabular}{lcccccc}
  \hline
 & L2Boost & MBoost & LADBoost & Robloss & SBoost & RRBoost \\ 
  \hline
$D_0$ & 4.75 (0.24) & 4.74 (0.25) & 4.96 (0.24) & 4.81 (0.24) & 5.26 (0.38) & 4.47 (0.25) \\ 
  $D_1 (10\%)$ & 8.20 (1.20) & 6.67 (1.41) & 5.22 (0.29) & 5.16 (0.29) & 5.21 (0.35) & 4.45 (0.20) \\ 
  $D_1 (20\%)$ & 10.60 (1.62) & 10.66 (1.40) & 5.63 (0.37) & 5.78 (0.50) & 4.92 (0.41) & 4.46 (0.26) \\ 
  $D_2 (10\%)$ & 10.33 (1.15) & 7.87 (2.20) & 5.40 (0.36) & 5.40 (0.40) & 5.39 (0.39) & 4.56 (0.25) \\ 
  $D_2 (20\%)$ & 16.48 (1.31) & 16.71 (1.38) & 6.29 (0.50) & 6.83 (0.60) & 5.19 (0.42) & 4.55 (0.25) \\ 
  $D_3$ & 5.71 (0.68) & 5.08 (0.27) & 5.26 (0.27) & 5.14 (0.22) & 5.44 (0.30) & 4.81 (0.26) \\ 
  $D_4$ & 212.54 (1373.08) & 6.17 (0.41) & 5.69 (0.38) & 5.79 (0.44) & 5.07 (0.42) & 4.80 (0.35) \\ 
   \hline
\end{tabular}
\caption{Summary statistics of RMSEs on the test sets by L2Boost, MBoost, LADBoost, Robloss, SBoost, and RRBoost applied with tree learners of $d$ = 2 for clean ($D_0$), symmetric gross error contaminated ($D_1$), asymmetric gross error contaminated ($D_2$),  skewed distributed ($D_3$),  and heavy-tailed distributed  ($D_4$) data generated from $g$ = $g_3$ S = $S_2$ $n$ = 300 $p$ = 10, displayed in the form of: mean (SD) calculated from 100 independent runs of the experiment.} 
\end{table}
\begin{table}[H]
\centering
\begin{tabular}{lccccc}
  \hline
 & L2Boost & MBoost & LADBoost & Robloss & RRBoost \\ 
  \hline
$D_0$ & 0.98 (0.06) & 0.98 (0.05) & 0.98 (0.05) & 0.98 (0.06) & 1.00 (0.02) \\ 
  $D_1 (10\%)$ & 0.48 (0.14) & 0.70 (0.23) & 0.97 (0.08) & 0.95 (0.10) & 1.00 (0.03) \\ 
  $D_1 (20\%)$ & 0.41 (0.16) & 0.39 (0.14) & 0.92 (0.11) & 0.86 (0.16) & 0.99 (0.03) \\ 
  $D_2 (10\%)$ & 0.51 (0.14) & 0.73 (0.21) & 0.96 (0.09) & 0.95 (0.11) & 1.00 (0.03) \\ 
  $D_2 (20\%)$ & 0.47 (0.16) & 0.51 (0.19) & 0.90 (0.15) & 0.81 (0.17) & 1.00 (0.02) \\ 
  $D_3$ & 0.82 (0.19) & 0.96 (0.08) & 0.98 (0.07) & 0.96 (0.09) & 0.99 (0.04) \\ 
  $D_4$ & 0.34 (0.20) & 0.68 (0.20) & 0.91 (0.11) & 0.83 (0.18) & 0.96 (0.08) \\ 
   \hline
\end{tabular}
\caption{Fractions of variables recovered by L2Boost, MBoost, LADBoost, Robloss, and RRBoost applied with tree learners of $d$ = 2 for clean ($D_0$), symmetric gross error contaminated ($D_1$), asymmetric gross error contaminated ($D_2$),  skewed distributed ($D_3$),  and heavy-tailed distributed  ($D_4$) data generated from $g$ = $g_3$ S = $S_2$ $n$ = 300 $p$ = 10, displayed in the form of: mean (SD) calculated from 100 independent runs of the experiment.} 
\end{table}
\begin{table}[H]
\centering
\begin{tabular}{lcccccc}
  \hline
 & L2Boost & MBoost & LADBoost & Robloss & SBoost & RRBoost \\ 
  \hline
$D_0$ & 4.93 (0.24) & 4.93 (0.22) & 5.24 (0.24) & 5.03 (0.22) & 5.48 (0.39) & 4.83 (0.25) \\ 
  $D_1 (10\%)$ & 9.95 (1.30) & 7.66 (2.01) & 5.64 (0.29) & 5.50 (0.29) & 5.49 (0.36) & 4.87 (0.28) \\ 
  $D_1 (20\%)$ & 13.34 (1.57) & 13.34 (1.36) & 6.07 (0.40) & 6.27 (0.45) & 5.35 (0.36) & 4.86 (0.27) \\ 
  $D_2 (10\%)$ & 11.59 (1.25) & 8.72 (2.66) & 5.78 (0.35) & 5.73 (0.41) & 5.58 (0.46) & 4.92 (0.25) \\ 
  $D_2 (20\%)$ & 17.92 (1.31) & 17.92 (1.38) & 6.84 (0.53) & 7.61 (0.68) & 5.68 (0.54) & 4.99 (0.25) \\ 
  $D_3$ & 5.76 (0.51) & 5.30 (0.27) & 5.50 (0.27) & 5.37 (0.28) & 5.67 (0.40) & 5.20 (0.25) \\ 
  $D_4$ & 215.55 (1372.79) & 6.57 (0.63) & 6.08 (0.38) & 6.05 (0.45) & 5.51 (0.33) & 5.29 (0.34) \\ 
   \hline
\end{tabular}
\caption{Summary statistics of RMSEs on the test sets by L2Boost, MBoost, LADBoost, Robloss, SBoost, and RRBoost applied with tree learners of $d$ = 3 for clean ($D_0$), symmetric gross error contaminated ($D_1$), asymmetric gross error contaminated ($D_2$),  skewed distributed ($D_3$),  and heavy-tailed distributed  ($D_4$) data generated from $g$ = $g_3$ S = $S_2$ $n$ = 300 $p$ = 10, displayed in the form of: mean (SD) calculated from 100 independent runs of the experiment.} 
\end{table}
\begin{table}[H]
\centering
\begin{tabular}{lccccc}
  \hline
 & L2Boost & MBoost & LADBoost & Robloss & RRBoost \\ 
  \hline
$D_0$ & 0.98 (0.07) & 0.98 (0.07) & 0.99 (0.05) & 0.99 (0.03) & 0.99 (0.04) \\ 
  $D_1 (10\%)$ & 0.63 (0.16) & 0.75 (0.17) & 0.97 (0.08) & 0.94 (0.11) & 0.99 (0.04) \\ 
  $D_1 (20\%)$ & 0.53 (0.16) & 0.51 (0.17) & 0.92 (0.12) & 0.83 (0.14) & 0.98 (0.06) \\ 
  $D_2 (10\%)$ & 0.66 (0.16) & 0.77 (0.19) & 0.96 (0.08) & 0.93 (0.12) & 0.97 (0.07) \\ 
  $D_2 (20\%)$ & 0.59 (0.15) & 0.57 (0.17) & 0.89 (0.14) & 0.82 (0.15) & 0.98 (0.05) \\ 
  $D_3$ & 0.87 (0.13) & 0.95 (0.09) & 0.97 (0.07) & 0.98 (0.07) & 0.98 (0.06) \\ 
  $D_4$ & 0.53 (0.19) & 0.79 (0.16) & 0.91 (0.12) & 0.87 (0.13) & 0.94 (0.11) \\ 
   \hline
\end{tabular}
\caption{Fractions of variables recovered by L2Boost, MBoost, LADBoost, Robloss, and RRBoost applied with tree learners of $d$ = 3 for clean ($D_0$), symmetric gross error contaminated ($D_1$), asymmetric gross error contaminated ($D_2$),  skewed distributed ($D_3$),  and heavy-tailed distributed  ($D_4$) data generated from $g$ = $g_3$ S = $S_2$ $n$ = 300 $p$ = 10, displayed in the form of: mean (SD) calculated from 100 independent runs of the experiment.} 
\end{table}
\begin{table}[H]
\centering
\begin{tabular}{lcccccc}
  \hline
 & L2Boost & MBoost & LADBoost & Robloss & SBoost & RRBoost \\ 
  \hline
$D_0$ & 3.56 (0.07) & 3.57 (0.07) & 3.61 (0.08) & 3.57 (0.08) & 3.86 (0.23) & 3.51 (0.07) \\ 
  $D_1 (10\%)$ & 5.18 (0.31) & 4.52 (0.73) & 3.67 (0.08) & 3.65 (0.08) & 3.73 (0.18) & 3.51 (0.08) \\ 
  $D_1 (20\%)$ & 5.84 (0.39) & 5.82 (0.30) & 3.75 (0.17) & 3.80 (0.31) & 3.64 (0.16) & 3.50 (0.08) \\ 
  $D_2 (10\%)$ & 8.43 (0.29) & 5.74 (1.65) & 3.72 (0.09) & 3.73 (0.10) & 3.74 (0.19) & 3.51 (0.07) \\ 
  $D_2 (20\%)$ & 14.59 (0.37) & 14.42 (0.47) & 4.07 (0.42) & 4.32 (0.49) & 3.65 (0.13) & 3.51 (0.08) \\ 
  $D_3$ & 3.95 (0.19) & 3.76 (0.08) & 4.09 (0.09) & 3.93 (0.09) & 4.58 (0.09) & 3.96 (0.09) \\ 
  $D_4$ & 37.41 (92.62) & 4.15 (0.13) & 3.76 (0.09) & 3.83 (0.09) & 3.65 (0.11) & 3.56 (0.08) \\ 
   \hline
\end{tabular}
\caption{Summary statistics of RMSEs on the test sets by L2Boost, MBoost, LADBoost, Robloss, SBoost, and RRBoost applied with tree learners of $d$ = 1 for clean ($D_0$), symmetric gross error contaminated ($D_1$), asymmetric gross error contaminated ($D_2$),  skewed distributed ($D_3$),  and heavy-tailed distributed  ($D_4$) data generated from $g$ = $g_3$ S = $S_2$ $n$ = 3000 $p$ = 10, displayed in the form of: mean (SD) calculated from 100 independent runs of the experiment.} 
\end{table}
\begin{table}[H]
\centering
\begin{tabular}{lccccc}
  \hline
 & L2Boost & MBoost & LADBoost & Robloss & RRBoost \\ 
  \hline
$D_0$ & 1.00 (0.00) & 1.00 (0.00) & 1.00 (0.00) & 1.00 (0.00) & 1.00 (0.00) \\ 
  $D_1 (10\%)$ & 0.92 (0.15) & 0.98 (0.07) & 1.00 (0.00) & 1.00 (0.00) & 1.00 (0.00) \\ 
  $D_1 (20\%)$ & 0.68 (0.19) & 0.68 (0.19) & 1.00 (0.00) & 1.00 (0.00) & 1.00 (0.00) \\ 
  $D_2 (10\%)$ & 0.95 (0.10) & 0.94 (0.09) & 1.00 (0.00) & 1.00 (0.00) & 1.00 (0.00) \\ 
  $D_2 (20\%)$ & 0.81 (0.20) & 0.67 (0.28) & 1.00 (0.02) & 0.99 (0.04) & 1.00 (0.00) \\ 
  $D_3$ & 1.00 (0.00) & 1.00 (0.00) & 1.00 (0.00) & 1.00 (0.00) & 1.00 (0.00) \\ 
  $D_4$ & 0.30 (0.32) & 1.00 (0.00) & 1.00 (0.00) & 1.00 (0.00) & 1.00 (0.00) \\ 
   \hline
\end{tabular}
\caption{Fractions of variables recovered by L2Boost, MBoost, LADBoost, Robloss, and RRBoost applied with tree learners of $d$ = 1 for clean ($D_0$), symmetric gross error contaminated ($D_1$), asymmetric gross error contaminated ($D_2$),  skewed distributed ($D_3$),  and heavy-tailed distributed  ($D_4$) data generated from $g$ = $g_3$ S = $S_2$ $n$ = 3000 $p$ = 10, displayed in the form of: mean (SD) calculated from 100 independent runs of the experiment.} 
\end{table}
\begin{table}[H]
\centering
\begin{tabular}{lcccccc}
  \hline
 & L2Boost & MBoost & LADBoost & Robloss & SBoost & RRBoost \\ 
  \hline
$D_0$ & 3.77 (0.11) & 3.80 (0.11) & 3.92 (0.11) & 3.85 (0.11) & 3.71 (0.12) & 3.54 (0.08) \\ 
  $D_1 (10\%)$ & 5.62 (0.30) & 4.86 (0.74) & 4.03 (0.12) & 3.95 (0.13) & 3.66 (0.11) & 3.55 (0.08) \\ 
  $D_1 (20\%)$ & 6.13 (0.33) & 6.18 (0.33) & 4.13 (0.15) & 4.10 (0.15) & 3.60 (0.10) & 3.54 (0.08) \\ 
  $D_2 (10\%)$ & 8.76 (0.31) & 6.24 (1.71) & 4.08 (0.12) & 4.06 (0.13) & 3.68 (0.10) & 3.55 (0.07) \\ 
  $D_2 (20\%)$ & 14.73 (0.37) & 14.70 (0.63) & 4.52 (0.17) & 4.59 (0.18) & 3.63 (0.09) & 3.55 (0.08) \\ 
  $D_3$ & 4.25 (0.20) & 3.99 (0.12) & 4.27 (0.11) & 4.15 (0.11) & 4.50 (0.12) & 4.03 (0.09) \\ 
  $D_4$ & 122.98 (371.19) & 4.50 (0.17) & 4.20 (0.16) & 4.14 (0.14) & 3.59 (0.10) & 3.58 (0.09) \\ 
   \hline
\end{tabular}
\caption{Summary statistics of RMSEs on the test sets by L2Boost, MBoost, LADBoost, Robloss, SBoost, and RRBoost applied with tree learners of $d$ = 2 for clean ($D_0$), symmetric gross error contaminated ($D_1$), asymmetric gross error contaminated ($D_2$),  skewed distributed ($D_3$),  and heavy-tailed distributed  ($D_4$) data generated from $g$ = $g_3$ S = $S_2$ $n$ = 3000 $p$ = 10, displayed in the form of: mean (SD) calculated from 100 independent runs of the experiment.} 
\end{table}
\begin{table}[H]
\centering
\begin{tabular}{lccccc}
  \hline
 & L2Boost & MBoost & LADBoost & Robloss & RRBoost \\ 
  \hline
$D_0$ & 1.00 (0.00) & 1.00 (0.00) & 1.00 (0.00) & 1.00 (0.00) & 1.00 (0.00) \\ 
  $D_1 (10\%)$ & 0.83 (0.18) & 0.94 (0.10) & 1.00 (0.00) & 1.00 (0.00) & 1.00 (0.00) \\ 
  $D_1 (20\%)$ & 0.62 (0.19) & 0.64 (0.18) & 1.00 (0.00) & 1.00 (0.00) & 1.00 (0.00) \\ 
  $D_2 (10\%)$ & 0.81 (0.19) & 0.96 (0.10) & 1.00 (0.00) & 1.00 (0.00) & 1.00 (0.00) \\ 
  $D_2 (20\%)$ & 0.68 (0.20) & 0.78 (0.15) & 1.00 (0.00) & 1.00 (0.00) & 1.00 (0.00) \\ 
  $D_3$ & 1.00 (0.00) & 1.00 (0.00) & 1.00 (0.00) & 1.00 (0.00) & 1.00 (0.00) \\ 
  $D_4$ & 0.30 (0.19) & 1.00 (0.00) & 1.00 (0.00) & 1.00 (0.00) & 1.00 (0.00) \\ 
   \hline
\end{tabular}
\caption{Fractions of variables recovered by L2Boost, MBoost, LADBoost, Robloss, and RRBoost applied with tree learners of $d$ = 2 for clean ($D_0$), symmetric gross error contaminated ($D_1$), asymmetric gross error contaminated ($D_2$),  skewed distributed ($D_3$),  and heavy-tailed distributed  ($D_4$) data generated from $g$ = $g_3$ S = $S_2$ $n$ = 3000 $p$ = 10, displayed in the form of: mean (SD) calculated from 100 independent runs of the experiment.} 
\end{table}
\begin{table}[H]
\centering
\begin{tabular}{lcccccc}
  \hline
 & L2Boost & MBoost & LADBoost & Robloss & SBoost & RRBoost \\ 
  \hline
$D_0$ & 3.94 (0.11) & 3.93 (0.11) & 4.21 (0.13) & 4.01 (0.11) & 4.03 (0.23) & 3.66 (0.08) \\ 
  $D_1 (10\%)$ & 5.69 (0.33) & 5.01 (0.61) & 4.32 (0.14) & 4.15 (0.12) & 3.81 (0.13) & 3.64 (0.08) \\ 
  $D_1 (20\%)$ & 6.33 (0.39) & 6.48 (0.44) & 4.44 (0.15) & 4.34 (0.15) & 3.67 (0.09) & 3.61 (0.07) \\ 
  $D_2 (10\%)$ & 8.73 (0.28) & 6.47 (1.58) & 4.36 (0.12) & 4.24 (0.13) & 3.88 (0.14) & 3.66 (0.07) \\ 
  $D_2 (20\%)$ & 14.79 (0.41) & 15.02 (0.60) & 4.75 (0.15) & 4.93 (0.16) & 3.78 (0.13) & 3.64 (0.09) \\ 
  $D_3$ & 4.48 (0.19) & 4.15 (0.13) & 4.47 (0.11) & 4.29 (0.12) & 4.58 (0.21) & 4.21 (0.09) \\ 
  $D_4$ & 160.99 (601.47) & 4.78 (0.17) & 4.45 (0.14) & 4.37 (0.15) & 3.66 (0.10) & 3.65 (0.09) \\ 
   \hline
\end{tabular}
\caption{Summary statistics of RMSEs on the test sets by L2Boost, MBoost, LADBoost, Robloss, SBoost, and RRBoost applied with tree learners of $d$ = 3 for clean ($D_0$), symmetric gross error contaminated ($D_1$), asymmetric gross error contaminated ($D_2$),  skewed distributed ($D_3$),  and heavy-tailed distributed  ($D_4$) data generated from $g$ = $g_3$ S = $S_2$ $n$ = 3000 $p$ = 10, displayed in the form of: mean (SD) calculated from 100 independent runs of the experiment.} 
\end{table}
\begin{table}[H]
\centering
\begin{tabular}{lccccc}
  \hline
 & L2Boost & MBoost & LADBoost & Robloss & RRBoost \\ 
  \hline
$D_0$ & 1.00 (0.00) & 1.00 (0.00) & 1.00 (0.00) & 1.00 (0.00) & 1.00 (0.00) \\ 
  $D_1 (10\%)$ & 0.85 (0.15) & 0.93 (0.12) & 1.00 (0.00) & 1.00 (0.00) & 1.00 (0.00) \\ 
  $D_1 (20\%)$ & 0.81 (0.15) & 0.78 (0.14) & 1.00 (0.00) & 1.00 (0.00) & 1.00 (0.00) \\ 
  $D_2 (10\%)$ & 0.83 (0.15) & 0.96 (0.09) & 1.00 (0.00) & 1.00 (0.00) & 1.00 (0.00) \\ 
  $D_2 (20\%)$ & 0.76 (0.15) & 0.80 (0.14) & 1.00 (0.00) & 1.00 (0.00) & 1.00 (0.00) \\ 
  $D_3$ & 1.00 (0.00) & 1.00 (0.00) & 1.00 (0.00) & 1.00 (0.00) & 1.00 (0.00) \\ 
  $D_4$ & 0.46 (0.19) & 1.00 (0.00) & 1.00 (0.00) & 1.00 (0.00) & 1.00 (0.00) \\ 
   \hline
\end{tabular}
\caption{Fractions of variables recovered by L2Boost, MBoost, LADBoost, Robloss, and RRBoost applied with tree learners of $d$ = 3 for clean ($D_0$), symmetric gross error contaminated ($D_1$), asymmetric gross error contaminated ($D_2$),  skewed distributed ($D_3$),  and heavy-tailed distributed  ($D_4$) data generated from $g$ = $g_3$ S = $S_2$ $n$ = 3000 $p$ = 10, displayed in the form of: mean (SD) calculated from 100 independent runs of the experiment.} 
\end{table}
\begin{table}[H]
\centering
\begin{tabular}{lcccccc}
  \hline
 & L2Boost & MBoost & LADBoost & Robloss & SBoost & RRBoost \\ 
  \hline
$D_0$ & 4.22 (0.28) & 4.23 (0.30) & 4.47 (0.31) & 4.31 (0.36) & 4.77 (0.46) & 3.77 (0.36) \\ 
  $D_1 (10\%)$ & 6.87 (0.92) & 5.71 (0.86) & 4.85 (0.41) & 4.75 (0.41) & 4.49 (0.39) & 3.74 (0.28) \\ 
  $D_1 (20\%)$ & 7.79 (1.33) & 7.90 (1.46) & 5.11 (0.38) & 5.27 (0.38) & 4.16 (0.37) & 3.71 (0.27) \\ 
  $D_2 (10\%)$ & 8.34 (0.74) & 6.71 (1.31) & 4.90 (0.37) & 4.88 (0.38) & 4.48 (0.46) & 3.71 (0.26) \\ 
  $D_2 (20\%)$ & 12.97 (1.31) & 13.15 (1.41) & 5.65 (0.44) & 6.09 (0.70) & 4.16 (0.42) & 3.69 (0.33) \\ 
  $D_3$ & 5.07 (0.47) & 4.54 (0.32) & 4.74 (0.35) & 4.57 (0.37) & 4.73 (0.53) & 3.93 (0.31) \\ 
  $D_4$ & 81.84 (241.80) & 5.47 (0.36) & 5.11 (0.37) & 5.15 (0.34) & 4.59 (0.55) & 4.07 (0.38) \\ 
   \hline
\end{tabular}
\caption{Summary statistics of RMSEs on the test sets by L2Boost, MBoost, LADBoost, Robloss, SBoost, and RRBoost applied with tree learners of $d$ = 1 for clean ($D_0$), symmetric gross error contaminated ($D_1$), asymmetric gross error contaminated ($D_2$),  skewed distributed ($D_3$),  and heavy-tailed distributed  ($D_4$) data generated from $g$ = $g_3$ S = $S_2$ $n$ = 300 $p$ = 400, displayed in the form of: mean (SD) calculated from 100 independent runs of the experiment.} 
\end{table}
\begin{table}[H]
\centering
\begin{tabular}{lccccc}
  \hline
 & L2Boost & MBoost & LADBoost & Robloss & RRBoost \\ 
  \hline
$D_0$ & 0.96 (0.08) & 0.97 (0.07) & 0.91 (0.12) & 0.95 (0.10) & 0.99 (0.04) \\ 
  $D_1 (10\%)$ & 0.33 (0.10) & 0.55 (0.22) & 0.80 (0.16) & 0.83 (0.17) & 0.99 (0.03) \\ 
  $D_1 (20\%)$ & 0.23 (0.11) & 0.24 (0.12) & 0.67 (0.15) & 0.64 (0.17) & 1.00 (0.03) \\ 
  $D_2 (10\%)$ & 0.33 (0.09) & 0.51 (0.19) & 0.79 (0.15) & 0.83 (0.15) & 1.00 (0.00) \\ 
  $D_2 (20\%)$ & 0.28 (0.10) & 0.29 (0.10) & 0.65 (0.15) & 0.57 (0.14) & 0.99 (0.04) \\ 
  $D_3$ & 0.68 (0.20) & 0.89 (0.14) & 0.87 (0.15) & 0.92 (0.13) & 0.99 (0.04) \\ 
  $D_4$ & 0.05 (0.11) & 0.53 (0.13) & 0.68 (0.17) & 0.66 (0.16) & 0.96 (0.09) \\ 
   \hline
\end{tabular}
\caption{Fractions of variables recovered by L2Boost, MBoost, LADBoost, Robloss, and RRBoost applied with tree learners of $d$ = 1 for clean ($D_0$), symmetric gross error contaminated ($D_1$), asymmetric gross error contaminated ($D_2$),  skewed distributed ($D_3$),  and heavy-tailed distributed  ($D_4$) data generated from $g$ = $g_3$ S = $S_2$ $n$ = 300 $p$ = 400, displayed in the form of: mean (SD) calculated from 100 independent runs of the experiment.} 
\end{table}
\begin{table}[H]
\centering
\begin{tabular}{lcccccc}
  \hline
 & L2Boost & MBoost & LADBoost & Robloss & SBoost & RRBoost \\ 
  \hline
$D_0$ & 4.58 (0.32) & 4.54 (0.33) & 5.14 (0.32) & 4.74 (0.32) & 5.57 (0.53) & 4.34 (0.34) \\ 
  $D_1 (10\%)$ & 8.90 (1.12) & 6.83 (1.44) & 5.53 (0.40) & 5.29 (0.37) & 5.59 (0.41) & 4.37 (0.39) \\ 
  $D_1 (20\%)$ & 10.89 (1.35) & 11.13 (1.40) & 5.98 (0.50) & 5.85 (0.57) & 5.34 (0.45) & 4.39 (0.35) \\ 
  $D_2 (10\%)$ & 10.31 (1.22) & 7.63 (1.69) & 5.79 (0.43) & 5.63 (0.45) & 5.77 (0.50) & 4.43 (0.41) \\ 
  $D_2 (20\%)$ & 14.67 (1.38) & 14.75 (1.35) & 7.21 (0.63) & 7.58 (0.89) & 5.87 (0.46) & 4.63 (0.32) \\ 
  $D_3$ & 5.42 (0.47) & 4.98 (0.31) & 5.34 (0.39) & 5.04 (0.33) & 5.60 (0.41) & 4.60 (0.37) \\ 
  $D_4$ & 65.23 (159.26) & 5.89 (0.53) & 5.93 (0.50) & 5.67 (0.46) & 5.72 (0.47) & 4.95 (0.42) \\ 
   \hline
\end{tabular}
\caption{Summary statistics of RMSEs on the test sets by L2Boost, MBoost, LADBoost, Robloss, SBoost, and RRBoost applied with tree learners of $d$ = 2 for clean ($D_0$), symmetric gross error contaminated ($D_1$), asymmetric gross error contaminated ($D_2$),  skewed distributed ($D_3$),  and heavy-tailed distributed  ($D_4$) data generated from $g$ = $g_3$ S = $S_2$ $n$ = 300 $p$ = 400, displayed in the form of: mean (SD) calculated from 100 independent runs of the experiment.} 
\end{table}
\begin{table}[H]
\centering
\begin{tabular}{lccccc}
  \hline
 & L2Boost & MBoost & LADBoost & Robloss & RRBoost \\ 
  \hline
$D_0$ & 0.89 (0.13) & 0.89 (0.13) & 0.90 (0.13) & 0.91 (0.11) & 0.98 (0.07) \\ 
  $D_1 (10\%)$ & 0.32 (0.12) & 0.52 (0.18) & 0.80 (0.14) & 0.77 (0.16) & 0.98 (0.06) \\ 
  $D_1 (20\%)$ & 0.24 (0.12) & 0.24 (0.12) & 0.70 (0.16) & 0.65 (0.18) & 0.97 (0.08) \\ 
  $D_2 (10\%)$ & 0.34 (0.10) & 0.50 (0.15) & 0.77 (0.15) & 0.76 (0.15) & 0.97 (0.07) \\ 
  $D_2 (20\%)$ & 0.27 (0.12) & 0.31 (0.10) & 0.60 (0.19) & 0.53 (0.15) & 0.95 (0.10) \\ 
  $D_3$ & 0.68 (0.19) & 0.82 (0.13) & 0.87 (0.14) & 0.87 (0.15) & 0.96 (0.09) \\ 
  $D_4$ & 0.10 (0.13) & 0.54 (0.14) & 0.71 (0.17) & 0.64 (0.17) & 0.89 (0.12) \\ 
   \hline
\end{tabular}
\caption{Fractions of variables recovered by L2Boost, MBoost, LADBoost, Robloss, and RRBoost applied with tree learners of $d$ = 2 for clean ($D_0$), symmetric gross error contaminated ($D_1$), asymmetric gross error contaminated ($D_2$),  skewed distributed ($D_3$),  and heavy-tailed distributed  ($D_4$) data generated from $g$ = $g_3$ S = $S_2$ $n$ = 300 $p$ = 400, displayed in the form of: mean (SD) calculated from 100 independent runs of the experiment.} 
\end{table}
\begin{table}[H]
\centering
\begin{tabular}{lcccccc}
  \hline
 & L2Boost & MBoost & LADBoost & Robloss & SBoost & RRBoost \\ 
  \hline
$D_0$ & 4.74 (0.23) & 4.73 (0.28) & 5.68 (0.34) & 4.90 (0.27) & 5.30 (0.52) & 4.90 (0.29) \\ 
  $D_1 (10\%)$ & 10.65 (1.05) & 8.16 (1.96) & 6.38 (0.38) & 5.95 (0.46) & 5.52 (0.50) & 5.01 (0.34) \\ 
  $D_1 (20\%)$ & 14.04 (1.35) & 14.02 (1.48) & 7.03 (0.49) & 6.94 (0.61) & 5.59 (0.37) & 5.15 (0.35) \\ 
  $D_2 (10\%)$ & 11.85 (1.23) & 8.99 (1.99) & 6.60 (0.49) & 6.37 (0.50) & 5.54 (0.48) & 5.04 (0.26) \\ 
  $D_2 (20\%)$ & 16.67 (1.48) & 16.54 (1.56) & 8.08 (0.60) & 8.73 (0.96) & 5.97 (0.71) & 5.39 (0.55) \\ 
  $D_3$ & 5.63 (0.58) & 5.19 (0.32) & 6.03 (0.30) & 5.28 (0.34) & 5.44 (0.47) & 5.16 (0.32) \\ 
  $D_4$ & 69.24 (166.07) & 6.75 (0.61) & 6.85 (0.44) & 6.29 (0.52) & 5.78 (0.48) & 5.57 (0.34) \\ 
   \hline
\end{tabular}
\caption{Summary statistics of RMSEs on the test sets by L2Boost, MBoost, LADBoost, Robloss, SBoost, and RRBoost applied with tree learners of $d$ = 3 for clean ($D_0$), symmetric gross error contaminated ($D_1$), asymmetric gross error contaminated ($D_2$),  skewed distributed ($D_3$),  and heavy-tailed distributed  ($D_4$) data generated from $g$ = $g_3$ S = $S_2$ $n$ = 300 $p$ = 400, displayed in the form of: mean (SD) calculated from 100 independent runs of the experiment.} 
\end{table}
\begin{table}[H]
\centering
\begin{tabular}{lccccc}
  \hline
 & L2Boost & MBoost & LADBoost & Robloss & RRBoost \\ 
  \hline
$D_0$ & 0.89 (0.13) & 0.90 (0.14) & 0.88 (0.14) & 0.92 (0.12) & 0.91 (0.12) \\ 
  $D_1 (10\%)$ & 0.36 (0.13) & 0.54 (0.17) & 0.78 (0.16) & 0.74 (0.17) & 0.90 (0.12) \\ 
  $D_1 (20\%)$ & 0.25 (0.12) & 0.25 (0.13) & 0.63 (0.18) & 0.58 (0.14) & 0.87 (0.15) \\ 
  $D_2 (10\%)$ & 0.40 (0.14) & 0.55 (0.15) & 0.75 (0.15) & 0.73 (0.16) & 0.89 (0.13) \\ 
  $D_2 (20\%)$ & 0.30 (0.13) & 0.32 (0.12) & 0.55 (0.14) & 0.54 (0.14) & 0.86 (0.14) \\ 
  $D_3$ & 0.78 (0.14) & 0.85 (0.14) & 0.84 (0.13) & 0.85 (0.15) & 0.88 (0.12) \\ 
  $D_4$ & 0.20 (0.17) & 0.63 (0.15) & 0.67 (0.15) & 0.68 (0.15) & 0.76 (0.15) \\ 
   \hline
\end{tabular}
\caption{Fractions of variables recovered by L2Boost, MBoost, LADBoost, Robloss, and RRBoost applied with tree learners of $d$ = 3 for clean ($D_0$), symmetric gross error contaminated ($D_1$), asymmetric gross error contaminated ($D_2$),  skewed distributed ($D_3$),  and heavy-tailed distributed  ($D_4$) data generated from $g$ = $g_3$ S = $S_2$ $n$ = 300 $p$ = 400, displayed in the form of: mean (SD) calculated from 100 independent runs of the experiment.} 
\end{table}
\begin{table}[H]
\centering
\begin{tabular}{lcccccc}
  \hline
 & L2Boost & MBoost & LADBoost & Robloss & SBoost & RRBoost \\ 
  \hline
$D_0$ & 3.02 (0.08) & 3.03 (0.08) & 3.12 (0.08) & 3.04 (0.08) & 3.05 (0.15) & 2.79 (0.06) \\ 
  $D_1 (10\%)$ & 4.77 (0.26) & 3.91 (0.67) & 3.17 (0.09) & 3.13 (0.09) & 2.98 (0.11) & 2.78 (0.07) \\ 
  $D_1 (20\%)$ & 5.25 (0.29) & 5.27 (0.29) & 3.28 (0.08) & 3.26 (0.10) & 2.93 (0.09) & 2.79 (0.06) \\ 
  $D_2 (10\%)$ & 7.04 (0.28) & 4.79 (1.26) & 3.24 (0.10) & 3.20 (0.16) & 2.98 (0.11) & 2.79 (0.07) \\ 
  $D_2 (20\%)$ & 11.59 (0.30) & 11.43 (0.36) & 3.53 (0.12) & 3.68 (0.23) & 2.92 (0.09) & 2.79 (0.06) \\ 
  $D_3$ & 3.62 (0.24) & 3.19 (0.08) & 3.37 (0.08) & 3.26 (0.08) & 3.55 (0.08) & 3.12 (0.07) \\ 
  $D_4$ & 308.06 (1512.48) & 3.68 (0.14) & 3.32 (0.10) & 3.33 (0.10) & 2.94 (0.07) & 2.87 (0.06) \\ 
   \hline
\end{tabular}
\caption{Summary statistics of RMSEs on the test sets by L2Boost, MBoost, LADBoost, Robloss, SBoost, and RRBoost applied with tree learners of $d$ = 1 for clean ($D_0$), symmetric gross error contaminated ($D_1$), asymmetric gross error contaminated ($D_2$),  skewed distributed ($D_3$),  and heavy-tailed distributed  ($D_4$) data generated from $g$ = $g_3$ S = $S_2$ $n$ = 3000 $p$ = 400, displayed in the form of: mean (SD) calculated from 100 independent runs of the experiment.} 
\end{table}
\begin{table}[H]
\centering
\begin{tabular}{lccccc}
  \hline
 & L2Boost & MBoost & LADBoost & Robloss & RRBoost \\ 
  \hline
$D_0$ & 1.00 (0.00) & 1.00 (0.00) & 1.00 (0.00) & 1.00 (0.00) & 1.00 (0.00) \\ 
  $D_1 (10\%)$ & 0.83 (0.15) & 0.96 (0.09) & 1.00 (0.00) & 1.00 (0.00) & 1.00 (0.00) \\ 
  $D_1 (20\%)$ & 0.57 (0.11) & 0.55 (0.10) & 1.00 (0.00) & 1.00 (0.00) & 1.00 (0.00) \\ 
  $D_2 (10\%)$ & 0.81 (0.15) & 0.95 (0.09) & 1.00 (0.00) & 1.00 (0.02) & 1.00 (0.00) \\ 
  $D_2 (20\%)$ & 0.64 (0.10) & 0.52 (0.17) & 1.00 (0.00) & 1.00 (0.03) & 1.00 (0.00) \\ 
  $D_3$ & 0.99 (0.06) & 1.00 (0.00) & 1.00 (0.00) & 1.00 (0.00) & 1.00 (0.00) \\ 
  $D_4$ & 0.05 (0.13) & 1.00 (0.00) & 1.00 (0.00) & 1.00 (0.00) & 1.00 (0.00) \\ 
   \hline
\end{tabular}
\caption{Fractions of variables recovered by L2Boost, MBoost, LADBoost, Robloss, and RRBoost applied with tree learners of $d$ = 1 for clean ($D_0$), symmetric gross error contaminated ($D_1$), asymmetric gross error contaminated ($D_2$),  skewed distributed ($D_3$),  and heavy-tailed distributed  ($D_4$) data generated from $g$ = $g_3$ S = $S_2$ $n$ = 3000 $p$ = 400, displayed in the form of: mean (SD) calculated from 100 independent runs of the experiment.} 
\end{table}
\begin{table}[H]
\centering
\begin{tabular}{lcccccc}
  \hline
 & L2Boost & MBoost & LADBoost & Robloss & SBoost & RRBoost \\ 
  \hline
$D_0$ & 3.18 (0.13) & 3.23 (0.12) & 3.54 (0.16) & 3.32 (0.09) & 3.37 (0.25) & 2.88 (0.06) \\ 
  $D_1 (10\%)$ & 5.18 (0.27) & 4.32 (0.76) & 3.70 (0.20) & 3.43 (0.13) & 3.10 (0.17) & 2.86 (0.07) \\ 
  $D_1 (20\%)$ & 5.54 (0.27) & 5.57 (0.28) & 3.83 (0.18) & 3.57 (0.18) & 2.99 (0.11) & 2.86 (0.06) \\ 
  $D_2 (10\%)$ & 7.34 (0.27) & 5.07 (1.18) & 3.77 (0.19) & 3.57 (0.17) & 3.18 (0.20) & 2.87 (0.08) \\ 
  $D_2 (20\%)$ & 11.76 (0.29) & 11.80 (0.54) & 4.29 (0.25) & 4.22 (0.29) & 3.00 (0.13) & 2.86 (0.07) \\ 
  $D_3$ & 3.86 (0.27) & 3.38 (0.14) & 3.75 (0.15) & 3.50 (0.12) & 3.81 (0.21) & 3.22 (0.08) \\ 
  $D_4$ & 288.80 (1367.36) & 3.96 (0.19) & 3.87 (0.20) & 3.66 (0.19) & 2.98 (0.10) & 2.93 (0.07) \\ 
   \hline
\end{tabular}
\caption{Summary statistics of RMSEs on the test sets by L2Boost, MBoost, LADBoost, Robloss, SBoost, and RRBoost applied with tree learners of $d$ = 2 for clean ($D_0$), symmetric gross error contaminated ($D_1$), asymmetric gross error contaminated ($D_2$),  skewed distributed ($D_3$),  and heavy-tailed distributed  ($D_4$) data generated from $g$ = $g_3$ S = $S_2$ $n$ = 3000 $p$ = 400, displayed in the form of: mean (SD) calculated from 100 independent runs of the experiment.} 
\end{table}
\begin{table}[H]
\centering
\begin{tabular}{lccccc}
  \hline
 & L2Boost & MBoost & LADBoost & Robloss & RRBoost \\ 
  \hline
$D_0$ & 1.00 (0.00) & 1.00 (0.00) & 1.00 (0.00) & 1.00 (0.00) & 1.00 (0.00) \\ 
  $D_1 (10\%)$ & 0.71 (0.18) & 0.96 (0.09) & 1.00 (0.00) & 1.00 (0.00) & 1.00 (0.00) \\ 
  $D_1 (20\%)$ & 0.53 (0.14) & 0.55 (0.14) & 1.00 (0.00) & 1.00 (0.00) & 1.00 (0.00) \\ 
  $D_2 (10\%)$ & 0.66 (0.19) & 0.97 (0.07) & 1.00 (0.00) & 1.00 (0.00) & 1.00 (0.00) \\ 
  $D_2 (20\%)$ & 0.57 (0.17) & 0.69 (0.12) & 1.00 (0.00) & 1.00 (0.00) & 1.00 (0.00) \\ 
  $D_3$ & 0.99 (0.04) & 1.00 (0.00) & 1.00 (0.00) & 1.00 (0.00) & 1.00 (0.00) \\ 
  $D_4$ & 0.04 (0.08) & 1.00 (0.00) & 1.00 (0.00) & 1.00 (0.00) & 1.00 (0.00) \\ 
   \hline
\end{tabular}
\caption{Fractions of variables recovered by L2Boost, MBoost, LADBoost, Robloss, and RRBoost applied with tree learners of $d$ = 2 for clean ($D_0$), symmetric gross error contaminated ($D_1$), asymmetric gross error contaminated ($D_2$),  skewed distributed ($D_3$),  and heavy-tailed distributed  ($D_4$) data generated from $g$ = $g_3$ S = $S_2$ $n$ = 3000 $p$ = 400, displayed in the form of: mean (SD) calculated from 100 independent runs of the experiment.} 
\end{table}
\begin{table}[H]
\centering
\begin{tabular}{lcccccc}
  \hline
 & L2Boost & MBoost & LADBoost & Robloss & SBoost & RRBoost \\ 
  \hline
$D_0$ & 3.34 (0.11) & 3.37 (0.12) & 3.85 (0.17) & 3.51 (0.12) & 4.07 (0.40) & 3.11 (0.16) \\ 
  $D_1 (10\%)$ & 5.27 (0.35) & 4.44 (0.53) & 4.00 (0.14) & 3.66 (0.15) & 3.80 (0.32) & 3.06 (0.08) \\ 
  $D_1 (20\%)$ & 6.12 (0.44) & 6.25 (0.44) & 4.12 (0.20) & 3.93 (0.15) & 3.55 (0.22) & 3.02 (0.08) \\ 
  $D_2 (10\%)$ & 7.44 (0.32) & 5.41 (1.05) & 4.02 (0.17) & 3.80 (0.15) & 3.77 (0.37) & 3.07 (0.09) \\ 
  $D_2 (20\%)$ & 11.96 (0.34) & 12.34 (0.59) & 4.49 (0.23) & 4.70 (0.20) & 3.50 (0.29) & 3.05 (0.09) \\ 
  $D_3$ & 4.11 (0.25) & 3.59 (0.14) & 3.96 (0.13) & 3.70 (0.12) & 4.35 (0.43) & 3.40 (0.10) \\ 
  $D_4$ & 300.70 (1347.86) & 4.26 (0.15) & 4.16 (0.17) & 3.94 (0.16) & 3.61 (0.24) & 3.15 (0.09) \\ 
   \hline
\end{tabular}
\caption{Summary statistics of RMSEs on the test sets by L2Boost, MBoost, LADBoost, Robloss, SBoost, and RRBoost applied with tree learners of $d$ = 3 for clean ($D_0$), symmetric gross error contaminated ($D_1$), asymmetric gross error contaminated ($D_2$),  skewed distributed ($D_3$),  and heavy-tailed distributed  ($D_4$) data generated from $g$ = $g_3$ S = $S_2$ $n$ = 3000 $p$ = 400, displayed in the form of: mean (SD) calculated from 100 independent runs of the experiment.} 
\end{table}
\begin{table}[H]
\centering
\begin{tabular}{lccccc}
  \hline
 & L2Boost & MBoost & LADBoost & Robloss & RRBoost \\ 
  \hline
$D_0$ & 1.00 (0.00) & 1.00 (0.00) & 1.00 (0.00) & 1.00 (0.00) & 1.00 (0.00) \\ 
  $D_1 (10\%)$ & 0.80 (0.14) & 0.92 (0.13) & 1.00 (0.00) & 1.00 (0.00) & 1.00 (0.00) \\ 
  $D_1 (20\%)$ & 0.68 (0.14) & 0.67 (0.14) & 1.00 (0.00) & 1.00 (0.00) & 1.00 (0.00) \\ 
  $D_2 (10\%)$ & 0.72 (0.13) & 0.92 (0.13) & 1.00 (0.00) & 1.00 (0.00) & 1.00 (0.00) \\ 
  $D_2 (20\%)$ & 0.64 (0.15) & 0.68 (0.15) & 1.00 (0.00) & 1.00 (0.02) & 1.00 (0.00) \\ 
  $D_3$ & 0.99 (0.05) & 1.00 (0.00) & 1.00 (0.00) & 1.00 (0.00) & 1.00 (0.00) \\ 
  $D_4$ & 0.06 (0.12) & 1.00 (0.03) & 1.00 (0.00) & 1.00 (0.00) & 1.00 (0.00) \\ 
   \hline
\end{tabular}
\caption{Fractions of variables recovered by L2Boost, MBoost, LADBoost, Robloss, and RRBoost applied with tree learners of $d$ = 3 for clean ($D_0$), symmetric gross error contaminated ($D_1$), asymmetric gross error contaminated ($D_2$),  skewed distributed ($D_3$),  and heavy-tailed distributed  ($D_4$) data generated from $g$ = $g_3$ S = $S_2$ $n$ = 3000 $p$ = 400, displayed in the form of: mean (SD) calculated from 100 independent runs of the experiment.} 
\end{table}

\subsection{Shrinkage}
\begin{figure}[H]
\centering
\includegraphics[scale = 0.6]{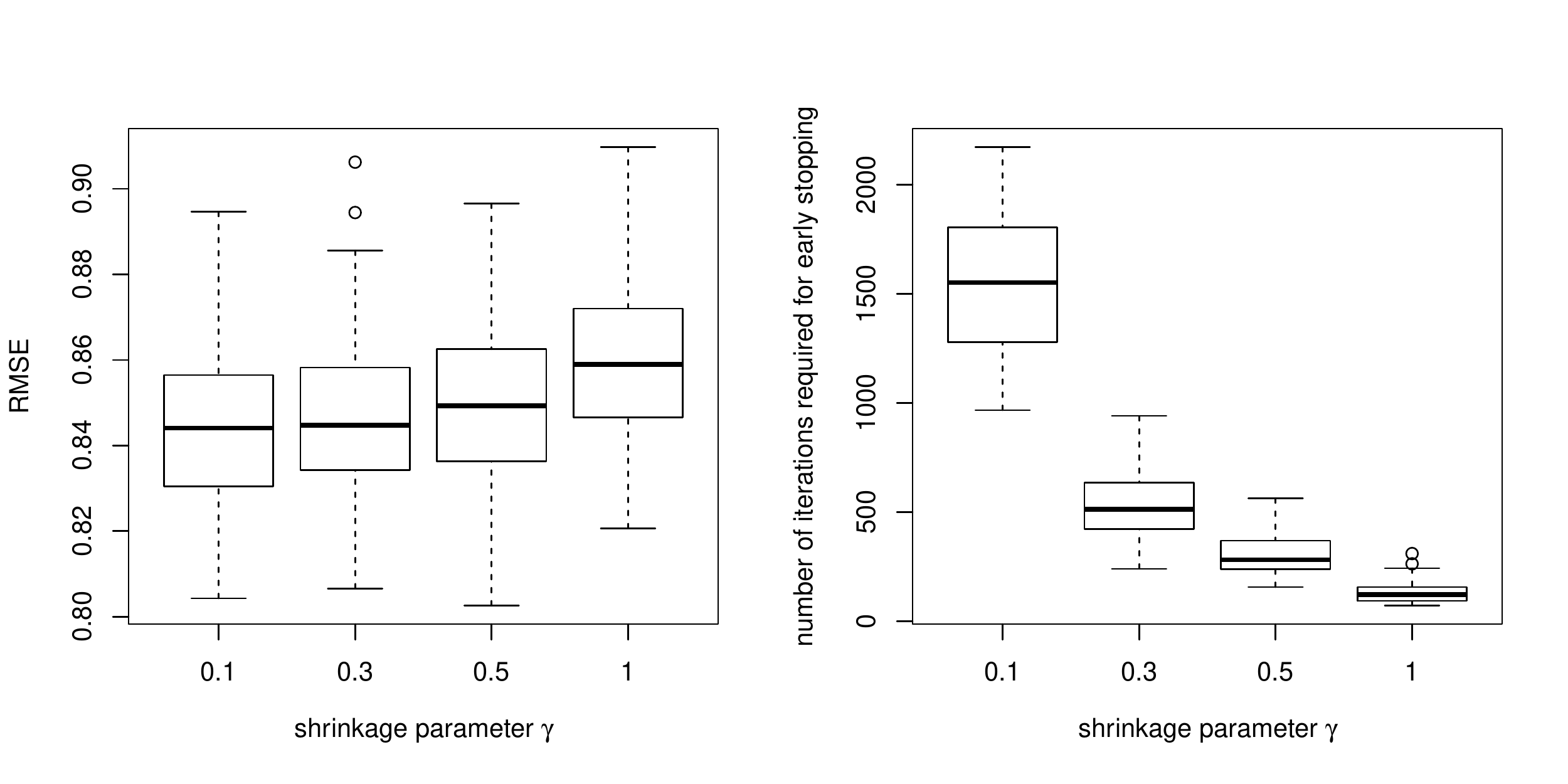}
\caption{Left panel: boxplots of RMSEs on the test sets  for RRBoost in 100 independent runs with  the shrinkage parameter $\gamma = 0.1,0.3,0.5,$ and 1; right panel: boxplots of number of iterations needed for early stopping for RRBoost in 100 independent runs with  the shrinkage parameter $\gamma = 0.1,0.3,0.5,$ and 1.}
\label{letter_2_2}
\end{figure}

	To illustrate the magnitude of the gain in prediction accuracy that 
	may be obtained using a shrinkage parameter, we include the
	results of a small experiment using 
	\texttt{RRBoost} in the following setting (please refer to Section 3.1 
	for the notation):
 $$g = g_2, S = S_2, d = 2, p = 10, n = 3000, D = D_2(10\%).$$
We considered shrinkage parameters $\gamma = 0.1,0.3,0.5,$ and $1$, and set
	$T_{1,\text{max}} = 1500$ and $T_{2,\text{max}} = 2000 $. 
	\Cref{letter_2_2} shows RMSEs on tests sets and
	the number of iterations required by \texttt{RRBoost} for early
	stopping, which adds early stopping times in two stages.
	It can be observed that a smaller value of $\gamma$ 
	results in a slight improvement of 
	prediction error at the cost of a significantly increased
	computational cost. 

\section{Empirical Applications}
\subsection{Average TRMSEs}
\subsubsection{Airfoil data}
\begin{table}[H]
\centering
\begin{tabular}{lccccc}
  \hline
 & L2Boost & MBoost & LADBoost & Robloss & RRBoost \\ 
  \hline
Original & 4.31(0.26) & 4.17(0.27) & 3.83(0.33) & 3.94(0.27) & 3.53(0.28) \\ 
  $D_1 (10\%)$ & 5.38(0.42) & 4.90(0.43) & 4.29(0.29) & 4.35(0.28) & 4.02(0.21) \\ 
  $D_1 (20\%)$ & 6.07(0.43) & 5.96(0.42) & 4.41(0.29) & 4.50(0.27) & 4.11(0.27) \\ 
  $D_2 (10\%)$ & 7.82(0.41) & 5.95(1.00) & 4.33(0.35) & 4.40(0.30) & 4.04(0.26) \\ 
  $D_2 (20\%)$ & 12.61(0.42) & 12.31(0.65) & 4.65(0.38) & 5.00(0.35) & 4.13(0.29) \\ 
  $D_3$ & 4.54(0.28) & 4.41(0.24) & 4.24(0.29) & 4.32(0.28) & 4.16(0.25) \\ 
  $D_4$ & 16.91(51.31) & 4.76(0.27) & 4.47(0.31) & 4.54(0.31) & 4.22(0.29) \\ 
   \hline
\end{tabular}
\caption{Summary statistics of average TRMSEs by L2Boost, MBoost, LADBoost, Robloss, and RRBoost applied with tree learners of $d$ = 1 for the original, symmetric gross error contaminated ($D_1$), asymmetric gross error contaminated ($D_2$),  skewed distributed ($D_3$),  and heavy-tailed distributed  ($D_4$) airfoil data displayed in the form of: mean (SD) calculated from 100 independent runs of the experiment.} 
\end{table}
\begin{table}[H]
\centering
\begin{tabular}{lccccc}
  \hline
 & L2Boost & MBoost & LADBoost & Robloss & RRBoost \\ 
  \hline
Original & 1.72(0.13) & 1.67(0.12) & 2.06(0.16) & 1.67(0.12) & 1.57(0.12) \\ 
  $D_1 (10\%)$ & 5.71(0.51) & 4.33(0.97) & 2.89(0.23) & 2.90(0.23) & 2.51(0.16) \\ 
  $D_1 (20\%)$ & 6.46(0.66) & 6.25(0.51) & 3.21(0.28) & 3.46(0.30) & 2.55(0.20) \\ 
  $D_2 (10\%)$ & 7.93(0.43) & 5.38(1.43) & 3.02(0.21) & 3.13(0.26) & 2.47(0.15) \\ 
  $D_2 (20\%)$ & 12.76(0.53) & 12.53(0.68) & 3.67(0.26) & 4.15(0.34) & 2.56(0.20) \\ 
  $D_3$ & 3.52(0.29) & 2.97(0.20) & 2.94(0.17) & 2.90(0.21) & 2.80(0.15) \\ 
  $D_4$ & 9.15(4.37) & 4.17(0.31) & 3.34(0.26) & 3.50(0.24) & 3.06(0.25) \\ 
   \hline
\end{tabular}
\caption{Summary statistics of average TRMSEs by L2Boost, MBoost, LADBoost, Robloss, and RRBoost applied with tree learners of $d$ = 2 for the original, symmetric gross error contaminated ($D_1$), asymmetric gross error contaminated ($D_2$),  skewed distributed ($D_3$),  and heavy-tailed distributed  ($D_4$) airfoil data displayed in the form of: mean (SD) calculated from 100 independent runs of the experiment.} 
\end{table}
\begin{table}[H]
\centering
\begin{tabular}{lccccc}
  \hline
 & L2Boost & MBoost & LADBoost & Robloss & RRBoost \\ 
  \hline
Original & 1.47(0.14) & 1.42(0.12) & 1.86(0.15) & 1.45(0.10) & 1.62(0.28) \\ 
  $D_1 (10\%)$ & 5.81(0.51) & 4.66(0.93) & 2.99(0.16) & 3.15(0.21) & 2.61(0.18) \\ 
  $D_1 (20\%)$ & 6.64(0.72) & 6.52(0.57) & 3.42(0.28) & 3.71(0.34) & 2.61(0.17) \\ 
  $D_2 (10\%)$ & 8.00(0.45) & 5.65(1.38) & 3.14(0.24) & 3.32(0.27) & 2.63(0.17) \\ 
  $D_2 (20\%)$ & 12.69(0.66) & 12.52(0.74) & 3.97(0.33) & 4.61(0.40) & 2.61(0.18) \\ 
  $D_3$ & 3.71(0.32) & 3.13(0.21) & 3.01(0.19) & 3.08(0.18) & 2.96(0.17) \\ 
  $D_4$ & 8.47(2.48) & 4.43(0.30) & 3.56(0.24) & 3.82(0.29) & 3.07(0.21) \\ 
   \hline
\end{tabular}
\caption{Summary statistics of average TRMSEs by L2Boost, MBoost, LADBoost, Robloss, and RRBoost applied with tree learners of $d$ = 3 for the original, symmetric gross error contaminated ($D_1$), asymmetric gross error contaminated ($D_2$),  skewed distributed ($D_3$),  and heavy-tailed distributed  ($D_4$) airfoil data displayed in the form of: mean (SD) calculated from 100 independent runs of the experiment.} 
\end{table}
\subsubsection{Abalone data}
\begin{table}[H]
\centering
\begin{tabular}{lccccc}
  \hline
 & L2Boost & MBoost & LADBoost & Robloss & RRBoost \\ 
  \hline
Original & 1.79(0.06) & 1.73(0.05) & 1.66(0.07) & 1.70(0.06) & 1.64(0.05) \\ 
  $D_1 (10\%)$ & 2.12(0.13) & 1.97(0.10) & 1.74(0.06) & 1.77(0.06) & 1.71(0.06) \\ 
  $D_1 (20\%)$ & 2.29(0.20) & 2.26(0.18) & 1.77(0.06) & 1.81(0.05) & 1.74(0.06) \\ 
  $D_2 (10\%)$ & 3.56(0.11) & 2.42(0.23) & 1.81(0.05) & 1.86(0.06) & 1.72(0.06) \\ 
  $D_2 (20\%)$ & 5.92(0.17) & 5.70(0.14) & 1.98(0.07) & 2.20(0.07) & 1.74(0.06) \\ 
  $D_3$ & 1.88(0.08) & 1.80(0.07) & 1.84(0.07) & 1.81(0.07) & 1.75(0.09) \\ 
  $D_4$ & 4.47(3.18) & 1.93(0.06) & 1.78(0.07) & 1.83(0.06) & 1.75(0.06) \\ 
   \hline
\end{tabular}
\caption{Summary statistics of average TRMSEs by L2Boost, MBoost, LADBoost, Robloss, and RRBoost applied with tree learners of $d$ = 1 for the original, symmetric gross error contaminated ($D_1$), asymmetric gross error contaminated ($D_2$),  skewed distributed ($D_3$),  and heavy-tailed distributed  ($D_4$) abalone data displayed in the form of: mean (SD) calculated from 100 independent runs of the experiment.} 
\end{table}
\begin{table}[H]
\centering
\begin{tabular}{lccccc}
  \hline
 & L2Boost & MBoost & LADBoost & Robloss & RRBoost \\ 
  \hline
Original & 1.81(0.07) & 1.75(0.06) & 1.68(0.05) & 1.71(0.06) & 1.65(0.07) \\ 
  $D_1 (10\%)$ & 2.08(0.16) & 1.98(0.11) & 1.79(0.09) & 1.82(0.08) & 1.71(0.06) \\ 
  $D_1 (20\%)$ & 2.30(0.25) & 2.22(0.22) & 1.82(0.09) & 1.86(0.09) & 1.73(0.08) \\ 
  $D_2 (10\%)$ & 3.59(0.12) & 2.59(0.30) & 1.87(0.09) & 1.90(0.10) & 1.75(0.07) \\ 
  $D_2 (20\%)$ & 5.95(0.15) & 5.68(0.15) & 2.07(0.09) & 2.23(0.07) & 1.73(0.07) \\ 
  $D_3$ & 1.92(0.09) & 1.85(0.09) & 1.89(0.08) & 1.85(0.08) & 1.79(0.08) \\ 
  $D_4$ & 3.23(2.06) & 1.95(0.10) & 1.83(0.09) & 1.88(0.09) & 1.77(0.06) \\ 
   \hline
\end{tabular}
\caption{Summary statistics of average TRMSEs by L2Boost, MBoost, LADBoost, Robloss, and RRBoost applied with tree learners of $d$ = 2 for the original, symmetric gross error contaminated ($D_1$), asymmetric gross error contaminated ($D_2$),  skewed distributed ($D_3$),  and heavy-tailed distributed  ($D_4$) abalone data displayed in the form of: mean (SD) calculated from 100 independent runs of the experiment.} 
\end{table}
\begin{table}[H]
\centering
\begin{tabular}{lccccc}
  \hline
 & L2Boost & MBoost & LADBoost & Robloss & RRBoost \\ 
  \hline
Original & 1.78(0.09) & 1.75(0.07) & 1.70(0.06) & 1.71(0.06) & 1.67(0.07) \\ 
  $D_1 (10\%)$ & 2.14(0.20) & 2.01(0.14) & 1.82(0.09) & 1.82(0.10) & 1.74(0.07) \\ 
  $D_1 (20\%)$ & 2.34(0.27) & 2.34(0.24) & 1.85(0.07) & 1.87(0.10) & 1.74(0.07) \\ 
  $D_2 (10\%)$ & 3.56(0.13) & 2.61(0.33) & 1.90(0.09) & 1.92(0.08) & 1.74(0.06) \\ 
  $D_2 (20\%)$ & 5.91(0.14) & 5.65(0.18) & 2.13(0.09) & 2.28(0.11) & 1.73(0.07) \\ 
  $D_3$ & 1.85(0.10) & 1.85(0.07) & 1.91(0.11) & 1.85(0.10) & 1.81(0.08) \\ 
  $D_4$ & 3.05(1.10) & 1.96(0.09) & 1.88(0.09) & 1.87(0.10) & 1.80(0.08) \\ 
   \hline
\end{tabular}
\caption{Summary statistics of average TRMSEs by L2Boost, MBoost, LADBoost, Robloss, and RRBoost applied with tree learners of $d$ = 3 for the original, symmetric gross error contaminated ($D_1$), asymmetric gross error contaminated ($D_2$),  skewed distributed ($D_3$),  and heavy-tailed distributed  ($D_4$) abalone data displayed in the form of: mean (SD) calculated from 100 independent runs of the experiment.} 
\end{table}
\subsubsection{Wage data}
\begin{table}[H]
\centering
\begin{tabular}{lccccc}
  \hline
 & L2Boost & MBoost & LADBoost & Robloss & RRBoost \\ 
  \hline
Original & 24.61(0.75) & 23.98(0.78) & 23.84(0.70) & 23.82(0.68) & 23.81(0.76) \\ 
  $D_1 (10\%)$ & 27.77(1.38) & 26.29(1.02) & 24.25(0.84) & 24.37(0.99) & 23.93(0.80) \\ 
  $D_1 (20\%)$ & 29.84(1.92) & 29.27(1.65) & 24.37(0.83) & 24.44(0.79) & 24.02(0.80) \\ 
  $D_2 (10\%)$ & 46.69(1.89) & 37.07(5.35) & 24.70(0.81) & 25.16(0.87) & 23.97(0.77) \\ 
  $D_2 (20\%)$ & 77.63(2.45) & 75.16(3.01) & 26.85(1.01) & 29.07(1.24) & 24.01(0.79) \\ 
  $D_3$ & 25.15(0.87) & 24.28(0.82) & 24.73(0.84) & 24.47(0.82) & 24.46(0.92) \\ 
  $D_4$ & 45.19(12.64) & 25.55(0.87) & 24.53(0.82) & 24.68(0.86) & 24.26(0.80) \\ 
   \hline
\end{tabular}
\caption{Summary statistics of average TRMSEs by L2Boost, MBoost, LADBoost, Robloss, and RRBoost applied with tree learners of $d$ = 1 for the original, symmetric gross error contaminated ($D_1$), asymmetric gross error contaminated ($D_2$),  skewed distributed ($D_3$),  and heavy-tailed distributed  ($D_4$) wage data displayed in the form of: mean (SD) calculated from 100 independent runs of the experiment.} 
\end{table}
\begin{table}[H]
\centering
\begin{tabular}{lccccc}
  \hline
 & L2Boost & MBoost & LADBoost & Robloss & RRBoost \\ 
  \hline
Original & 25.21(0.87) & 24.37(0.84) & 24.34(0.91) & 24.26(0.82) & 23.69(0.79) \\ 
  $D_1 (10\%)$ & 29.46(2.01) & 27.54(1.20) & 24.93(0.95) & 25.26(0.90) & 23.92(0.77) \\ 
  $D_1 (20\%)$ & 31.61(2.59) & 31.01(2.05) & 25.19(0.76) & 25.42(1.03) & 24.04(0.76) \\ 
  $D_2 (10\%)$ & 47.26(2.30) & 37.69(5.08) & 25.61(0.93) & 26.15(0.95) & 23.97(0.85) \\ 
  $D_2 (20\%)$ & 77.98(2.90) & 76.18(2.99) & 28.06(1.18) & 30.22(1.34) & 24.06(0.78) \\ 
  $D_3$ & 26.10(0.94) & 25.05(0.82) & 25.46(0.88) & 25.20(0.95) & 24.55(0.85) \\ 
  $D_4$ & 44.96(14.14) & 26.83(0.92) & 25.64(1.09) & 25.69(0.99) & 24.36(0.84) \\ 
   \hline
\end{tabular}
\caption{Summary statistics of average TRMSEs by L2Boost, MBoost, LADBoost, Robloss, and RRBoost applied with tree learners of $d$ = 2 for the original, symmetric gross error contaminated ($D_1$), asymmetric gross error contaminated ($D_2$),  skewed distributed ($D_3$),  and heavy-tailed distributed  ($D_4$) wage data displayed in the form of: mean (SD) calculated from 100 independent runs of the experiment.} 
\end{table}
\begin{table}[H]
\centering
\begin{tabular}{lccccc}
  \hline
 & L2Boost & MBoost & LADBoost & Robloss & RRBoost \\ 
  \hline
Original & 25.53(0.81) & 24.66(0.85) & 24.50(0.71) & 24.57(0.65) & 23.81(0.73) \\ 
  $D_1 (10\%)$ & 29.99(1.72) & 28.20(1.46) & 25.45(0.89) & 25.37(1.04) & 24.03(0.83) \\ 
  $D_1 (20\%)$ & 32.91(3.57) & 32.55(2.66) & 25.61(0.96) & 25.92(0.85) & 24.16(0.81) \\ 
  $D_2 (10\%)$ & 47.93(2.19) & 39.12(5.30) & 25.98(1.00) & 26.47(0.89) & 24.05(0.81) \\ 
  $D_2 (20\%)$ & 78.32(2.76) & 76.73(2.64) & 28.66(1.25) & 30.74(1.43) & 24.11(0.80) \\ 
  $D_3$ & 26.38(0.99) & 25.62(0.74) & 25.85(1.03) & 25.58(0.84) & 24.94(0.95) \\ 
  $D_4$ & 41.40(10.11) & 27.29(0.95) & 25.73(0.90) & 26.05(1.11) & 24.50(0.79) \\ 
   \hline
\end{tabular}
\caption{Summary statistics of average TRMSEs by L2Boost, MBoost, LADBoost, Robloss, and RRBoost applied with tree learners of $d$ = 3 for the original, symmetric gross error contaminated ($D_1$), asymmetric gross error contaminated ($D_2$),  skewed distributed ($D_3$),  and heavy-tailed distributed  ($D_4$) wage data displayed in the form of: mean (SD) calculated from 100 independent runs of the experiment.} 
\end{table}
\subsubsection{Nir data}
\begin{table}[H]
\centering
\begin{tabular}{lccccc}
  \hline
 & L2Boost & MBoost & LADBoost & Robloss & RRBoost \\ 
  \hline
Original & 3.89(1.01) & 3.95(1.09) & 4.42(1.23) & 3.97(0.93) & 3.77(1.00) \\ 
  $D_1 (10\%)$ & 9.73(2.20) & 7.30(2.69) & 5.33(1.33) & 5.62(1.12) & 4.71(1.00) \\ 
  $D_1 (20\%)$ & 13.69(4.50) & 12.68(3.64) & 5.58(1.46) & 5.99(1.34) & 4.74(1.18) \\ 
  $D_2 (10\%)$ & 11.25(2.92) & 8.36(3.58) & 5.57(1.12) & 5.79(1.27) & 4.85(1.13) \\ 
  $D_2 (20\%)$ & 18.49(3.16) & 16.16(3.41) & 6.58(1.62) & 7.76(2.55) & 5.00(1.28) \\ 
  $D_3$ & 5.71(1.52) & 5.38(1.65) & 5.27(1.01) & 5.21(1.33) & 5.15(1.09) \\ 
  $D_4$ & 10.44(3.10) & 6.99(1.54) & 5.52(1.65) & 6.12(1.38) & 5.41(1.32) \\ 
   \hline
\end{tabular}
\caption{Summary statistics of average TRMSEs by L2Boost, MBoost, LADBoost, Robloss, and RRBoost applied with tree learners of $d$ = 1 for the original, symmetric gross error contaminated ($D_1$), asymmetric gross error contaminated ($D_2$),  skewed distributed ($D_3$),  and heavy-tailed distributed  ($D_4$) nir data displayed in the form of: mean (SD) calculated from 100 independent runs of the experiment.} 
\end{table}
\begin{table}[H]
\centering
\begin{tabular}{lccccc}
  \hline
 & L2Boost & MBoost & LADBoost & Robloss & RRBoost \\ 
  \hline
Original & 4.17(1.01) & 4.15(1.10) & 4.53(1.19) & 4.17(0.92) & 4.47(1.17) \\ 
  $D_1 (10\%)$ & 11.02(3.27) & 8.49(3.66) & 5.70(1.31) & 5.53(1.30) & 5.24(1.02) \\ 
  $D_1 (20\%)$ & 16.52(5.21) & 15.31(5.99) & 6.32(1.49) & 6.52(1.57) & 5.41(1.24) \\ 
  $D_2 (10\%)$ & 12.74(4.70) & 8.63(4.26) & 5.81(1.23) & 5.81(1.37) & 5.34(1.09) \\ 
  $D_2 (20\%)$ & 19.41(5.66) & 16.85(6.96) & 7.15(1.45) & 8.74(2.44) & 5.09(1.23) \\ 
  $D_3$ & 5.52(1.49) & 5.23(1.26) & 5.51(1.12) & 5.09(1.21) & 5.16(1.23) \\ 
  $D_4$ & 11.67(5.22) & 7.59(2.18) & 6.34(1.78) & 6.55(1.76) & 5.64(1.42) \\ 
   \hline
\end{tabular}
\caption{Summary statistics of average TRMSEs by L2Boost, MBoost, LADBoost, Robloss, and RRBoost applied with tree learners of $d$ = 2 for the original, symmetric gross error contaminated ($D_1$), asymmetric gross error contaminated ($D_2$),  skewed distributed ($D_3$),  and heavy-tailed distributed  ($D_4$) nir data displayed in the form of: mean (SD) calculated from 100 independent runs of the experiment.} 
\end{table}
\begin{table}[H]
\centering
\begin{tabular}{lccccc}
  \hline
 & L2Boost & MBoost & LADBoost & Robloss & RRBoost \\ 
  \hline
Original & 4.10(1.02) & 4.31(1.06) & 4.73(1.12) & 4.15(1.02) & 4.42(1.12) \\ 
  $D_1 (10\%)$ & 12.04(4.19) & 8.97(3.93) & 5.81(1.32) & 5.64(1.62) & 5.11(0.93) \\ 
  $D_1 (20\%)$ & 16.97(6.44) & 16.31(6.56) & 6.92(1.77) & 6.99(1.65) & 5.35(1.15) \\ 
  $D_2 (10\%)$ & 12.17(4.75) & 9.18(5.46) & 5.96(1.35) & 5.93(1.40) & 5.20(1.11) \\ 
  $D_2 (20\%)$ & 21.24(7.57) & 18.59(7.18) & 8.18(1.87) & 9.64(3.09) & 5.48(1.14) \\ 
  $D_3$ & 5.60(1.43) & 5.27(1.23) & 5.61(1.05) & 5.01(1.22) & 5.54(1.07) \\ 
  $D_4$ & 11.81(7.24) & 8.23(2.32) & 6.40(1.38) & 6.64(2.10) & 5.92(1.52) \\ 
   \hline
\end{tabular}
\caption{Summary statistics of average TRMSEs by L2Boost, MBoost, LADBoost, Robloss, and RRBoost applied with tree learners of $d$ = 3 for the original, symmetric gross error contaminated ($D_1$), asymmetric gross error contaminated ($D_2$),  skewed distributed ($D_3$),  and heavy-tailed distributed  ($D_4$) nir data displayed in the form of: mean (SD) calculated from 100 independent runs of the experiment.} 
\end{table}
\subsubsection{Glass data}
\begin{table}[H]
\centering
\begin{tabular}{lccccc}
  \hline
 & L2Boost & MBoost & LADBoost & Robloss & RRBoost \\ 
  \hline
Original & 0.10(0.04) & 0.11(0.04) & 0.11(0.02) & 0.12(0.03) & 0.10(0.02) \\ 
  $D_1 (10\%)$ & 0.34(0.38) & 0.19(0.19) & 0.23(0.08) & 0.22(0.12) & 0.21(0.08) \\ 
  $D_1 (20\%)$ & 0.44(0.46) & 0.33(0.31) & 0.28(0.12) & 0.23(0.13) & 0.22(0.07) \\ 
  $D_2 (10\%)$ & 0.61(0.39) & 0.29(0.30) & 0.25(0.10) & 0.22(0.13) & 0.20(0.08) \\ 
  $D_2 (20\%)$ & 1.17(0.48) & 1.11(0.62) & 0.30(0.14) & 0.22(0.17) & 0.22(0.07) \\ 
  $D_3$ & 0.17(0.15) & 0.19(0.11) & 0.26(0.04) & 0.25(0.09) & 0.27(0.05) \\ 
  $D_4$ & 0.45(0.49) & 0.20(0.15) & 0.28(0.13) & 0.24(0.15) & 0.28(0.14) \\ 
   \hline
\end{tabular}
\caption{Summary statistics of average TRMSEs by L2Boost, MBoost, LADBoost, Robloss, and RRBoost applied with tree learners of $d$ = 1 for the original, symmetric gross error contaminated ($D_1$), asymmetric gross error contaminated ($D_2$),  skewed distributed ($D_3$),  and heavy-tailed distributed  ($D_4$) glass data displayed in the form of: mean (SD) calculated from 100 independent runs of the experiment.} 
\end{table}

\begin{table}[H]
\centering
\begin{tabular}{lccccc}
  \hline
 & L2Boost & MBoost & LADBoost & Robloss & RRBoost \\ 
  \hline
Original & 0.11(0.05) & 0.14(0.05) & 0.11(0.02) & 0.14(0.03) & 0.10(0.02) \\ 
  $D_1 (10\%)$ & 0.39(0.40) & 0.22(0.14) & 0.27(0.08) & 0.30(0.14) & 0.21(0.06) \\ 
  $D_1 (20\%)$ & 0.76(0.77) & 0.57(0.60) & 0.31(0.10) & 0.35(0.17) & 0.22(0.06) \\ 
  $D_2 (10\%)$ & 0.49(0.50) & 0.32(0.33) & 0.31(0.09) & 0.31(0.13) & 0.19(0.06) \\ 
  $D_2 (20\%)$ & 1.10(0.70) & 1.12(0.78) & 0.41(0.18) & 0.39(0.28) & 0.23(0.06) \\ 
  $D_3$ & 0.19(0.15) & 0.28(0.17) & 0.29(0.06) & 0.33(0.08) & 0.29(0.04) \\ 
  $D_4$ & 0.57(0.84) & 0.33(0.28) & 0.33(0.12) & 0.36(0.23) & 0.27(0.10) \\ 
   \hline
\end{tabular}
\caption{Summary statistics of average TRMSEs by L2Boost, MBoost, LADBoost, Robloss, and RRBoost applied with tree learners of $d$ = 2 for the original, symmetric gross error contaminated ($D_1$), asymmetric gross error contaminated ($D_2$),  skewed distributed ($D_3$),  and heavy-tailed distributed  ($D_4$) glass data displayed in the form of: mean (SD) calculated from 100 independent runs of the experiment.} 
\end{table}
\begin{table}[H]
\centering
\begin{tabular}{lccccc}
  \hline
 & L2Boost & MBoost & LADBoost & Robloss & RRBoost \\ 
  \hline
Original & 0.13(0.06) & 0.14(0.05) & 0.11(0.02) & 0.14(0.02) & 0.09(0.02) \\ 
  $D_1 (10\%)$ & 0.49(0.55) & 0.32(0.39) & 0.27(0.07) & 0.36(0.13) & 0.22(0.07) \\ 
  $D_1 (20\%)$ & 0.79(0.78) & 0.72(0.71) & 0.31(0.08) & 0.42(0.15) & 0.25(0.07) \\ 
  $D_2 (10\%)$ & 0.43(0.52) & 0.31(0.39) & 0.30(0.09) & 0.37(0.10) & 0.22(0.06) \\ 
  $D_2 (20\%)$ & 1.04(0.82) & 1.11(0.93) & 0.40(0.11) & 0.50(0.28) & 0.25(0.06) \\ 
  $D_3$ & 0.23(0.10) & 0.32(0.11) & 0.29(0.05) & 0.36(0.08) & 0.29(0.05) \\ 
  $D_4$ & 0.62(0.75) & 0.39(0.33) & 0.36(0.10) & 0.48(0.18) & 0.28(0.10) \\ 
   \hline
\end{tabular}
\caption{Summary statistics of average TRMSEs by L2Boost, MBoost, LADBoost, Robloss, and RRBoost applied with tree learners of $d$ = 3 for the original, symmetric gross error contaminated ($D_1$), asymmetric gross error contaminated ($D_2$),  skewed distributed ($D_3$),  and heavy-tailed distributed  ($D_4$) glass data displayed in the form of: mean (SD) calculated from 100 independent runs of the experiment.} 
\end{table}
\subsection{Variable importance}
\begin{table}[H]
\centering
\begin{tabular}{llcccccc}
  \hline
Method & Error & Field & 1st & 2nd & 3rd & 4th & 5th \\ 
  \hline
L2Boost & Original & Variable & velocity & angle & frequency & thickness & chord length \\ 
   &   & Rank & 2.34(1.24) & 2.40(1.32) & 2.92(1.31) & 3.04(1.21) & 4.30(1.09) \\ 
   & $D_1 (20\%)$ & Variable & thickness & chord length & velocity & frequency & angle \\ 
   &   & Rank & 1.34(0.63) & 2.18(0.72) & 2.48(0.65) & 4.50(0.51) & 4.50(0.51) \\ 
   & $D_2 (20\%)$ & Variable & thickness & chord length & velocity & angle & frequency \\ 
   &   & Rank & 1.16(0.42) & 2.34(0.63) & 2.50(0.61) & 4.36(0.48) & 4.64(0.48) \\ 
   & $D_4$ & Variable & thickness & chord length & velocity & frequency & angle \\ 
   &  & Rank & 2.48(1.09) & 2.66(1.06) & 2.90(1.07) & 3.08(1.98) & 3.88(1.27) \\ 
  MBoost & Original & Variable & angle & frequency & velocity & thickness & chord length \\ 
   &   & Rank & 2.00(1.05) & 2.66(1.36) & 2.72(1.09) & 3.20(1.39) & 4.42(0.91) \\ 
   & $D_1 (20\%)$ & Variable & velocity & thickness & chord length & frequency & angle \\ 
   &   & Rank & 2.08(1.28) & 2.36(1.05) & 2.56(1.03) & 3.74(1.29) & 4.26(0.93) \\ 
   & $D_2 (20\%)$ & Variable & thickness & chord length & velocity & angle & frequency \\ 
   &   & Rank & 2.08(1.07) & 2.76(1.24) & 2.90(1.37) & 3.48(1.42) & 3.78(1.31) \\ 
   & $D_4$ & Variable & velocity & chord length & angle & thickness & frequency \\ 
   &  & Rank & 2.56(1.30) & 2.84(1.02) & 2.90(1.31) & 3.12(1.30) & 3.58(1.85) \\ 
  LADBoost & Original & Variable & angle & frequency & velocity & thickness & chord length \\ 
   &   & Rank & 2.12(1.19) & 2.36(1.17) & 2.52(1.18) & 3.70(1.11) & 4.30(1.05) \\ 
   & $D_1 (20\%)$ & Variable & thickness & velocity & chord length & frequency & angle \\ 
   &   & Rank & 1.58(0.78) & 2.20(0.81) & 2.34(0.75) & 4.30(0.77) & 4.58(0.37) \\ 
   & $D_2 (20\%)$ & Variable & thickness & chord length & velocity & angle & frequency \\ 
   &   & Rank & 1.30(0.58) & 2.36(0.69) & 2.52(0.79) & 4.27(0.83) & 4.55(0.47) \\ 
   & $D_4$ & Variable & thickness & chord length & velocity & angle & frequency \\ 
   &  & Rank & 2.40(1.18) & 2.42(1.07) & 2.72(1.13) & 3.44(1.26) & 4.02(1.66) \\ 
  Robloss & Original & Variable & angle & frequency & velocity & thickness & chord length \\ 
   &   & Rank & 1.80(0.93) & 2.26(1.07) & 2.50(1.13) & 4.00(0.86) & 4.44(0.88) \\ 
   & $D_1 (20\%)$ & Variable & thickness & chord length & velocity & frequency & angle \\ 
   &   & Rank & 2.04(1.14) & 2.24(0.94) & 2.40(1.01) & 4.05(0.97) & 4.27(1.07) \\ 
   & $D_2 (20\%)$ & Variable & thickness & chord length & velocity & angle & frequency \\ 
   &   & Rank & 2.06(1.02) & 2.34(0.92) & 2.44(1.23) & 3.83(1.22) & 4.33(1.01) \\ 
   & $D_4$ & Variable & thickness & angle & velocity & chord length & frequency \\ 
   &  & Rank & 2.52(1.27) & 2.68(1.24) & 2.76(1.29) & 2.92(1.03) & 4.12(1.64) \\ 
  RRBoost & Original & Variable & thickness & frequency & velocity & angle & chord length \\ 
   &   & Rank & 2.12(1.22) & 2.84(1.35) & 3.04(1.40) & 3.06(1.24) & 3.94(1.30) \\ 
   & $D_1 (20\%)$ & Variable & thickness & velocity & chord length & frequency & angle \\ 
   &   & Rank & 1.70(0.84) & 2.28(0.90) & 2.38(0.95) & 4.22(0.95) & 4.42(0.81) \\ 
   & $D_2 (20\%)$ & Variable & thickness & velocity & chord length & angle & frequency \\ 
   &   & Rank & 1.46(0.68) & 2.36(0.80) & 2.44(0.88) & 4.11(0.95) & 4.63(0.52) \\ 
   & $D_4$ & Variable & velocity & thickness & chord length & frequency & angle \\ 
   &  & Rank & 2.74(1.31) & 2.74(1.08) & 2.82(1.34) & 3.04(1.91) & 3.66(1.14) \\ 
   \hline
\end{tabular}
\caption{Top 5 most important variables in terms of the average rank generated by L2Boost, MBoost, LADBoost, Robloss, and RRBoost respectively,  applied with tree learners of $d$ = 2 for the original, symmetric gross error contaminated ($D_1$), asymmetric gross error contaminated ($D_2$),  skewed distributed ($D_3$),  
                         and heavy-tailed distributed  ($D_4$) airfoil data; rank displayed in the form of: mean (SD) calculated from 50 random splits of the data.} 
\end{table}
\begin{table}[H]
\hspace{-1cm}
\begin{tabular}{llcccccc}
  \hline
Method & Error & Field & 1st & 2nd & 3rd & 4th & 5th \\ 
  \hline
L2Boost & Original & Variable & viscera weight & height & shucked weight & length & diameter \\ 
   &   & Rank & 1.52(0.58) & 1.68(0.71) & 2.80(0.45) & 4.54(0.58) & 4.58(0.64) \\ 
   & $D_1 (20\%)$ & Variable & shucked weight & viscera weight & shell weight & diameter & height \\ 
   &   & Rank & 1.12(0.33) & 1.88(0.33) & 3.90(0.93) & 4.92(1.30) & 5.29(1.47) \\ 
   & $D_2 (20\%)$ & Variable & shucked weight & viscera weight & shell weight & height & diameter \\ 
   &   & Rank & 1.20(0.45) & 1.90(0.42) & 4.00(1.03) & 4.27(1.63) & 5.36(1.20) \\ 
   & $D_4$ & Variable & shucked weight & viscera weight & height & shell weight & diameter \\ 
   &  & Rank & 1.50(0.74) & 2.16(0.79) & 3.94(1.38) & 4.58(1.03) & 4.79(2.06) \\ 
  MBoost & Original & Variable & height & diameter & shell weight & shucked weight & length \\ 
   &   & Rank & 1.00(0.00) & 2.18(0.39) & 3.08(1.05) & 4.44(0.91) & 5.18(0.90) \\ 
   & $D_1 (20\%)$ & Variable & diameter & shell weight & height & whole weight & shucked weight \\ 
   &   & Rank & 1.63(1.01) & 2.20(1.47) & 3.16(1.38) & 4.78(1.02) & 5.05(1.27) \\ 
   & $D_2 (20\%)$ & Variable & diameter & shell weight & height & whole weight & shucked weight \\ 
   &   & Rank & 2.05(1.28) & 2.30(1.49) & 2.80(1.55) & 4.50(1.50) & 5.03(1.11) \\ 
   & $D_4$ & Variable & height & shell weight & whole weight & shucked weight & viscera weight \\ 
   &  & Rank & 1.32(0.55) & 3.50(1.50) & 4.22(1.47) & 4.44(1.21) & 4.55(1.42) \\ 
  LADBoost & Original & Variable & viscera weight & shell weight & diameter & length & height \\ 
   &   & Rank & 1.48(0.50) & 1.52(0.50) & 3.54(0.81) & 4.26(0.94) & 4.42(0.78) \\ 
   & $D_1 (20\%)$ & Variable & viscera weight & shell weight & whole weight & length & shucked weight \\ 
   &   & Rank & 1.38(0.49) & 1.62(0.49) & 4.42(1.28) & 4.85(1.53) & 5.00(1.06) \\ 
   & $D_2 (20\%)$ & Variable & viscera weight & shell weight & whole weight & height & diameter \\ 
   &   & Rank & 1.40(0.49) & 1.60(0.49) & 4.25(1.42) & 4.97(1.37) & 5.01(1.39) \\ 
   & $D_4$ & Variable & viscera weight & shell weight & whole weight & diameter & length \\ 
   &  & Rank & 1.24(0.48) & 1.98(0.55) & 4.59(1.30) & 4.78(1.26) & 4.94(2.13) \\ 
  Robloss & Original & Variable & length & height & whole weight & shucked weight & diameter \\ 
   &   & Rank & 1.08(0.27) & 1.92(0.27) & 3.56(1.25) & 4.74(1.10) & 5.12(1.12) \\ 
   & $D_1 (20\%)$ & Variable & whole weight & height & shucked weight & viscera weight & length \\ 
   &   & Rank & 1.60(0.53) & 1.61(0.85) & 4.22(1.20) & 4.52(1.11) & 4.93(1.88) \\ 
   & $D_2 (20\%)$ & Variable & whole weight & height & shucked weight & viscera weight & shell weight \\ 
   &   & Rank & 1.34(0.48) & 1.84(0.82) & 4.42(1.37) & 4.72(1.25) & 5.06(1.11) \\ 
   & $D_4$ & Variable & height & whole weight & shucked weight & length & diameter \\ 
   &  & Rank & 1.18(0.39) & 2.18(0.44) & 4.32(1.24) & 4.88(2.33) & 5.09(1.39) \\ 
  RRBoost & Original & Variable & length & shucked weight & shell weight & diameter & viscera weight \\ 
   &   & Rank & 1.04(0.20) & 2.38(0.60) & 2.60(0.49) & 4.08(0.34) & 4.98(0.43) \\ 
   & $D_1 (20\%)$ & Variable & shell weight & shucked weight & length & diameter & whole weight \\ 
   &   & Rank & 1.08(0.27) & 3.17(1.38) & 3.29(1.78) & 4.63(1.83) & 5.25(1.35) \\ 
   & $D_2 (20\%)$ & Variable & shell weight & length & shucked weight & diameter & height \\ 
   &   & Rank & 1.06(0.24) & 3.01(1.21) & 3.32(1.69) & 4.83(1.61) & 4.96(1.51) \\ 
   & $D_4$ & Variable & shell weight & shucked weight & diameter & height & whole weight \\ 
   &  & Rank & 1.22(0.42) & 2.95(1.54) & 3.76(1.27) & 4.86(1.49) & 4.96(1.32) \\ 
   \hline
\end{tabular}
\caption{Top 5 most important variables ordered in terms of the average rank generated by L2Boost, MBoost, LADBoost, Robloss, and RRBoost respectively,  applied with tree learners of $d$ = 1 for the original, symmetric gross error contaminated ($D_1$), asymmetric gross error contaminated ($D_2$),  skewed distributed ($D_3$),  
                         and heavy-tailed distributed  ($D_4$) abalone data; rank displayed in the form of: mean (SD) calculated from 50 random splits of the data.} 
\end{table}
\begin{table}[H]
\centering
\begin{tabular}{llcccccc}
  \hline
Method & Error & Field & 1st & 2nd & 3rd & 4th & 5th \\ 
  \hline
L2Boost & Original & Variable & health & health ins & year & maritl & age \\ 
   &   & Rank & 1.00(0.00) & 2.08(0.27) & 3.38(0.53) & 3.54(0.61) & 5.62(0.99) \\ 
   & $D_1 (20\%)$ & Variable & health & health ins & year & maritl & education \\ 
   &   & Rank & 1.00(0.00) & 2.46(0.68) & 3.12(0.82) & 3.76(1.02) & 6.28(0.98) \\ 
   & $D_2 (20\%)$ & Variable & health & health ins & year & maritl & education \\ 
   &   & Rank & 1.00(0.00) & 2.32(0.59) & 3.26(0.83) & 3.54(0.76) & 6.26(0.99) \\ 
   & $D_4$ & Variable & health & health ins & year & maritl & age \\ 
   &  & Rank & 1.06(0.24) & 2.64(0.75) & 3.38(0.92) & 3.60(1.09) & 5.82(1.37) \\ 
  MBoost & Original & Variable & race & year & education & jobclass & maritl \\ 
   &   & Rank & 1.00(0.00) & 2.08(0.27) & 3.40(0.61) & 3.56(0.61) & 5.36(0.92) \\ 
   & $D_1 (20\%)$ & Variable & race & year & jobclass & maritl & education \\ 
   &   & Rank & 1.25(1.14) & 2.46(0.50) & 4.15(2.17) & 4.98(1.63) & 5.41(2.01) \\ 
   & $D_2 (20\%)$ & Variable & race & year & jobclass & education & maritl \\ 
   &   & Rank & 1.16(0.58) & 2.08(0.40) & 3.91(1.91) & 5.27(1.63) & 5.38(1.46) \\ 
   & $D_4$ & Variable & year & jobclass & race & health ins & maritl \\ 
   &  & Rank & 1.66(0.66) & 2.76(1.58) & 4.40(2.80) & 4.50(1.31) & 4.63(1.88) \\ 
  LADBoost & Original & Variable & age & jobclass & health ins & health & education \\ 
   &   & Rank & 1.00(0.00) & 2.04(0.20) & 3.40(0.57) & 3.56(0.50) & 5.51(0.94) \\ 
   & $D_1 (20\%)$ & Variable & age & health ins & health & jobclass & year \\ 
   &   & Rank & 1.06(0.24) & 3.06(0.82) & 3.27(1.21) & 4.04(2.40) & 5.01(1.04) \\ 
   & $D_2 (20\%)$ & Variable & age & health ins & health & jobclass & year \\ 
   &   & Rank & 1.02(0.14) & 2.92(0.75) & 3.21(1.41) & 3.95(1.88) & 5.11(1.23) \\ 
   & $D_4$ & Variable & age & jobclass & health ins & health & year \\ 
   &  & Rank & 1.12(0.33) & 2.52(0.76) & 3.68(0.89) & 3.70(1.27) & 5.63(1.05) \\ 
  Robloss & Original & Variable & maritl & health & race & year & health ins \\ 
   &   & Rank & 1.00(0.00) & 2.04(0.20) & 3.26(0.56) & 3.72(0.45) & 5.44(0.86) \\ 
   & $D_1 (20\%)$ & Variable & maritl & health & year & race & health ins \\ 
   &   & Rank & 1.00(0.00) & 2.48(0.54) & 3.72(0.83) & 3.79(2.04) & 5.25(1.07) \\ 
   & $D_2 (20\%)$ & Variable & maritl & health & year & race & health ins \\ 
   &   & Rank & 1.00(0.00) & 2.28(0.64) & 3.34(0.63) & 4.46(1.82) & 5.62(1.41) \\ 
   & $D_4$ & Variable & maritl & health & year & race & health ins \\ 
   &  & Rank & 1.30(0.46) & 2.50(0.54) & 3.60(0.83) & 4.33(2.72) & 5.30(1.56) \\ 
  RRBoost & Original & Variable & education & race & jobclass & age & year \\ 
   &   & Rank & 1.00(0.00) & 2.20(0.49) & 3.30(0.61) & 3.50(0.65) & 5.74(0.90) \\ 
   & $D_1 (20\%)$ & Variable & education & age & race & year & health \\ 
   &   & Rank & 1.04(0.20) & 3.11(1.21) & 3.47(2.00) & 5.05(1.45) & 5.09(1.29) \\ 
   & $D_2 (20\%)$ & Variable & education & age & race & jobclass & year \\ 
   &   & Rank & 1.14(0.45) & 2.90(1.46) & 4.10(2.01) & 4.77(1.87) & 5.12(1.78) \\ 
   & $D_4$ & Variable & education & jobclass & age & health & year \\ 
   &  & Rank & 1.02(0.14) & 2.66(0.75) & 2.80(0.97) & 5.27(1.33) & 5.48(1.71) \\ 
   \hline
\end{tabular}
\caption{Top 5 most important variables ordered in terms of the average rank generated by L2Boost, MBoost, LADBoost, Robloss, and RRBoost respectively, applied with tree learners of $d$ = 1 for the original, symmetric gross error contaminated ($D_1$), asymmetric gross error contaminated ($D_2$),  skewed distributed ($D_3$),  
                         and heavy-tailed distributed  ($D_4$) wage data; rank displayed in the form of: mean (SD) calculated from 50 random splits of the data.} 
\end{table}
\begin{table}[H]
\centering
\begin{tabular}{llcccccc}
  \hline
Method & Error & Field & 1st & 2nd & 3rd & 4th & 5th \\ 
  \hline
L2Boost & Original & Variable & V73 & V78 & V92 & V21 & V40 \\ 
   &   & Rank & 28.93(29.79) & 31.59(22.93) & 32.50(31.65) & 32.67(29.08) & 34.36(28.59) \\ 
   & $D_1 (20\%)$ & Variable & V73 & V93 & V92 & V54 & V96 \\ 
   &   & Rank & 37.81(21.59) & 40.84(24.93) & 41.05(26.74) & 41.68(21.08) & 42.24(22.73) \\ 
   & $D_2 (20\%)$ & Variable & V97 & V54 & V73 & V51 & V92 \\ 
   &   & Rank & 38.87(24.28) & 39.40(22.80) & 41.00(20.52) & 41.46(25.92) & 42.14(21.02) \\ 
   & $D_4$ & Variable & V51 & V73 & V54 & V92 & V97 \\ 
   &  & Rank & 39.22(24.93) & 39.66(19.99) & 40.41(22.20) & 40.43(26.86) & 40.90(21.84) \\ 
  MBoost & Original & Variable & V96 & V92 & V40 & V74 & V73 \\ 
   &   & Rank & 29.49(29.35) & 30.78(29.93) & 31.23(25.89) & 33.33(23.57) & 33.48(27.35) \\ 
   & $D_1 (20\%)$ & Variable & V73 & V54 & V58 & V102 & V95 \\ 
   &   & Rank & 39.63(22.57) & 40.08(24.11) & 40.80(19.73) & 41.68(21.24) & 41.87(20.67) \\ 
   & $D_2 (20\%)$ & Variable & V54 & V92 & V96 & V94 & V84 \\ 
   &   & Rank & 33.10(27.23) & 33.41(23.47) & 39.53(29.94) & 41.22(25.47) & 41.81(26.02) \\ 
   & $D_4$ & Variable & V54 & V96 & V73 & V93 & V59 \\ 
   &  & Rank & 35.73(24.51) & 37.02(25.78) & 38.40(20.82) & 40.98(25.88) & 41.85(17.55) \\ 
  LADBoost & Original & Variable & V96 & V54 & V92 & V13 & V35 \\ 
   &   & Rank & 28.90(29.73) & 33.64(28.15) & 34.73(29.18) & 34.89(26.45) & 34.90(28.86) \\ 
   & $D_1 (20\%)$ & Variable & V96 & V92 & V95 & V91 & V77 \\ 
   &   & Rank & 32.16(22.94) & 36.02(25.39) & 39.62(27.28) & 41.26(23.37) & 41.32(20.08) \\ 
   & $D_2 (20\%)$ & Variable & V96 & V98 & V92 & V91 & V97 \\ 
   &   & Rank & 36.61(25.86) & 38.14(24.61) & 40.55(22.29) & 41.92(21.20) & 41.97(20.37) \\ 
   & $D_4$ & Variable & V96 & V16 & V54 & V35 & V92 \\ 
   &  & Rank & 36.79(28.68) & 41.73(17.84) & 41.74(25.10) & 42.18(17.68) & 43.32(25.73) \\ 
  Robloss & Original & Variable & V77 & V88 & V16 & V96 & V35 \\ 
   &   & Rank & 28.28(25.19) & 31.45(28.89) & 34.25(22.99) & 34.91(26.14) & 36.18(23.36) \\ 
   & $D_1 (20\%)$ & Variable & V88 & V91 & V94 & V90 & V96 \\ 
   &   & Rank & 31.08(30.25) & 38.21(26.12) & 40.07(23.95) & 40.58(20.03) & 40.73(23.25) \\ 
   & $D_2 (20\%)$ & Variable & V91 & V92 & V88 & V58 & V95 \\ 
   &   & Rank & 33.88(28.35) & 38.82(23.24) & 39.49(31.24) & 41.57(21.83) & 42.26(26.05) \\ 
   & $D_4$ & Variable & V88 & V35 & V98 & V77 & V96 \\ 
   &  & Rank & 33.72(27.57) & 37.56(22.61) & 40.33(19.91) & 40.74(18.51) & 40.82(22.04) \\ 
  RRBoost & Original & Variable & V96 & V91 & V58 & V69 & V39 \\ 
   &   & Rank & 28.34(22.97) & 31.50(32.86) & 32.21(29.56) & 32.55(28.97) & 35.38(31.28) \\ 
   & $D_1 (20\%)$ & Variable & V91 & V58 & V69 & V95 & V77 \\ 
   &   & Rank & 28.66(26.26) & 37.66(23.56) & 40.50(18.37) & 42.30(22.75) & 42.33(19.38) \\ 
   & $D_2 (20\%)$ & Variable & V91 & V102 & V96 & V69 & V54 \\ 
   &   & Rank & 31.50(26.00) & 37.51(26.11) & 39.87(18.60) & 41.24(21.38) & 41.72(20.46) \\ 
   & $D_4$ & Variable & V91 & V95 & V92 & V97 & V16 \\ 
   &  & Rank & 36.93(24.23) & 39.80(18.65) & 40.73(19.98) & 42.10(18.69) & 42.46(17.18) \\ 
   \hline
\end{tabular}
\caption{Top 5 most important variables ordered in terms of the average rank generated by L2Boost, MBoost, LADBoost, Robloss, and RRBoost respectively, applied with tree learners of $d$ = 1 for the original, symmetric gross error contaminated ($D_1$), asymmetric gross error contaminated ($D_2$),  skewed distributed ($D_3$),  
                         and heavy-tailed distributed  ($D_4$) nir data; rank displayed in the form of: mean (SD) calculated from 50 random splits of the data.} 
\end{table}
\begin{table}[H]
\hspace{-1cm}
\begin{tabular}{llcccccc}
  \hline
Method & Error & Field & 1st & 2nd & 3rd & 4th & 5th \\ 
  \hline
L2Boost & Original & Variable & V447 & V253 & V254 & V59 & V155 \\ 
   &   & Rank & 173.24(122.28) & 173.97(129.48) & 193.77(129.64) & 197.07(98.96) & 197.29(98.77) \\ 
   & $D_1 (20\%)$ & Variable & V254 & V293 & V223 & V253 & V418 \\ 
   &   & Rank & 220.47(72.70) & 220.59(71.94) & 225.33(65.51) & 225.59(81.02) & 229.81(58.41) \\ 
   & $D_2 (20\%)$ & Variable & V350 & V351 & V253 & V490 & V455 \\ 
   &   & Rank & 215.77(79.63) & 220.73(73.94) & 225.65(66.34) & 225.80(65.57) & 225.87(65.29) \\ 
   & $D_4$ & Variable & V448 & V495 & V155 & V351 & V304 \\ 
   &  & Rank & 216.16(79.41) & 216.17(79.41) & 220.87(73.59) & 221.09(73.20) & 221.20(72.50) \\ 
  MBoost & Original & Variable & V156 & V254 & V213 & V304 & V252 \\ 
   &   & Rank & 133.38(123.16) & 187.78(169.94) & 190.69(119.17) & 203.96(136.24) & 206.55(169.86) \\ 
   & $D_1 (20\%)$ & Variable & V350 & V446 & V449 & V298 & V252 \\ 
   &   & Rank & 225.29(66.53) & 225.31(66.38) & 225.32(66.39) & 225.58(65.36) & 225.61(81.52) \\ 
   & $D_2 (20\%)$ & Variable & V252 & V490 & V411 & V303 & V94 \\ 
   &   & Rank & 220.20(73.48) & 225.44(80.08) & 225.67(65.23) & 225.91(80.35) & 229.90(58.34) \\ 
   & $D_4$ & Variable & V407 & V491 & V254 & V102 & V304 \\ 
   &  & Rank & 211.50(83.55) & 216.35(91.15) & 220.95(72.93) & 225.62(66.26) & 225.76(81.12) \\ 
  LADBoost & Original & Variable & V349 & V251 & V253 & V252 & V346 \\ 
   &   & Rank & 177.98(120.48) & 192.69(161.47) & 203.14(149.30) & 208.11(120.56) & 208.61(87.42) \\ 
   & $D_1 (20\%)$ & Variable & V250 & V484 & V489 & V254 & V252 \\ 
   &   & Rank & 225.33(66.09) & 229.89(75.47) & 230.03(57.95) & 230.12(75.00) & 230.18(89.53) \\ 
   & $D_2 (20\%)$ & Variable & V445 & V288 & V298 & V489 & V249 \\ 
   &   & Rank & 220.86(87.17) & 220.88(72.18) & 225.87(64.48) & 226.02(65.44) & 226.07(63.69) \\ 
   & $D_4$ & Variable & V349 & V253 & V302 & V497 & V252 \\ 
   &  & Rank & 215.85(79.73) & 220.73(73.33) & 220.90(72.39) & 221.14(86.30) & 225.51(66.30) \\ 
  Robloss & Original & Variable & V252 & V157 & V253 & V254 & V350 \\ 
   &   & Rank & 144.73(175.26) & 191.78(111.56) & 192.24(147.24) & 202.15(155.24) & 202.54(92.71) \\ 
   & $D_1 (20\%)$ & Variable & V350 & V254 & V59 & V422 & V493 \\ 
   &   & Rank & 215.54(79.91) & 225.21(66.65) & 230.09(58.48) & 230.15(58.15) & 230.22(57.83) \\ 
   & $D_2 (20\%)$ & Variable & V250 & V483 & V252 & V350 & V348 \\ 
   &   & Rank & 216.12(78.29) & 220.79(87.15) & 229.90(74.27) & 230.21(58.42) & 230.26(58.52) \\ 
   & $D_4$ & Variable & V500 & V251 & V298 & V350 & V491 \\ 
   &  & Rank & 216.12(78.65) & 225.84(65.37) & 226.03(64.60) & 230.37(76.16) & 230.38(74.79) \\ 
  RRBoost & Original & Variable & V349 & V252 & V499 & V350 & V250 \\ 
   &   & Rank & 114.55(140.17) & 178.85(136.55) & 193.04(110.97) & 197.44(98.55) & 199.43(126.06) \\ 
   & $D_1 (20\%)$ & Variable & V350 & V293 & V252 & V444 & V254 \\ 
   &   & Rank & 225.35(66.63) & 225.50(65.86) & 225.52(65.95) & 225.57(66.10) & 230.16(57.82) \\ 
   & $D_2 (20\%)$ & Variable & V254 & V447 & V252 & V495 & V299 \\ 
   &   & Rank & 215.68(93.36) & 225.37(66.85) & 225.59(66.18) & 225.74(65.55) & 225.86(65.15) \\ 
   & $D_4$ & Variable & V492 & V254 & V407 & V491 & V214 \\ 
   &  & Rank & 221.03(71.68) & 221.12(72.14) & 225.78(65.42) & 225.78(64.69) & 226.02(64.40) \\ 
   \hline
\end{tabular}
\caption{Top 5 most important variables ordered in terms of the average rank generated by L2Boost, MBoost, LADBoost, Robloss, and RRBoost respectively, applied with tree learners of $d$ = 1 for the original, symmetric gross error contaminated ($D_1$), asymmetric gross error contaminated ($D_2$),  skewed distributed ($D_3$),  
                         and heavy-tailed distributed  ($D_4$)  glass data; rank displayed in the form of: mean (SD) calculated from 50 random splits of the data.} 
\end{table}